\documentclass[aps,prx,twocolumn,nofootinbib,longbibliography,notitlepage,floatfix,10pt]{revtex4-2}
\usepackage{graphicx,outlines,subfigure}
\usepackage{physics,graphicx,diagbox,markdown}
\usepackage{tabularx}
\usepackage{amsmath,mathtools}
\usepackage{amstext,cuted}
\usepackage{amssymb}
\usepackage{xfrac}
\usepackage[colorlinks,citecolor=magenta]{hyperref}
\usepackage{graphicx,makecell}
\usepackage{amsmath}
\usepackage{amstext}
\usepackage{amssymb}
\usepackage{longtable,booktabs}
\usepackage{hyperref}
\usepackage{url}
\usepackage{subfigure}
\usepackage{dsfont}
\usepackage{booktabs}
\usepackage{amsbsy}
\usepackage{dcolumn}
\usepackage{amsthm}
\usepackage{times}
\usepackage{bm}
\usepackage{esint}
\usepackage{multirow}
\usepackage{hyperref}
\usepackage{cleveref}
\usepackage{amsmath}
\usepackage{amssymb}
\usepackage{mathrsfs}
\usepackage{amsbsy}
\usepackage{dcolumn}
\usepackage{bm}
\usepackage{multirow}
\usepackage{color}
\usepackage{extarrows}
\usepackage{datetime}
\usepackage{comment}
\usepackage[super]{nth}
\usepackage{tikz}
\usetikzlibrary{arrows.meta}
\hypersetup{
	colorlinks=true,
	linkcolor=blue,
	filecolor=blue,
	urlcolor=blue,
}

\tikzset{bold/.style={color=blue, line width=2pt}}
\tikzset{redop/.style={circle,fill=red}}
\tikzset{blueop/.style={circle,fill=blue}}
\tikzset{greenop/.style={circle,fill=teal}}

\renewcommand{\i}{\mathrm{i}}

\def\Y{\hat{Y}^{(a/b)}_{ij}}

\def\YY{\hat{Y}^{(a/b)}_{i'j'}}
\def\za{\hat{Z}^{(a)}_{ij}}
\def\xa{\hat{X}^{(a)}_{ij}}
\def\ya{\hat{Y}^{(a)}_{ij}}
\def\zb{\hat{Z}^{(b)}_{ij}}
\def\xb{\hat{X}^{(b)}_{ij}}
\def\yb{\hat{Y}^{(b)}_{ij}}
\def\zza{\hat{Z}^{(a)}_{i'j'}}
\def\xxa{\hat{X}^{(a)}_{i'j'}}

\def\zzb{\hat{Z}^{(b)}_{i'j'}}
\def\xxb{\hat{X}^{(b)}_{i'j'}}

\usepackage{datetime}
\usepackage{comment}
\usepackage{lineno}
\graphicspath{{Pictures/}}
\usepackage{xcolor}

\begin{document}
	\title{Higher-Order Cellular Automata Generated Symmetry-Protected Topological Phases and Detection Through Multi-Point Strange Correlators}
	\author{Jie-Yu Zhang}
	\author{Meng-Yuan Li}
	\author{Peng Ye}
	\email{yepeng5@mail.sysu.edu.cn}
	\affiliation{Guangdong Provincial Key Laboratory of Magnetoelectric Physics and Devices, State Key Laboratory of Optoelectronic Materials and Technologies, and School of Physics,   Sun Yat-sen University, Guangzhou, 510275, China}
	\date{\today}	
	\begin{abstract}
In computer and system sciences, higher-order cellular automata (HOCA) are a type of cellular automata that evolve over multiple time steps and generate complex patterns, which have various applications such as secret sharing schemes, data compression, and image encryption. In this paper, we introduce HOCA to quantum many-body physics and   construct a series of symmetry-protected topological (SPT) phases of matter, in which symmetries are supported on a great variety of subsystems embbeded in the SPT bulk. We call these phases HOCA-generated SPT (HGSPT) phases. Specifically, we show that HOCA can generate not only well-understood SPTs with symmetries supported on either regular (e.g., line-like subsystems in the 2D cluster model) or fractal subsystems, but also a large class of   unexplored SPTs with symmetries supported on more choices of subsystems. One example is \textit{mixed-subsystem SPT} that has either fractal and line-like subsystem symmetries simultaneously or two distinct types of fractal symmetries simultaneously. Another example is \textit{chaotic-subsystem SPT}  in which   chaotic-looking symmetries are significantly different from and thus cannot reduce to  fractal or regular subsystem symmetries. We also introduce a new notation system to characterize HGSPTs. We prove that all possible subsystem symmetries in square lattice can be locally simulated by an HOCA generated symmetry. As the usual two-point strange correlators are trivial in most HGSPTs, we find that the nontrivial SPT orders can be detected by what we call \textit{multi-point strange correlators}. We propose a universal procedure to design the spatial configuration of the multi-point strange correlators for a given HGSPT phase. Specifically, we find deep connections between multi-point strange correlators and the spurious topological entanglement entropy (STEE), both exhibiting long range behavior in a short range entangled state. Our HOCA approaches and multi-point strange correlators pave the way for a unified paradigm to design, classify, and detect phases of matter with symmetries supported on a great variety of subsystems, and also provide potential useful perspective in surpassing the computational irreducibility of HOCA in a quantum mechanical way. 
	\end{abstract}
	
	\maketitle
	\section{Introduction}

Cellular automata (CA) are dynamic systems that evolve in discrete time steps, which have been wided used in comptuer and system sciences~\cite{wolfram_cellular_1984}. CA have rather simple evolution rules, but produce rich structures. Because of  their ability to model a wide range of phenomena, they have been used to model various real-world systems and can be used for prediction and simulation~\cite{Blecic2013CellularAS,Guan2010AGP}. 
\begin{table*}[htb]
		\label{table1}	\caption{Representative examples of SPT phases generated by HOCA. Strange correlators of model-IVc and model-Vb can also be calculated by the same procedure given in the paper, but is not explicitly shown in the table as these models are not the main focus of this paper.}
	\begin{ruledtabular}
		\begin{tabular}{cccccc}
			SPT phases& Lattice  models& CA order& Symmetry description  &Strange correlators\\
			\hline
			I-MSPT&   I (Eq.~(\ref{MSPT1}))&2&    a mixture of line-like and fractal-like symmetry (Fig.~\ref{mixCApic})  &Sec.~\ref{D-I-MSPT}\\
			II-MSPT& II (Eq.~(\ref{MSPT2}))&3&     a mixture of two types of fractal-like symmetry (Fig.~\ref{mixCApic2}) &Sec.~\ref{D-II-MSPT}\\
			CSPT& III (Eq.~(\ref{MSPT3}))&3&   chaotic-looking, neither line-like nor fractal-like symmetry (Fig.~\ref{otherpic}) &Sec.~\ref{D-CSPT}\\
			RSPT& IVa (Eq.~(\ref{MSPT4}))&2&    regular (e.g., line-like, membrane-like) subsystem symmetry (Fig.~\ref{periodic pic}) \cite{zhou_detecting_2022,you2018a} &Sec.~\ref{D-RSPT}\\
            RSPT&IVb (Eq.(\ref{h_ivb}))&3&line-like symmetry, chaotic-looking symmetry (Fig.~\ref{pic_ivb})&Sec.~\ref{D-RSPT2}\\
            RSPT&IVc \footnote{Previously known as subsystem SPT (SSPT), see also footnote~\ref{footnote_SSPT}}(Eq.(\ref{clusterRule}))&2&line-like symmetry, a deformed 2D cluster model\footnote{Though a lot of SPT ordered states considered in this paper belong to the family of 2D cluster states, for convenience, we reserve the terminology ``2D cluster model'' for the Hamiltonian Fig.~\ref{clusterH} with SPT order protected by linear subsystem symmetries.}&/\\
			FSPT& Va (Eq.~(\ref{FSPT}))&1&  fractal-like symmetry \cite{devakul_fractal_2019}&Sec.~\ref{D-FSPT}\\
   FSPT& Vb (Eq.~(\ref{fiborule}))&1&  fractal-like symmetry \cite{devakul_fractal_2019}&/
		\end{tabular}
	\end{ruledtabular}
\end{table*}
\begin{figure*}[htb]
	\includegraphics[width=0.9\linewidth]{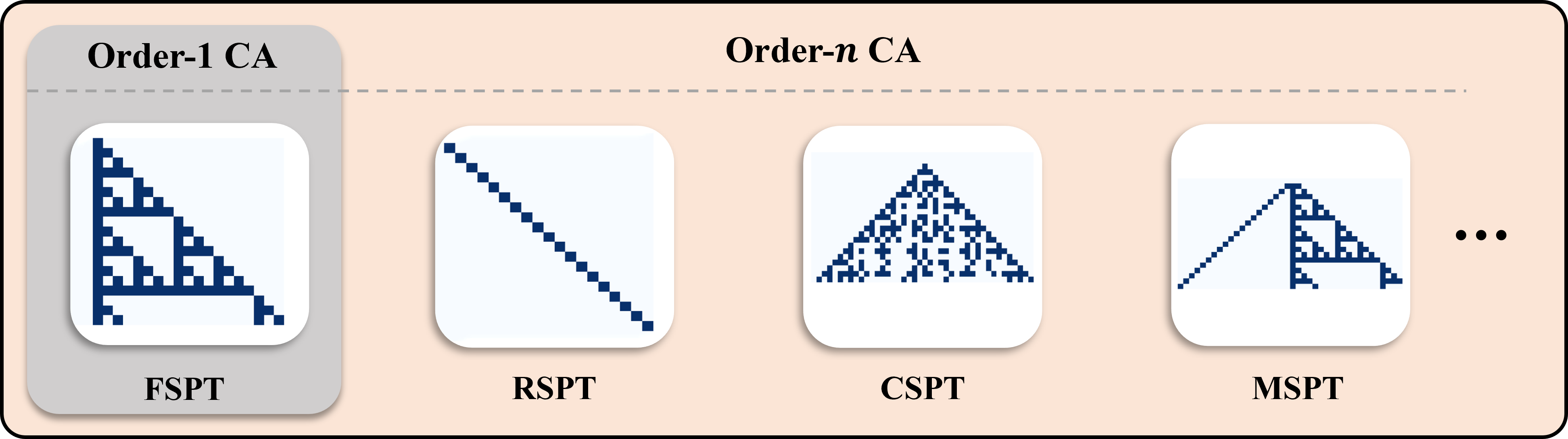}
	\caption{A brief schematic introduction to the types of SPT models produced by (linear) CA of different orders. The order-1 CA can only produce fractal patterns (see Appendix~\ref{order1} for detailed argument). It would require HOCA, i.e., CA with order $n\geq 2$, to create SPT models like RSPT, CSPT, and MSPT. These are SPT models whose symmetries are supported on regular subsets of lattice, chaotic-looking subsets of lattice, and more than one type of subsets of lattice, respectively. }
	\label{relation}
\end{figure*}

In the field of condensed matter physics, CA are often adopted to simulate   dynamical properties of systems, such as Ref.~\cite{CA1}. An example is the quantum cellular automata (QCA), originating from von Neumann and Feynman~\cite{neumann1966a,feynman1982a,feynman1986a}. QCA consist of arrays of identical finite-dimensional quantum systems that evolve in discrete-time steps by iterating a unitary operator $U$ \cite{arrighi_overview_2019}.  QCA are useful for simulating quantum systems and processes, such as quantum walks, quantum circuits, and quantum phase transitions \cite{arnault2016a,i2017a,giuseppe2016a,molfetta2014a,QCA1}.
Apart from simulation, CA also play a role in the study of symmetry-protected topological (SPT) order. Let us review some basic facts of SPT physics. 
SPT phases are short-range entangled states that cannot be smoothly deformed into trivial states without breaking some symmetries~\cite{haldane1983a,affleck1987a,su1979a,schuch2011a,turner2011a,fidkowski2011a,pollmann2012a,chen2011a,chen2011b,kane2005a,moore2007a,fu2007a,chen2012a,levin_braiding_2012,vishwanath2013a,yao2013a,gu2014a,qi2013a,cheng2018a,gaiotto2016a,thorngren2018a,else2014a,yoshida2016a}. 
SPT phases have been extensively explored through various methods, sparking interests from fields like condensed matter physics, mathematical physics and quantum information. These methods include group cohomology~\cite{chen_symmetry_2013}, cobordism groups~\cite{Kapustin2014FermionicSP,Kapustin2014SymmetryPT}, non-linear sigma models (NLSM)~\cite{PhysRevB.91.134404,PhysRevB.93.245135}, topological field theories~\cite{FT1,FT2,FT3,FT4}, conformal field theories (CFT)~\cite{CFT1,CFT2,CFT3}, decoration construction~\cite{chen2014a}, topological response theory~\cite{tr1,tr2,tr3,tr4,tr5,tr6,tr7}, projective or parton construction~\cite{pc1,pc2,pc3,pc4,pc5,pc6}, and braiding statistics \cite{bs1,bs2,bs3,bs4}. 
However, all the above SPTs  are limited to the cases where the symmetries are global, meaning that they act uniformly on the whole system. Recently, partially motivated by the field of fracton physics~\cite{f1,f2,f3,f4,f5,f6,f7,f8,f9,f10,f11,f12,f13,f14,f15,f16,f17,f18,f19,f20,f21,f22,f23,f24,f25,f26,f27,f28,f29,f30,f31,f32,f33,f34,f35,f36,f37}, people have realized that not only symmetries themselves but also where exactly symmetries act on the system matters, catalyzing the research of symmetry protected topological phases protected by symmetries supported on either  regular (e.g., line-like, membrane-like)~\cite{you2018a,you2020a,stephen2019a,schmitz2019a,miguel2021a,burnell2022a,devakul2018a,you2018a,zhou_detecting_2022} or fractal~\cite{devakul_fractal_2019,FSPT1,FSPT2} \textit{subsystems}. However, systematically designing the Hamiltonian in order to get the model protected by symmetries acting on a specific kind of subsets of the system remains a challenging problem. On one hand, the Hamiltonian of a lattice model usually involves some products of local operators; on the other hand, we need to control the subsets that the symmetries act on (usually the subsets are non-local and spread through the system).

Remarkably, CA coincidentally have simple local rules but exhibit complex behaviors, making them an ideal paradigm to design the Hamiltonian of condensed matter physics. So far, order-$1$ linear cellular automata have been  used to construct FSPT models whose nontrivial edge states are protected by fractal subsystem symmetries~\cite{devakul_fractal_2019}. Although fractal geometry in physics is also an interesting topic \cite{fr1,fr2,fr3,fr4,fr5,fr6,fr7,fr8,fr9,fr10,fr11,fr12,fr13,fr14,fr15,fr16,fr17},  order-$1$  linear CA cannot   produce subsystem symmetries without self-similarity (see Appendix \ref{order1}). Therefore,  it is natural to ask whether we can include SPT orders protected by symmetries supported on regular subsystems in the framework of CA\footnote{Strictly speaking, SPT protected by line-like symmetries can be produced by linear order-1 CA if  there is only one term in the update rule. The resulting model can be recognized as the trivial stacking of many 1D SPT phases, which was called    ``weak'' SSPT   in Ref.~\cite{you2018a}. Linear order-1 CA cannot produce ``strong'' SSPT in 2D, which is what people usually mean by SSPT, a ``genuine'' SSPT.}, and whether there are other possible subsets supporting symmetry action. 

In this paper, we go beyond the linear order-$1$ CA by using linear higher-order cellular automata (HOCA) to generate SPTs with various kinds of subsystem symmetries. We call these phases HOCA-generated SPT (HGSPT) phases. HOCA are cellular automata whose evolution involves multiple time steps~\cite{toffoli1977a}, and are widely used in computer science \cite{gu2000a,rey2005a,bruyn1991a,HOCA-C}. HOCA produce a rich variety of subsystem patterns in the spacetime lattice, including line-like and fractal patterns. HOCA have local update rules that make them useful for constructing Hamiltonians\footnote{We only consider linear versions of CA of any order, unless otherwise specified.}.  By using HOCA, we obtain a series of exactly solvable lattice models with various types of subsystem symmetries, as shown in Table~\ref{table1} and Fig.~\ref{relation}. These models include not only SPT models with symmetries supported on regular or fractal subsystems, but also more peculiar models. Therefore, in this paper we propose a notation for the types of SPT orders protected by symmetries supported on subsystems, which includes \textit{regular(-subsystems) SPT} (RSPT), \textit{fractal(-subsystem) SPT} (FSPT), \textit{mixed(-subsystem) SPT} (MSPT) and \textit{chaotic(-subsystem) SPT} (SPT) orders. For example, the 2D cluster model with linear subsystem symmetries discussed in Ref.~\cite{you2018a} and order-$1$ CA generated models with fractal subsystem symmetries discussed in Ref.~\cite{devakul_fractal_2019} are respectively classified into RSPT \footnote{The term ``subsystem symmetry protected topological phases'' and the acronym ``SSPT'' were introduced in previous works~\cite{you2018a}, where ``subsystem'' means lattice subsets with lower-dimensional regular shapes, such as lines in 2D systems and planes in 3D systems. In this paper, we introduce a slightly different terminology system for clarity, since we define ``symmetries'' on a great variety of subsystems. We use ``subsystem symmetry'' to mean a symmetry supported on any spatial subsystems with rigid shapes, including line-like, fractal, chaotic subsystem symmetries and other exotic cases that we will introduce in this paper. Therefore, the term ``SSPT'' in the previous work corresponds to the special case that we call ``RSPT'' in this paper, and we will avoid using ``SSPT'' to prevent confusion.\label{footnote_SSPT}} and FSPT orders. Besides, we also have MSPT models with both line-like and fractal-like subsystem symmetries, MSPT models with two distinct types of fractal-like subsystem symmetries and CSPT models with chaotic-looking subsystem symmetries. A more detailed and technical definition of these types of SPT orders is given in Sec.~\ref{notation}. We also introduce a new notation system to characterize these newly discovered subsystem symmetries, claiming that HOCA patterns can be labeled and HOCA rules can be classified by the patterns they produce. This notation plays the role of an attempt towards constructing a universal classification system of all possible configurations of subsystem symmetries, as the spatial form of symmetry elements has not been mathematically labeled as has symmetry group itself. We also discuss how universal the HOCA symmetries are, i.e. if any kind of subsystem symmetries can be understood as an HOCA symmetry. Nevertheless, it is particularly worth noticing that the CSPT is a class of SPT models supported on symmetries with highly exotic spatial distribution, which is very challenging to extract the mathematical properties of these symmetry patterns and labeling them. Because of the application of HOCA in the realm of data processing and encryption, we deem that CSPT models are also applicable in quantum computation, being an resource of quantum encryption algorithm, which is left to future exploration.


To detect the nontriviality of a given SPT ordered ground state, one may use   \textit{strange correlators}~\cite{you_wave_2014}.  By definition, a usual strange correlator is introduced as a two-point correlation function in which bra- and ket- wave functions are a symmetric short-range entangled state and the state to be diagnosed, respectively. If the state to be diagnosed is indeed SPT-ordered, the strange correlator will either saturate to a constant or decay algebraically at long distances. While nontrivial phenomena of SPT order are fully characterized by the boundary with 't Hooft anomaly, strange correlators enable us to   detect the nontrivial SPT order directly from the bulk, which removes potential analytic and numerical complexity induced by intricate boundary conditions.
So far, strange correlators have been successfully applied to  many SPT phases~\cite{you_wave_2014,zhou_detecting_2022,sc2,sc3,sc4,sc5,sc6,sc7}, and have also been applied in   intrinsic topological orders and conformal field theories (CFT)~\cite{c2,c5,c7,c8,c10,c13,c14,c15}. In particular, in Ref.~\cite{zhou_detecting_2022},  this tool has  been successfully applied to a 2D SPT order protected by line-like subsystem symmetries, i.e., a RSPT order following the nomenclature of the present paper. Therefore, one may wonder how to detect SPT phases with other types of subsystem symmetries, such as FSPT, MSPT, and CSPT discussed in the present paper. It is also interesting to ask whether or not the usual definition of two-point strange correlators is sufficient to  detect  all HGSPTs. 

In this paper, we find that HGSPTs  can be efficiently detected by what we call \textit{multi-point strange correlators} (MPSC). More concretely, in some HGSPT models, $2$-point strange correlators are insufficient for probing SPT orders, which leads us to generalize the usual strange correlators to multi-point. This approach reveals the complexity of SPT physics induced by HOCA and also expands the research scope of strange correlators. To explore this topic, we design a general procedure to detect the nontrivial SPT orders in the HGSPT models and to determine the class of HOCA update rules for a given HGSPT phase. We have shown that there are models that can only be detected by MPSC with more than 2 points by rigorous mathematical proof. We explicitly present the multi-point strange correlators designed for the models discussed in this paper. By generalizing the strange correlator to multi-point strange correlator, we find that the spatial properties of the symmetry can be reflected by the configuration of multi-point strange correlator, exhibiting the complexity of the HOCA evolution. 

It is also worth noticing that we discover MPSC, as the long range behavior in a short range entangled ground state, is inextricably connected to the \textit{spurious topological entanglement entropy} (STEE) \cite{spurious}, a ``spurious'' long range behavior in SSPT models devoid of topological order. We show that the nonlocal stabilizers that can run along the boundary in boundary geometries like Levin-Wen prescription \cite{LW} or dumbbell-like tripartition in \cite{spurious}, have the exact form of MPSC in these HGSPT models. We demonstrate this claim by many concrete models and mathematical proof in the paper. We discover a large variety of MPSC that can serve as the nonlocal stabilizers giving STEE in multiple HGSPT models, showing that STEE is a common character of most SPT orders with subsystem symmetry, beyond the scope of models protected by line-like symmetries only. To enable spurious contributions of these stabilizers,  one must consider more generalized boundary geometries, e.g. staggered boundary and even detached boundary. MPSC and STEE are both long range behaviors in a short range entangled state, between which the relation is a profound topic.

Hopefully, the multi-point strange correlator can potentially serve as a quantum mechanical approach to surpass the computational irreducibility~\cite{wolfram_cellular_1984, wolfram1984d, wolfram2002a} of HOCA, since we can effectively verify whether an arbitrary step of HOCA evolution is the given configuration by measuring the corresponding multi-point strange correlator (see also Sec.~\ref{D-CSPT}). These emergent phenomena naturally urge us to establish a more general and fundamental theory of strange correlators, shedding light on its underlying physics and explaining its efficacy in probing SPT phases, which is left to future exploration.

The rest of this  paper is organized as follows. In Sec.~\ref{hoca}, we provide some basic knowledge of HOCA. In Sec.~\ref{hgps}, we present the details of HGSPTs (see also Table~\ref{table1}), including the edge states, symmetry protection, duality, and concrete examples of models. In Sec.~\ref{notation}, we introduce a notation system to label HOCA rules and patterns, and we give a technical definition of the types of SPT orders protected by subsystems symmetries base on these notations. Some typical examples are summarized in Table~\ref{table_label}. In Sec.~\ref{sc}, we find that the phases mentioned above can be detected via multi-point strange correlators, and we propose a general procedure to design multi-point strange correlators for a given HGSPT model. We apply the procedure  to all models discussed in Sec.~\ref{hgps} and show the results in  Sec.~\ref{detection}. In Sec.~\ref{MPSC+STEE} we show the relation between MPSC and STEE by some concrete examples. In Appendix~\ref{order1},  we prove that order-1 CA cannot produce genuine RSPT phases as HOCA can, demonstrating the need for introducing HOCA. In Appendix~\ref{diff} we compare the model-IVa and the commonly known 2D cluster model, showing their subtle differences. In Appendix~\ref{calculation}, we demonstrate that for an FSPT model, strange correlators with $2$ onsite operators are all trivial. And as a special case, in Appendix~\ref{proof} we prove that there must be at least 3 points in the multi-point strange correlator to detect the nontriviality of an FSPT model. We further add a part comparing different CA approaches in constructing subsystem symmetries in Appendix~\ref{ca_review}, and a part that gives a brief review of calculating several dynamical properties of HOCA in Appendix~\ref{HOCA_prop}. The mathematical discussion of 2 criteria in Sec.~\ref{sc_definition} on how to detect the class of HGSPT phases using MPSC is given in Appendix~\ref{sc_proof}. Finally, a mathematical discussion on the universality of HGSPT phase is given in Appendix~\ref{universality}.

	\section{Higher-order cellular automata (HOCA)}	\label{hoca}

 \subsection{Preliminaries of CA and HOCA}\label{preliminary}
	
	Cellular automata (CA), first introduced by von Neumann~\cite{neumann1966a}, have been recognized as a good dynamical system for simulating complex physical systems. CA have a simple structure but exhibit a great variety of complex behaviors, and are used to model phenomena with local, uniform, and synchronous processing~\cite{dennunzio_dynamical_2019}. Formally speaking, a CA consists of an infinite set of identical finite automata placed over a lattice and all taking a state from a finite set called the \emph{alphabet} of the CA.
	
There are many possible variants of CA. People have explored CA in higher dimensions to model systems with multiple degrees of freedom, and higher-order CA with memory size $n>1$, which is the main focus of this paper. A higher-order cellular automaton (HOCA) is a discrete dynamic system whose evolution involves multiple time steps, first introduced by Toffoli in \cite{toffoli1977a}. While ordinary linear cellular automata  always generate self-similar patterns (e.g., fractal patterns in the spacetime lattice, which can be proved by the \emph{Freshman's Dream} theorem of a polynomial over $\mathbb{F}_p$ \cite{noauthor_freshmans_2023}, see also Appendix \ref{order1}), (linear) HOCA produce many peculiar patterns besides fractal patterns\footnote{Linear CA rules can be written in terms of polynomial representations, making it possible to write Hamiltonians with respect to these rules.}. For example, HOCA can exhibit chaotic behaviors, which are often used in secret sharing schemes \cite{rey2005a,bruyn1991a}, data compression and image encryption \cite{gu2000a}. The encryption algorithm based on HOCA can be efficiently implemented in hardware due to the simple structure of CA, and is hard to decipher due to the chaotic behavior of the HOCA.

Despite the above interesting applications in computer science, the understanding of the dynamic behavior of HOCA is still at an early stage, and few results are known for linear HOCA~\cite{dennunzio_dynamical_2019}. More properties and applications of HOCA are still to be studied and explored. Furthermore, to the best of our knowledge, HOCA have not been used in the realm of physics so far, and this paper will serve as an attempt to explore the interdisciplinary amalgamation of HOCA and condensed matter physics. 

Now we introduce some basic notations of HOCA. Consider a set of $1D$ lattice sites $\{i\},~i\in\mathbb  Z$ with \textit{alphabet} $a_i\in \{0,1,\cdots,p-1\}= \mathbb{F}_p$ evolving with time $j$; the state of any given site at any given time may be expressed as $a_i(j)$. We introduce the \textit{polynomial representations} to simplify our notation. By doing the substitution
\begin{equation}
	a_i(j)\to a_{ij}x^iy^j, \text{ where } a_{ij}\equiv a_i(j)\in \mathbb{F}_p,
\end{equation}
we express the spacetime configuration of all lattice sites by a polynomial:
\begin{equation}\label{eq:HOCA_config}
	\mathscr{F}(x,y)=\sum_{i=-\infty}^{\infty}\sum_{j=0}^\infty a_{ij}x^iy^j.
\end{equation}
Also, we define site configuration at time $j_0$ with respect to $x$ as
\begin{equation}\label{rx}
	 r_{j_0}(x)\equiv\sum_{i=-\infty}^{\infty}a_{ij_0}x^i
\end{equation}
by picking all terms with $y$-exponent equals to $j_0$.
Notice that our model is defined on a semi-infinite plane here, which shows the entire evolution of the HOCA rule. The HOCA model can be also defined on an open slab by truncation, which we will introduce later in subsection \ref{general_discussion}.
We will use $r_j(x)$ to denote the configuration at time $j$ from now on. Now we introduce the concept of higher-order cellular automata (HOCA), which are extensions of traditional cellular automata that involve interactions across multiple time steps. In an order-$n$ CA \footnote{The order of CA is also referred to as the \textit{memory size} of the CA in computer science.}, the state of a site at time $j_0$ is determined by the states of a neighborhood of sites at times $j_0-1,j_0-2,\cdots, j_0-n$.
From now on, we focus on HOCA defined on $\mathbb{F}_2=\{0,1\}$. Every HOCA rule mentioned below is defined on $\mathbb{F}_2$. If $a_{i_0j_0}$ can be written as translationally invariant sums of elements in $\{a_{ij}|i\in\mathbb{Z},j\in\{j_0-1,j_0-2,\cdots,j_0-n\}\}$, then the HOCA is defined to be \textit{linear}, meaning that
\begin{equation}
	a_{i_0,j_0}=\sum_{q=-n}^{-1}\sum_{p=-R}^{R} c_{pq}a_{i_0+p,j_0+q},
\end{equation}
where $c_{pq}\in \mathbb{F}_2$ are coefficients, $R$ is radius, a constant describing the maximal range of $p$, which does not scale with the system size, making the rule local. We concentrate on linear HOCA because update rule of a linear order-$n$ HOCA can be represented by $n$ polynomials, which enables us to construct Hamiltonians with decorated defect construction using these update rules.  We demand $R<\infty$ to make sure the HOCA rule is local, which means that the effect of the HOCA rule (i.e. change of $a_{ij}$ due to the HOCA rule) will not propagate faster than the speed $R$. We denote an HOCA rule (i.e. update rule) by an $n$-row vector $\mathbf{f}$, dubbed as the \textit{update rule} of the HOCA:
\begin{equation}
\label{eq:update_rule}
	\mathbf{f}(x)\equiv\begin{pmatrix} 
		f_1(x),f_2(x),
		\cdots,
		f_n(x)
	\end{pmatrix}^T,
\end{equation}
where the superscript $T$ denotes the transpose of the vector.
When $n=1$, the HOCA returns to the normal CA as discussed in \cite{devakul_fractal_2019}. The time evolution of local linear HOCA can be denoted by a single formula (here we assume $j_0>n$):
\begin{equation}\label{evolve}
	r_{j}(x)=\sum_{k=1}^n r_{j-k}(x)f_k(x),
\end{equation}
where $r_{j}(x)$ is defined in Eq.~(\ref{rx}).

To ascertain the whole time evolution process of all lattice sites of an order-$n$ CA, one needs to manually specify the configurations of first $n$ time steps $r_0,r_1,...,r_{n-1}$, which is called the \textit{initial condition} of the system. It can also be denoted by an $n$-row vector $\mathbf{q}(x)$:

\begin{equation}
	\mathbf{q}(x)\equiv \begin{pmatrix}
		r_0(x),r_1(x),\cdots,r_{n-1}(x)
	\end{pmatrix}^T.
\end{equation}

By specifying an HOCA rule $\mathbf{f}$ and an initial condition $\mathbf{q}$, the whole spacetime pattern can be uniquely defined. We define
\begin{equation}
	\begin{aligned}
		\mathcal{E}^{(1)}(\mathbf{f})&\equiv \begin{pmatrix}
			f_n,f_{n-1},\cdots,f_1
		\end{pmatrix}^T,\\
		\mathcal{E}^{(2)}(\mathbf{f})&\equiv \begin{pmatrix}
			f_1f_n,f_n+f_1f_{n-1},\cdots,f_2+(f_1)^2
		\end{pmatrix}^T
	\end{aligned}
\end{equation}
and so on, such that $r_{n-1+i}(x)=\mathbf{q}^T(x)\cdot\mathcal{E}^{(i)}(\mathbf{f})$.
$\mathcal{E}^{(i)}$ is dubbed as the \textit{evolution operator}, which can be calculated using Eq.~(\ref{evolve}).
We can always write $r_{j}(x),~j\ge n$ as sum of each row in the initial condition multiplied by some update rules.  It follows that the whole spacetime pattern can be expressed as
\begin{equation}\label{pattern}
	\begin{aligned}
		\mathscr{F}(x,y)&=\mathbf{q}^T(x)\cdot \mathbf{ y}_{0,n}+\sum_{k=1}^\infty y^{n-1+k}\mathbf{q}^T(x)\cdot  \mathcal{E}^{(k)}(\mathbf{f})\\
		&=\mathbf{q}^T(x)\cdot \left[ \mathbf{ y}_{0,n} +\sum_{k=1}^\infty y^{n-1+k}  \mathcal{E}^{(k)}(\mathbf{f})\right]\\
		&\equiv \mathbf{q}^T(x)\cdot \mathbf{F}(x,y),
	\end{aligned}
\end{equation}
where $\mathbf{q}$ and $\mathbf{F}$ capture the effect of the initial condition and the update rule separately, and the label $\mathbf{y}_{p,q}$ is defined as 
\begin{equation}
\label{eq:vector_y}
	\mathbf{y}_{p,q}=\begin{pmatrix} 
		y^p,
		y^{p+1},
		\cdots,
		y^{p+q-1}
	\end{pmatrix}^T.
\end{equation}
Eq.~(\ref{pattern}) is useful in the calculation of commutation polynomial (Eq.~(\ref{compol})), which is important to the discussion of the symmetry elements in the HOCA generated SPT phases (to be discussed in section \ref{general_discussion}).
If we treat the time axis as another spatial dimension, we can view the whole time evolution of the given HOCA $\mathscr{F}(x,y)$ as a static pattern in a $2D$ semi-infinite plane. Any given HOCA rule $\mathbf{f}$ can generate infinite number of patterns by adjusting initial condition $\mathbf{q}(x)$. 
		
 \subsection{Spin (qubit) model in terms of polynomial representations}
	
	Polynomial representations can also express spin systems by identifying $a_{ij}$ with the state of the spin located at site $x^iy^j$. Then the whole HOCA pattern $\mathscr{  F}(x,y)$ (given in Eq.~(\ref{pattern})) naturally expresses the spin configuration in the lattice. By introducing polynomial representations, we naturally transplant HOCA into the realm of spin systems. Consider a spin model defined on a $d$-dimensional square lattice with $\alpha$ sublattices (i.e. each site contains $\alpha$ independent degrees of freedom, which don't have to equal the order of the HOCA). One spin is placed on each site of the sublattice. We introduce the following conventions
	\begin{outline}[itemize]
		\1 The coordinate of site $s=(i_1,i_2,...,i_d)$ are represented by a monomial $m$ with respect to $x_1,\cdots, x_d$:  $
			m=x_1^{i_1}\cdots x_d^{i_d}$. 
		In this paper, we focus on $d=2$ case, and we use the notation $i_1\equiv i,\ i_2\equiv j,\ x_1\equiv x, x_2 \equiv y$.
		\1 Previously defined $a_{ij}$ expresses the state of spin located at site $x^iy^j$ (in a specific sublattice). $a_{ij}=0$ represents that the spin at $x^iy^j$ is at the state $\ket{0}$, and $a_{ij}=1$ represents $\ket{1}$. The sublattice that the spin belongs to will be defined in the next point.
		\1 If the onsite Pauli operators Pauli $\hat X$, Pauli $\hat Y$ and Pauli $\hat Z$ operators are represented by $O=X,Y,Z$ respectively, a many-body Pauli operator $\mathscr  O$ can be  denoted as
		\begin{align}
		 \mathscr{O}:=O\begin{pmatrix} 
				m^{(1)}_1+m^{(1)}_2\cdots +m^{(1)}_{k_1}\\
				\vdots\\
				m^{(\alpha)}_1+m^{(\alpha)}_2\cdots +m^{(\alpha)}_{k_\alpha}
				\end{pmatrix}\!=\!O\begin{pmatrix} 
				P_1\\ \vdots \\ P_\alpha
				 \end{pmatrix},
		\end{align}
		where $m^{(i)}_k$ denotes the position of site $s_k$ in sublattice $i$ that the operator $\mathscr{O}$ acts nontrivially on, and $P_k$ denotes a polynomial with respect to $x$.
		\1 We define the coefficients of monomials $ m^{(i)}_k $ to be in $\mathbb{Z}_2$, and then we can naturally obtain
		\begin{equation}
		\mathscr{O}_1\mathscr{O}_2\!=\!O\!\!\begin{pmatrix} 
				P^{\mathscr{O}_1}_1+P^{\mathscr{O}_2}_1\\
				\vdots\\
				P^{\mathscr{O}_1}_\alpha+P^{\mathscr{O}_2}_\alpha
				 \end{pmatrix}\!\!=\!O\!\!\left[\!\begin{pmatrix} 
				 P^{\mathscr{O}_1}_1\\
				 \vdots\\
				 P^{\mathscr{O}_1}_\alpha
				 \end{pmatrix}\!+\!\begin{pmatrix} 
				 P^{\mathscr{O}_2}_1\\
				 \vdots\\
				 P^{\mathscr{O}_2}_\alpha
				 \end{pmatrix}\!\right].
		\end{equation}
	\end{outline}

	\section{HOCA generated symmetry-protected topological phases }
	\label{hgps}

	\subsection{Lattice models, short-range entanglement, symmetry, and symmetry protected edge states}\label{general_discussion}
In this section we construct symmetry-protected topological (SPT) phases protected by HOCA generated symmetry. Due to the great variety of the HOCA behavior, we can naturally obtain models with fractal symmetries, line-like symmetries, both of the above, and even chaotic symmetries, which are respectively classified into FSPT, RSPT, MSPT and CSPT orders according to our notation. And as a summary, in Sec.~\ref{notation}, we give the technical definition of all these types of SPT orders based on the concrete examples demonstrated in this section. 
We'll begin with some basics and notations.

Through HOCA and decorated defect construction, we can obtain symmetry-protected topological (SPT) order with two types of subsystem symmetry, which we refer to as \textit{mixed-subsystem SPT} (MSPT). Among all MSPT models, there are models with both fractal-like and line-like subsystem symmetries (abbreviated as I-MSPT), and models with 2 different fractal symmetries (abbreviated as II-MSPT). Additionally, there are models with only chaotic-looking symmetries (dubbed as chaotic SPT) and models with line-like and membrane-like symmetries (resembling the previously known SSPT model). These HOCA generated models are all defined on a 2D square lattice with 2 sublattices $(a)$ and $(b)$, and the Hamiltonians can be generally written as
\begin{equation}\label{MSPT}
	\begin{aligned}
		\mathscr{H}=&-\sum_{i j} Z\left(\begin{array}{c}
			x^i y^j(1+\bar{\mathbf{f}}\cdot \mathbf{\bar{ y}}_{1,n}) \\
			x^i y^j
		\end{array}\right)\\
		&-\sum_{i j} X\left(\begin{array}{c}
			x^i y^j \\
			x^i y^j(1+\mathbf{f}\cdot \mathbf{y}_{1,n})
		\end{array}\right) 
	\end{aligned}
\end{equation}
where $\mathbf{f}$ is an HOCA update rule (see Eq.~(\ref{eq:update_rule})), $\mathbf{\bar{y}}$ is a vector composed of monomials of $y$ (see Eq.~(\ref{eq:vector_y})), and the notation $\bar{\mathbf{f}}$ means $\bar{\mathbf{f}}(x):=\mathbf{f}(\bar x):=\mathbf{f}(x^{-1})$.

\begin{figure*}[htbp]
	\centering
	\subfigure{\includegraphics[width=0.35\linewidth]{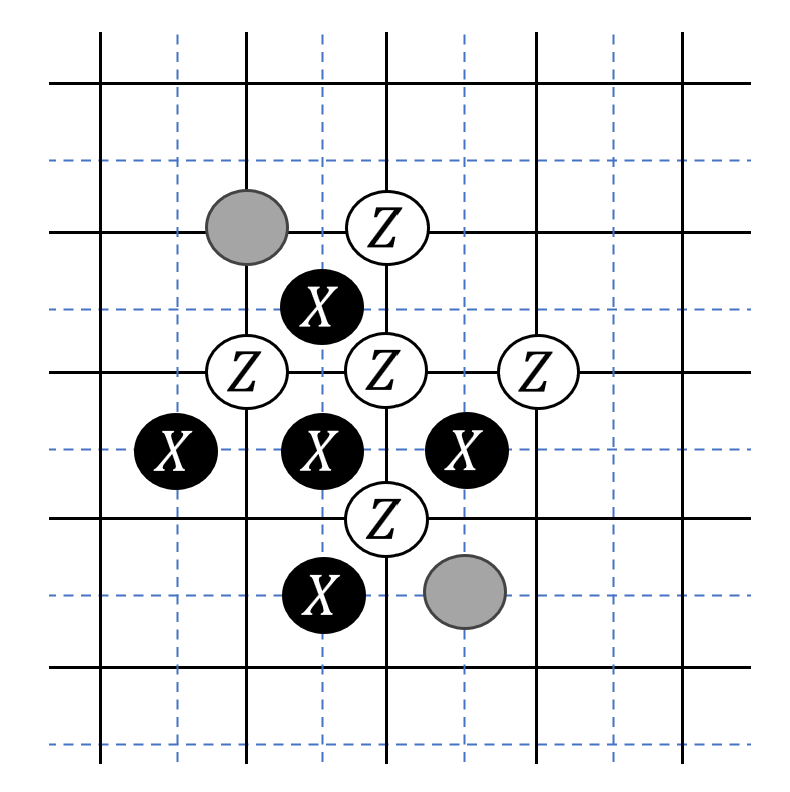}}\qquad\qquad\qquad\qquad\qquad
	\subfigure{\includegraphics[width=0.35\linewidth]{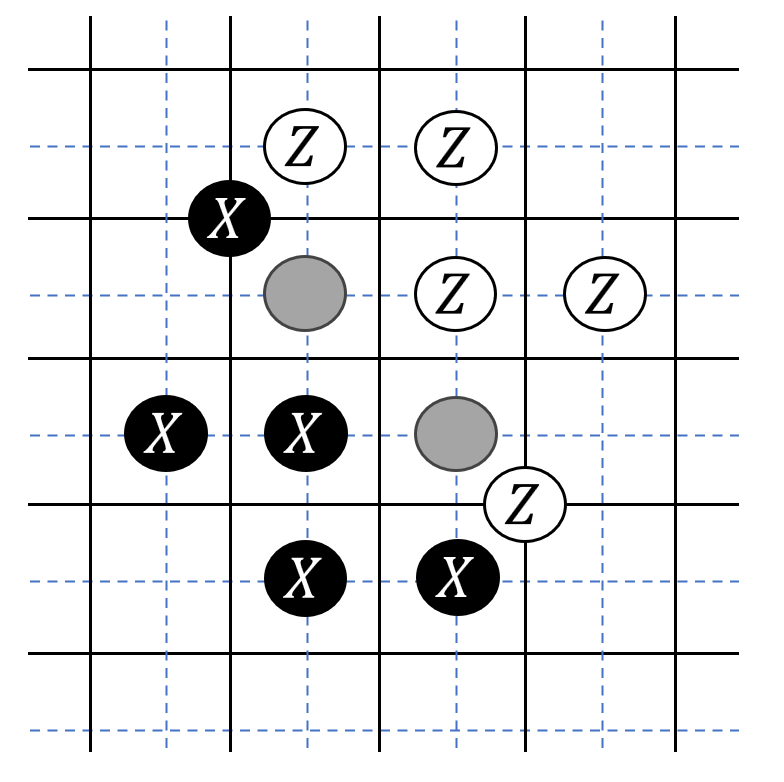}}
	\caption{Two possible overlapping ways of Hamiltonian terms of I-MSPT model (Eq.~(\ref{MSPT1})). Gray circles are the overlapping $X$ and $Z$ Pauli matrices from two terms. Black lattice and blue dashed lattice denote sublattices $(a)$ and $(b)$.}
	\label{pic-1-1}
\end{figure*}

The Hamiltonian in Eq.~(\ref{MSPT}) describes an exactly solvable cluster model with a short-range entangled unique ground state on a torus, similar to the usual cluster states. In fact, we can obtain the commonly seen $1D$ cluster model (which is an SPT phase protected by global $\mathbb{Z}_2\times \mathbb{Z}_2$ symmetry) by taking $\mathbf{f}=1$, and define it on a lattice with $L_x=1$. The Hamiltonian we get are equivalent to that of the $1D$ cluster model up to a change of basis ($Z\leftrightarrow X$). The exact solvability of the model can be proved by noting that there are always $0$ or $2$ overlapping operators between two terms in the Hamiltonian, ensuring that every Hamiltonian term commutes with each other. Two examples are shown in Fig.~\ref{pic-1-1}. 
The property of the ground state can be verified by noting that there are $2L_xL_y$ qubits and $2L_xL_y$ Hamiltonian terms in a model defined on a $L_x\times L_y$ torus. Notice that each Hamiltonian term corresponds to a unique onsite Pauli operator, resulting that all Hamiltonian terms are independent with each other. For example, each term $Z\left(\begin{array}{c}
	x^i y^j(1+\bar{\mathbf{f}}\cdot \mathbf{\bar{ y}}_{1,n}) \\
	x^i y^j
\end{array}\right)$ corresponds to operator $Z\left(\begin{array}{c}
0 \\
x^i y^j
\end{array}\right)$ , and each term $X\left(\begin{array}{c}
x^i y^j \\
x^i y^j(1+\mathbf{f}\cdot \mathbf{y}_{1,n})
\end{array}\right)$ correspond to operator $X\left(\begin{array}{c}
x^i y^j \\
0
\end{array}\right)$.
 With no other constraints being present, the ground state subspace has dimension $2^{2L_xL_y}/2^{2L_xL_y}=2^0=1$, giving a unique ground state on the torus.

Now, let's investigate deeper into the symmetry elements of the model. Suppose these models are defined on an open slab with a size of $L_x\times L_y$, and all Hamiltonian terms with operators outside of the boundary are excluded. Here, $L_x$ is the length in the $i$ direction and $L_y$ is the length in the $j$ direction. For a SPT model generated by an order-$n$ CA, the open slab should satisfy
$	L_x\ge p_{\text{min} }+p_{\text{max} } $ and $ L_y> n
 $ , where $p_{\text{max} }$ and $-p_{\text{min} }$ are respetively the largest and smallest power of $x$ in $\mathbf{f}(x)$ (if $p_{\text{max} }$ or $p_{\text{min} }$ are less than zero, then it is defined to be zero), to ensure there is at least one valid Hamiltonian term in the model.

We now focus on Hamiltonian terms whose coordinate $x^iy^j$ is in the slab but contains sites outside of the slab. Assuming the coordinate axis is taken as in Fig.~\ref{pic-1}, then for sublattice $(a)$ there are $n$ rows of such Hamiltonian terms excluded at the top edge of the system, $p_{\text{max} }$ terms at the left edge, and $p_{\text{min} }$ terms at the right edge. Each excluded term with coordinate $x^iy^j$ plays the role of a lost constraint on the ground state manifold, producing a free spin at site $x^iy^j$. Similarly, we can obtain extra degrees of freedom at sublattice $(b)$, with everything reversed (top$\leftrightarrow$down, left$\leftrightarrow$right, etc.). Suppose there are $k$ such excluded Hamiltonian terms in the model, then the ground state degeneracy of the model will be $2^k$.

We are now able to flip free spins at the edge without changing the energy of the system. Flipping spins at the edge will generally affect spins in the bulk following the HOCA update rule, producing symmetry elements in the shape of the HOCA pattern. The operations that flip spins in these HOCA patterns commute with the Hamiltonian of the model, being the symmetries that protect the degenerate edge state.

These model has $2$ sets of subsystem symmetries $S^{(a)}$ and $S^{(b)}$, each set per sublattice:
		\begin{align}\!\! S^{(a)}(\mathbf q)=X\begin{pmatrix}\mathscr{\tilde F}(x,y)\\0\end{pmatrix}\,,\,\, S^{(b)}(\mathbf q)=Z\begin{pmatrix}0\\ \tilde{\bar{\mathscr{  F}}}(x,y)\end{pmatrix},\label{eq_symm}
	\end{align}
	where $\mathscr{\tilde F}(x,y)$ is the truncated HOCA pattern $\mathscr{F}(x,y)$ (Eq.~(\ref{pattern})) specified by an HOCA rule $ \mathbf{f}$ and an initial condition $\mathbf{q}$, and all terms which are not fully in the slab are excluded. The HOCA rule $\mathbf{f}$ controls which type of subsystem symmetry can be found in this model, and $\mathbf{q}$ controls the specific pattern of the symmetry. We can enumerate these symmetry elements in an HOCA generated SPT model by counting all possible different initial conditions that can be defined in the slab. 
	
	These symmetry elements commute with the Hamiltonian terms, which can be verified by examining the \textit{commutation polynomial}. The commutation polynomial with respect to two polynomials $\alpha  , \beta  $ is defined as:
 $		P( \alpha  , \beta  )\equiv \alpha  \bar  \beta$. 
	If the coefficient of $x^0y^0$ in $P( \alpha  , \beta  )$ is zero, $X( \alpha  )$ and $Z( \beta  )$ commute with each other \cite{devakul_fractal_2019}.
	
	Now we verify the commutation relation in sublattice $(a)$. We will do the calculation in the semi-infinite plane ($-\infty<i<\infty,\ j\ge0$) and do the truncation afterwards. In sublattice $(a)$, the symmetry writes (sublattice (b) are not represented below)
 $		X(\mathscr{F}(x,y))$
	and a general Hamiltonian term (we consider the $Z$-term only) writes
	$Z(x^iy^j(1+\mathbf{\bar f}\cdot \mathbf{\bar y}_{1,n})).
	 $ 	The commutation polynomial is
	\begin{equation}\label{compol}
		\begin{aligned}
			&P(\mathscr{F}(x,y),x^iy^j(1+\mathbf{\bar f}\cdot \mathbf{\bar y}_{1,n}))\\
			=&\mathscr{F}(x,y)x^{-i}y^{-j}(1+\mathbf{f}\cdot \mathbf{y}_{1,n})\\
			=&x^{-i}y^{-j}(1+\mathbf{f}\cdot \mathbf{y}_{1,n})\sum_{k=0}^{\infty}y^kr_k(x)\\
			=&x^{-i}y^{-j}\left[\sum_{k=0}^{\infty}y^kr_k(x)+\sum_{k=n}^{\infty}y^kr_k(x)+\sum_{k=0}^{n-1}y^k\tilde{r}(x)\right]\\
			=&x^{-i}y^{-j}\left[\sum_{k=0}^{n-1}y^k{r}(x)+\sum_{k=0}^{n-1}y^k\tilde{r}(x)\right].
		\end{aligned}
	\end{equation}
Here, $\tilde{r}(x)$ is defined as follows:
\begin{equation}
	\sum_{k=0}^{n-1}y^k\tilde{r}(x)\equiv\mathbf{f}\cdot \mathbf{y}_{1,n}\sum_{k=0}^{\infty}y^kr_k(x)-\sum_{k=n}^{\infty}y^kr_k(x),
\end{equation}
where all terms have a power of $y$ that is lower than $n$. 
Note that $i\in \mathbb{Z}$ and $j\in\mathbb{N}$. The definition of $\mathscr{  F}$ can be found in Eq.~(\ref{pattern}). While the exponent of $y$ is given by $k-j$ in Eq.~(\ref{compol}), and $k=0,1,...,n-1$, which means that all terms in the commutation polynomial have a $y$-power less than $n$. 
According to our convention, Hamiltonian terms with $j<n$ are all excluded. Hamiltonian terms with $j\ge n$ have a null commutation polynomial since the $y$-power of all terms in the commutation polynomial are less than zero, which are also outside of the slab. Thus, we proved that such symmetry elements indeed commute with the Hamiltonian (Eq.~(\ref{MSPT})). 

The edge states are protected by the above symmetry elements. These symmetry elements are all in the shape of an HOCA pattern (and their superposition after translation), which can be generated by choosing an initial condition $\mathbf{q}$ and truncating the resulting HOCA pattern to fit the open slab. The visual property of the symmetry element is controlled by both initial condition $\mathbf{q}$ and HOCA rule $\mathbf{f}$. To explore edge physics, we can define a series of new Pauli operators at the edge. Taking sublattice (a) as an example, the edge Pauli matrices are written as:
	\begin{align}
		&\mathscr{X}_{ij}^{(a)} = X\begin{pmatrix}0\\x^iy^j\end{pmatrix};\mathscr{Z}_{ij}^{(a)} = Z\begin{pmatrix}x^iy^j(1+\bar{\mathbf{f}}\cdot \bar{\mathbf{y}}_{1,n})\\x^iy^j\end{pmatrix},\nonumber\\ &\mathscr{Y}_{ij}^{(a)} = Z\begin{pmatrix}x^iy^j(1+\bar{\mathbf{f}}\cdot \bar{\mathbf{y}}_{1,n})\\0\end{pmatrix}Y\begin{pmatrix}0\\x^iy^j\end{pmatrix}, \nonumber 
	\end{align}
where these operators are truncated to the slab by default. Here, edge states in sublattice $(a)$ are distributed along the top, left and right edges of the slab. These three matrices all commute with remaining Hamiltonian terms and they form a Pauli algebra. To open the gap of a degenerate edge state, we can add a magnetic field to an edge free spin. This operation must violate approximately $2^{k-1}$ symmetry elements which act nontrivially on this edge spin, reducing the ground state degeneracy by half. 

Now we give a brief picture of the duality of HGSPT model. On open slab, each ground state of an HGSPT model can be mapped to a symmetry breaking ground state of the dual model. By redefining Pauli operators, an HGSPT Hamiltonian becomes 2 decoupled copies of symmetry breaking orders with the same HOCA generated symmetry, each showing a phase transition at magnetic field $h=1$ via Kramers-Wannier duality. So for an HGSPT model, we can add magnetic field in two different directions ($h_x$ and $h_z$), and transition will happen at $h_x=1$ and $h_z=1$. If both magnetic fields are smaller than 1, the model remains in the HGSPT order. If one of $h_x$ and $h_z$ is bigger than $1$, the model becomes symmetry breaking phase in one sublattice. If $h_x,\ h_z$ are both bigger than 1, the system are in the trivial paramagnetic phase.

We will give several examples of HGSPT models in the next few subsections.

	\subsection{Model-I: I-MSPT generated by order-2 CA}
\begin{figure}[htbp]
	 	\includegraphics[width=\linewidth]{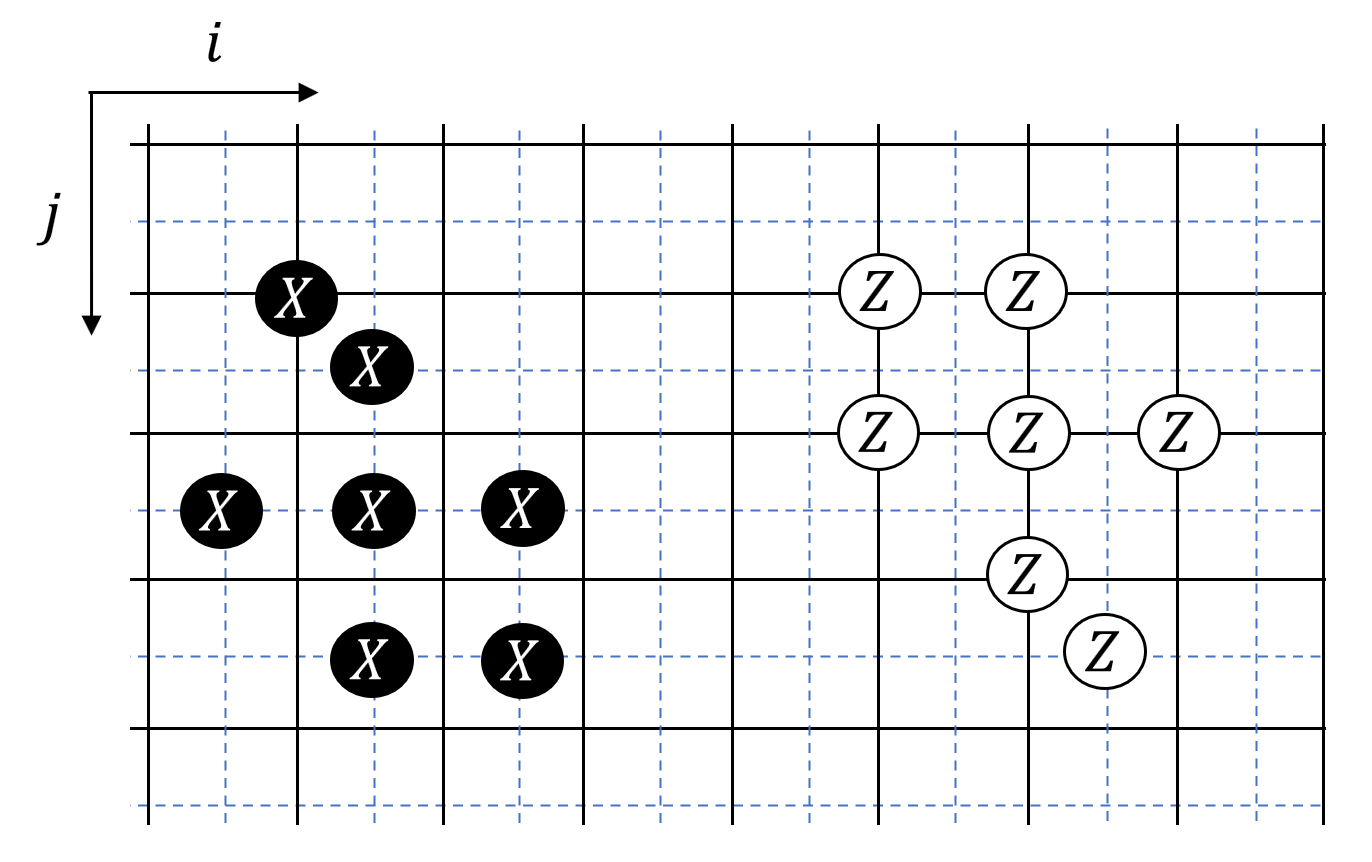}
	 	\caption{Pictorial illustration of two typical Hamiltonian terms of I-MSPT model (Eq.~(\ref{MSPT1})). Black lattice and blue dashed lattice denote sublattices $(a)$ and $(b)$.}
	 	\label{pic-1}
	 \end{figure}

	An example of I-MSPT Hamiltonian writes
	\begin{equation}\label{MSPT1}
		\begin{aligned}
			&\mathscr{H}=\\&-\sum_{i j} Z\left(\begin{array}{c}
				x^i y^j[1+y^{-1}(x^{-1}+1+x)+y^{-2}(1+x^{-1})] \\
				x^i y^j
			\end{array}\right)\\
			&-\sum_{i j} X\left(\begin{array}{c}
				x^i y^j \\
				x^i y^j[1+y(x^{-1}+1+x)+y^2(1+x)]
			\end{array}\right),
		\end{aligned}
	\end{equation}
	generated by an order-2 CA with update rule
		\begin{equation}\label{fl-MSPT1}
		\mathbf{f}(x)=\begin{pmatrix}
			x^{-1}+1+x\\
			1+x
		\end{pmatrix}.
		\end{equation}
	The Hamiltonian is pictorially shown in Fig.~\ref{pic-1}.

\begin{figure*}[htbp]
		\subfigure{\includegraphics[width=0.38\linewidth]{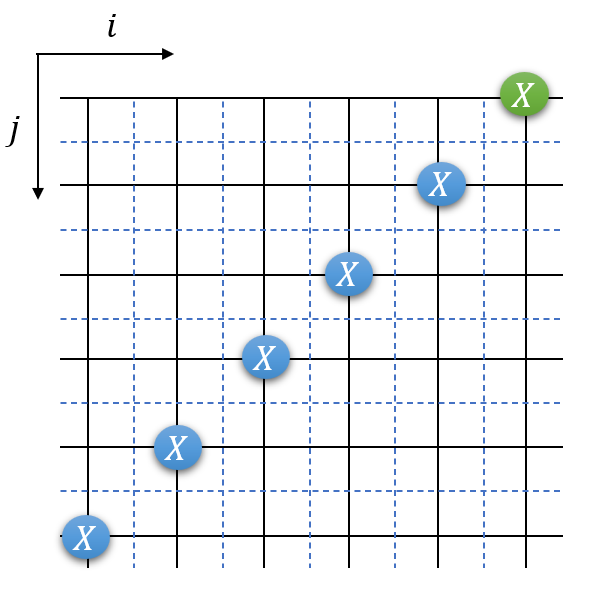}}
		\subfigure{\includegraphics[width=0.38\linewidth]{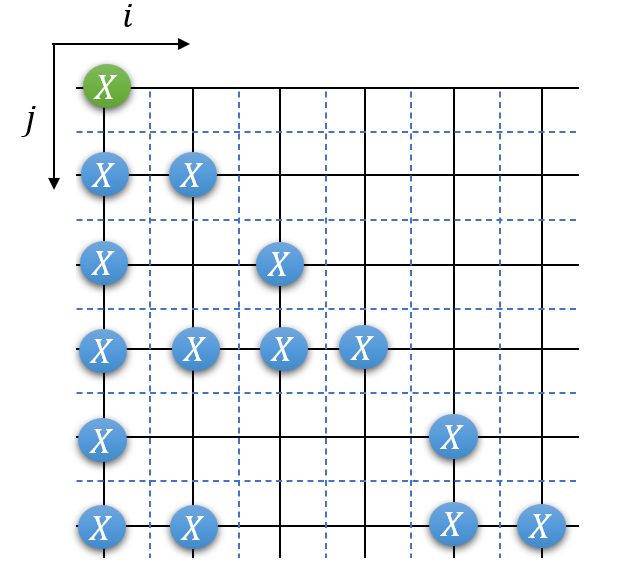}}
		\caption{Two examples of symmetries that protect the edge free spin located at green circle at the top edge of the system. The Hamiltonian is (Eq.~(\ref{MSPT1})). If we manually break the degenerate edge mode at green circle (e.g. by adding a Zeeman term), such modification of the Hamiltonian will anticommute with two symmetry elements shown in the figure (and many other terms that act nontrivially on this site), showing that the edge mode is indeed protected by our HOCA generated symmetry.}
		\label{symprotect}
	\end{figure*}

	 The order of the HOCA is $n=2$, so we need to specify 2 rows of initial conditions by a $2$-row vector $\mathbf{q}$. Now we focus on symmetry elements in sublattice $(a)$, and symmetries in sublattice $(b)$ can be obtained similarly (by reversing everything, as mentioned in subsection~\ref{general_discussion}). First, take an initial condition $\mathbf{q}$ and plug it into Eq.~(\ref{pattern}) to get the symmetry pattern. Then, if the model is defined on an $L_x\times L_y$ open slab, we place the first row of symmetry pattern on top of the slab (row with $j=0$) and exclude parts that are not in the slab. Since the HOCA rule is translationally invariant in $x$-axis, we can move our pattern in the $x$-direction with terms outside of the slab being excluded. The operation above gives us $\tilde{\mathscr{F}}$, which we plug into Eq.~(\ref{eq_symm}) to get the symmetry element. In the case of sublattice $(a)$, the symmetry elements are made up to Pauli-$X$ operators in the shape of $\tilde{\mathscr{F}}$, just like we defined in Eq.~(\ref{eq_symm}). Consider symmetry elements generated by the following $4$ initial conditions:
\allowdisplaybreaks[4]
	 \begin{subequations}
		\begin{align}
			\mathbf{q}_1(x)&=\begin{pmatrix}
				x\\
				1+x
			\end{pmatrix}\label{ic1},\\
			\mathbf{q}_2(x)&=\begin{pmatrix}
				1\\
				x
			\end{pmatrix}\label{ic2},\\
			\mathbf{q}_3(x)&=\begin{pmatrix}
				0\\
				x^{-1}+1+x\\
			\end{pmatrix}\label{ic3},\\
			\mathbf{q}_4(x)&=\begin{pmatrix}
				0\\
				1
			\end{pmatrix}\label{ic4}.
		\end{align}
	\end{subequations}
	\begin{figure*}[htbp]
		\centering
		\subfigure[$\mathbf{q}_1(x)$\label{p1}]{\includegraphics[width=0.48\linewidth]{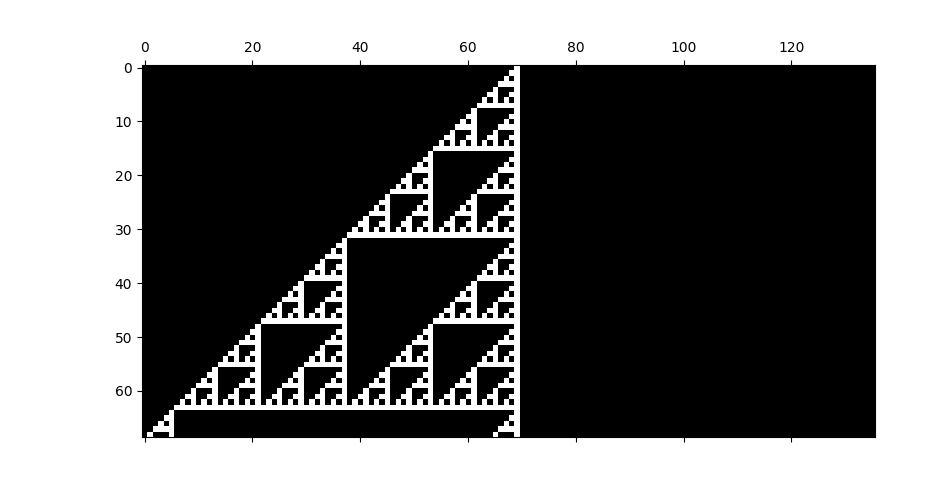}}
		\subfigure[$\mathbf{q}_2(x)$\label{p2}]{\includegraphics[width=0.48\linewidth]{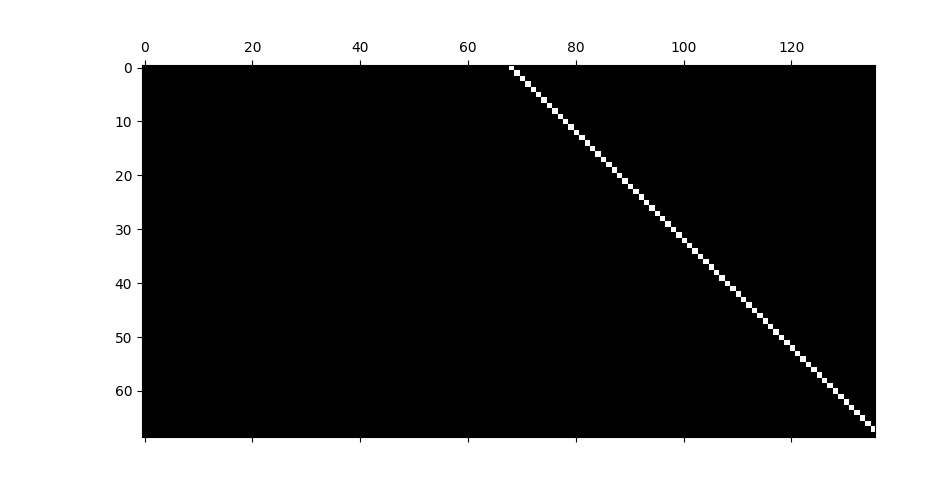}}\\
		\subfigure[$\mathbf{q}_3(x)$\label{p3}]{\includegraphics[width=0.48\linewidth]{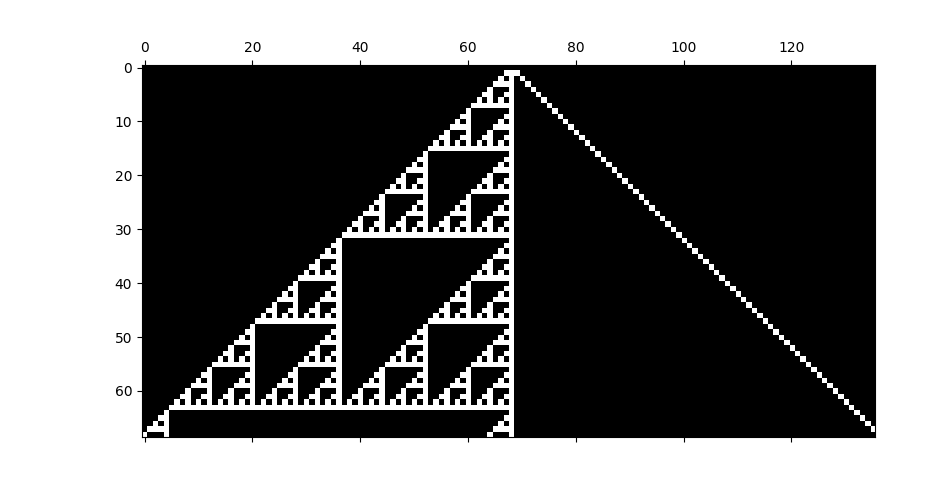}}
		\subfigure[$\mathbf{q}_4(x)$]{\includegraphics[width=0.48\linewidth]{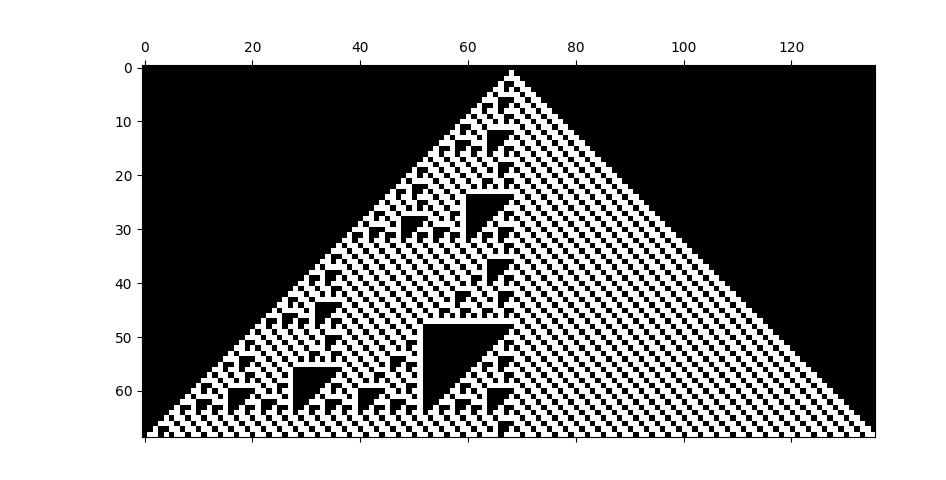}}
		\caption{4 subsystem symmetries of I-MSPT model (Eq.~(\ref{MSPT1})) in sublattice $(a)$. The initial condition are shown in Eq.~(\ref{ic1}), Eq.~(\ref{ic2}), Eq.~(\ref{ic3}), Eq.~(\ref{ic4}). White pixels are spins that the Pauli-$X$ operator acts nontrivially on. The first 2 rows in each figure are determined by the initial condition, and the rest is determined by HOCA rule.}

		\label{mixCApic}
	\end{figure*}
The overall results are shown in Fig.~\ref{mixCApic}. It can be seen clearly that a Sierpinski triangle (Fig.~\ref{p1}) and a line (Fig.~\ref{p2}) can both be the symmetry element, and there are symmetry elements that look like the attachment of two patterns (Fig.~\ref{p3}). Although we only consider the initial condition with the absolute values of exponent of $x$ (denoted as $P$) in each row less or equal than $1$, any $\mathbf{q}$ can be chosen in principle if it can fit into the size of the open slab. However, doing so does not bring us extra peculiar phenomenon. So far we have not found guiding principles of ascertaining $\mathbf{q}$ for a given type of symmetry element (e.g. line-like or fractal), and 4 initial conditions mentioned above are found by computer enumeration.
	The edge states of the model are protected by the symmetries we generated above (and other possible HOCA generated symmetries). In Fig.~\ref{symprotect} we explicitly show an example of symmetry protection for model (Eq.~(\ref{MSPT1})). In the figure we show that to open the gap of an edge spin must violate two symmetries: a line-like symmetry and a fractal-like symmetry (and many other HOCA generated symmetries that act nontrivially on this spin). The only way to modify edge spins while keeping all commutation relation with symmetries is to couple edge spins at different edges, which is either non-local or located at the corner of the system.

	\subsection{Model-II: II-MSPT generated by order-3 CA}
	There are also models with 2 different fractal symmetries. Consider a Hamiltonian
	\begin{widetext}
		\begin{equation}\label{MSPT2}
			\begin{aligned}
				\mathscr{H}=&-\sum_{i j} Z\left(\begin{array}{c}
					x^i y^j[1+y^{-1}(x^{-1}+1+x)+y^{-2}x+y^{-3}(x^{-1}+1)] \\
					x^i y^j
				\end{array}\right)\\
				&-\sum_{i j} X\left(\begin{array}{c}
					x^i y^j \\
					x^i y^j[1+y(x^{-1}+1+x)+y^2x^{-1}+y^3(1+x)]
				\end{array}\right)
			\end{aligned}
		\end{equation}
	\end{widetext}
 	
 	generated by an order-3 HOCA rule
	\begin{equation}\label{ff-MSPT}
		\mathbf{f}(x)=\begin{pmatrix} 
			x^{-1}+1+x\\x^{-1}\\1+x
			\end{pmatrix}.
	\end{equation}
	
\begin{figure}[htbp]
		\includegraphics[width=\linewidth]{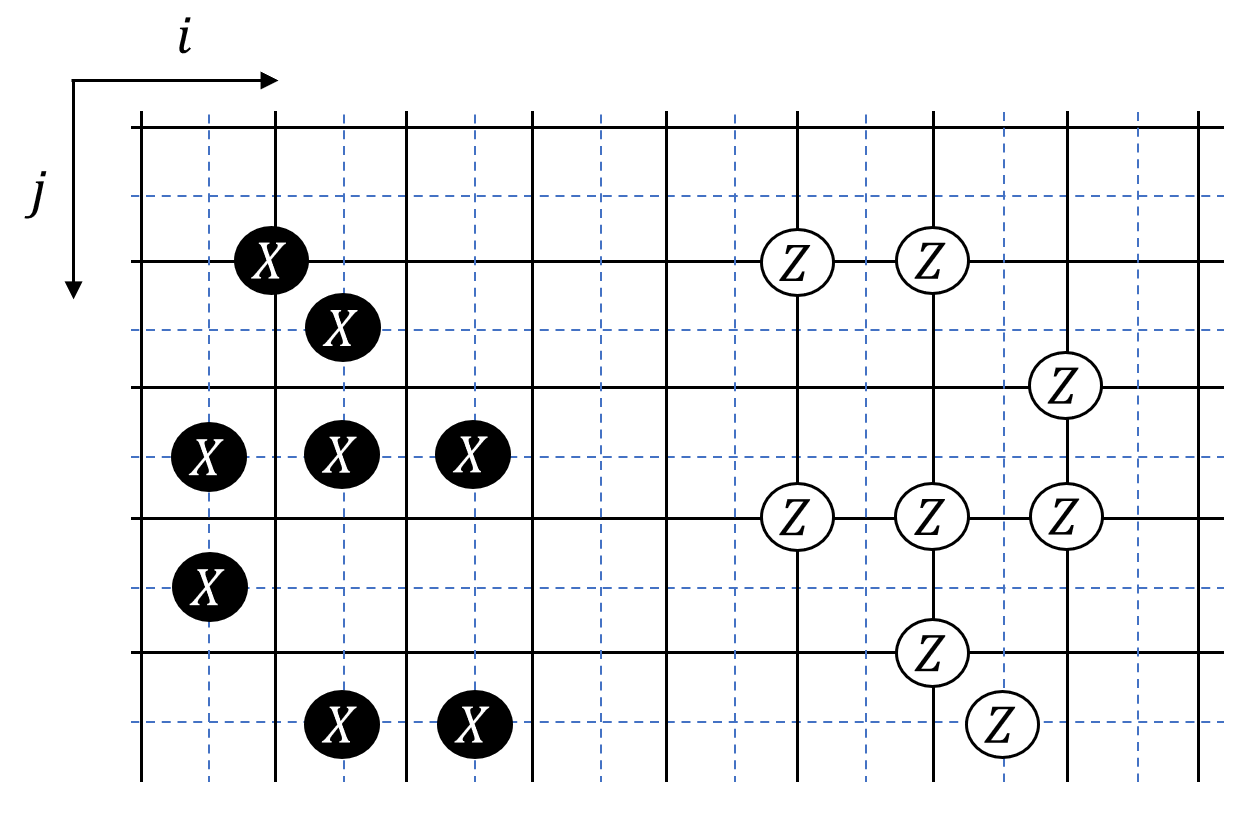}
		\caption{Pictorial illustration of two typical Hamiltonian terms of II-MSPT model (Eq.~(\ref{MSPT2})). Black lattice and blue dashed lattice denote 2 sublattices.}
		\label{pic-2}
	\end{figure}

	The Hamiltonian is pictorially shown in Fig.~\ref{pic-2}.
	
\begin{figure*}[htbp]
		\centering
		\subfigure[$\mathbf{q}_1(x)$\label{p4}]{\includegraphics[width=0.48\linewidth]{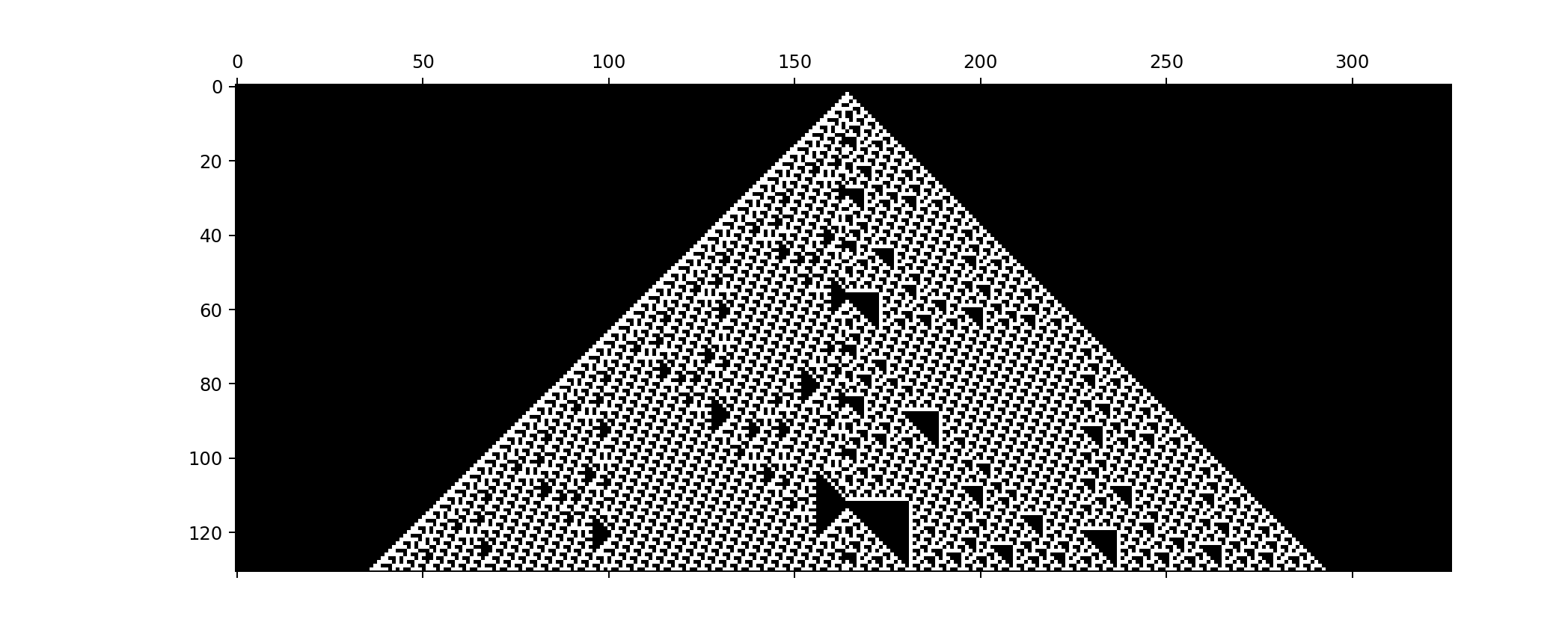}}
		\subfigure[$\mathbf{q}_2(x)$\label{p5}]{\includegraphics[width=0.48\linewidth]{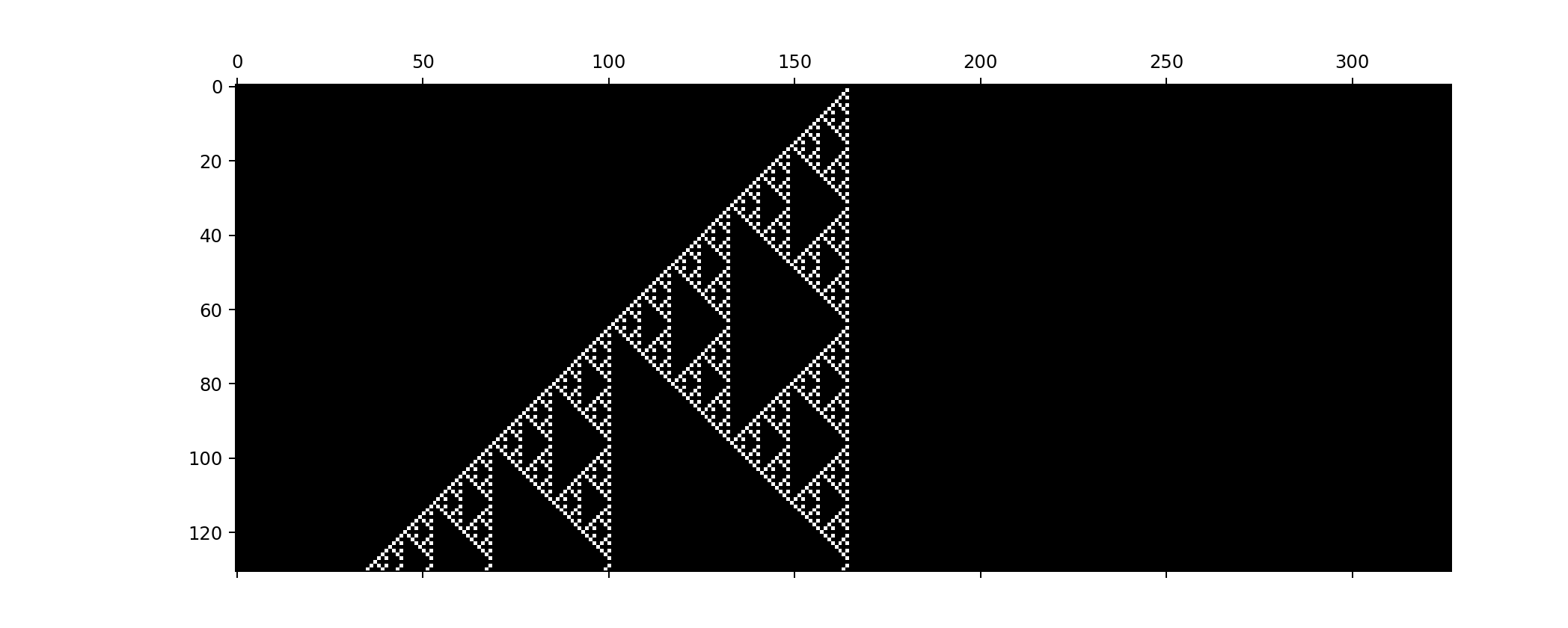}}\\
		\subfigure[$\mathbf{q}_3(x)$\label{p6}]{\includegraphics[width=0.48\linewidth]{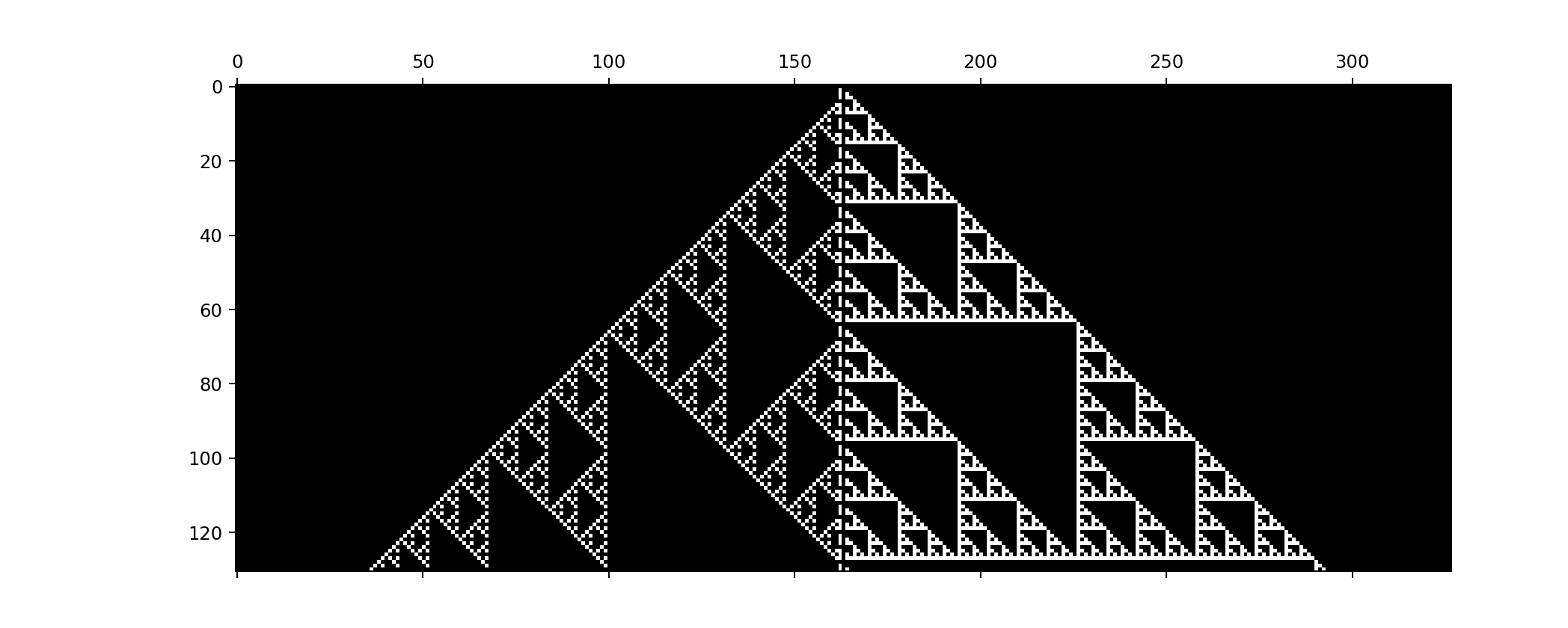}}
		\subfigure[$\mathbf{q}_4(x)$\label{p7}]{\includegraphics[width=0.48\linewidth]{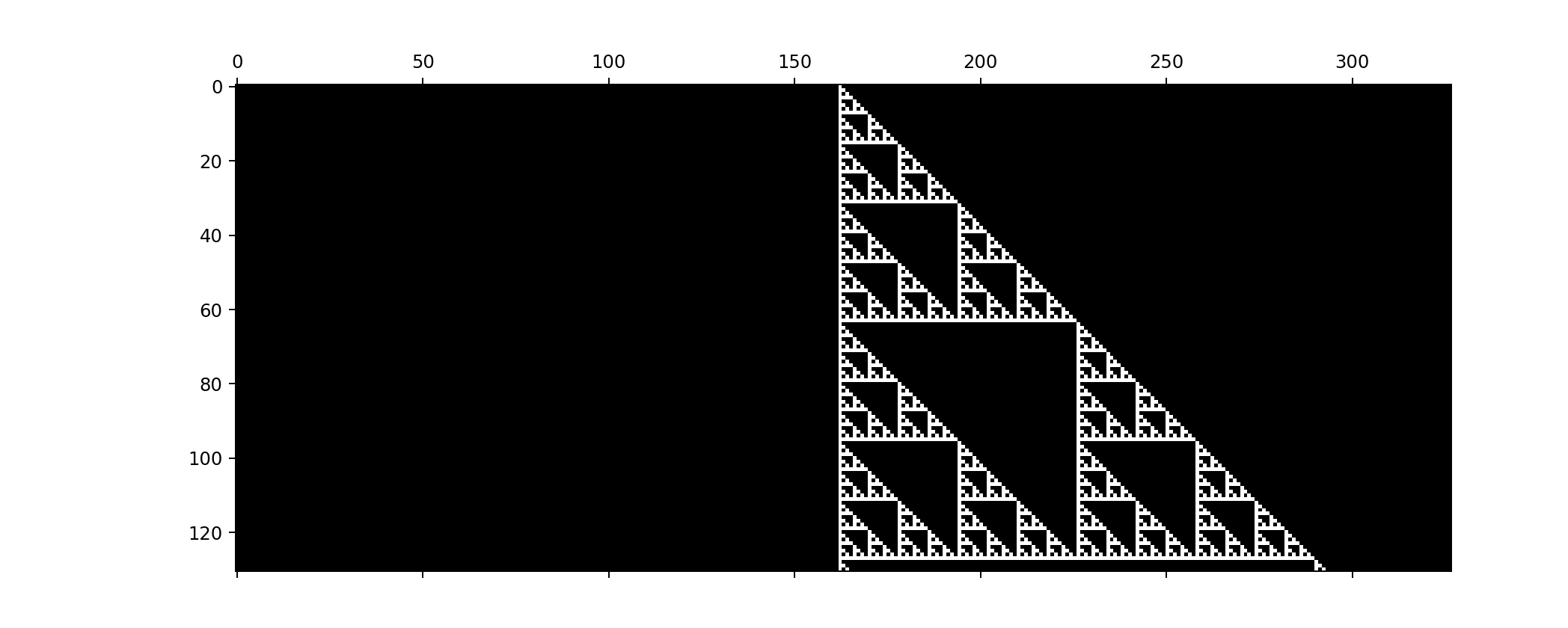}}
		\caption{4 subsystem symmetries of II-MSPT model (Eq.~(\ref{MSPT2})) in sublattice $(a)$. The initial condition are shown in Eq.~(\ref{ff1}), Eq.~(\ref{ff2}), Eq.~(\ref{ff3}), Eq.~(\ref{ff4}).  White pixels are spins that the Pauli-$X$ operator acts nontrivially on. The first 3 rows in each figure are determined by the initial condition, and the rest is determined by HOCA rule.}
		\label{mixCApic2}
	\end{figure*}

Examine symmetries generated by the following initial conditions:
{\allowdisplaybreaks[4]
	\begin{subequations}
		\begin{align}
			\mathbf{q}_1(x)&=\begin{pmatrix}
				0\\
				0\\
				1
			\end{pmatrix}\label{ff1},\\
			\mathbf{q}_2(x)&=\begin{pmatrix}
				0\\
				x\\
				1
			\end{pmatrix}\label{ff2},\\
			\mathbf{q}_3(x)&=\begin{pmatrix}
				0\\
				x^{-1}\\
				x^{-1}+x
			\end{pmatrix}\label{ff3},\\
			\mathbf{q}_4(x)&=\begin{pmatrix}
				x^{-1}\\
				x^{-1}+1\\
				x^{-1}+x
			\end{pmatrix}\label{ff4}.
		\end{align}
	\end{subequations}
}
The overall results are pictorially shown in Fig.~\ref{mixCApic2}. When there is only one flipped spin in the initial condition, the HOCA rule gives an chaotic pattern (shown in Fig.~\ref{p4}). Fig.~\ref{p5} and Fig.~\ref{p7} are two fractal subsystem symmetries of the model (Eq.~(\ref{MSPT2})). They are both Sierpinski triangles but with different shapes and orientation. Fig.~\ref{p6} can be viewed as the attachment of two fractal symmetries (with small modifications in the middle).
	
	\subsection{Model-III: CSPT generated by order-3 CA}
\begin{figure}[htbp]
		\includegraphics[width=\linewidth]{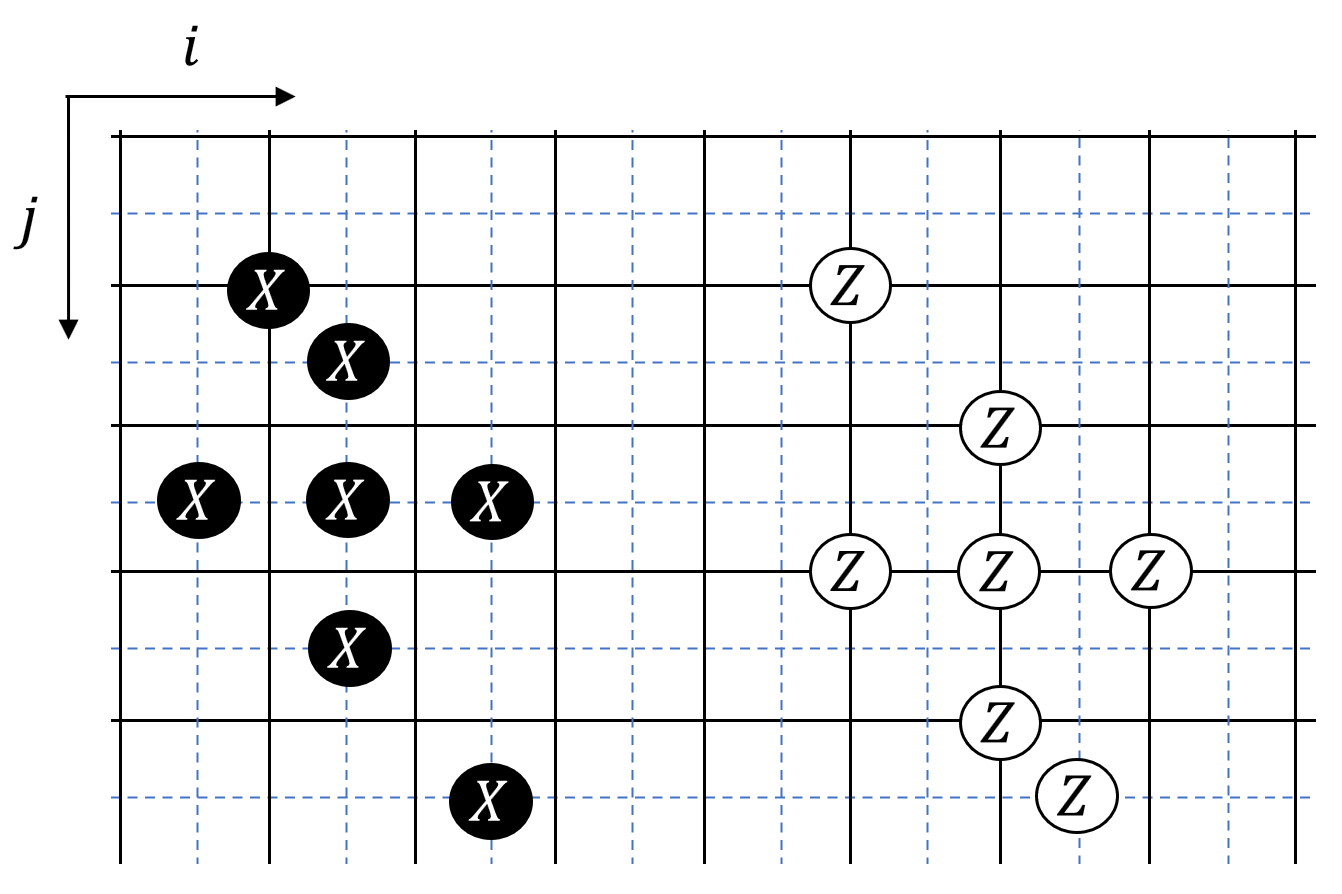}
		\caption{Pictorial illustration of two typical Hamiltonian terms of  model (Eq.~(\ref{MSPT3})). Black lattice and blue dashed lattice denote 2 sublattices.}
		\label{pic-3}
	\end{figure}

		Chaotic SPT (CSPT) models only contain subsystem symmetries in chaotic patterns. Under various initial conditions, the symmetry elements majorly show chaotic patterns. HOCA rules producing chaotic patterns are often used for encryption algorithm in computer science, as minor change in the initial condition may produce entirely different chaotic pattern. An example writes
		
			\begin{equation}\label{MSPT3}
		\begin{aligned}
			&\mathscr{H}=\\&-\sum_{i j} Z\left(\begin{array}{c}
				x^i y^j[1+y^{-1}(x^{-1}+1+x)+y^{-2}+y^{-3}x^{-1}] \\
				x^i y^j
			\end{array}\right)\\
			&-\sum_{i j} X\left(\begin{array}{c}
				x^i y^j \\
				x^i y^j[1+y(x^{-1}+1+x)+y^2+y^3x]
			\end{array}\right)
		\end{aligned}
	\end{equation}
	 with HOCA rule
	 
	\begin{equation}\label{irregular_rule}
		\mathbf{f}(x)=\begin{pmatrix} 
			x^{-1}+1+x\\
			1\\
			x
		\end{pmatrix}.
	\end{equation}
	The Hamiltonian is pictorially shown in Fig.~\ref{pic-3}.
	
	Given 4 initial conditions
	\begin{subequations}
		\begin{align}
			\mathbf{q}_1(x)&=\begin{pmatrix}
				0\\
				0\\
				1
			\end{pmatrix},\label{ir1}\\
			\mathbf{q}_2(x)&=\begin{pmatrix}
				0\\
				0\\
				1+x
			\end{pmatrix},\label{ir2}\\
			\mathbf{q}_3(x)&=\begin{pmatrix}
				1\\
				x^{-1}+1\\
				x
			\end{pmatrix},\label{ir3}\\
			\mathbf{q}_4(x)&=\begin{pmatrix}
				x^{-1}+1+x\\
				1\\
				x
			\end{pmatrix}.\label{ir4}
		\end{align}
	\end{subequations}
	
\begin{figure*}[htbp]
		\centering
		\subfigure[$\mathbf{q}_1(x)$]{\includegraphics[width=0.45\linewidth]{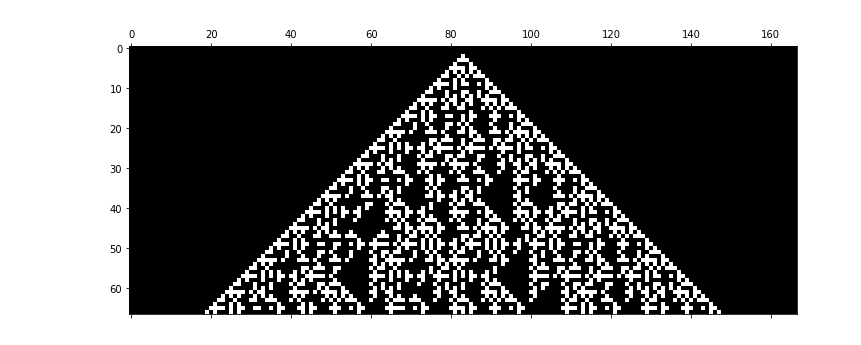}}
		\subfigure[$\mathbf{q}_2(x)$]{\includegraphics[width=0.45\linewidth]{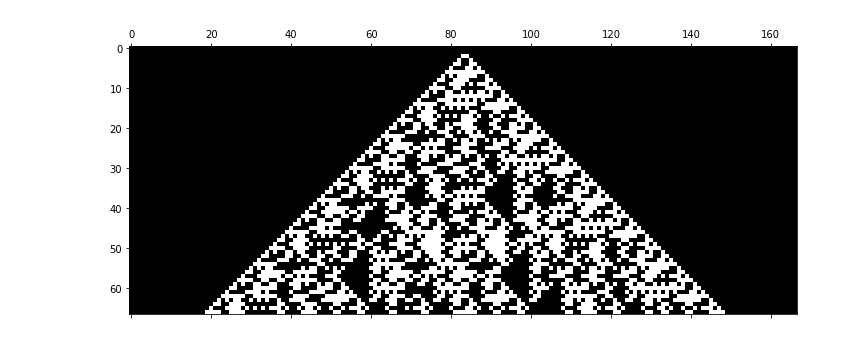}}\\
		\subfigure[$\mathbf{q}_3(x)$]{\includegraphics[width=0.45\linewidth]{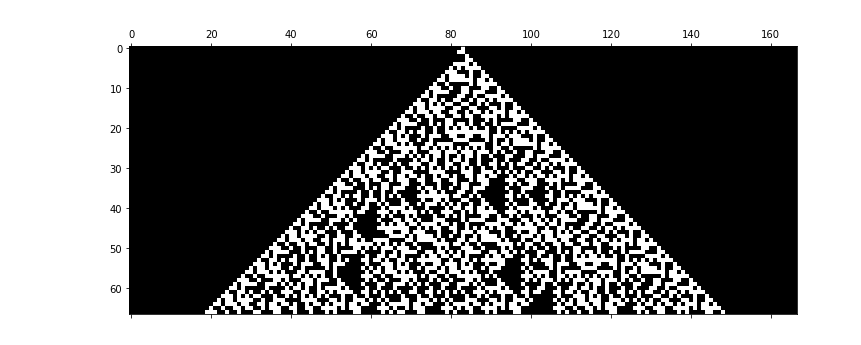}}
		\subfigure[$\mathbf{q}_4(x)$]{\includegraphics[width=0.45\linewidth]{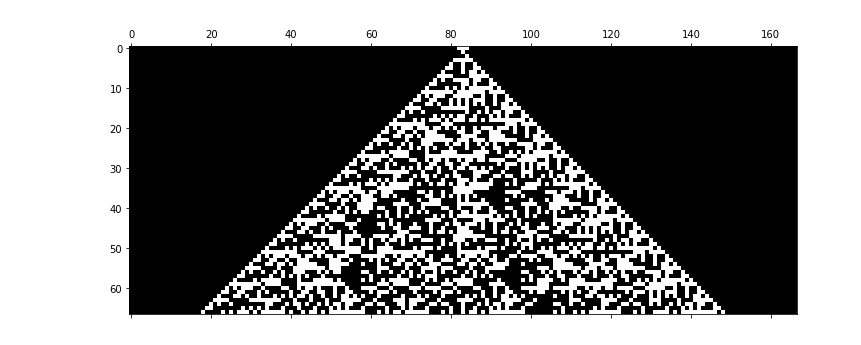}}
		\caption{4 symmetry elements generated by order-3 CA (Eq.~(\ref{irregular_rule})) in sublattice $(a)$. The initial conditions are shown in Eq.~(\ref{ir1}), Eq.~(\ref{ir2}), Eq.~(\ref{ir3}), Eq.~(\ref{ir4}).  White pixels are spins that the Pauli-$X$ operator acts nontrivially on. The first 3 rows in each figure are determined by the initial condition, and the rest is determined by HOCA rule.}
		\label{otherpic}
	\end{figure*}

	The resulting subsystem symmetries are shown in Fig.~\ref{otherpic}.

	\subsection{Model-IVa: RSPT generated by order-2 CA}
\begin{figure*}[htbp]
		\centering
		\subfigure[$\mathbf{q}_1(x)$]{\includegraphics[width=0.45\linewidth]{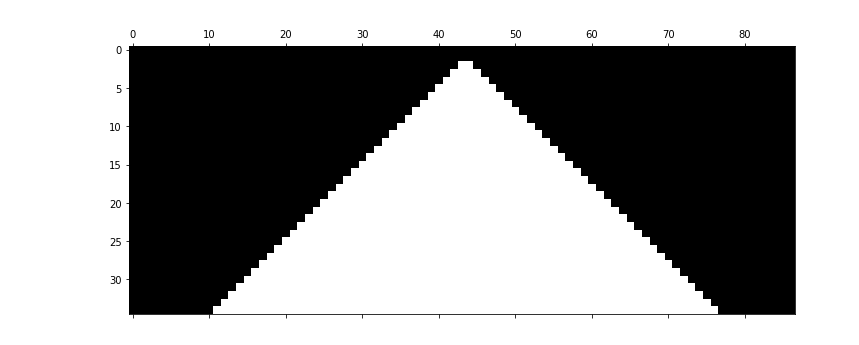}}
		\subfigure[$\mathbf{q}_2(x)$\label{line-like}]{\includegraphics[width=0.45\linewidth]{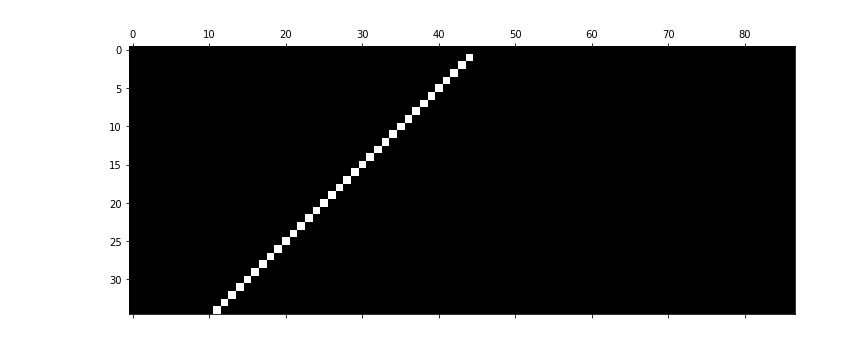}}\\
		\subfigure[$\mathbf{q}_3(x)$]{\includegraphics[width=0.45\linewidth]{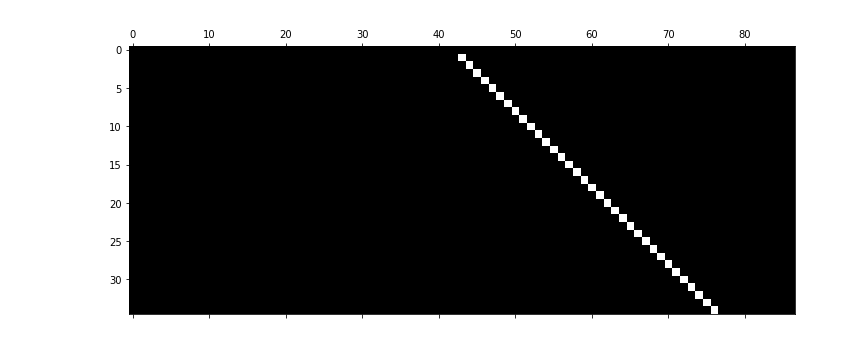}}
		\subfigure[$\mathbf{q}_4(x)$\label{membrane}]{\includegraphics[width=0.45\linewidth]{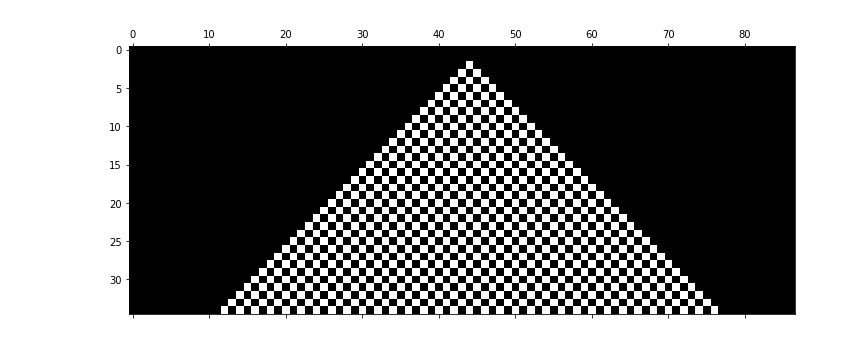}}
		\caption{4 patterns generated by order-2 CA (Eq.~(\ref{periodic_rule})) in sublattice $(a)$. The initial condition are shown in Eq.~(\ref{pr2}), Eq.~(\ref{pr3}), Eq.~(\ref{pr4}), Eq.~(\ref{pr5}).  White pixels are spins that the Pauli-$X$ operator acts nontrivially on. The first 2 rows in each figure are determined by the initial condition, and the rest is determined by HOCA rule.}
		\label{periodic pic}
	\end{figure*}

	It has been known that 2D regular SPT (RSPT, previously referred to as SSPT in Ref.~\cite{you2018a}) can be generated by a cluster model. Now we want to show that our HOCA framework also includes quantum models with line-like and membrane-like symmetry elements just like $\mathbb{Z}_2^{sub}$ strong SSPT discussed in \cite{you2018a}. Our model is different from the model discussed in \cite{you2018a}, but some of their subsystem symmetries share the same type. A rigorous proof of this statement is shown in Appendix~\ref{diff}. In addition, there are also checkerboard-like membrane symmetries (Fig.~\ref{membrane}) in our model, which is different from the previously defined SSPT model. A typical update rule of such an RSPT model writes
	\begin{equation}\label{periodic_rule}
		\mathbf{f}(x)=\begin{pmatrix}
			x^{-1}+x\\
			1
		\end{pmatrix},
	\end{equation} 
	which generates the Hamiltonian 
 \begin{widetext}
		\begin{equation}\label{MSPT4}
		\begin{aligned}
			\mathscr{H}=&-\sum_{i j} Z\left(\begin{array}{c}
				x^i y^j[1+y^{-1}(x^{-1}+x)+y^{-2}] \\
				x^i y^j
			\end{array}\right)
			-\sum_{i j} X\left(\begin{array}{c}
				x^i y^j \\
				x^i y^j[1+y(x^{-1}+x)+y^2]
			\end{array}\right).
		\end{aligned}
	\end{equation}
 \end{widetext}

\begin{figure}[htbp]
		\includegraphics[width=\linewidth]{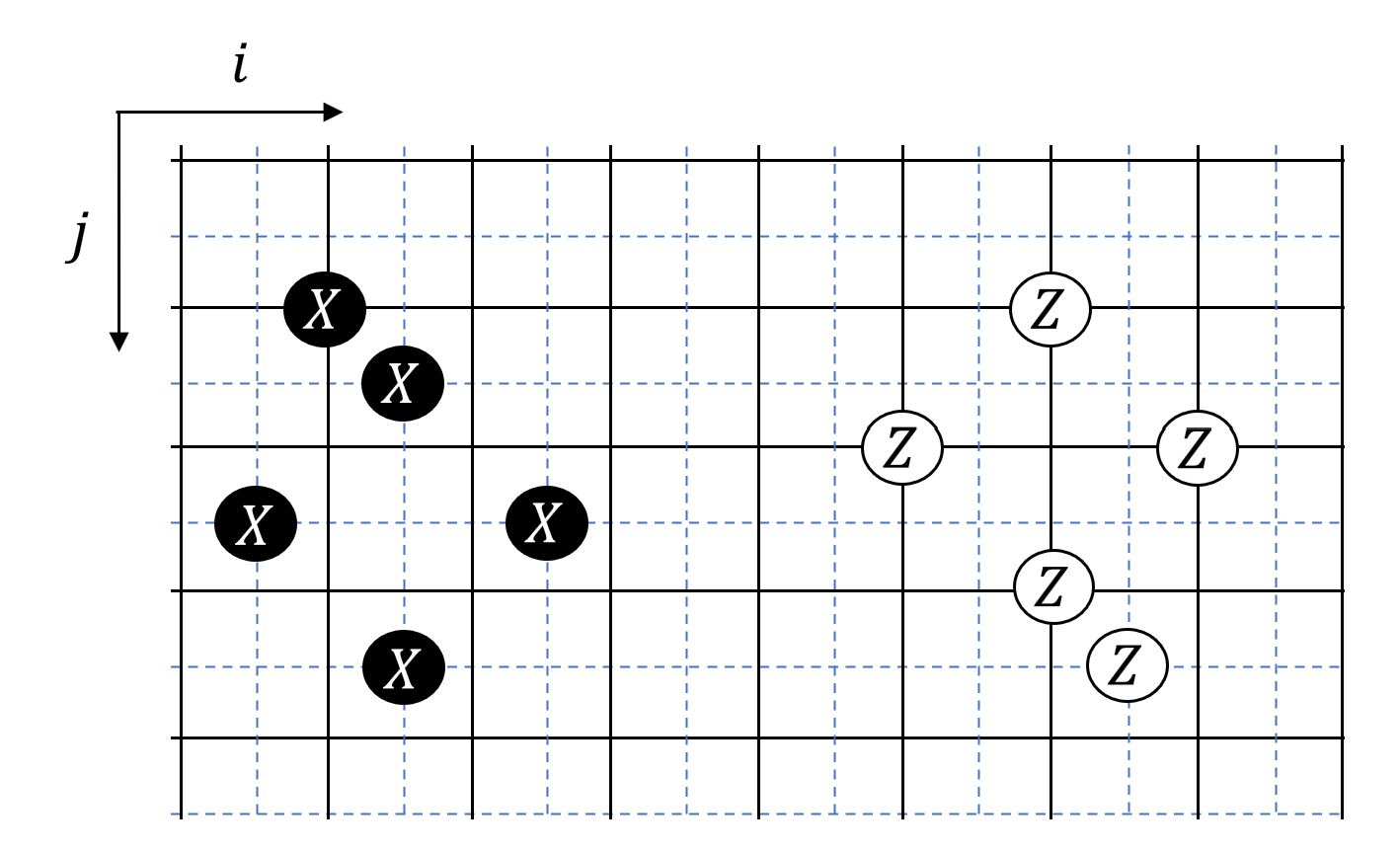}
		\caption{Pictorial illustration of two typical Hamiltonian terms of  model (Eq.~(\ref{periodic_rule})). Black lattice and blue dashed lattice denote 2 sublattices.}
		\label{HGSPT4}
	\end{figure}

	Hamiltonian generated by this rule are pictorially shown in Fig.~\ref{HGSPT4}.
	
	Consider following initial conditions:
	\begin{subequations}
		\begin{align}
			\mathbf{q}_1(x)&=\begin{pmatrix} 
				0\\
				1+x\\
			\end{pmatrix},\label{pr2}\\
			\mathbf{q}_2(x)&=\begin{pmatrix} 
				x\\
				1
			\end{pmatrix},\label{pr3}\\
			\mathbf{q}_3(x)&=\begin{pmatrix} 
				1\\
				x
			\end{pmatrix},\label{pr4}\\
			\mathbf{q}_4(x)&=\begin{pmatrix} 
				1\\
				0
			\end{pmatrix}.\label{pr5}
		\end{align}
	\end{subequations}
	There are line-like and membrane-like symmetry elements present in the model, as shown in Fig.~\ref{periodic pic}.

  \subsection{Model-IVb: RSPT generated by order-3 CA}\label{model_ivb}
\begin{figure}
    \centering
    \includegraphics[width=0.85\linewidth]{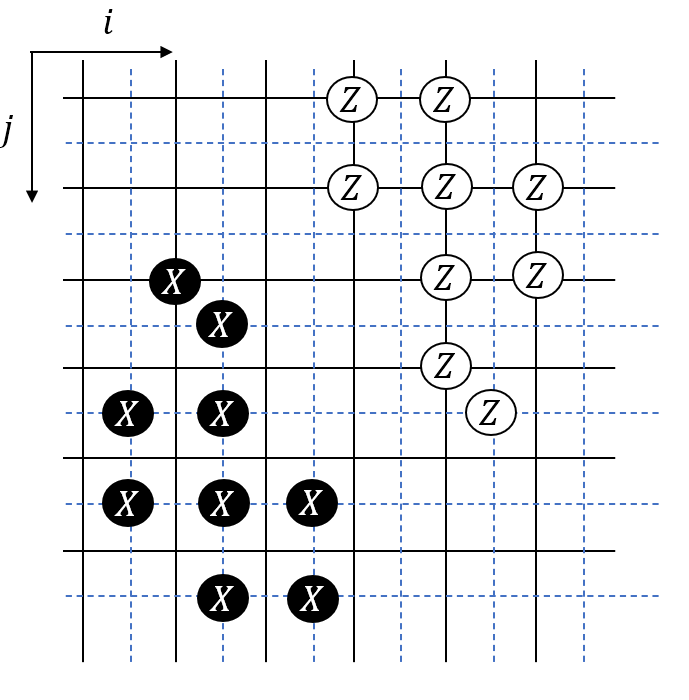}
    \caption{Pictorial illustration of two typical Hamiltonian terms of  model (Eq.~(\ref{rule_ivb})). Black lattice and blue dashed lattice denote 2 sublattices.}
    \label{H_ivb}
\end{figure}

  Now let us consider a more nontrivial RSPT model generated by an order-3 CA:
\begin{equation}\label{rule_ivb}
		\mathbf{f}(x)=\begin{pmatrix}
			x^{-1}+1\\
			x^{-1}+1+x\\
        1+x
		\end{pmatrix},
	\end{equation} 
	which generates the Hamiltonian 
 \begin{widetext}
		\begin{equation}\label{h_ivb}
		\begin{aligned}
			\mathscr{H}=&-\sum_{i j} Z\left(\begin{array}{c}
				x^i y^j[1+y^{-1}(1+x)+y^{-2}(x^{-1}+1+x)+y^{-3}(x^{-1}+1)] \\
				x^i y^j
			\end{array}\right)\\
			&-\sum_{i j} X\left(\begin{array}{c}
				x^i y^j \\
				x^i y^j[1+y(x^{-1}+1)+y^2(x^{-1}+1+x)+y^3(1+x)]
			\end{array}\right).
		\end{aligned}
	\end{equation}
 \end{widetext}
The Hamiltonian of Model-IVb is shown pictorially in Fig.~\ref{H_ivb}.

\begin{figure*}[htbp]
		\centering
		\subfigure[$\mathbf{q}_1(x)$]{\includegraphics[width=0.45\linewidth]{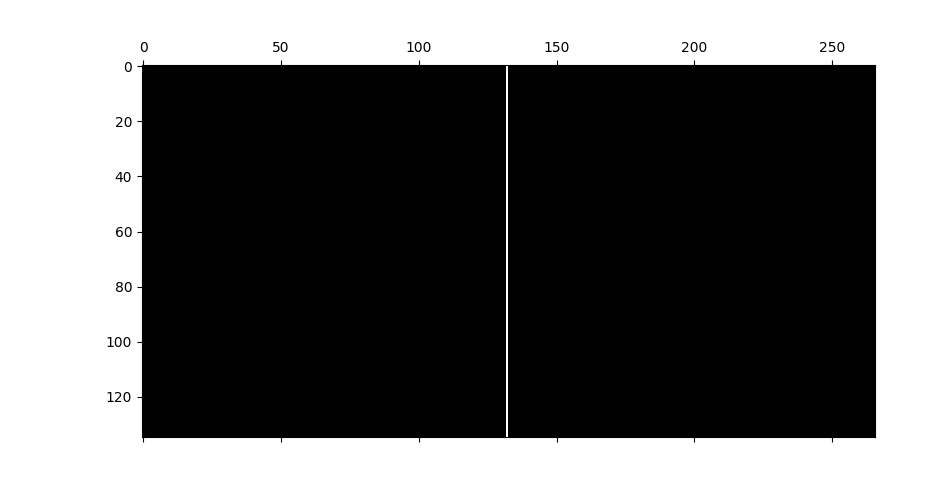}}
		\subfigure[$\mathbf{q}_2(x)$]{\includegraphics[width=0.45\linewidth]{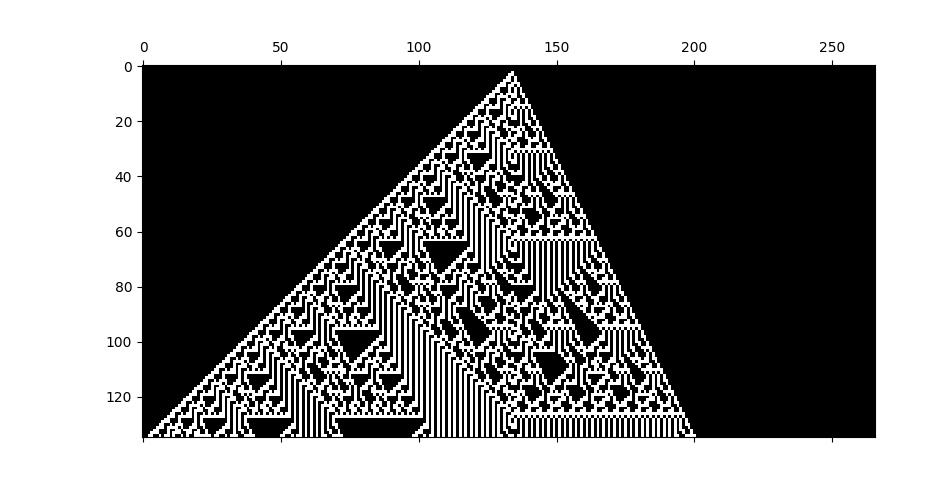}}\\
		\subfigure[$\mathbf{q}_3(x)$]{\includegraphics[width=0.45\linewidth]{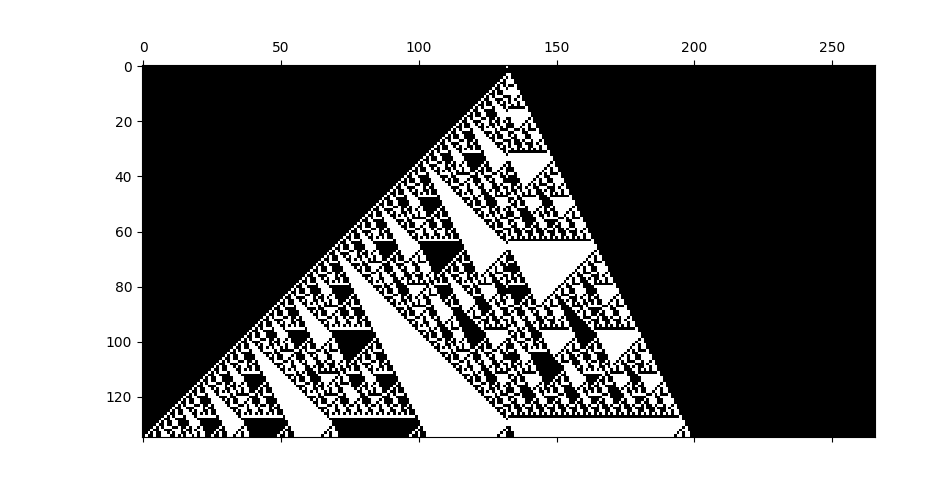}}
		\subfigure[$\mathbf{q}_4(x)$\label{ivb_p4}]{\includegraphics[width=0.45\linewidth]{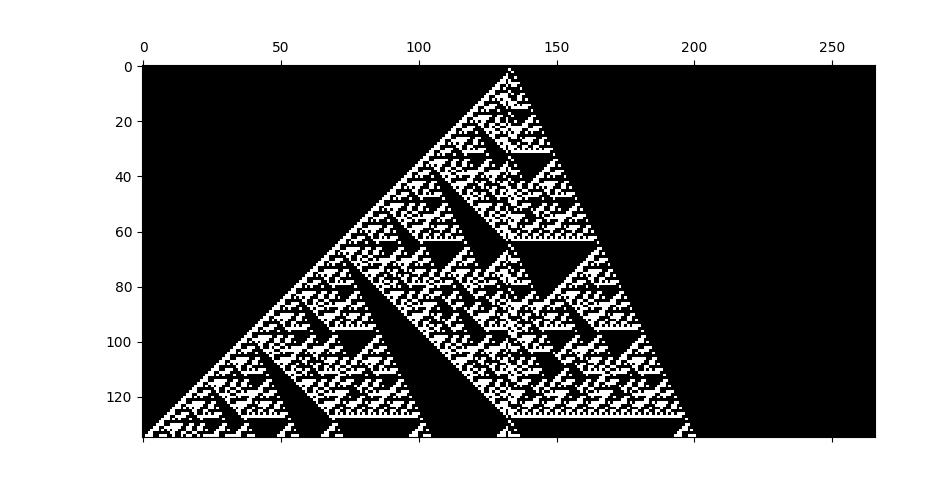}}
		\caption{4 patterns generated by order-2 CA (Eq.~(\ref{rule_ivb})) in sublattice $(a)$. The initial condition are shown in Eq.~(\ref{ivb1}), Eq.~(\ref{ivb2}), Eq.~(\ref{ivb3}), Eq.~(\ref{ivb4}).  White pixels are spins that the Pauli-$X$ operator acts nontrivially on. The first 3 rows in each figure are determined by the initial condition, and the rest is determined by HOCA rule.}
		\label{pic_ivb}
	\end{figure*}

Given 4 initial conditions:
  \begin{subequations}
		\begin{align}
			\mathbf{q}_1(x)&=\begin{pmatrix} 
				1\\
				1\\
                1
			\end{pmatrix},\label{ivb1}\\
			\mathbf{q}_2(x)&=\begin{pmatrix} 
				0\\
				0\\
                1\\
			\end{pmatrix},\label{ivb2}\\
			\mathbf{q}_3(x)&=\begin{pmatrix} 
				1\\
				0\\
                0
			\end{pmatrix},\label{ivb3}\\
			\mathbf{q}_4(x)&=\begin{pmatrix} 
                0\\
				1\\
				x^{-1}
			\end{pmatrix},\label{ivb4}
		\end{align}
	\end{subequations}
  we obtain 4 different subsystem symmetries shown in Fig.~\ref{pic_ivb}. Different from Model-IVa which can only generate regular patterns, there are some chaotic-looking symmetries in Model-IVb. At first look some symmetries (e.g. Fig.~\ref{ivb_p4}) may seem to possess a fractal structure, but they do not actually have a rigorous self-similarity (see Sec.~\ref{notation} for more detailed discussions). 

	\subsection{A notation system for labeling the higher-order cellular automata}\label{notation}

Now, we want to introduce new notations to characterize various subsystem patterns generated by HOCA. As we have shown in the previous sections, HOCA can generate various types of patterns in the spacetime lattice, such as fractal patterns with rigorous self-similarity (e.g., Sierpinski triangle), patterns that consist of some periodic repetition of basic structures (e.g., checkerboard), and even patterns that look like a mixture of the two above. We also found these subsystems have various dimensions (in the sense of Hausdorff dimension). We claim that these properties of an HOCA pattern $\mathscr{F}$ (defined on the semi-infinite plane, see Eq.~(\ref{pattern})) generated by a finite initial condition (i.e. there are finite terms in the initial condition) can be captured by a mathematical object
 $X(\mathscr{F})=(d(\mathscr{F}),M(\mathscr{F}))$. Here, $d$ is defined as the Hausdorff dimension of the pattern with infinite time evolution steps, which can be approached numerically by box dimension. If we denote the number of evolution time steps by $t$, and the number of sites with state $1$ from time 0 to time $t$ by $a(t)$, then we have
 $	d=\lim_{t\to \infty}\frac{\ln a(t)}{\ln t}$. 

For a Sierpinski triangle with Hausdorff dimension $d_H=\ln3/\ln2\approx 1.5850$, the numerical result with $t=256$ gives $d=1.5830$, which is quite close to the exact result. Another quantity, $M$, is dubbed as mix rate, describing how fractal or periodic the pattern is. For an order-$n$ HOCA pattern, $M$ is mathematically defined as
\begin{equation}\label{M1}
	M=\frac{S_u-S_d}{S_u},
\end{equation}
where
\begin{equation}\label{M2}
	S_u(n)=\limsup_{k\to\infty} \frac{\sum_{i=k}^{k+n-1}A(i)}{k}
\end{equation}
and
\begin{equation}\label{M3}
	S_d(n)=\liminf_{k\to\infty}\frac{\sum_{i=k}^{k+n-1}A(i)}{k}, 
\end{equation}
where $A(i)$ is the number of cells in state 1 of the $i$-th row. 

The definition of $M$ comes from following observations. Now we have observed two possible local behaviors of a HOCA evolution patterns for all HOCA rules with radius $r\leq 3/2$ and $n\leq 3$:
\begin{outline}[enumerate]
	\1 Self-similar fractal structure: Some parts of a HOCA pattern tend to appear recurringly while we increase the time of evolution (i.e. zooming out the pattern). While the whole pattern may not be fully self-similar in general, there are recognizable fractal structure in many patterns. An example is Fig.~\ref{p7}.
	\1 Regular structure: Some parts of a HOCA pattern may appear to be filled by some local repeating structures, like Fig.~\ref{periodic pic}. 
\end{outline}
While in general a HOCA pattern may not be a fully fractal or regular pattern, a large subset of HOCA patterns can be viewed as some mixture of fractal and regular patterns. This visual observation can be clearly seen for almost all HOCA patterns with $r=3/2$ and $n\leq3$, while it is subtle to argue whether HOCA patterns like Fig.~\ref{otherpic} can be viewed as this kind of mixture just by watching. To give a more quantitative and rigorous description of this mixing behavior, we observe that fractal patterns and regular patterns are distinguishable in terms of counting cells with state $1$ in every row. We denote the number of cells with state $1$ in row $t$ to be $A(t)$ as in Eq.~\ref{M2},~\ref{M3}, then we have following qualitative observations: 
\begin{outline}[enumerate]
	\1 In a fully fractal pattern, there always exist a infinite sequence $\{t_i\}$ such that $A(t_i)/t_i\to 0$. This observation is obvious because of the self-similar essence of fractal. Self-similarity means there are infinitely many rows can be represented by simply scaling a single row configurations. An example of this are the sequence $\{t=2^n-1:n \in \mathbb{Z^+}\}$ of Fig.~\ref{p7}. All rows with index in this sequence have 2 cells in state $1$, scaled by different proportions. 
	\1 In a fully fractal pattern, there is always a sequence $\{t_i\}$ such that the sequence $\{A(t_i)\}$ grows to infinity. A heuristic explanation of this observation are fractal patterns always tend to grow bigger as $t$ increases. A rigorous proof of this statement would involve the sensitivity of HOCA rule, which will be introduced in Section~\ref{sensitivity}. In short, this statement is always true for a fractal pattern. 
   \1 In a fully regular pattern, where every row configuration can be viewed as some repetitions of a specific local structure (e.g. each row in Fig.~\ref{membrane} can be viewed as the repetitions of a white-black checkerboard structure), the number of sites with state $1$ in each row either grows linearly (for a membrane-like pattern) , or remains a constant (for a line-like pattern), or oscillate between two constants (for a zipper-like pattern). In all situations the upper limit and the lower limit of the sequence $\{A(t)\}$ when $t\to \infty$ will be at the same order, being both $O(t)$ or $O(1)$. 
\end{outline}

\begin{figure*}[htbp]
    \centering
    \subfigure[Membrane-like regular pattern\label{MP1}]{\includegraphics[width=0.55\linewidth]{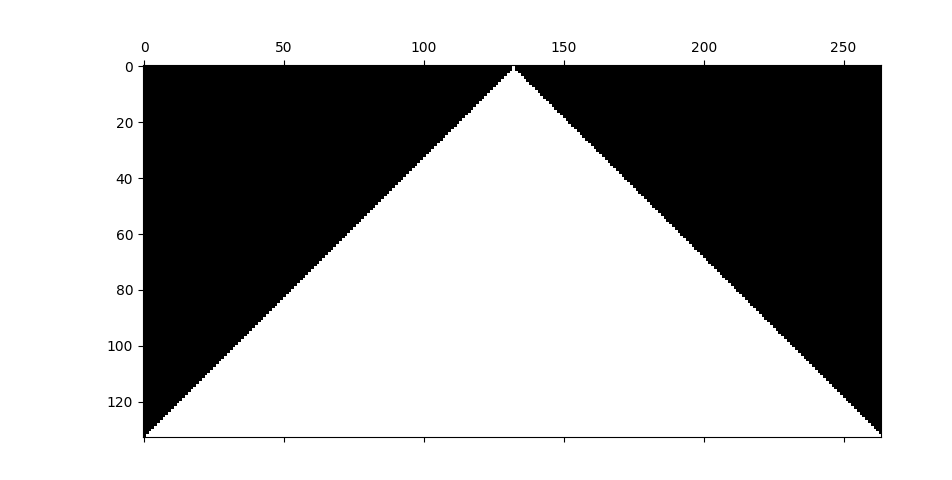}}
    \subfigure[$M=0$\label{MP2}]{\includegraphics[width=0.4\linewidth]{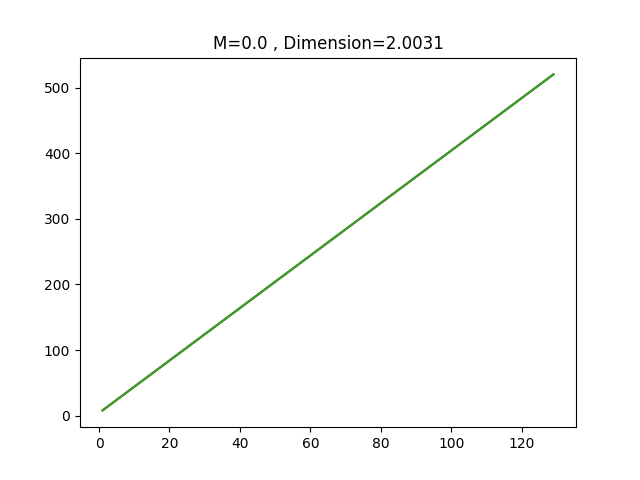}}
    \subfigure[Checkerboard-like regular pattern\label{MP3}]{\includegraphics[width=0.55\linewidth]{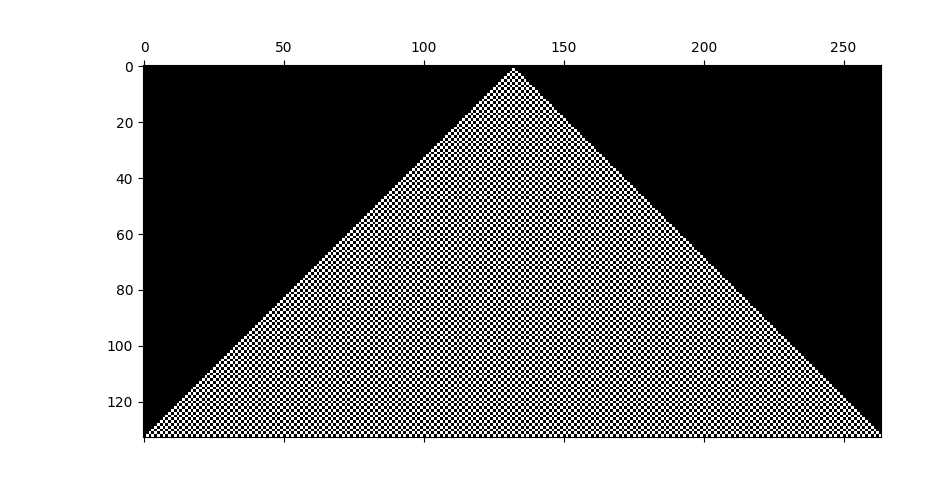}}
    \subfigure[$M=0$\label{MP4}]{\includegraphics[width=0.4\linewidth]{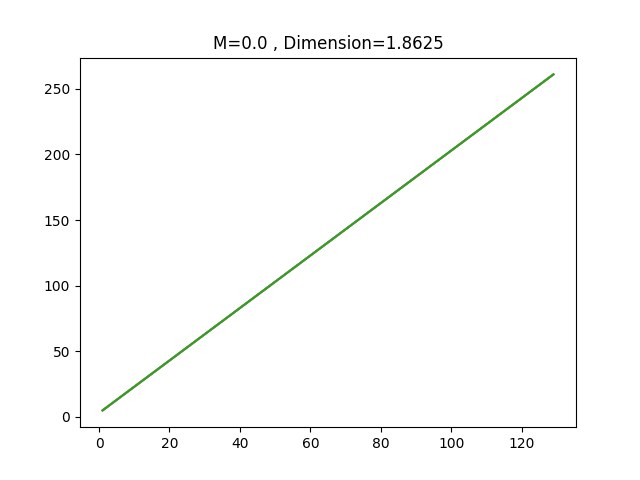}}
    \subfigure[Line-like regular pattern\label{MP5}]{\includegraphics[width=0.55\linewidth]{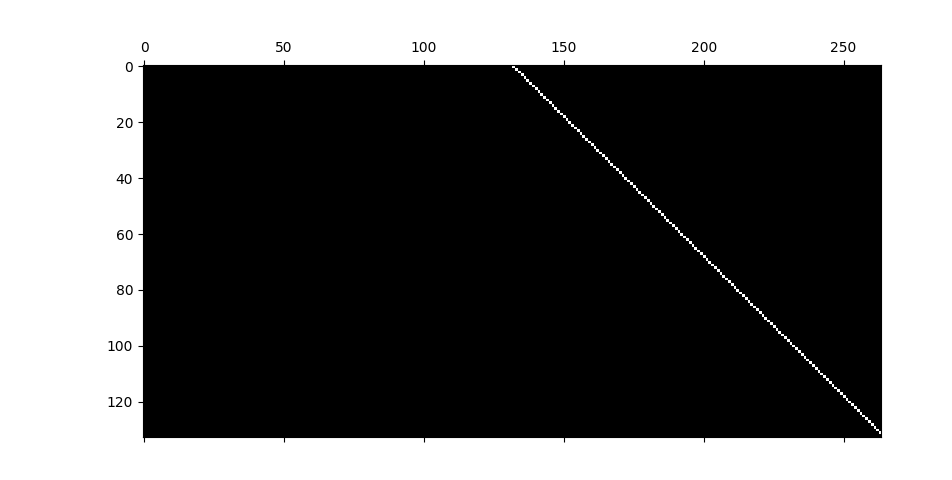}}
    \subfigure[$M=0$\label{MP6}]{\includegraphics[width=0.4\linewidth]{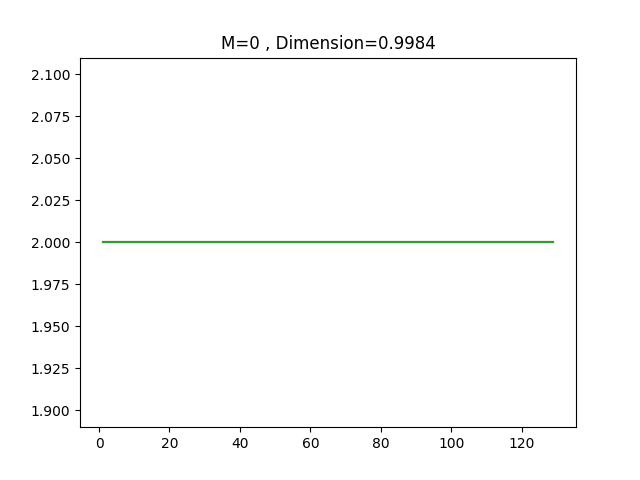}}
       \caption{Example 1 (2 in total) illustrating the validity of $M$ in classifying HOCA patterns. Fig.~\ref{MP1}, \ref{MP3}, \ref{MP5}, \ref{MP7}, \ref{MP9}, \ref{MP11} show 6 examples of HOCA evolution pattern. Fig.~\ref{MP2}, \ref{MP4}, \ref{MP6}, \ref{MP8}, \ref{MP10}, \ref{MP12} show how $\sum_{k=i}^{i+n}A(k)$ grows with $i$, where the numerical results of $d, M$ is shown above each figure. $S_u$ and $S_d$ can be understood as the slope of green and orange straight lines in the subfigures in the right column. For regular patterns two slopes always equal, while for fractal pattern the slope of orange lines are always zero, for chaotic-looking patterns two straight lines have different nonzero slopes. }
    \label{fig:M_def1}
    \end{figure*}
    
\begin{figure*}[htbp]
    \subfigure[Fibonacci fractal pattern\label{MP7}]{\includegraphics[width=0.55\linewidth]{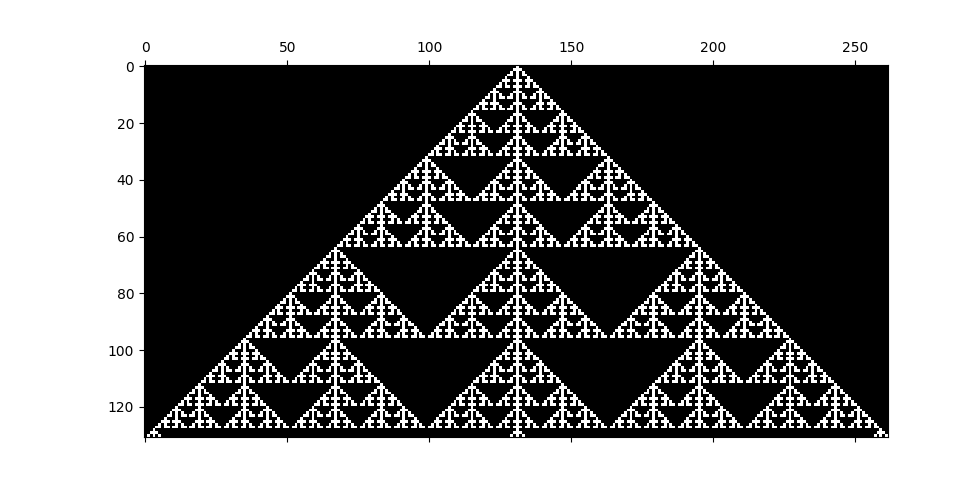}}
    \subfigure[$M=1$\label{MP8}]{\includegraphics[width=0.4\linewidth]{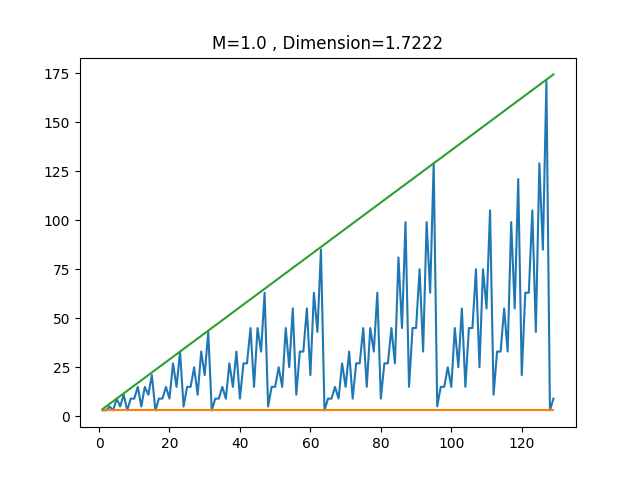}}
    \subfigure[Sierpinski fractal pattern\label{MP9}]{\includegraphics[width=0.55\linewidth]{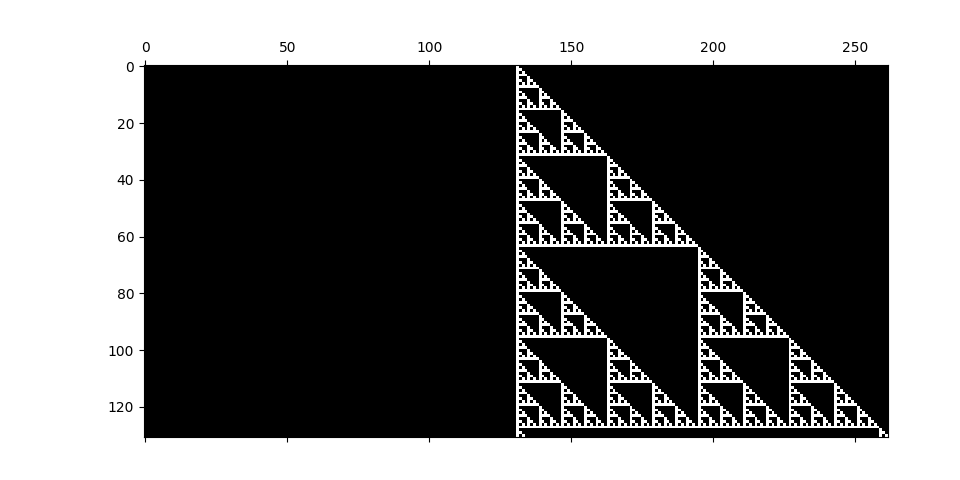}}
    \subfigure[$M=1$\label{MP10}]{\includegraphics[width=0.4\linewidth]{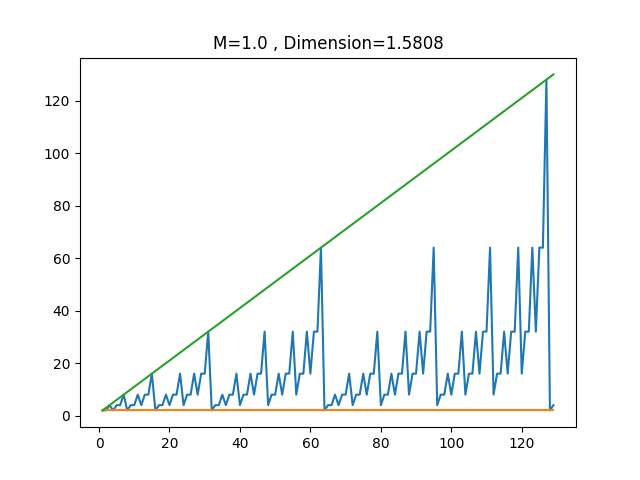}}
    \subfigure[Chaotic-looking pattern\label{MP11}]{\includegraphics[width=0.55\linewidth]{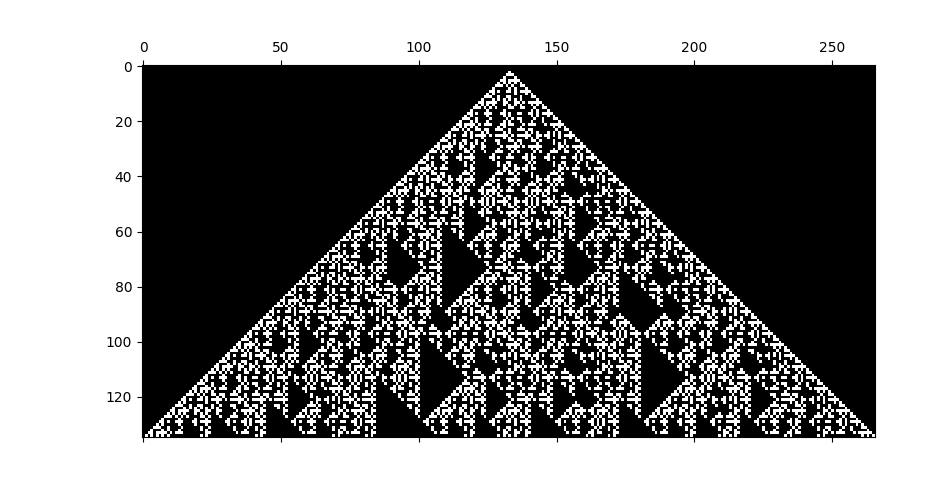}}
    \subfigure[$M\approx0.54$\label{MP12}]{\includegraphics[width=0.4\linewidth]{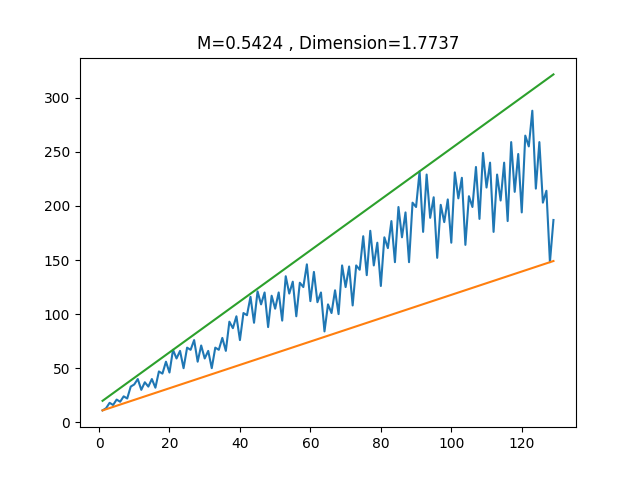}}
    \caption{Example 2 (2 in total) illustrating the validity of $M$ in classifying HOCA patterns. Fig.~\ref{MP1}, \ref{MP3}, \ref{MP5}, \ref{MP7}, \ref{MP9}, \ref{MP11} show 6 examples of HOCA evolution pattern. Fig.~\ref{MP2}, \ref{MP4}, \ref{MP6}, \ref{MP8}, \ref{MP10}, \ref{MP12} show how $\sum_{k=i}^{i+n}A(k)$ grows with $i$, where the numerical results of $d, M$ is shown above each figure. $S_u$ and $S_d$ can be understood as the slope of green and orange straight lines in the subfigures in the right column. For regular patterns two slopes always equal, while for fractal pattern the slope of orange lines are always zero, for chaotic-looking patterns two straight lines have different nonzero slopes. }
    \label{fig:M_def2}
\end{figure*}

The observation above enables us to define the ``mix rate'' $M$ of two types of structures in a single HOCA pattern. As seen in the definition of $M$ (Eq.~(\ref{M1}), (\ref{M2}),( \ref{M3})), for a fully fractal pattern there will be a constant upper limit of $\{A(t)/t\}$, and a zero lower limit of $\{A(t)/t\}$ (all constant lower limits of $\{A(t)\}$ will be suppressed by the $1/t$ factor). Thus, $M=1$ is always true for the fractal case. As for the case of regular pattern, the upper and lower limit of $\{A(t)/t\}$ will either be equal to an identical constant (membrane-like pattern) or equal to zero (line-like, zipper-like case). Both cases lead to the result that $M=0$, being the character of a regular pattern. Generally, $M$ is in $[0,1]$ for a general HOCA pattern, and this property is true for any finite HOCA pattern generated by finite initial condition (only sites in a limited area have nonzero values). Specifically, the sum in the numerator of Eq.(~\ref{M2}), (\ref{M3}) is taken from $i=k$ to $i=k+n$ because the fact that there will be at most $n-1$ consecutive empty rows in a pattern generated by an order-$n$ HOCA, which can cause some non-fractal pattern to obtain a zero lower limit of $\{A(t)\}$, affecting the effectiveness of $M$. As the research on the dynamical properties of HOCA is still at early stages, there are still no mathematical paper to give rigorous classifcations of these properties. Therefore, we have done an exhaustive and case-by-case verification of the validity of $M$ for all $r\leq 3/2$ and $n\leq 3$ HOCA rules (512 rules in total), and have found no counterexamples of our classification based on our current observation. While there are certainly mathematical foundations of this quantity, the topic is beyond the scope of this research and is left as a part of future works.


We provide a pictorial description of various HOCA patterns:
\begin{outline}[enumerate]\label{M_expl}
\1 $M=1$: It is a fractal pattern exhibiting a self-similar structure, like Fig.~\ref{p7}. Also there are mixed patterns that can be considered as the attachment of $(d ,1)$ fractal pattern and a $(1 ,0)$ pattern, as shown in Fig.~\ref{p3}.
\1  $M=0$: $(d ,0)$ pattern can be considered as spatial repeating of some minimal structures or some stable patterns. The overall pattern can extend in the 2D plane (Fig.~\ref{membrane}) or propagate along some 1D subsystem of the plane (Fig.~\ref{line-like}). 
\1  $0<M<1$: Chaotic patterns can show recognizable repeating structures locally or appear to be irregular, but it does not fit into the classification above, as shown in Fig.~\ref{otherpic}. 
\end{outline}
More explicit examples of the validity of $M$ in classifying HOCA is shown in Fig.~\ref{fig:M_def1}, \ref{fig:M_def2}.

An HOCA rule can generate infinite patterns by varying the initial condition $\mathbf{q}(x)$. However, if we collect all possible patterns generated by an HOCA rule, we find that different rules may produce different types of patterns, dividing HOCA rules into 4 classes. We use $\mathcal{X}(\mathbf{f})=(X_r,X_f)$ to denote the classes $[\mathbf{f}]$, where
\begin{equation}\label{eq:class_label}
	\begin{aligned}
		X_r&=1-\lceil \min\{M\}\rceil,\\
		X_f&=\lfloor\max\{M\}\rfloor,
	\end{aligned}
\end{equation}
where $\lceil~\rceil$ and $\lfloor~\rfloor$ are ceil and floor functions, respectively. $\{M\}$ represents the set of all possible $M$ generated by the given HOCA rule. Heuristically, $X_r$ and $X_f$ describe whether a certain HOCA rule can generate regular pattern or fractal pattern. For example, $X_r=1$ means that there are at least one regular pattern (e.g. line, membrane, checkerboard, etc.) can be obtained from the HOCA rule by varying the initial condition, and vice versa.

Typical examples of patterns above can be found in Fig.~\ref{fig:M_def1}, \ref{fig:M_def2}. Given a specific update rule $\mathbf{f}$, different patterns can emerge when we adjust the initial conditions. Thus, we can classify different update rules by the patterns they can produce, as shown below:
\begin{outline}[itemize]
	\1 $\mathcal{X}(\mathbf{f})=(0,0)$: These HOCA rules
 only produces chaotic patterns like symmetry elements presented in chaotic SPT. Neither fractal nor like-like patterns can be found in this class.
	\1 $\mathcal{X}(\mathbf{f})=(0,1)$: HOCA rules in this class can produces fractal patterns but not periodic patterns. Sierpinski FSPT \cite{devakul_fractal_2019}, and previously mentioned II-MSPT can be generated by CAs in this class.
	\1 $\mathcal{X}(\mathbf{f})=(1,0)$: HOCA rules in this class produces periodic patterns, including line-like, membrane-like patterns. SSPT \cite{you2018a} can by generated by HOCA rules in this class.
	\1 $\mathcal{X}(\mathbf{f})=(1,1)$: HOCA rules in this class produces both fractal and periodic patterns. These rule can generate I-MSPT phases.
\end{outline}

To capture finer details of an HOCA rule, we define two sets of \textit{characteristic dimension} of an HOCA rule $\mathbf{f}$:
\begin{equation}
	D_r=\{d(\mathscr{F})|M(\mathscr{F})=0\},
\end{equation}
and
\begin{equation}
	D_f=\{d(\mathscr{F})|M(\mathscr{F})=1\},
\end{equation}
where $\mathscr{  F}$ denotes any possible HOCA pattern generated by the HOCA rule $\mathbf{f}$, and $d(\mathscr F)$ denotes the box dimension of the given HOCA pattern. 
By examining $D_f$ and $D_r$, one can quickly ascertain the types of patterns that a given HOCA rule can produce. For example, if $D_f$ is empty, then the HOCA rule cannot produce any fractal patterns. If $D_r$ contains only one element, then the HOCA rule can only produce periodic patterns with the same dimension. Note that different fractal patterns can share the same dimension, an example of which is shown in our II-MSPT model (Eq.~(\ref{MSPT2})). Two HOCA symmetries in this model are Sierpinski triangles facing different directions, thus sharing the same dimension. With this notation in hand, we can give a technical definition of RSPT, FSPT, MSPT and CSPT orders as follows:
\begin{itemize}
    \item \textit{Regular(-subsystem) SPT} (RSPT) is the SPT phases protected by subsystem symmetries that necessarily \textit{(i)} include regular subsystem symmetries (e.g. line-like symmetry) and \textit{(ii)} exclude fractal subsystem symmetries. For HGSPT models, it means that RSPT models correspond to HOCA rules $\mathbf{f}$ satisfying $\mathcal{X}(\mathbf{f})=(1,0)$.
    \item \textit{Fractal(-subsystem) SPT (FSPT)} is the SPT phases protected by subsystem symmetries that necessarily \textit{(i)} exclude regular subsystem symmetries (e.g. line-like symmetry) and \textit{(ii)} include fractal subsystem symmetries. For HGSPT models, it means that FSPT models correspond to HOCA  rules$\mathbf{f}$ satisfying $\mathcal{X}(\mathbf{f})=(0,1)$.
    \item \textit{Type-I mixed(-subsystem) SPT} (I-MSPT) is the SPT phases protected by subsystem symmetries that necessarily \textit{(i)} include regular subsystem symmetries (e.g. line-like symmetry) and \textit{(ii)} include fractal subsystem symmetries. For HGSPT models, it means that I-MSPT models correspond to HOCA rules $\mathbf{f}$ satisfying $\mathcal{X}(\mathbf{f})=(1,1)$.
    \item \textit{Chaotic(-subsystem) SPT} (CSPT) is the SPT phases protected by subsystem symmetries that necessarily \textit{(i)} exclude regular subsystem symmetries (e.g. line-like symmetry) and \textit{(ii)} exclude fractal subsystem symmetries. For HGSPT models, it means that CSPT models correspond to HOCA rules $\mathbf{f}$ satisfying $\mathcal{X}(\mathbf{f})=(0,0)$.
\end{itemize}
Given the technical definitions above, there are still points need further clarification. Firstly, we can notice that these definitions are not completely intuitive: for example, when regular symmetries and chaotic-looking symmetries exist simultaneously in one model and fractal symmetries do not, the model is classified as RSPT phase as well. Besides, II-MSPT orders with two different kinds of fractal subsystem symmetries do not have a specific position in this classification, as purely according to the $\mathcal{X}(\mathbf{f})$ of HOCA rules they would be classified into FSPT orders. Furthermore, for a HOCA rule $\mathbf{f}$ a rigorous proof between $\mathcal{X}(\mathbf{f})=(0,0)$ and chaos is still lacked, although in our observation $\mathcal{X}(\mathbf{f})=(0,0)$ always implies the HOCA rule can generate chaotic-looking patterns. A finer classification of HGSPT orders naturally depends on a more complete and sophisticated understanding of the dynamics of HOCA, thus it is beyond the scope of this paper, but we expect it to be an important future direction which may lead to further understanding of subsystem symmetries.

Here we list some SPT phases characterized by our notation in Table~\ref{table_label}:
\begin{table}[htbp]\caption{Typical SPT phases denoted by the new notation system. Here, model I and II are respectively I-MSPT and II-MSPT models, model III is a CSPT model, model IVa, IVb, IVc are all RSPT models, model Va and Vb are both FSPT models. Specially, we can notice that though model IVa, IVb and IVc are all classified as 2D RSPT models, their behavior can be very different.}\label{table_label}

\begin{ruledtabular}
\begin{tabular}{c|cccc}
	Model Number & $X_r$ & $X_f$ & $D_r$ & $D_f$\\
	\hline
	I (Eq.~(\ref{MSPT1})) & 1&1&$\{1\}$ & $\{\ln3/\ln2\}$\\
	II (Eq.~(\ref{MSPT2})) & 0&1&$\varnothing$ & $\{\ln3/\ln2\}$\\
	III (Eq.~(\ref{MSPT3})) & 0& 0 &$\varnothing$ &$\varnothing$	\\
	IVa (Eq.~(\ref{MSPT4})) & 1&0& $\{1,2\}$ &$\varnothing$\\
 IVb (Eq.~(\ref{h_ivb})) & 1&0& $\{1\}$ &$\varnothing$\\
 IVc (Eq.~(\ref{clusterRule})) & 1&0& $\{1,2\}$ &$\varnothing$\\
	Va (Eq.~(\ref{FSPT})) & 0 &1& $\varnothing$ & $\{\ln3/\ln2\}$\\
 Vb (Eq.~(\ref{fiborule})) & 0 &1& $\varnothing$ & $\{1+\log_2\left(\frac{1+\sqrt 5}{2}\right)\}$\\
\end{tabular}
\end{ruledtabular}
\end{table}

	\section{Multi-point strange correlator detection}
	\label{sc}
	Originally proposed in \cite{you_wave_2014}, ``strange correlator'' is a powerful tool to detect nontrivial short-range entangled states. Recently,  strange correlators have been used to detect the nontriviality of the RSPT state \cite{zhou_detecting_2022} (see footnote~\ref{footnote_SSPT}). This work shows that RSPT state with line-like subsystem symmetries can be detected by strange correlators with two operators $\phi$ being in the same straight line corresponding to the anisotropy of the subsystem symmetries. This naturally motivates us to detect the nontriviality of HGSPT phases through strange correlator, with the hope that the configuration of the operators inside the strange correlator will reflect the property of the HOCA generated symmetry of the model.
	
	In the previous sections, we have shown that HOCA are able to successfully generate SPTs protected by various kinds of subsystem symmetries. In the following, given a specific HGSPT, we want to detect its nontriviality and the class that its HOCA rule belongs to. We will show  that this task can be completed by what we call ``multi-point strange correlator'' (MPSC). 
	\subsection{Definition}\label{sc_definition}
	The strange correlator is defined as follows in Ref. ~\cite{you_wave_2014}:
	\begin{equation}
		C(r,r')=\frac{\bra{ \Omega}\phi(r) \phi(r')\ket{\Psi}}{\braket{\Omega}{\Psi}},
	\end{equation}
	where $ \ket{\Psi} $ is the short range entangled (SRE) state to be diagnosed, $ \ket{\Omega} $ is the trivial disordered state in the same Hilbert space as $ \ket{\Psi} $, $ \phi $ is some local operator. For nontrivial SRE states, the strange correlator will saturate to a constant or undergo a power law decay for specific $ \phi   $, while that of trivial SRE states will decay exponentially or become null.
	The strange correlator defined above involves 2 local operators, and the definition can be extended to the case of $\mathfrak{n}$ local operators, dubbed as multi-point strange correlator:
	\begin{equation}
		C(r_1,r_2,...,r_\mathfrak{n})=\frac{\bra{ \Omega}\phi(r_1) \phi(r_2)\cdots \phi  (r_\mathfrak{n}) \ket{\Psi}}{\braket{\Omega}{\Psi}}.
	\end{equation}
	Here, $\mathfrak{n}$ are dubbed as the \textit{correlation number} of the multi-point strange correlator. We also introduce the multi-point normal correlator, which can be regarded as the strange correlator of the trivial disordered state, serving as the ``background'' to be subtracted from the strange correlator:
	\begin{equation}
		N(r_1,r_2,...,r_\mathfrak{n})=\frac{\bra{ \Omega}\phi(r_1) \phi(r_2)\cdots \phi  (r_\mathfrak{n}) \ket{\Omega}}{\braket{\Omega}{\Omega}}.
	\end{equation}
For a given $\phi$ and a spatial configuration $\{r_i\}$, we say the strange correlator $C(\{r_i\})$ gives nontrivial result if and only if $C(\{r_i\})-N(\{r_i\})$ saturate to a constant or decay algebraically. Since if $C(\{r_i\})$ is (quasi-)long range ordered but $C(\{r_i\})-N(\{r_i\})$ is not, it would mean that we cannot distinguish non-trivial SPT ordered states from the trivial symmetric state with this strange correlator.
	In this work we demand $\phi  $ to be onsite Pauli operators. By means of multi-point strange correlator, we construct a general procedure that detects the nontriviality of the HGSPT ground state.

 It is also worth noticing that using the duality relation between the SPT model and the symmetry breaking model, the MPSC can be mapped to the membrane-like order parameters in the symmetry breaking models, some examples have previously shown in \cite{devakul_fractal_2019,zhou2021fractalquantumphasetransitions,doherty_identifying_2009}. However, it has not been discussed up to our knowledge that whether the ground state of SPT models with fractal symmetries can only be detected by strange correlators with more than 2 local operators, i.e. if there are ``intrinsic'' multi-point nature in these models. In the following texts, we explore systematically the behavior of MPSC in various HGSPT models and proved that there is indeed SPT models that can only be detected by strange correlators with more than 2 points, as shown in Appendix~\ref{proof}.

	\subsection{Detection through multi-point strange correlators}\label{detection}
In this subsection we are to probe the nontrivial ground states of HGSPT models. The aim of this subsection is to raise a universal approach that distinguishes HGSPT models in different classes. Now that we have demonstrated that all HGSPT models have degenerate edge states on an open slab in the previous section, it is guaranteed that we can find a specific $\phi$ and a particular spatial configuration of $\phi$ to produce nontrivial results. The point here is to find the $\phi$ and configuration that give nontrivial results while reflecting the symmetry properties of the system. 

If we denote the position of operators in the multi-point strange correlator by $\{r_i\}$, then for a HGSPT generated by an order-$n$ CA, we claim that the nontrivial ground state of the HGSPT model can be detected by the multi-point strange correlator in the following configuration:
\begin{equation}
		C(\mathbf{q},\mathbf{f},L;\{r_i\})=\frac{\bra{ \Omega}X \begin{pmatrix} 
				0\\
					D_{L}(\mathbf{q},\mathbf{f};x,y)
			\end{pmatrix} \ket{\Psi}}{\braket{\Omega}{\Psi}},
	\end{equation}
 where
	\begin{equation}\label{scconfig}
			D_{L}(\mathbf{q},\mathbf{f};x,y)=[\mathbf{\tilde{q}}(L;x)+\mathbf{m}(L;x)]^T\cdot \mathbf{y}_{1,2n+L},
	\end{equation}
	\begin{equation}\label{eq:qx}
		\mathbf{\tilde{q}}(x)=\begin{pmatrix} 
							&q_{0}(x)\\
							&q_{1}(x)\\
							&\vdots\\
							&q_{n-1}(x)\\
							&\phantom{=}\begin{rcases}
								&0\\
								&\vdots\\
								&0
								\end{rcases} n+L \text{ rows}\\
							 \end{pmatrix},
	\end{equation}
	\begin{equation}\label{eq:mx}
		\mathbf{m}(x)=\begin{pmatrix} 
			m_{0}(x)\\
			m_{1}(x)\\
			\vdots\\
			m_{n-1}(x)\\
			\qquad \phantom{==}\begin{rcases}
				0\\
				\vdots\\
				0
			\end{rcases} L \text{ rows}\\
			p_0(x)\\
			p_1(x)\\
			\vdots\\
			p_{n-1}(x)
		\end{pmatrix},
	\end{equation}
	\begin{equation}
		\mathbf{y}_{1,2n+L}=\begin{pmatrix} 
			1\\
			y\\
			y^2\\
			\vdots\\
			y^{2n+L-1}
		\end{pmatrix},
	\end{equation}
and we choose the trivial symmetric state $\ket{\Omega}=\ket{\hat X^{(a)}=\hat Z^{(b)}=1}$. Specifically, $m_0(x)=0$, and $m_i(x),\ i>0$ can be calculated by
	\begin{equation}
		m_i(x)=\sum_{k=1}^{i}q_{i-k}(x)f_k(x),
	\end{equation}
	and
	\begin{equation}
		p_i(x)=\sum_{k=i+1}^{n}r_{n+L+i-k}(x)f_k(x).
	\end{equation}
All calculations of polynomials above can be easily done by computer.

$L$ plays the role of ``distance'' in this  configuration and takes value in $\mathbb N$, and is named \textit{evolution distance} hereafter. Given the HOCA rule of the HGSPT model, we can construct a series of multi-point strange correlator by fixing an initial condition $\mathbf q$ (as long as the initial condition can be defined in the given bulk) and increase $L$. Different series of multi-point strange correlators will behave variously depending on the $\mathcal X(\mathbf f)$ and $\mathbf q$. Specifically,  we define the \textit{correlation number} $\mathfrak n$ of a strange correlator to be the number of onsite operators $\phi$ included in the correlator $\mathfrak{n}[D_{L}(\mathbf{q},\mathbf{f};x,y)]$. Then, we claim that by observing how $\mathfrak n$ grows with $L$ will be helpful to determine the $\mathcal X(\mathbf f)$ (Sec. \ref{notation}),  the class of the HOCA rule for the given HGSPT phase.
	\begin{figure*}[htbp]
	\subfigure[Model-I (Eq.~(\ref{MSPT1})), initial condition are given by Eq.~(\ref{ic2}), $L=14$, correlation number $\mathfrak{n}=6$.\label{min1}]{\includegraphics[width=0.4\linewidth]{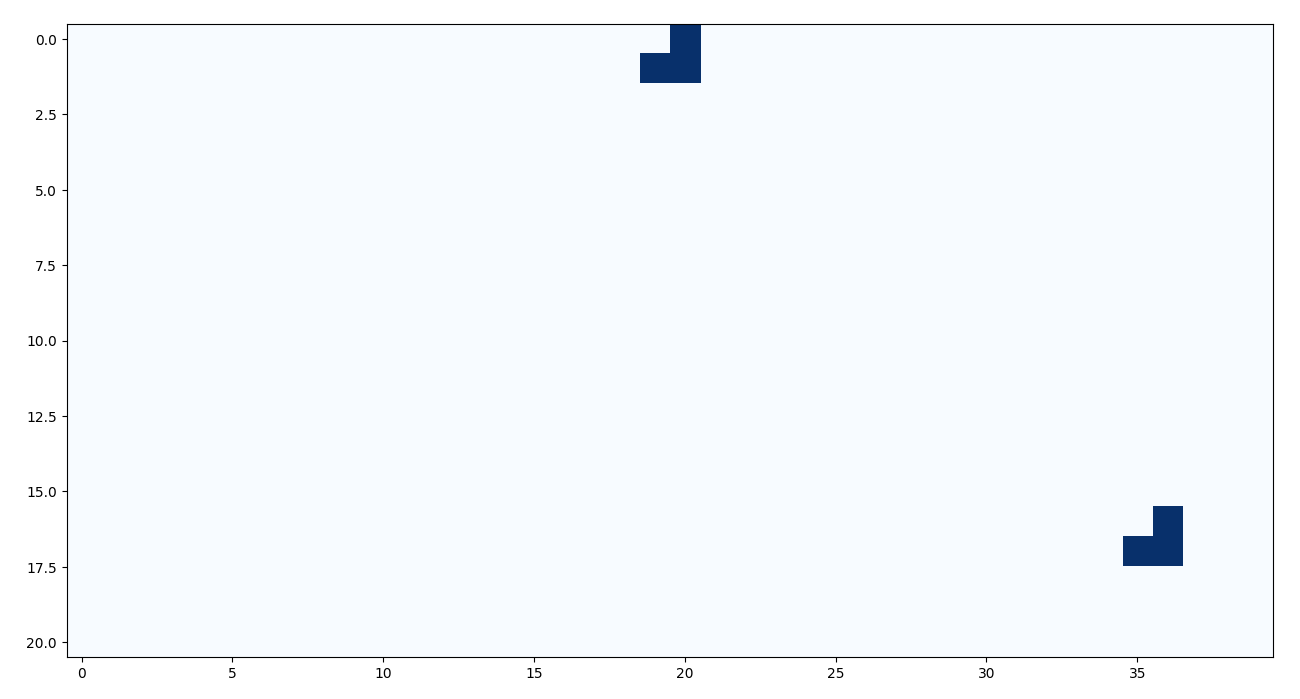}}
	\subfigure[Model-I (Eq.~(\ref{MSPT1})), initial condition are given by Eq.~(\ref{ic3}), $L=14$, correlation number $\mathfrak{n}=10$.\label{min4}]{\includegraphics[width=0.4\linewidth]{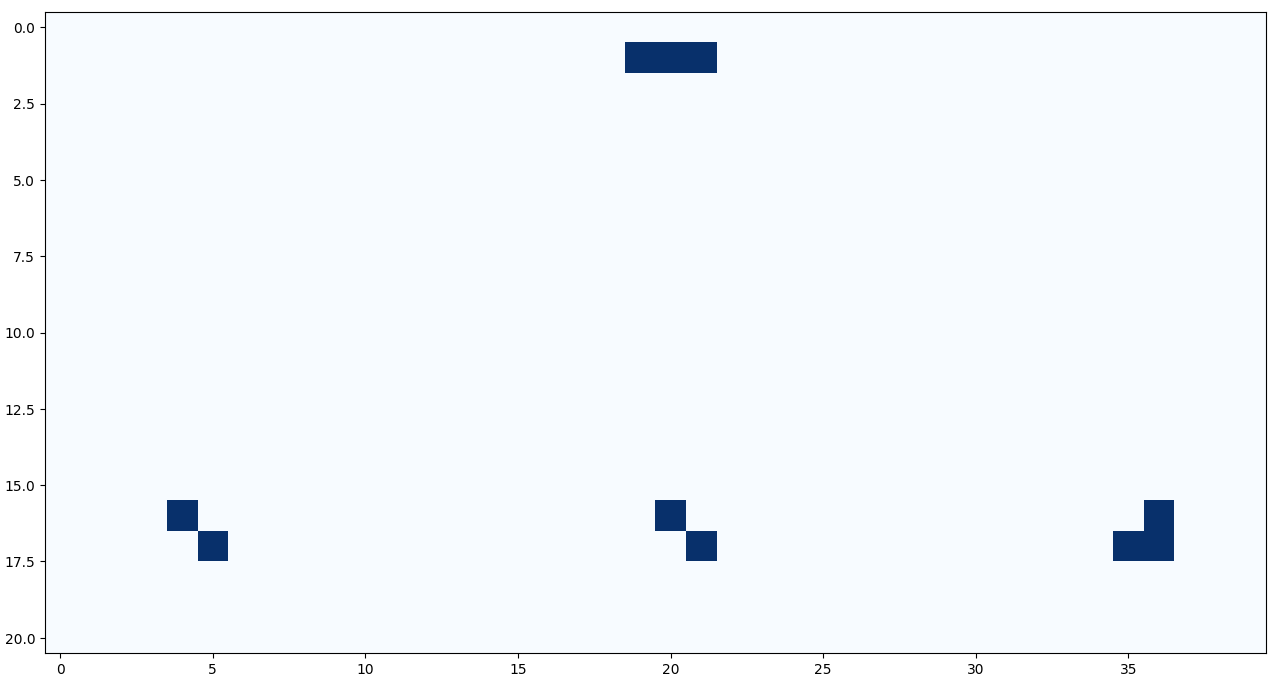}}
	\subfigure[Model-II (Eq.~(\ref{MSPT2})), Initial condition are given by Eq.~(\ref{ff4}), $L=13$, correlation number $\mathfrak{n}=9$.\label{min2}]{\includegraphics[width=0.4\linewidth]{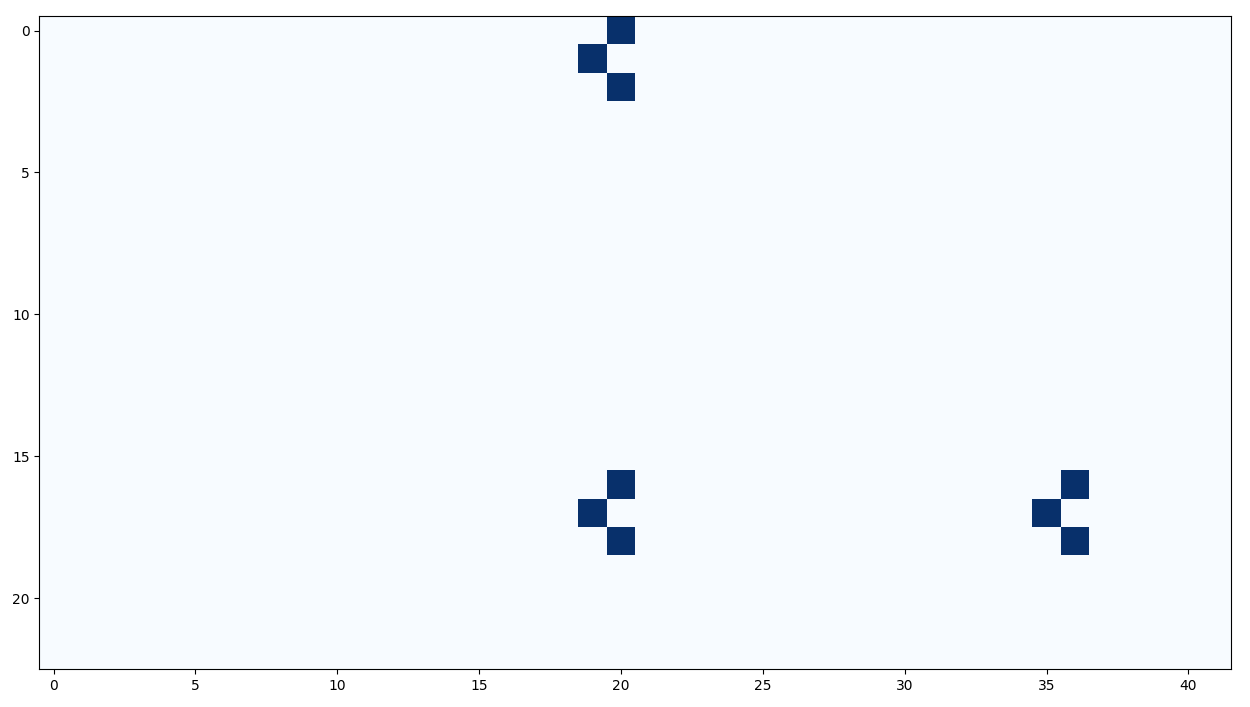}}
	\subfigure[Model-IV (Eq.~(\ref{MSPT4})), Initial condition are given by Eq.~(\ref{pr4}), $L=16$, correlation number $\mathfrak{n}=4$.\label{min3}]{\includegraphics[width=0.4\linewidth]{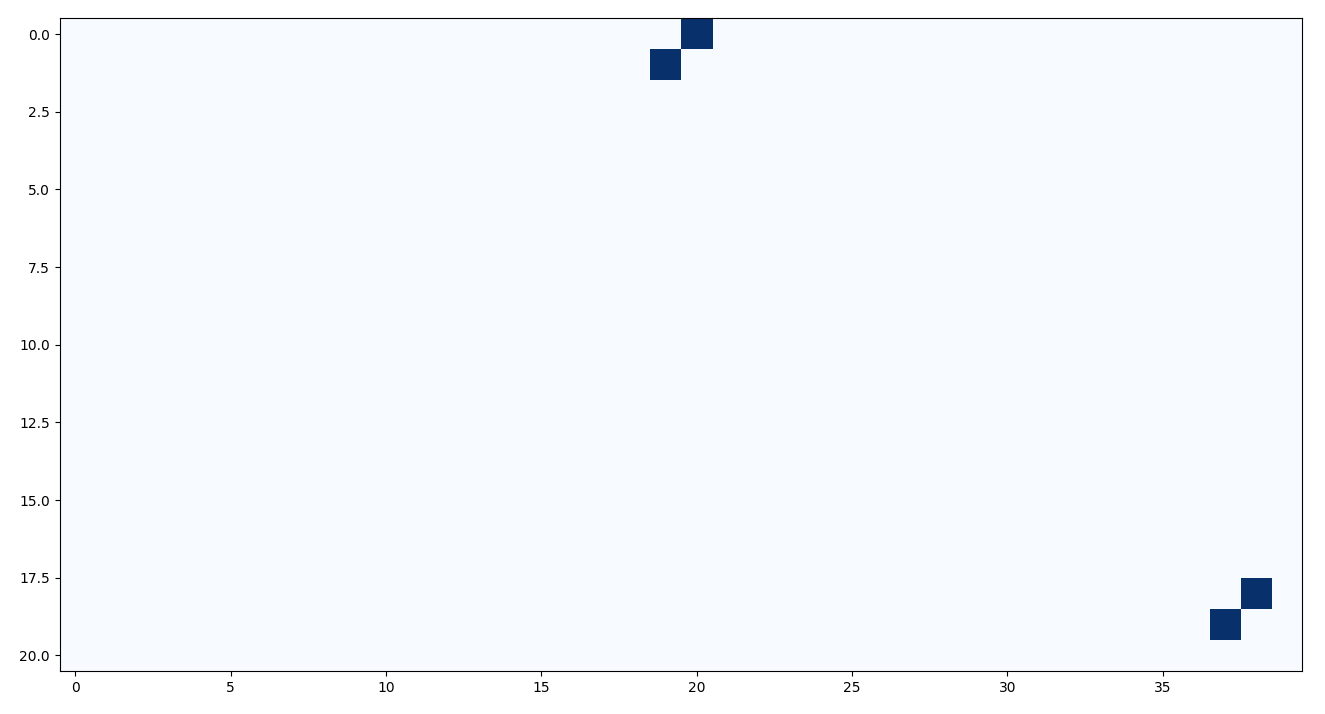}}
	\caption{Examples of $\mathbf{q}$ and $\mathbf{f}$ that make $\mathscr{N}(\mathbf{q},\mathbf{f})$ satisfies criterion 1 or 2. The configuration $D_{n,L}(\mathbf{q};x,y)$ generated by the given $\mathbf{q}$ and $\mathbf{f}$ are shown in the figure above. Detailed results are shown in Table \ref{results}.}
	\label{result2}
\end{figure*}

	The seemingly complicated configuration can be interpreted below: $\mathbf{\tilde q}(x)$ can be recognized as the initial condition $\mathbf{q}(x)$ that generates the whole configuration. Since generally the strange correlator is acted in the bulk, so the configuration $q(x)$ may possibly violate the HOCA update rule, resulting in trivial result of strange correlator, forcing us to introduce the term $\mathbf{m}$. The term $\mathbf{m}(x) $ is added for two reasons: (i) $m_i(x)$ is added to each $q_i(x)$ to make sure that the product of operators commute with the symmetry, and is determined by $q(x) $; (ii) $p_{i}(x)$ has the form of HOCA pattern generated by $q(x)$ in the corresponding rows to meet the commutation relations. Such construction ensures that $X \begin{pmatrix} 0\\D_{L}(\mathbf{q},\mathbf{f};x,y)\end{pmatrix}$ (take sublattice $(b)$ as an example) act equivalently as products of Hamiltonian terms in the given sublattice, giving a trivial action on the ground state $ \ket{ \Psi}$. This guarantees that the multi-point strange correlator $C(\mathbf{q},\mathbf{f},L;\{r_i\})$ gives nontrivial results since
	\begin{equation}\label{sc_derivation}
		\begin{aligned}
			C(\mathbf{q},\mathbf{f},L;\{r_i\})&=\frac{\bra{ \Omega}X \begin{pmatrix} 
   0\\
					D_{L}(\mathbf{q},\mathbf{f};x,y)
				\end{pmatrix} \ket{\Psi}}{\braket{\Omega}{\Psi}}\\
			&=\frac{\bra{ \Omega} \prod_{\alpha,\beta} X\begin{pmatrix} 
					x^\alpha y^\beta\\0
				\end{pmatrix} \prod_{\alpha,\beta} H^X_{\alpha, \beta} \ket{\Psi}}{\braket{\Omega}{\Psi}}\\
			&=1.
		\end{aligned}
	\end{equation}
	
	Here $H^X_{\alpha, \beta}$ is the Hamiltonian term defined on the site denoted by $\alpha, \beta$ and composed of $X$ operators, where $\alpha, \beta$ sum over all the sites with nontrivial values in $D_L(\mathbf{q},\mathbf{f};x,y)$. 
	
Now we give some comments of this construction of configuration. There are two main goals that we consider while designing this specific configuration. First, we want to find out the simplest configuration of $\phi$ needed to show the nontrivial results in an HGSPT ground state. Second, we hope that the configuration we design will reflect the symmetry property of the given HGSPT model. The second goal is well-achieved in our construction in all models discussed in this paper, while the first goal is not always easy to satisfy. Generally, we conjecture that configuration that meets the first goal are always included in this configuration. This claim is proved rigorously in the case of model-Va (Eq.~(\ref{FSPT})), an FSPT model.

	For an order-$n$ HOCA generated SPT, we can determine the class $[\mathbf{f}]$ by examining the behavior of correlation number $\mathfrak{n}$ of the strange correlator in the configuration Eq.~(\ref{scconfig}) as $L\to\infty$. 
	
Based on the definition of correlation number$\mathfrak{n}[D_{L}(\mathbf{q},\mathbf{f};x,y)]$, we can further define the following quantity:
	\begin{equation}\label{eq:N}
		 \mathscr{N}(\mathbf{q},\mathbf{f})\equiv \frac{\mathfrak{n}_{\text{inf} }}{\mathfrak{n}_{\text{sup}}}\in[0,1],
	\end{equation}
 where
 \begin{equation}
 \mathfrak{n}_{\text{inf} }\equiv\liminf_{L\to \infty} \mathfrak{n}[D_{L}(\mathbf{q},\mathbf{f};x,y)]
\end{equation}
 and
 \begin{equation}
     \mathfrak{n}_{\text{sup}}\equiv\limsup_{L\to \infty} \mathfrak{n}[D_{L}(\mathbf{q},\mathbf{f};x,y)].
 \end{equation}
Then the following two criteria holds:

\begin{outline}[itemize]

	\1 Criterion 1:
	
	\begin{equation}
		\begin{aligned}
			&\text{If }\exists \mathbf{q}(x)\ne (0,0,\cdots,0 )^T \text{ such that }\\
			& \mathscr{N}(\mathbf{q},\mathbf{f})=1,\\
			&\text{then } X_r=1.
		\end{aligned}
	\end{equation}

	\1 Criterion 2:	

	\begin{equation}\label{crit2}
		\begin{aligned}
			&\text{If }\exists \mathbf{q}(x)\ne (0,0,\cdots,0 )^T \text{ such that }\\
			&\mathscr{N}(\mathbf{q},\mathbf{f})=0 ,\\
			&\text{then } X_f=1.
		\end{aligned}
	\end{equation}
\end{outline}	 
Fig.~\ref{result2} shows some explicit examples of these two criteria. So far the search for the initial conditions that satisfy criterion 1 or 2 is done by computers, and hopefully an analytical way will be found in future works. The detailed mathematical discussion and concrete examples of these two criteria are shown in Appendix~\ref{sc_proof}.

Now we apply these criteria to HOCA generated SPT phases that we have discussed before:
\begin{table}[htbp]    

    \begin{ruledtabular}
        \begin{tabular}{c|cccc}
            Model Number & $X_r$ & $X_f$ & Criterion 1 & Criterion 2 \\
            \hline
            I (Eq.~(\ref{MSPT1})) & 1&1&Yes (Fig.~\ref{min1}) & Yes (Fig.~\ref{min4})\\
            II (Eq.~(\ref{MSPT2})) & 0&1&No & Yes (Fig.~\ref{min2})\\
            III (Eq.~(\ref{MSPT3})) & 0& 0 &No &No\\
            IVa (Eq.~(\ref{MSPT4})) & 1&0& Yes (Fig.~\ref{min3}) &No\\
            IVb (Eq.~(\ref{h_ivb})) &1 &0 &Yes (Fig.~\ref{sc_ivb}) &No\\
            Va (Eq.~(\ref{FSPT})) & 0 &1& No & Yes (Appendix \ref{proof})
        \end{tabular}
    \end{ruledtabular}

    \caption{Result of detecting HOCA generated SPT phases by multi-point strange correlators. Model-IVc and model-Vb is not included in this table, as these are not the models that we focus on.
    }
    \label{results}
\end{table}

It can be seen in Table \ref{results} that the class $[\mathbf{f}]$ can be detected by multi-point strange correlator. Examples that meet the criteria in the table are shown in Fig.~\ref{result2}.

 \subsubsection{Detecting FSPT (Model-Va) generated by order-1 CA}\label{D-FSPT}
In this section, we will probe the nontriviality of the Sierpinski FSPT ground state via the strange correlator. The Sierpinski FSPT model has degenerate edge modes, which are localized states at the boundary of the system that are protected by the fractal symmetry of the model. These edge modes are argued in detail in \cite{devakul_fractal_2019}, and they indicate that the ground state of the model must be a nontrivial short-range entangled (SRE) state, which is a quantum state that cannot be transformed into a product state by local unitary operations. We take the Sierpinski model as a model generated by order-1 CA, and explore its nontriviality by means of multi-point strange correlator. The update rule of the Sierpinski FSPT model \cite{devakul_fractal_2019} is

\begin{equation}\label{sierrule}
	\mathbf{f}(x)=1+x,
\end{equation}

	\begin{figure*}[htbp]
	\subfigure[Configuration of $D_{0}(\mathbf{q},\mathbf{f};x,y)$  \label{FS1}]{\includegraphics[width=0.4\linewidth]{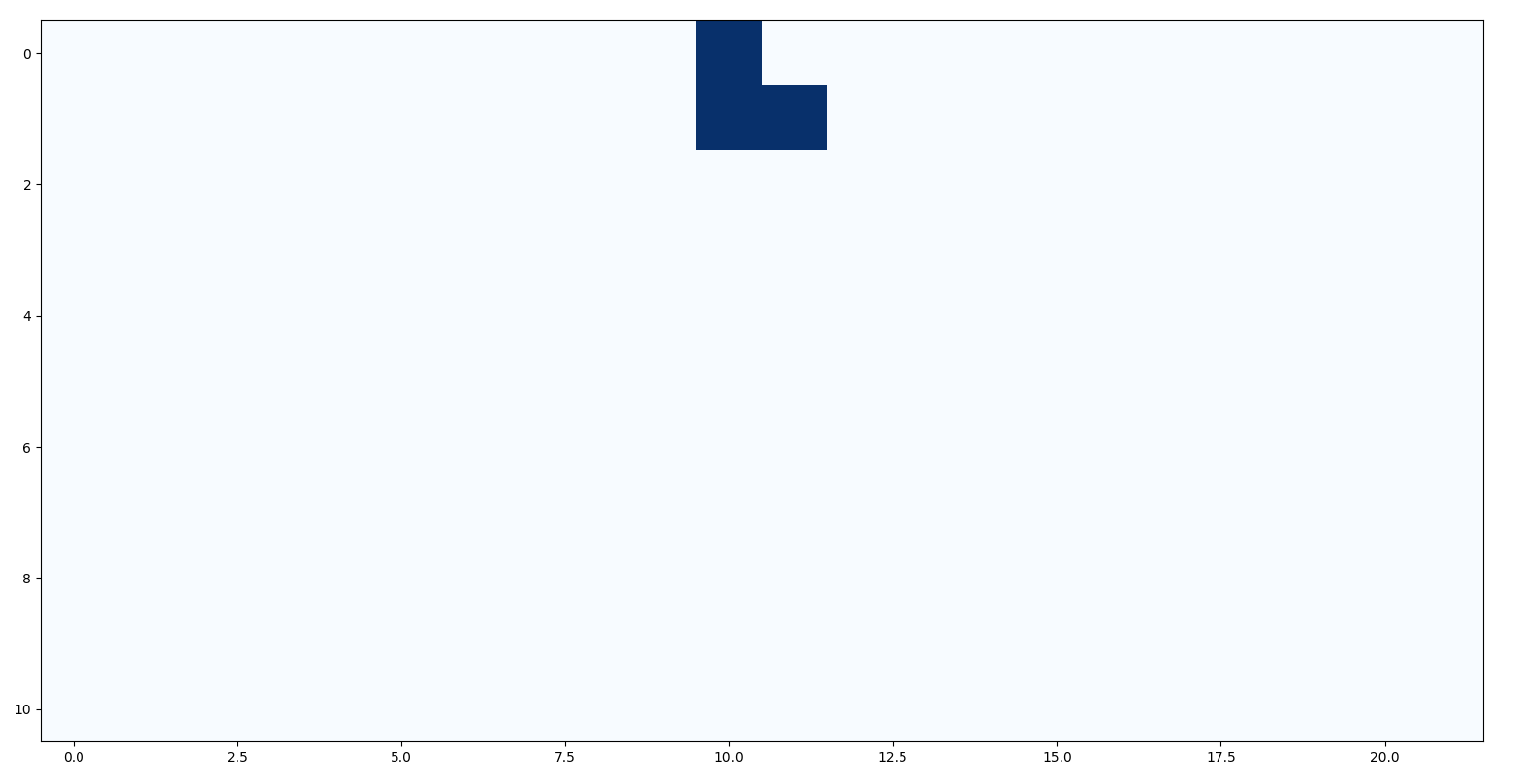}}
	\subfigure[Configuration of $D_{1}(\mathbf{q},\mathbf{f};x,y)$ \label{FS2}]{\includegraphics[width=0.4\linewidth]{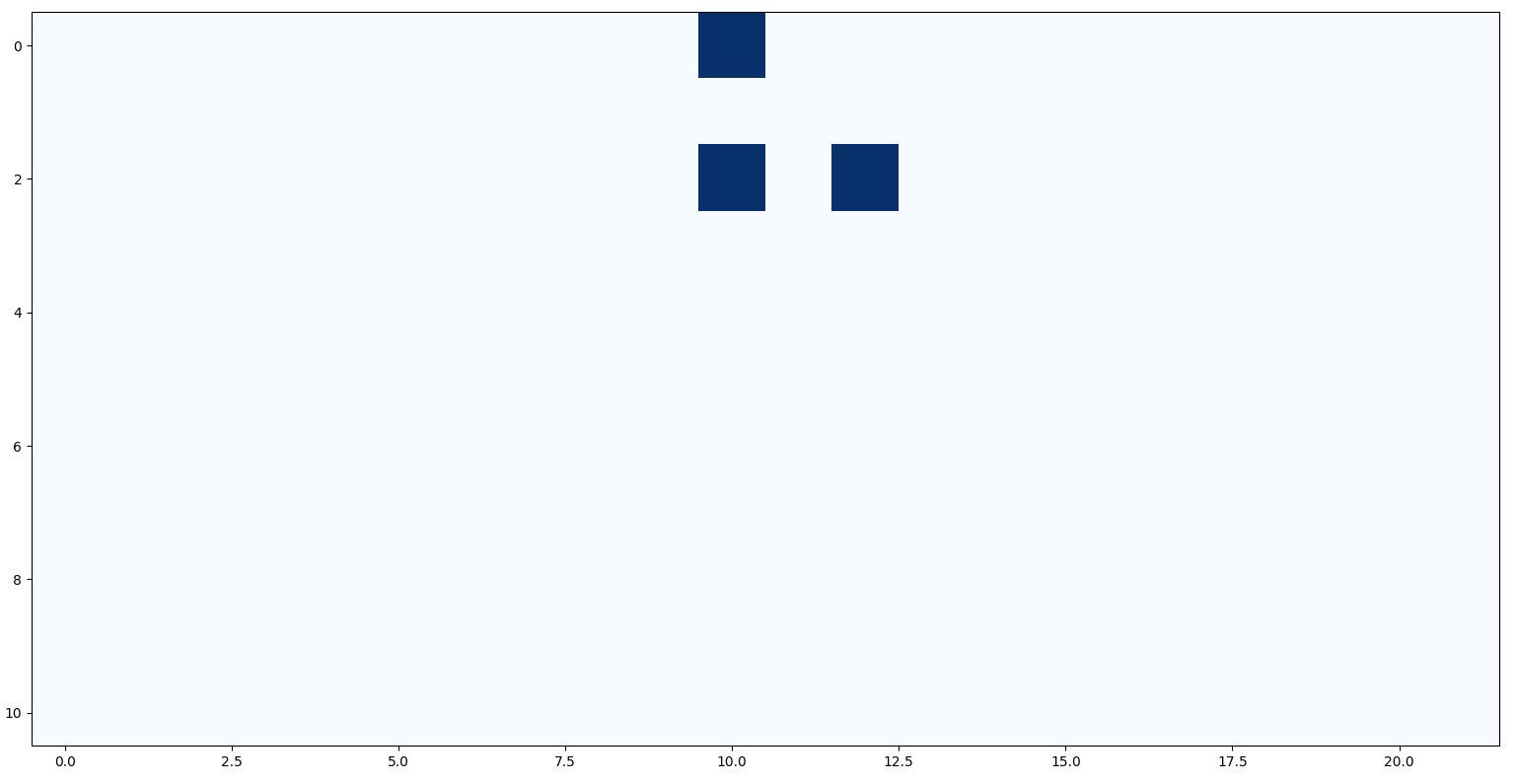}}\\
	\subfigure[Configuration of $D_{3}(\mathbf{q},\mathbf{f};x,y)$ \label{FS3}]{\includegraphics[width=0.4\linewidth]{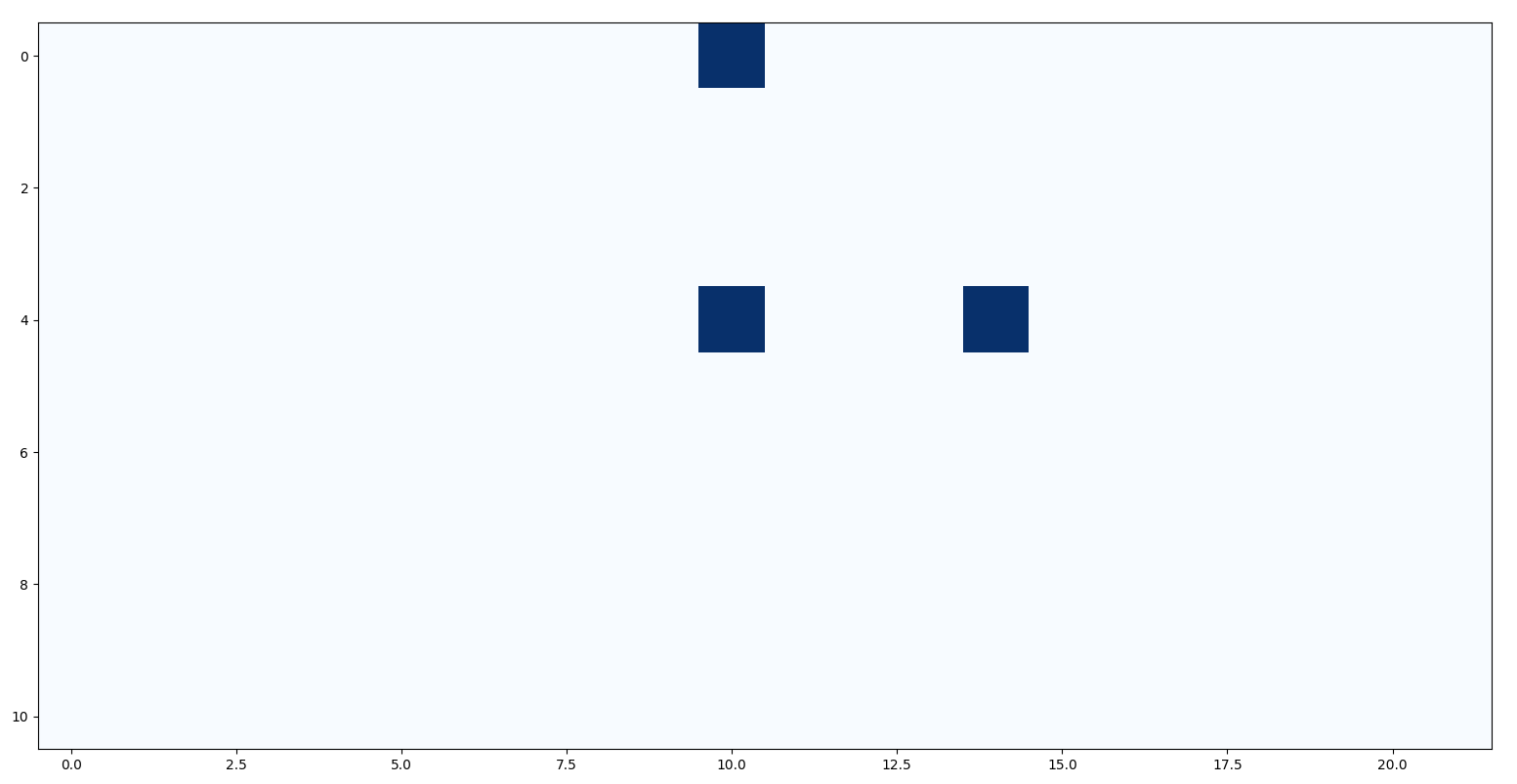}}
	\subfigure[Configuration of $D_{7}(\mathbf{q},\mathbf{f};x,y)$  \label{FS4}]{\includegraphics[width=0.4\linewidth]{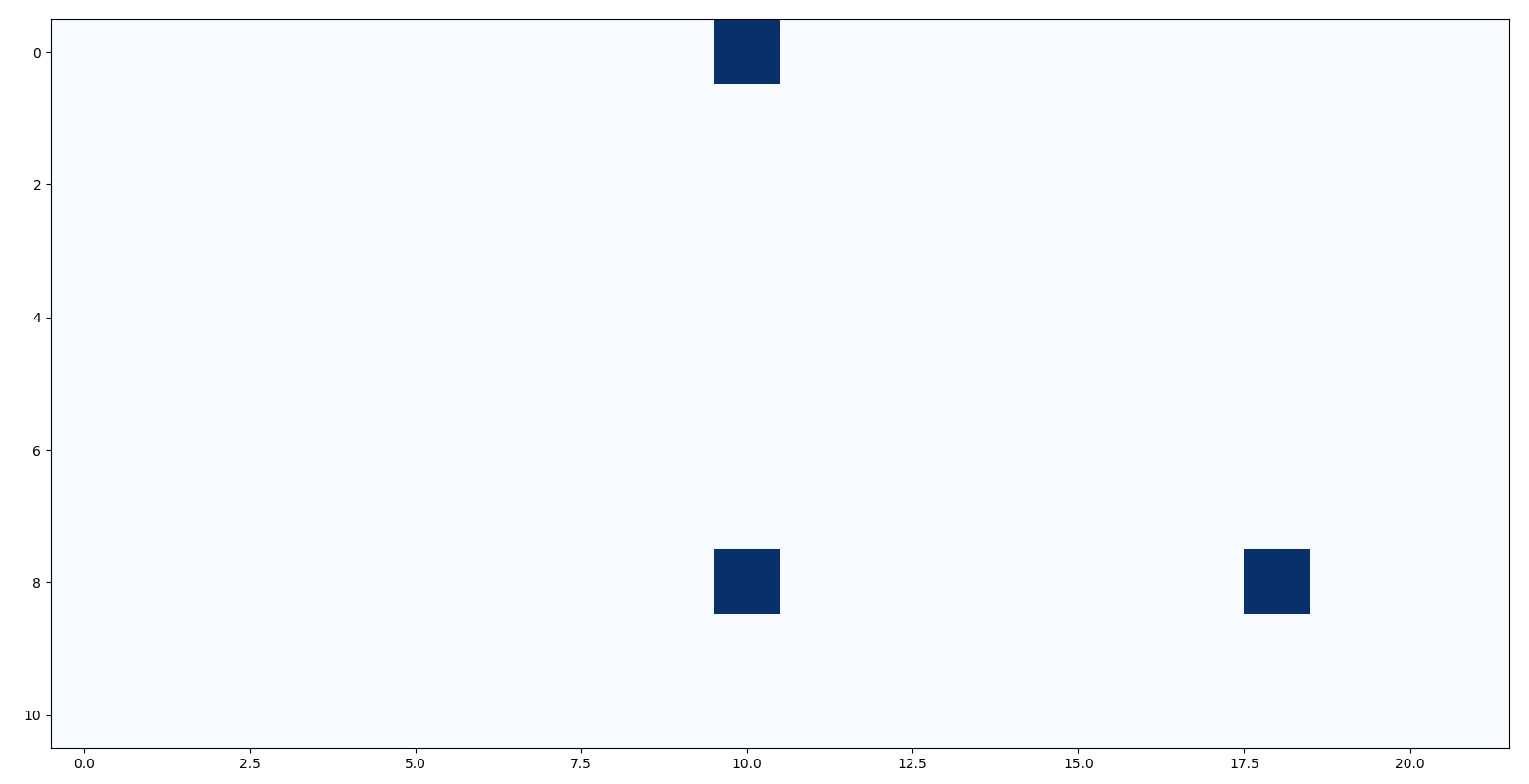}}
	\caption{Pictorial illustration of configuration of multi-point strange correlator $D_{L}(\mathbf{q},\mathbf{f};x,y)$ of update rule Eq.~(\ref{sierrule}), which generates an FSPT phase (nontrivial terms are shown in blue cubes in Fig.~\ref{FS1}, Fig.~\ref{FS2}, Fig.~\ref{FS3}, Fig.~\ref{FS4}). 4 configurations above have $\mathbf{q}(x)=1$ and $\mathfrak{n}=3$.}
	\label{minimal2}
\end{figure*}

and the Hamiltonian is written as

\begin{equation}\label{FSPT}
	\begin{aligned}
		\mathscr{H}=-\sum_{i j}&Z\begin{pmatrix}
		    x^iy^j(1+y^{-1}+x^{-1}y^{-1})\\x^iy^j
		\end{pmatrix}\\&-X\begin{pmatrix}
		    x^iy^j\\x^iy^j(1+y+xy)
		\end{pmatrix},
	\end{aligned}
\end{equation}
Applying our procedure to the model (Eq.~(\ref{FSPT})), we find that when
\begin{equation}\label{sc_FSPT}
    C(\{r_i\})=\frac{\bra{ \Omega}X \begin{pmatrix} 
    0\\
				D_{2^k-1}(\mathbf{q},\mathbf{f};x,y)
			\end{pmatrix} \ket{\Psi}}{\braket{\Omega}{\Psi}},~k\in N_+,
\end{equation}
where $\mathbf{q}=1$ and $\mathbf{f}=1+x$,
then criterion 2 is satisfied:
\begin{equation}
    \mathscr{N}(\mathbf{q},\mathbf{f})=0,
\end{equation}
Also, there are no possible configuration that satisfies criterion 1, which can be proved by the self-similarity nature of the order-1 CA, or simply by enumearting all possible initial conditions on an open slab. Thus, we indeed find out that the Sierpinski rule (Eq.~(\ref{sierrule})) is in the $(0,1)$ class.
This configuration reflects the fractal symmetry of the Sierpinski triangle. At the same time, this configuration possesses the minimal correlation number among all possible strange correlators made up of Pauli matrices in this FSPT model. It can be proved that all 2-point strange correlators show trivial results, giving the same result as the normal correlator gives. Detailed calculation can be found in appendix \ref{calculation}.  It is natural to ask what is the minimal $n$ that gives nontrivial multi-point strange correlators (giving different results from what multi-point normal correlator gives). It is proved that 
\begin{equation}
	\min(\mathfrak{n})=3
\end{equation}
in the case of the Sierpinski FSPT model, and these 3 Pauli matrices must be placed at the 3 corners of a Sierpinski triangle in the lattice. This claim is proved in appendix \ref{proof}.

\subsubsection{Detecting I-MSPT (Model-I) generated by order-2 CA }\label{D-I-MSPT}
\begin{figure*}[htb]
	\subfigure[HOCA pattern  \label{mini1}]{\includegraphics[width=0.4\linewidth]{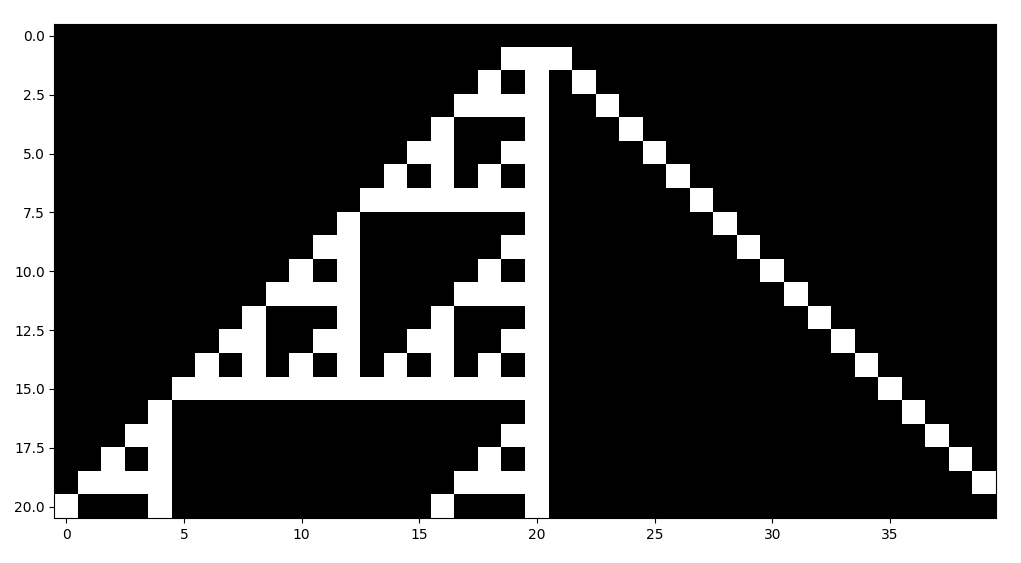}}
	\subfigure[Configuration of $D_{2}(\mathbf{q},\mathbf{f};x,y)$ \label{mini2}]{\includegraphics[width=0.4\linewidth]{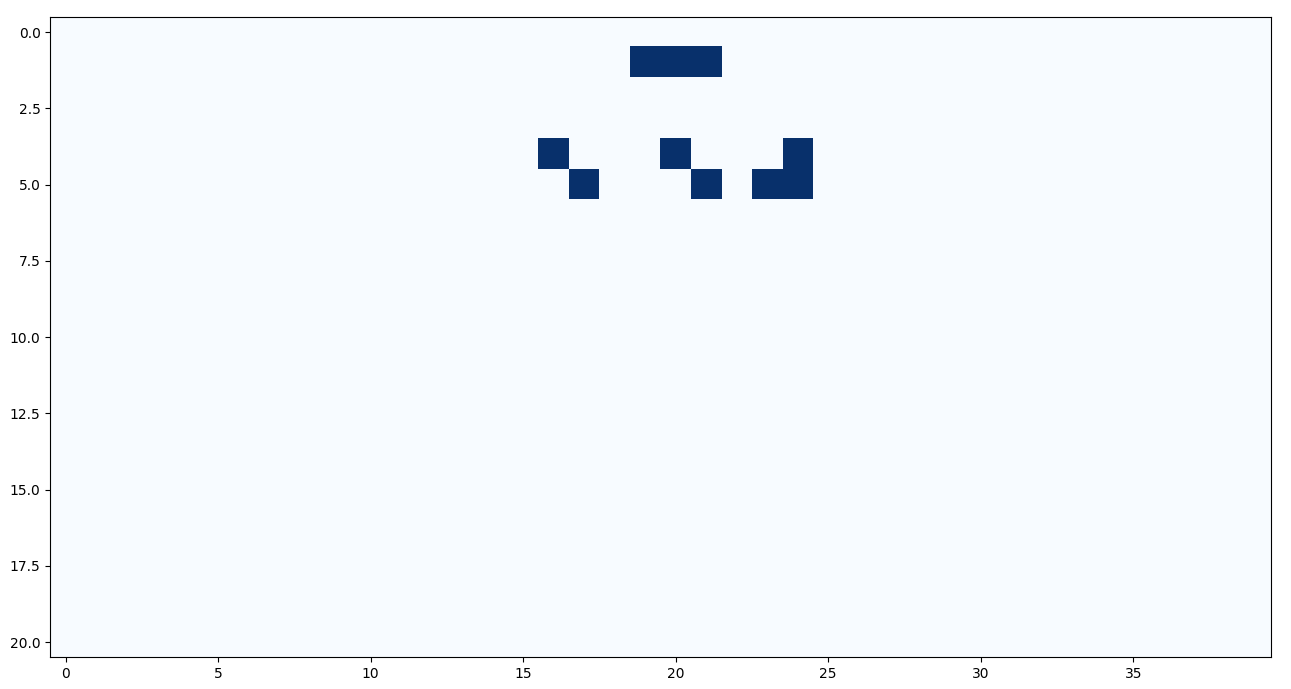}}\\
	\subfigure[Configuration of $D_{6}(\mathbf{q},\mathbf{f};x,y)$ \label{mini3}]{\includegraphics[width=0.4\linewidth]{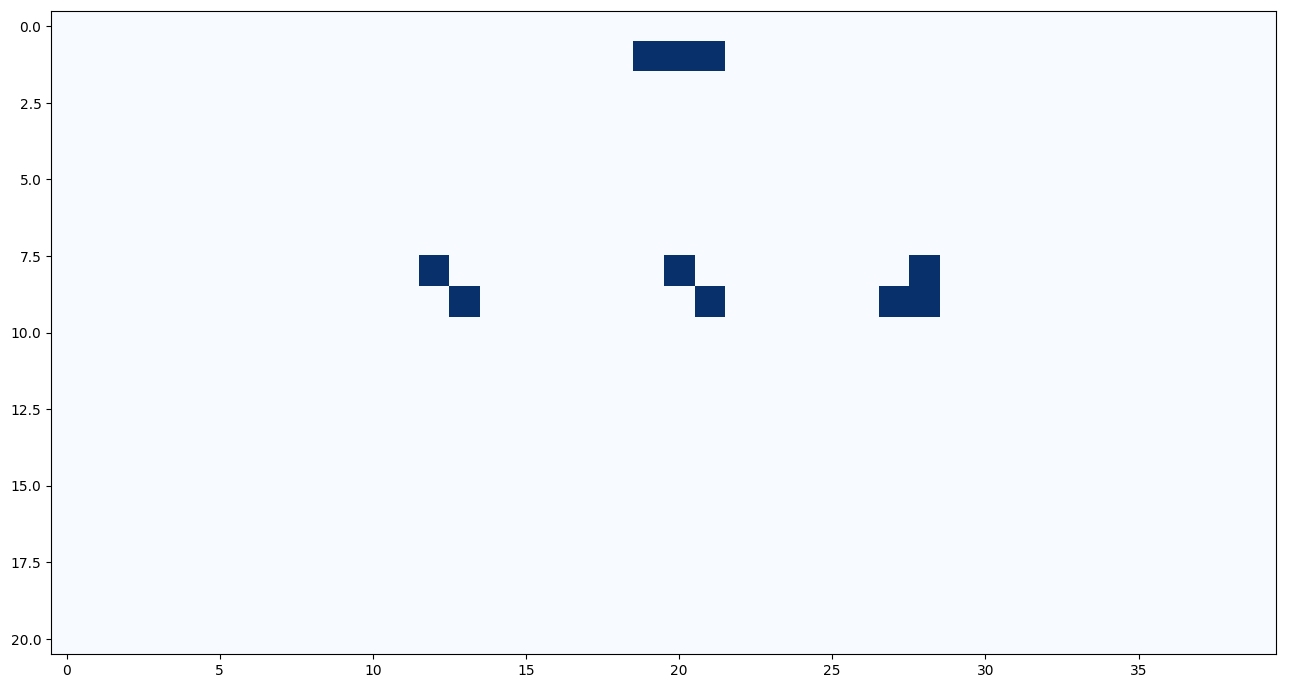}}
	\subfigure[Configuration of $D_{14}(\mathbf{q},\mathbf{f};x,y)$  \label{mini4}]{\includegraphics[width=0.4\linewidth]{minimal2}}
	\caption{Pictorial illustration of configuration of multi-point strange correlator $D_{L}(\mathbf{q},\mathbf{f};x,y)$ of update rule Eq.~(\ref{fl-MSPT1}), which generates an I-MSPT phase (nontrivial terms are shown in blue cubes in Fig.~\ref{mini2}, Fig.~\ref{mini3}, Fig.~\ref{mini4}). 3 configurations above have $\mathbf{q}(x)=\begin{pmatrix} 0\\x^{-1}+1+x	\end{pmatrix}$ and $\mathfrak{n}=10$. Fig.~\ref{mini1} shows the HOCA pattern generated by the initial condition above. It can be seen in the figure that 3 MPSC with different $L$ share the same correlation number $\mathfrak{n}$, showing the scaling invariance of this MPSC.}
	\label{minimal}
\end{figure*}

For model-I (Eq.~(\ref{MSPT1})), we expect that both criteria above can be satisfied by controlling the initial condition. We observe that the multi-point strange correlator
\begin{equation}\label{sc_MSPT1}
    C(\{r_i\})=\frac{\bra{ \Omega}X \begin{pmatrix}
    0\\
				D_{2^k-2}(\mathbf{q},\mathbf{f};x,y)
			\end{pmatrix} \ket{\Psi}}{\braket{\Omega}{\Psi}},~k\in N_+,
\end{equation}
satisfies
\begin{equation}
     \mathscr{N}(\mathbf{q},\mathbf{f})=0,
\end{equation}
when the initial condition $\mathrm{q}$ is set to be Eq.~(\ref{ic3}). The correlation number of the configurations above are
\begin{equation}
	\mathfrak{n}=10,
\end{equation} 
of which 3 examples are shown in Fig.~\ref{mini2}, Fig.~\ref{mini3}, Fig.~\ref{mini4}.
Also, we find that the multi-point strange correlator
\begin{equation}\label{sc_MSPT2}
	C(\{r_i\})=\frac{\bra{ \Omega}X \begin{pmatrix} 
 0\\
			D_{k}(\mathbf{q},\mathbf{f};x,y)
		\end{pmatrix} \ket{\Psi}}{\braket{\Omega}{\Psi}},~k\in N_+,
\end{equation}
satisfies
\begin{equation}
	\mathscr{N}(\mathbf{q},\mathbf{f})=0,
\end{equation}
when the initial condition $\mathrm{q}$ is set to be Eq.~(\ref{ic2}). The correlation number of the configurations above are
\begin{equation}
	\mathfrak{n}=6,
\end{equation}
of which an example is shown in Fig.~\ref{min1}.
\begin{figure*}[htb]
	\subfigure[Configuration of $D_{0}(\mathbf{q},\mathbf{f};x,y)$  \label{CS1}]{\includegraphics[width=0.4\linewidth]{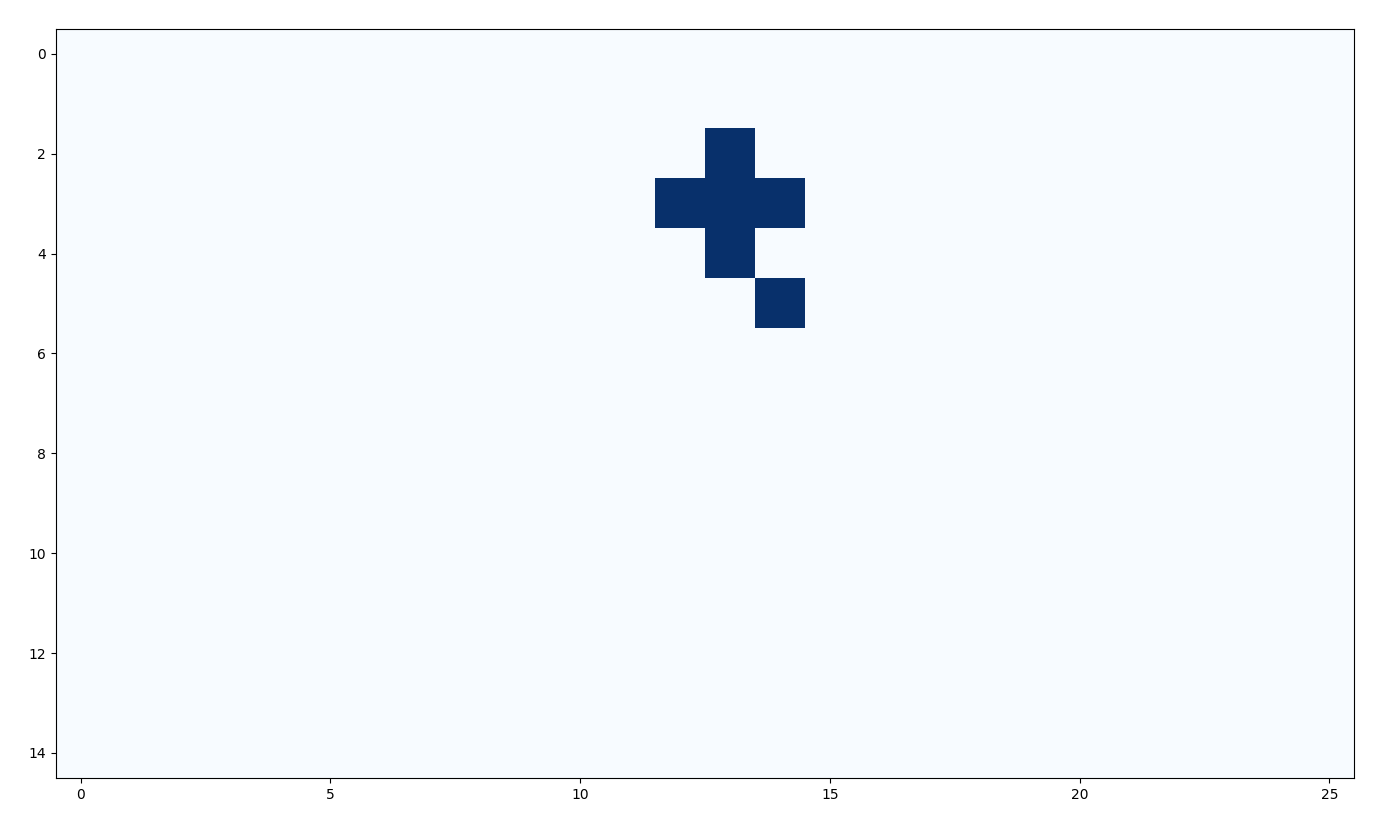}}
	\subfigure[Configuration of $D_{2}(\mathbf{q},\mathbf{f};x,y)$ \label{CS2}]{\includegraphics[width=0.4\linewidth]{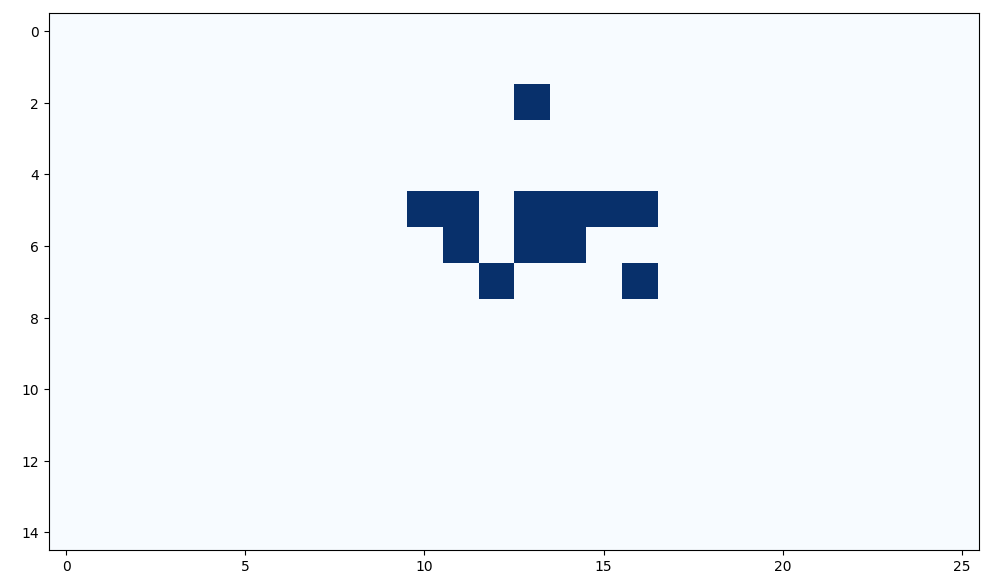}}\\
	\subfigure[Configuration of $D_{4}(\mathbf{q},\mathbf{f};x,y)$ \label{CS3}]{\includegraphics[width=0.4\linewidth]{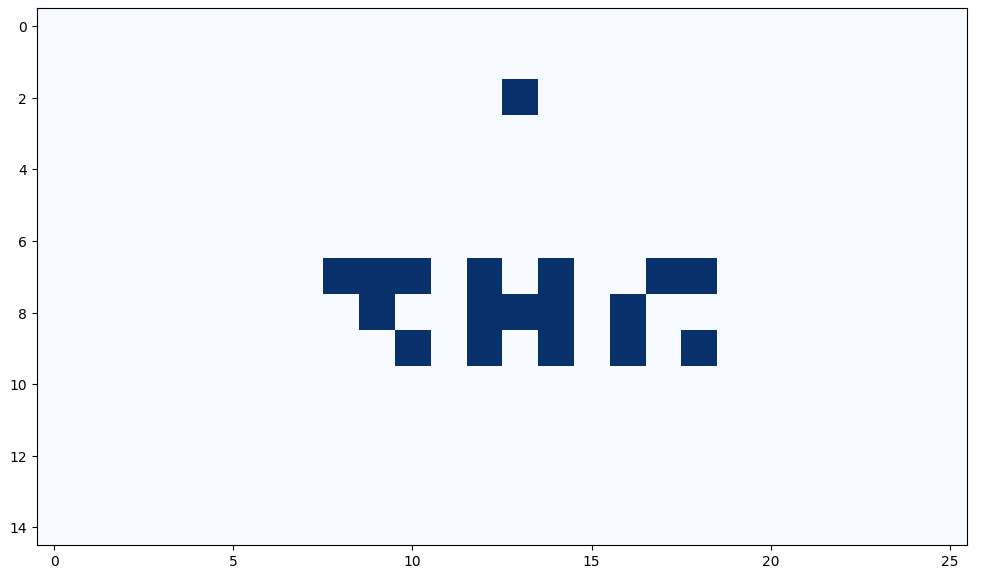}}
	\subfigure[Configuration of $D_{8}(\mathbf{q},\mathbf{f};x,y)$  \label{CS4}]{\includegraphics[width=0.4\linewidth]{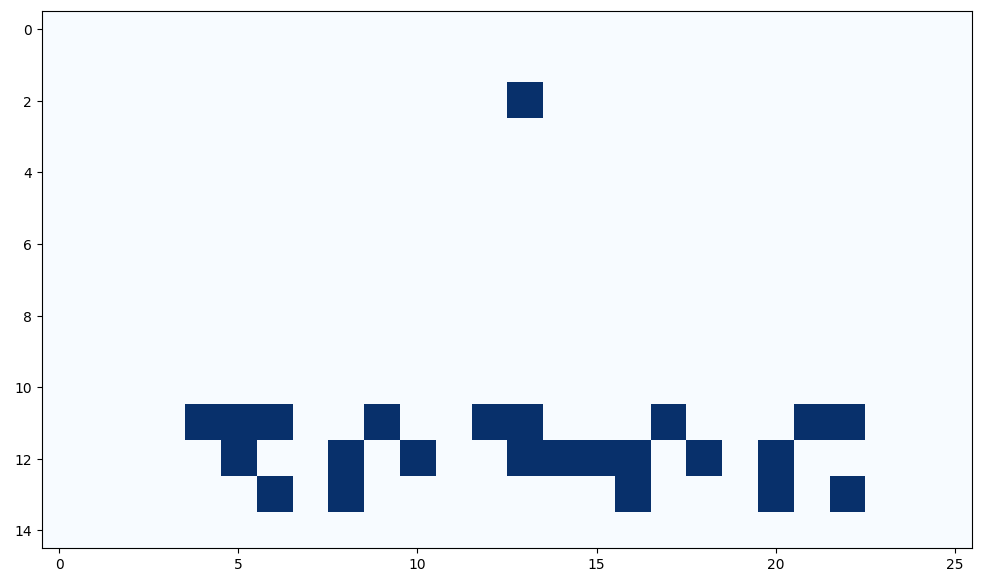}}
	\caption{Pictorial illustration of configuration of multi-point strange correlator $D_{L}(\mathbf{q},\mathbf{f};x,y)$ of update rule Eq.~(\ref{irregular_rule}), which generates an CSPT phase (nontrivial terms are shown in blue cubes in Fig.~\ref{CS1}, Fig.~\ref{CS2}, Fig.~\ref{CS3}, Fig.~\ref{CS4}). The initial condition of 4 configurations above is given by Eq.~(\ref{ir1}).}
	\label{CSPT_SC}
\end{figure*}

\subsubsection{Detecting II-MSPT (Model-II) generated by order-3 CA }\label{D-II-MSPT}
For model-II (Eq.~(\ref{MSPT2})), we expect that criterion 2 can be satisfied by controlling the initial condition. We observe that the multi-point strange correlator
\begin{equation}\label{sc_MSPT3}
	C(\{r_i\})=\frac{\bra{ \Omega}X \begin{pmatrix}
 0\\
			D_{2^k-3}(\mathbf{q},\mathbf{f};x,y)
		\end{pmatrix} \ket{\Psi}}{\braket{\Omega}{\Psi}},~k\in N_+,
\end{equation}
satisfies
\begin{equation}
	\mathscr{N}(\mathbf{q},\mathbf{f})=0,
\end{equation}
when the initial condition $\mathrm{q}$ is set to be Eq.~(\ref{ff4}). The correlation number of the configurations above are
\begin{equation}
	\mathfrak{n}=9,
\end{equation} 
of which an example is shown in Fig.~\ref{min2}.
\subsubsection{Detecting CSPT (Model-III) generated by order-3 CA }\label{D-CSPT}
For model-IV (Eq.~(\ref{MSPT3})), we expect that no criterion can be satisfied. This claim is confirmed by computational search on initial conditions with size $L_x\le 50$, and increasing the size generally do not give any new phenomenon. For all possible configurations we observe the correlation number $\mathfrak{n}$ generally increase with $L$. Unlike other models mentioned in this paper, fixing any initial condition $\mathbf{q}$, we will never obtain an infinite sequence of $L$ that makes the multi-point strange correlators share the same correlation number $\mathfrak{n}$. Among all strange correlators in this model, the one with minimal correlation number writes
\begin{equation}\label{CSPT}
	C(\{r_i\})=\frac{\bra{ \Omega}X \begin{pmatrix}
 0\\
			D_{0}(\mathbf{q},\mathbf{f};x,y)
		\end{pmatrix} \ket{\Psi}}{\braket{\Omega}{\Psi}},
\end{equation}
where the initial condition is set to be Eq.~(\ref{ir1}), of which the figure is shown in Fig.~\ref{CSPT_SC}.

We notice that multi-point strange correlators in CSPT order seem to give a promising procedure to overcome the computation irreducibility of CA. While the computational irreducibility states that we cannot directly calculate an arbitrary step in CA evolution (for CA showing complex behaviors, not CA with regular and predicable patterns, e.g. HOCA rules in CSPT models) without calculating steps before the wanted step, in principle we can efficiently prepare the ground state of this model in an array of qubits and measure the strange correlator by a series of quantum operations in this qubit array. Only the multi-point strange correlator with the correct configuration will show nontrivial result. That is to say, we can verify whether the result of an arbitrary step is a given configuration with zero knowledge about the intermediate steps, which can potentially serve as an quantum approach to surpass the well-known principle of computational irreducibility~\cite{wolfram_cellular_1984, wolfram1984d, wolfram2002a}.

\subsubsection{Detecting RSPT (Model-IVa) generated by order-2 CA}\label{D-RSPT}
For model-IVa (Eq.~(\ref{MSPT4})), we expect that criterion 1 can be satisfied. We observe that the multi-point strange correlator
\begin{equation}\label{sc_RSPT1}
	C(\{r_i\})=\frac{\bra{ \Omega}X \begin{pmatrix}
 0\\
			D_{k}(\mathbf{q},\mathbf{f};x,y)
		\end{pmatrix} \ket{\Psi}}{\braket{\Omega}{\Psi}},~k\in N_+,
\end{equation}
satisfies
\begin{equation}
	\mathscr{N}(\mathbf{q},\mathbf{f})=1,
\end{equation}
when the initial condition $\mathbf{q}$ is set to be Eq.~(\ref{pr2}). The correlation number of the configurations above are
\begin{equation}
	\mathfrak{n}=4,
\end{equation} 
of which an example is shown in Fig.~\ref{min3}.

\subsubsection{Detecting RSPT (Model-IVb) generated by order-3 CA}\label{D-RSPT2}
\begin{figure}
    \centering
    \includegraphics[width=1\linewidth]{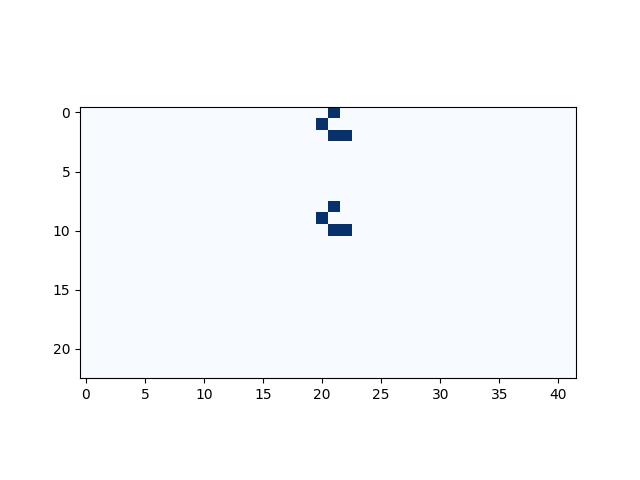}
    \caption{Pictorial illustration of configuration of multi-point strange correlator $D_{L}(\mathbf{q},\mathbf{f};x,y)$ of update rule Eq.~(\ref{rule_ivb}), which generates an RSPT phase (nontrivial terms are shown in blue cubes). The initial condition is given by Eq.~(\ref{ivb1}).}
    \label{sc_ivb}
\end{figure}

For model-IVb (Eq.~(\ref{rule_ivb})), we expect that criterion 1 can be satisfied. We observe that the multi-point strange correlator
\begin{equation}\label{sc_RSPT2}
	C(\{r_i\})=\frac{\bra{ \Omega}X \begin{pmatrix}
 0\\
			D_{k}(\mathbf{q},\mathbf{f};x,y)
		\end{pmatrix} \ket{\Psi}}{\braket{\Omega}{\Psi}},~k\in N_+,
\end{equation}
satisfies
\begin{equation}
	\mathscr{N}(\mathbf{q},\mathbf{f})=1,
\end{equation}
when the initial condition $\mathbf{q}$ is set to be Eq.~(\ref{ivb1}). The correlation number of the configurations above are
\begin{equation}
	\mathfrak{n}=10,
\end{equation} 
of which an example is shown in Fig.~\ref{sc_ivb}.

\section{Multi-point strange correlator and spurious topological entanglement entropy}\label{MPSC+STEE}
Firstly studied in \cite{STEEa}, it has been pointed out in Ref.~\cite{spurious} that the extraction of \textit{topological entanglement entropy} (TEE) $ -  \gamma $ via $S_{\text{topo}}$ (e.g. prescriptions proposed by Levin-Wen \cite{LW} and Kitaev-Preskill \cite{KP}) can suffer from spurious contributions from the nonlocal string order in the SSPT order, and this spurious contribution is extensively studied \cite{STEEb,STEEc,STEEd}. Authors of Ref.~\cite{spurious} have shown explicitly the string-like nonlocal stabilizer generators that spread across the boundary of subregions can contribute to the $S_{\text{topo}}$ in 2D cluster model, and have proposed a quantity $S_{\text{dumb}}$ to detect this spurious contribution. 

In this work, we want to show that MPSC are closely related to the spurious contributions in these models, and the spurious contributions in calculating TEE exist in a large variety of SPT orderes protected by subsystem symmetry. We will show how spurious topological entanglement entropy (STEE) appears in the HGSPT order, and how the configurations of the nonlocal string-like stabilizers that contribute to STEE are exactly mapped to the spatial distributions of local operators in MPSC that can detect the nontrivial SRE ground state of an HGSPT order. 

In the next following sections, we will discuss what kind of nonlocal stabilizers can contribute to STEE in Sec.~\ref{gen_disc}, explore the connection between STEE and MPSC in Sec.~\ref{con_mpsc}, and show concrete models in Sec.~\ref{model_study}.

\subsection{Nonlocal stabilizers contributing to $S_{\text{topo}}$}\label{gen_disc}
In this subsection we will discuss what kind of nonlocal stabilizers can finally contribute to the calculation of $S_{\text{topo}}$.

While extracting the topological entanglement entropy of a given physical model by means of tripartitions (e.g. Kitaev-Preskill and Levin-Wen), we argue that the calculation can be massively simplified by counting only a special type of nonlocal stabilizer generators. First we start with the entanglement entropy of a specific subregion $A$ in the system, which is given by

$$
S_{A}=-\text{tr } \rho_{A}\log_{2}\rho_{A}=N_{A}-\log_{2}|G_{A}|,
$$
where $N_{A}$ is the number of qubits contained in region $A$, $G_{A}$ is the stabilizer group whose elements are fully supported in $A$. In gapped spin systems, the entropy scales with

$$
S_{A}\sim c|\partial A|-\gamma.
$$
Here $c$ is some non-universal constant, and the $-\gamma$ term is the so-called \emph{topological entanglement entropy} (TEE), which is a universal constant. Kitaev and Preskill as well as Levin and Wen proposed two tripartitions and used linear combinations of entanglement entropy of these parts to cancel out the non-universal constant and extract the constant term $-\gamma$ in the scaling behavior of the entanglement entropy. The corresponding quantity is named \emph{topological entropy} $S_{\text{topo}}$, defined as

$$
S_{\text{topo}}\equiv S_{A}+S_{B}+S_{C}-S_{AB}-S_{AC}-S_{BC}+S_{ABC}.
$$
It is argued that the combinations in $S_{\text{topo}}$ managed to cancel out all the boundary and corner contributions in the entanglement entropy, leaving

$$
S_{\text{topo}}=-\gamma.
$$
However it is pointed out in \cite{spurious} that $S_{\text{topo}}$ is not fully topologically invariant, being sensitive to the deformation of the region boundaries in the subsystem symmetry-protected topological (SSPT) phase. In SSPT phase, this extraction process may suffer from unwanted contributions due to long-range string order, giving a nonzero $S_{\text{topo}}$ in SSPT phase, being devoid of the topological order, which is different from our expectation that $-\gamma=0$ in the SSPT phase.

Here we want to systematically explore the origin of the spurious contribution $\Delta S\equiv S_{\text{topo}}+\gamma$ in the realm of the lattice model.

First we observe that

$$
\begin{aligned}
S_{AB}&=N_{AB}-\log_{2}|G_{AB}|\\ 
&=N_{A}+N_{B}-\log_{2}|G_{A}|-\log_{2}|G_{B}|\\
&\quad -\log_{2}|G_{A\wedge B}|+\log_{2}|G_{A \vee B}|\\ 
&=N_{A}+N_{B}-O_{A}-O_{B}-O_{A \wedge B}+O_{A \vee B}.
\end{aligned}
$$
$N_{A \wedge B}$ is the number of qubits that exist in $AB$ but not in $A$ or $B$. $O_{A}$ is the number of independent stabilizer generators in $A$, and $O_{A \wedge B}$ is the number of independent stabilizer generators that exists in $AB$ but not in $A$ or $B$. $O_{A \vee B}$ is the number of independent stabilizer generators that exists in $A$ or $B$, but become no longer independent in $AB$. The following relationship holds:

$$
O_{AB}=O_{A}+O_{B}+O_{A \wedge B}-O_{A \vee B}.
$$
Similar observations can be found for $S_{ABC}$:

$$
\begin{aligned}
S_{ABC}&=N_{ABC}-\log_{2}|G_{ABC}| \\
&=N_{AB}+N_{C}-O_{AB}-O_{C}-O_{AB \wedge C}+O_{AB \vee C} \\
&= N_{A}+N_{B}+N_{C}-O_{A}-O_{B}-O_{A \wedge B}\\
&\quad +O_{A \vee B}-O_{C}-O_{AB \wedge C}+ O_{AB \vee C}.
\end{aligned}
$$
Adding up each term in $S_{\text{topo}}$, we can explicitly observe what terms are cancelled out in the linear combination. The calculation process are shown below:

$$
\begin{aligned}
S_{\text{topo}}&=S_{A}+S_{B}+S_{C}-S_{AB}-S_{AC}-S_{BC}+S_{ABC} \\ 
&=-O_{A}-O_{B}-O_{C}+O_{AB}+O_{AC}+O_{BC}-O_{ABC}  \\
&= -(O_{AB \wedge C}-O_{A \wedge C}-O_{B \wedge C})\\
&\quad +(O_{AB \vee C}-O_{A \vee C}- O_{B \vee C}). \\
\end{aligned}
$$
To explore what kind of operators can contribute to $S_{\text{topo}}$ eventually, we introduce some notation to keep track of terms that a specific stabilizer generator enters.
The distribution vector $\mathbf{D}(g)$ of a valid generator (to be explained below) $g$ is defined as

$$
\mathbf{D}(g)\equiv (T(g),P(g)),
$$
where $T(g)$ denotes the number of basic partitions (i.e. the unions of partitions are not included here) that $g$ as a total has a support in. $P(g)$ denotes the number of basic partitions that supports at least one local Hamiltonian term making up the generator. From now on, we will denote the support of the operator as \textit{total support} and the union of supports of Hamiltonian terms that make up the operator as \textit{partial support}. To specify the contribution of $g$ in $S_{\text{topo}}$, we further define the contribution vector $\mathbf{C}(g)$, which writes:

$$
\mathbf{C}(g)\equiv (C_{1}(g),C_{2}(g),\dots,C_{N}(g)),
$$
where $N$ is the number of partitions and $C_{i}(g)$ denotes how many times does $g$ appear as the minimal generators in $i$-th order region. An $i$-th order region is the union of $i$ basic partitions (e.g. $ABC$ is a $3$-rd order region), and a minimal generator in $i$-th order region is a generator that cannot be written as products of generators in $j$-th ($j<i$) regions with less area of support.
We see explicitly in the definition of $\mathbf{C}(g)$ that the contribution of $g$ in $S_{\text{topo}}$ is

$$
\Delta S(g)=\sum_{i=1}^N (-)^{i}C_{i}(g).
$$
Through enumerations we find that for $N=3$ case (including LW,KP prescription):

\begin{equation}\label{stopo2}
|\{g:T(g)=i,~P(g)=3\}|=|\{g:C_{j}(g)=\delta_{ij}\}|.
\end{equation}
And by calculation we observe that only the stabilizer generators with $P(g)=3$ have a nonzero $\Delta S(g)$ in $N=3$ case (LW, KP prescription). We define the power of set $|\{g:T(g)=i,~P(g)=j\}|$ as $Q(i,j)$. Then we obtain:

\begin{equation}\label{stopo}
S_{\text{topo}}^{(3)}=\sum_{i=1}^3 (-)^i Q(i,3).
\end{equation}
This quantity shows immediate potential to be generalized to $N$-partitions.

Eq.~(\ref{stopo2}, \ref{stopo}) can be interpreted as following: The stablizers that finally contribute to the calculation of $S_{\text{topo}}$ are the ones that have a local support on all 3 subregions, and their contributions depend on how many subregions their global support have. With this in hand, we can massively simplify the calculation of $S_{\text{topo}}$ by counting these special nonlocal stabilizers only.

\subsection{Connection to multi-Point strange correlator}\label{con_mpsc}
In this subsection we want to show that the nonlocal stabilizers will exactly be the operator that gives nontrivial results in the multi-point strange correlator.

In Eq.~(\ref{stopo}) we conclude that the stabilizers which can contribute to the spurious value of TEE have partial support that span across all three partitions. This kind of stabilizer either appears at the triple intersection point of partitions (KP case), or serves as nonlocal stabilizers running along the boundaries (both LW and KP case). Next we will show that \textit{the stabilizer in these cases will exactly be the operator that gives nontrivial results in the multi-point strange correlator}. We will start with some explicit examples. In the following part of this section, we will show some nonlocal stabilizers that exists in some HGSPT models. These nonlocal stabilizers have nonzero contributions to the value of $S_{\text{topo}}$ in certain partition geometries. While it was pointed out in \cite{spurious} that the string-like nonlocal stabilizers in 2D cluster model can be detected by calculating a tripartite topological entropy $S_{\text{dumb}}$  in a dumbbell-like tripartition, we want to show the following facts:
\begin{outline}[enumerate]
    \1 In HGSPT models with generally more exotic subsystem symmetries, there will be nonlocal stabilizers that contribute to the spurious values of $S_\text{topo}$ that cannot be detected by the original definition of $S_{\text{dumb}}$.
    \1 There will also be string-like nonlocal stabilizers in the HGSPT models devoid of any line-like symmetries, which can also be detected by $S_{\text{dumb}}$. So there is no general correspondence between the presence of line-like symmetries and nonlocal long range string order proposed in \cite{spurious}.
\end{outline}

And it is worth noticing that these nonlocal stabilizers happen to be the configuration of operators that can detect the nontrivial SRE ground state of the corresponding HGSPT phase. It is natural to notice that both STEE and MPSC point out the fact that there are hidden long range order in the SPT phases protected by subsystem symmetries: STEE is the unexpected contribution to the TEE in the absence of the topological order, while MPSC is the hidden long range correlation behavior in a short range entangled ground state. By this exact relation we see that two quantities share the same physical origin.

In the following texts, we show some nonlocal stabilizers that can contribute to the tripartite topological entropy
\begin{equation}
    S_{\text{topo}}=S_A+S_B+S_C-S_{AB}-S_{AC}-S_{BC}+S_{ABC},
\end{equation}
where the tripartitions $A,B,C$ are denoted by blue, green, red areas respectively in the figure. 

As a reminder, we would like to clarify that when we say a nonlocal stabilizer is generated by a certain symmetry, we are actually saying that we pick certain rows from the symmetry pattern (Eq.~(\ref{eq_symm})) to be the initial condition that generates the MPSC (Eq.~(\ref{scconfig})), and the resulting local operator configuration in the MPSC is the nonlocal stabilizer generated by this symmetry. From the discussion above it is clear that a certain symmetry pattern can generate a huge amount of MPSC, but not all MPSC can serve as the nonlocal stabilizers that contribute to $S_{\text{topo}}$ calculation. Only the ones that fit the boundary geometry of the tripartition can potentially have a spurious contribution. So we will explicitly draw the boundary geometry that admits spurious contributions of nonlocal stabilizers in the following texts. For the sake of simplifying the picture, we exerted a coarse-graining procedure on the lattice, combining two sublattices. Now the model is defined on a 2D square lattice with 2 qubits per site. A general Pauli operator acting on a site will be represented as $O:=O_1O_2$, where $O_1$ denotes the operator acting on the sublattice $(a)$ and $O_2$ denotes the operator acting on sublattice $(b)$, respectively.

The reason why there exists such correspondence between the configuration of MPSC and the nonlocal stabilizers with spurious TEE contributions can be explained as follows:
\begin{outline}[enumerate]
    \1 Nonlocal stabilizers in the context of STEE are always made up of products of Hamiltonian terms along the boundary, which contain Hamiltonian terms with supports outside of the tripartition. However, the nonlocal stabilizer as a whole does not have support outside of the tripartition, making it an independent stabilizer generator of stabilizer group of area $ABC$ (denoted as $G_{ABC}$). Details are discussed in Section~\ref{gen_disc}.
    \1 Such nonlocal stabilzers naturally act trivially on the SRE ground state of the HGSPT phase as products of Hamiltonian terms, as shown in the second row of Eq.~(\ref{sc_derivation}).
\end{outline}
\subsection{Model study}\label{model_study}
From the observation above we can see that any product of operators with the form designed in Eq.~(\ref{scconfig}) automatically have the form of product of Hamiltonian terms therefore having the potential to contribute to STEE. The only thing we need to do is to analyze the boundary geometry that can admit such nonlocal stabilizers. In the following sections, we will show some concrete examples of nonlocal stabilizers together with the corresponding boundary geometries of tripartition. The possible nonlocal stabilizers that may appear in these models extend beyond the examples we will show below, so we will demonstrate some typical examples only.

\begin{table*}[htb]
    \centering
    \begin{tabular}{c|ccccc}
    \hline 
    \hline
        Nonlocal Stabilizer&Boundary Geometry &Orientation & HOCA rule $\mathbf{f}$  &Initial Condition $\mathbf{q}$ & Evolution Distance $L$\\
         \hline
        Fig.~\ref{ns_I_1}&Smooth&Horizontal & Eq.~(\ref{fl-MSPT1})  & $\begin{pmatrix}
            x+x^3+x^5+x^7\\
            1+x+x^2+x^3+x^4+x^5+x^6+x^7+x^8
        \end{pmatrix}$ & 0 \\
         \hline
         Fig.~\ref{ns_I_2}&Smooth&Diagonal & Eq.~(\ref{fl-MSPT1})  & $\begin{pmatrix}
             1\\x
         \end{pmatrix}$ & 5\\
         \hline
         Fig.~\ref{ns_II_1}&Smooth&Horizontal&Eq.~(\ref{ff-MSPT})&$\begin{pmatrix}
             x+x^2+x^5+x^6\\1+x+x^3+x^4+x^5+x^7\\x+x^2+x^3+x^4+x^5+x^6+x^7+x^8
         \end{pmatrix}$&0\\
         \hline
         Fig.~\ref{ns_iva_1}&Detached horizontally&Horizontal&Eq.~(\ref{periodic_rule})&$\begin{pmatrix}
             0\\1+x^2+x^4+x^6
         \end{pmatrix}$&0\\
         \hline
         Fig.~\ref{ns_iva_2}&Detached vertically&Horizontal&Eq.~(\ref{periodic_rule})&$\begin{pmatrix}
             0\\1+x^2+x^4+x^6
         \end{pmatrix}$&0\\
         \hline
         Fig.~\ref{ns_iva_3}& Smooth&Diagonal&Eq.~(\ref{periodic_rule})&$\begin{pmatrix}
             1\\x
         \end{pmatrix}$&5\\
         \hline
         Fig.~\ref{ns_ivb_1}&\makecell{Mostly smooth,\\locally deformed}&Vertical&Eq.~(\ref{rule_ivb})&$\begin{pmatrix}
             1\\1\\1
         \end{pmatrix}$&4\\
         \hline
         Fig.~\ref{ns_ivb_2}&Smooth&Horizontal&Eq.~(\ref{rule_ivb})&$\begin{pmatrix}
             x+x^2+x^5+x^6\\x+x^3+x^5+x^7\\1+x+x^2+x^3+x^4+x^5+x^6+x^7
         \end{pmatrix}$&0\\
         \hline
         Fig.~\ref{ns_va_1}&Smooth&Horizontal&Eq.~(\ref{sierrule})&$1+x+x^2+x^3+x^4+x^5+x^6+x^7$&0\\
         \hline
         Fig.~\ref{ns_vb_1}&Smooth&Horizontal&Eq.~(\ref{fiborule})&$1+x+x^3+x^4+x^6+x^7$&0\\
         \hline
         Fig.~\ref{ns_vb_2}&Staggered&Horizontal&Eq.~(\ref{fiborule})&$1+x^2+x^4+x^6$&0\\
         \hline
         \hline
    \end{tabular}
    \caption{Nonlocal stabilizers in the HGSPT models that gives spurious contributions to $S_{\text{topo}}$. The MPSC configurations $D_L(\mathbf{q},\mathbf{f};x,y)$ that correspond to the nonlocal stabilizers are shown in last three columns of the table.}
    \label{tab:i}
\end{table*}

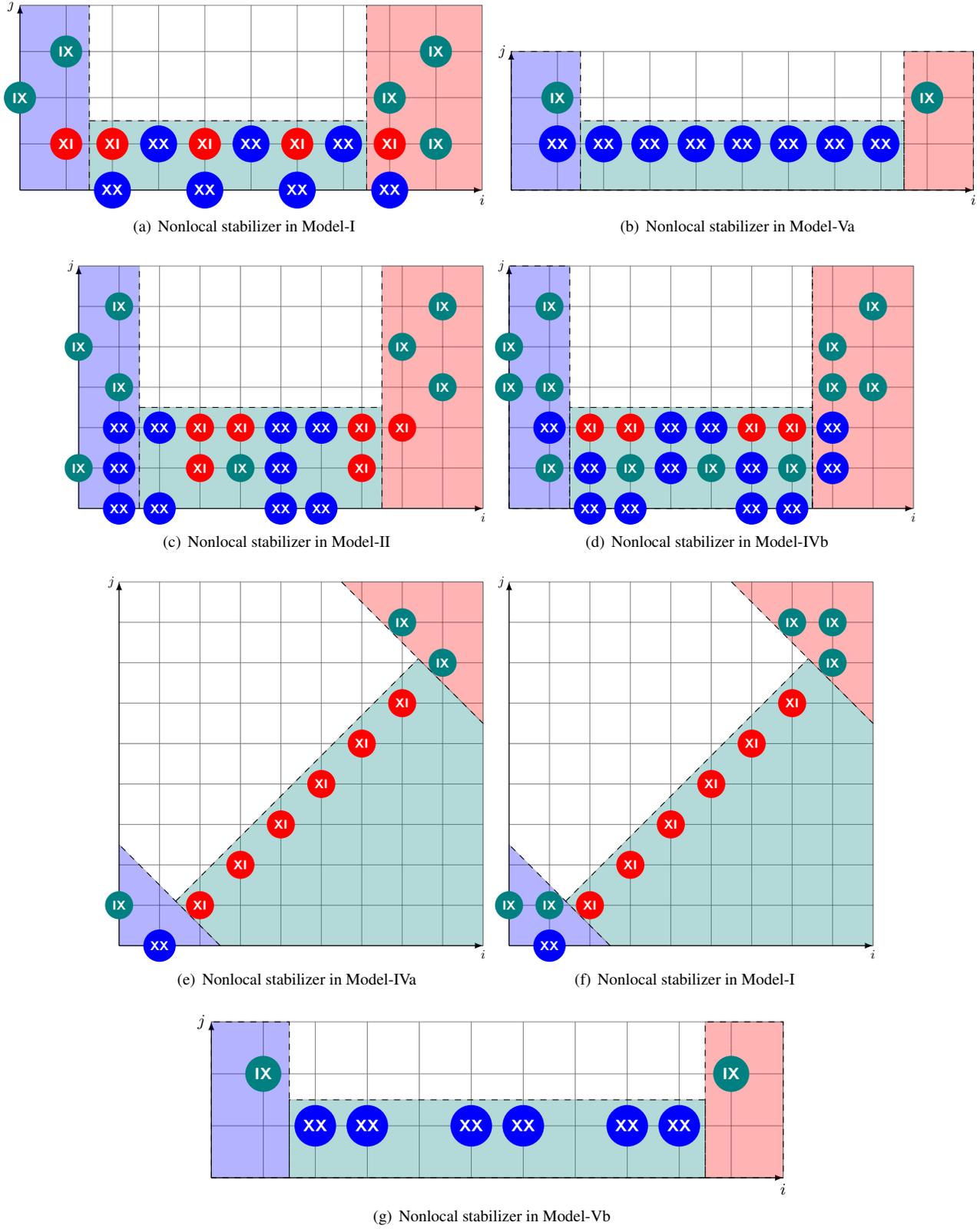
\begin{figure*}[htb]
    \centering
    \subfigure[Nonlocal stabilizer in Model-I \label{ns_I_1}]{\begin{tikzpicture}[scale=0.8,every node/.style={scale=0.8}]
\draw[help lines,step=1] (0,0) grid (10,4);
\draw[-latex] (0,0) -- (10,0);
\draw[-latex] (0,0) -- (0,4);
\draw (10,0)coordinate (A)node[below] {$i$};
\draw (0,4)coordinate (A)node[left] {$j$};
\filldraw[fill=blue,opacity=0.3](0,0)rectangle(1.5,4);
\filldraw[fill=teal,opacity=0.3](1.5,0)rectangle(7.5,1.5);
\filldraw[fill=red,opacity=0.3](7.5,0)rectangle(10,4);
\node[greenop] (a) at (0,2)
{\color{white}$\bm{\mathsf{IX}}$};
\node[redop] (a) at (1,1)
{\color{white}$\bm{\mathsf{XI}}$};
\node[greenop] (a) at (1,3)
{\color{white}$\bm{\mathsf{IX}}$};
\node[blueop] (a) at (2,0)
{\color{white}$\bm{\mathsf{XX}}$};
\node[redop] (a) at (2,1)
{\color{white}$\bm{\mathsf{XI}}$};
\node[blueop] (a) at (3,1)
{\color{white}$\bm{\mathsf{XX}}$};
\node[blueop] (a) at (4,0)
{\color{white}$\bm{\mathsf{XX}}$};
\node[redop] (a) at (4,1)
{\color{white}$\bm{\mathsf{XI}}$};
\node[blueop] (a) at (5,1)
{\color{white}$\bm{\mathsf{XX}}$};
\node[blueop] (a) at (6,0)
{\color{white}$\bm{\mathsf{XX}}$};
\node[redop] (a) at (6,1)
{\color{white}$\bm{\mathsf{XI}}$};
\node[blueop] (a) at (7,1)
{\color{white}$\bm{\mathsf{XX}}$};
\node[blueop] (a) at (8,0)
{\color{white}$\bm{\mathsf{XX}}$};
\node[redop] (a) at (8,1)
{\color{white}$\bm{\mathsf{XI}}$};
\node[greenop] (a) at (8,2)
{\color{white}$\bm{\mathsf{IX}}$};
\node[greenop] (a) at (9,1)
{\color{white}$\bm{\mathsf{IX}}$};
\node[greenop] (a) at (9,3)
{\color{white}$\bm{\mathsf{IX}}$};
\draw[dashed] (1.5,0)--(1.5,4);
\draw[dashed] (1.5,1.5)--(7.5,1.5);
\draw[dashed] (7.5,0)--(7.5,4);
\end{tikzpicture}}
    \subfigure[Nonlocal stabilizer in Model-Va\label{ns_va_1}]{\begin{tikzpicture}[scale=0.8,every node/.style={scale=0.8}]
\draw[help lines,step=1] (-1,-1) grid (9,2);
\draw[-latex] (-1,-1) -- (9,-1);
\draw[-latex] (-1,-1) -- (-1,2);
\draw[dashed] (-1,-1)rectangle(0.5,2);
\draw[dashed] (0.5,-1)rectangle(7.5,0.5);
\draw[dashed] (7.5,-1)rectangle(9,2);
\filldraw[fill=blue,opacity=0.3] (-1,-1)rectangle(0.5,2);
\filldraw[fill=teal,opacity=0.3] (0.5,-1)rectangle(7.5,0.5);
\filldraw[fill=red,opacity=0.3] (7.5,-1)rectangle(9,2);
\draw (9,-1)coordinate (A)node[below] {$i$};
\draw (-1,2)coordinate (A)node[left] {$j$};\node[blueop] (a) at (0,0)
{\color{white}$\bm{\mathsf{XX}}$};
\node[greenop] (a) at (0,1)
{\color{white}$\bm{\mathsf{IX}}$};
\node[blueop] (a) at (1,0)
{\color{white}$\bm{\mathsf{XX}}$};
\node[blueop] (a) at (2,0)
{\color{white}$\bm{\mathsf{XX}}$};
\node[blueop] (a) at (3,0)
{\color{white}$\bm{\mathsf{XX}}$};
\node[blueop] (a) at (4,0)
{\color{white}$\bm{\mathsf{XX}}$};
\node[blueop] (a) at (5,0)
{\color{white}$\bm{\mathsf{XX}}$};
\node[blueop] (a) at (6,0)
{\color{white}$\bm{\mathsf{XX}}$};
\node[blueop] (a) at (7,0)
{\color{white}$\bm{\mathsf{XX}}$};
\node[greenop] (a) at (8,1)
{\color{white}$\bm{\mathsf{IX}}$};
\end{tikzpicture}}
    \subfigure[Nonlocal stabilizer in Model-II\label{ns_II_1}]{\begin{tikzpicture}[scale=0.7,every node/.style={scale=0.7}]
\draw[help lines,step=1] (0,0) grid (10,6);
\draw[-latex] (0,0) -- (10,0);
\draw[-latex] (0,0) -- (0,6);
\draw (10,0)coordinate (A)node[below] {$i$};
\draw (0,6)coordinate (A)node[left] {$j$};
\draw[dashed] (1.5,0)--(1.5,6);
\draw[dashed] (1.5,2.5)--(7.5,2.5);
\draw[dashed] (7.5,0)--(7.5,6);
\filldraw[fill=blue,opacity=0.3](0,0)rectangle(1.5,6);
\filldraw[fill=teal,opacity=0.3](1.5,0)rectangle(7.5,2.5);
\filldraw[fill=red,opacity=0.3](7.5,0)rectangle(10,6);
\node[greenop] (a) at (0,1)
{\color{white}$\bm{\mathsf{IX}}$};
\node[greenop] (a) at (0,4)
{\color{white}$\bm{\mathsf{IX}}$};
\node[blueop] (a) at (1,0)
{\color{white}$\bm{\mathsf{XX}}$};
\node[blueop] (a) at (1,1)
{\color{white}$\bm{\mathsf{XX}}$};
\node[blueop] (a) at (1,2)
{\color{white}$\bm{\mathsf{XX}}$};
\node[greenop] (a) at (1,3)
{\color{white}$\bm{\mathsf{IX}}$};
\node[greenop] (a) at (1,5)
{\color{white}$\bm{\mathsf{IX}}$};
\node[blueop] (a) at (2,0)
{\color{white}$\bm{\mathsf{XX}}$};
\node[blueop] (a) at (2,2)
{\color{white}$\bm{\mathsf{XX}}$};
\node[redop] (a) at (3,1)
{\color{white}$\bm{\mathsf{XI}}$};
\node[redop] (a) at (3,2)
{\color{white}$\bm{\mathsf{XI}}$};
\node[greenop] (a) at (4,1)
{\color{white}$\bm{\mathsf{IX}}$};
\node[redop] (a) at (4,2)
{\color{white}$\bm{\mathsf{XI}}$};
\node[blueop] (a) at (5,0)
{\color{white}$\bm{\mathsf{XX}}$};
\node[blueop] (a) at (5,1)
{\color{white}$\bm{\mathsf{XX}}$};
\node[blueop] (a) at (5,2)
{\color{white}$\bm{\mathsf{XX}}$};
\node[blueop] (a) at (6,0)
{\color{white}$\bm{\mathsf{XX}}$};
\node[blueop] (a) at (6,2)
{\color{white}$\bm{\mathsf{XX}}$};
\node[redop] (a) at (7,1)
{\color{white}$\bm{\mathsf{XI}}$};
\node[redop] (a) at (7,2)
{\color{white}$\bm{\mathsf{XI}}$};
\node[redop] (a) at (8,2)
{\color{white}$\bm{\mathsf{XI}}$};
\node[greenop] (a) at (8,4)
{\color{white}$\bm{\mathsf{IX}}$};
\node[greenop] (a) at (9,3)
{\color{white}$\bm{\mathsf{IX}}$};
\node[greenop] (a) at (9,5)
{\color{white}$\bm{\mathsf{IX}}$};
\end{tikzpicture}}
    \subfigure[Nonlocal stabilizer in Model-IVb\label{ns_ivb_2}]{\begin{tikzpicture}[scale=0.7,every node/.style={scale=0.7}]
\draw[help lines,step=1] (0,0) grid (10,6);
\draw[-latex] (0,0) -- (10,0);
\draw[-latex] (0,0) -- (0,6);
\draw (10,0)coordinate (A)node[below] {$i$};
\draw (0,6)coordinate (A)node[left] {$j$};
\draw[dashed] (0,0)rectangle(1.5,6);
\draw[dashed] (1.5,0)rectangle(7.5,2.5);
\draw[dashed] (7.5,0)rectangle(7.5,6);
\filldraw[fill=blue,opacity=0.3] (0,0)rectangle(1.5,6);
\filldraw[fill=teal,opacity=0.3] (1.5,0)rectangle(7.5,2.5);
\filldraw[fill=red,opacity=0.3] (7.5,0)rectangle(10,6);
\node[greenop] (a) at (0,3)
{\color{white}$\bm{\mathsf{IX}}$};
\node[greenop] (a) at (0,4)
{\color{white}$\bm{\mathsf{IX}}$};
\node[greenop] (a) at (1,1)
{\color{white}$\bm{\mathsf{IX}}$};
\node[blueop] (a) at (1,2)
{\color{white}$\bm{\mathsf{XX}}$};
\node[greenop] (a) at (1,3)
{\color{white}$\bm{\mathsf{IX}}$};
\node[greenop] (a) at (1,5)
{\color{white}$\bm{\mathsf{IX}}$};
\node[blueop] (a) at (2,0)
{\color{white}$\bm{\mathsf{XX}}$};
\node[blueop] (a) at (2,1)
{\color{white}$\bm{\mathsf{XX}}$};
\node[redop] (a) at (2,2)
{\color{white}$\bm{\mathsf{XI}}$};
\node[blueop] (a) at (3,0)
{\color{white}$\bm{\mathsf{XX}}$};
\node[greenop] (a) at (3,1)
{\color{white}$\bm{\mathsf{IX}}$};
\node[redop] (a) at (3,2)
{\color{white}$\bm{\mathsf{XI}}$};
\node[blueop] (a) at (4,1)
{\color{white}$\bm{\mathsf{XX}}$};
\node[blueop] (a) at (4,2)
{\color{white}$\bm{\mathsf{XX}}$};
\node[greenop] (a) at (5,1)
{\color{white}$\bm{\mathsf{IX}}$};
\node[blueop] (a) at (5,2)
{\color{white}$\bm{\mathsf{XX}}$};
\node[blueop] (a) at (6,0)
{\color{white}$\bm{\mathsf{XX}}$};
\node[blueop] (a) at (6,1)
{\color{white}$\bm{\mathsf{XX}}$};
\node[redop] (a) at (6,2)
{\color{white}$\bm{\mathsf{XI}}$};
\node[blueop] (a) at (7,0)
{\color{white}$\bm{\mathsf{XX}}$};
\node[greenop] (a) at (7,1)
{\color{white}$\bm{\mathsf{IX}}$};
\node[redop] (a) at (7,2)
{\color{white}$\bm{\mathsf{XI}}$};
\node[blueop] (a) at (8,1)
{\color{white}$\bm{\mathsf{XX}}$};
\node[blueop] (a) at (8,2)
{\color{white}$\bm{\mathsf{XX}}$};
\node[greenop] (a) at (8,3)
{\color{white}$\bm{\mathsf{IX}}$};
\node[greenop] (a) at (8,4)
{\color{white}$\bm{\mathsf{IX}}$};
\node[greenop] (a) at (9,3)
{\color{white}$\bm{\mathsf{IX}}$};
\node[greenop] (a) at (9,5)
{\color{white}$\bm{\mathsf{IX}}$};
\end{tikzpicture}}
    \subfigure[Nonlocal stabilizer in Model-IVa\label{ns_iva_3}]{\begin{tikzpicture}[scale=0.7,every node/.style={scale=0.7}]
\draw[help lines,step=1] (0,0) grid (9,9);
\draw[-latex] (0,0) -- (9,0);
\draw[-latex] (0,0) -- (0,9);
\draw (9,0)coordinate (A)node[below] {$i$};
\draw (0,9)coordinate (A)node[left] {$j$};
\draw[dashed] (2.5,0)--(0,2.5);
\draw[dashed] (1.4,1.1)--(7.4,7.1);
\draw[dashed] (5.5,9)--(9,5.5);
\filldraw[fill=blue,opacity=0.3] (0,0)--(2.5,0)--(0,2.5)--(0,0);
\filldraw[fill=teal,opacity=0.3] (2.5,0)--(1.4,1.1)--(7.4,7.1)--(9,5.5)--(9,0)--(2.5,0);
\filldraw[fill=red,opacity=0.3] (5.5,9)--(9,5.5)--(9,9)--(5.5,9);
\node[greenop] (a) at (0,1)
{\color{white}$\bm{\mathsf{IX}}$};
\node[blueop] (a) at (1,0)
{\color{white}$\bm{\mathsf{XX}}$};
\node[redop] (a) at (2,1)
{\color{white}$\bm{\mathsf{XI}}$};
\node[redop] (a) at (3,2)
{\color{white}$\bm{\mathsf{XI}}$};
\node[redop] (a) at (4,3)
{\color{white}$\bm{\mathsf{XI}}$};
\node[redop] (a) at (5,4)
{\color{white}$\bm{\mathsf{XI}}$};
\node[redop] (a) at (6,5)
{\color{white}$\bm{\mathsf{XI}}$};
\node[redop] (a) at (7,6)
{\color{white}$\bm{\mathsf{XI}}$};
\node[greenop] (a) at (7,8)
{\color{white}$\bm{\mathsf{IX}}$};
\node[greenop] (a) at (8,7)
{\color{white}$\bm{\mathsf{IX}}$};
\end{tikzpicture}}
    \subfigure[Nonlocal stabilizer in Model-I \label{ns_I_2}]{\begin{tikzpicture}[scale=0.7,every node/.style={scale=0.7}]
\draw[help lines,step=1] (0,0) grid (9,9);
\draw[-latex] (0,0) -- (9,0);
\draw[-latex] (0,0) -- (0,9);
\draw (9,0)coordinate (A)node[below] {$i$};
\draw (0,9)coordinate (A)node[left] {$j$};
\draw[dashed] (2.5,0)--(0,2.5);
\draw[dashed] (1.4,1.1)--(7.4,7.1);
\draw[dashed] (5.5,9)--(9,5.5);
\filldraw[fill=blue,opacity=0.3] (0,0)--(2.5,0)--(0,2.5)--(0,0);
\filldraw[fill=teal,opacity=0.3] (2.5,0)--(1.4,1.1)--(7.4,7.1)--(9,5.5)--(9,0)--(2.5,0);
\filldraw[fill=red,opacity=0.3] (5.5,9)--(9,5.5)--(9,9)--(5.5,9);
\node[greenop] (a) at (0,1)
{\color{white}$\bm{\mathsf{IX}}$};
\node[blueop] (a) at (1,0)
{\color{white}$\bm{\mathsf{XX}}$};
\node[greenop] (a) at (1,1)
{\color{white}$\bm{\mathsf{IX}}$};
\node[redop] (a) at (2,1)
{\color{white}$\bm{\mathsf{XI}}$};
\node[redop] (a) at (3,2)
{\color{white}$\bm{\mathsf{XI}}$};
\node[redop] (a) at (4,3)
{\color{white}$\bm{\mathsf{XI}}$};
\node[redop] (a) at (5,4)
{\color{white}$\bm{\mathsf{XI}}$};
\node[redop] (a) at (6,5)
{\color{white}$\bm{\mathsf{XI}}$};
\node[redop] (a) at (7,6)
{\color{white}$\bm{\mathsf{XI}}$};
\node[greenop] (a) at (7,8)
{\color{white}$\bm{\mathsf{IX}}$};
\node[greenop] (a) at (8,7)
{\color{white}$\bm{\mathsf{IX}}$};
\node[greenop] (a) at (8,8)
{\color{white}$\bm{\mathsf{IX}}$};
\end{tikzpicture}}
    \subfigure[Nonlocal stabilizer in Model-Vb\label{ns_vb_1}]{\begin{tikzpicture}[scale=0.9,every node/.style={scale=0.9}]
\draw[help lines,step=1] (-1,-1) grid (10,2);
\draw[-latex] (-1,-1) -- (10,-1);
\draw[-latex] (-1,-1) -- (-1,2);
\draw[dashed] (-1,-1)rectangle(0.5,2);
\draw[dashed] (0.5,-1)rectangle(8.5,0.5);
\draw[dashed] (8.5,-1)rectangle(10,2);
\filldraw[fill=blue,opacity=0.3] (-1,-1)rectangle(0.5,2);
\filldraw[fill=teal,opacity=0.3] (0.5,-1)rectangle(8.5,0.5);
\filldraw[fill=red,opacity=0.3] (8.5,-1)rectangle(10,2);
\draw (10,-1)coordinate (A)node[below] {$i$};
\draw (-1,2)coordinate (A)node[left] {$j$};\node[greenop] (a) at (0,1)
{\color{white}$\bm{\mathsf{IX}}$};
\node[blueop] (a) at (1,0)
{\color{white}$\bm{\mathsf{XX}}$};
\node[blueop] (a) at (2,0)
{\color{white}$\bm{\mathsf{XX}}$};
\node[blueop] (a) at (4,0)
{\color{white}$\bm{\mathsf{XX}}$};
\node[blueop] (a) at (5,0)
{\color{white}$\bm{\mathsf{XX}}$};
\node[blueop] (a) at (7,0)
{\color{white}$\bm{\mathsf{XX}}$};
\node[blueop] (a) at (8,0)
{\color{white}$\bm{\mathsf{XX}}$};
\node[greenop] (a) at (9,1)
{\color{white}$\bm{\mathsf{IX}}$};
\end{tikzpicture}}
 \caption{Nonlocal stabilizers that can contribute to $S_\text{topo}$ in tripartition given in the figure. Red, green, and blue area is respectively the partition $A,B,C$. Such products of operators all have the same configuration with configuration of local operators in some certain MPSC. Nonlocal stabilizers in this figure can contribute to a smooth linear boundary geometry (dashed line surrounding green area).}
  \label{fig:STEE1}
\end{figure*}

\begin{figure*}[ht]
\centering
    \!\!\!\!\!\!\!\!\!\!\!\!\!\!\!\!\!\!\!\!\!\!\!\!\!\!\!\!\!\!\!\!\!\!\!\!\!\!\!\!\!\!\!\!\!\!\!\!\subfigure[Nonlocal stabilizer in Model-IVa\label{ns_iva_1}]{\begin{tikzpicture}[scale=0.75,every node/.style={scale=0.75}]
\draw[help lines,step=1] (-1,-1) grid (9,3);
\draw[-latex] (-1,-1) -- (9,-1);
\draw[-latex] (-1,-1) -- (-1,3);
\draw (9,-1)coordinate (A)node[below] {$i$};
\draw (-1,3)coordinate (A)node[left] {$j$};
\draw[dashed] (-1,-1)rectangle(0.5,3);
\draw[dashed] (2.5,3)rectangle(3.5,-1) (0.5,3)rectangle(1.5,-1);
\draw[dashed] (4.5,3)rectangle(5.5,-1) (6.5,3)rectangle(7.5,-1);
\draw[dashed] (7.5,3)rectangle(9,-1);
\filldraw[fill=blue,opacity=0.3](-1,-1)rectangle(0.5,3);
\filldraw[fill=teal,opacity=0.3](2.5,3)rectangle(3.5,-1) (0.5,3)rectangle(1.5,-1);
\filldraw[fill=teal,opacity=0.3](4.5,3)rectangle(5.5,-1) (6.5,3)rectangle(7.5,-1);
\filldraw[fill=red,opacity=0.3](7.5,3)rectangle(9,-1);

\node[greenop] (a) at (0,1)
{\color{white}$\bm{\mathsf{IX}}$};
\node[blueop] (a) at (1,0)
{\color{white}$\bm{\mathsf{XX}}$};
\node[greenop] (a) at (1,2)
{\color{white}$\bm{\mathsf{IX}}$};
\node[blueop] (a) at (3,0)
{\color{white}$\bm{\mathsf{XX}}$};
\node[greenop] (a) at (3,2)
{\color{white}$\bm{\mathsf{IX}}$};
\node[blueop] (a) at (5,0)
{\color{white}$\bm{\mathsf{XX}}$};
\node[greenop] (a) at (5,2)
{\color{white}$\bm{\mathsf{IX}}$};
\node[blueop] (a) at (7,0)
{\color{white}$\bm{\mathsf{XX}}$};
\node[greenop] (a) at (7,2)
{\color{white}$\bm{\mathsf{IX}}$};
\node[greenop] (a) at (8,1)
{\color{white}$\bm{\mathsf{IX}}$};
\end{tikzpicture}}
    \qquad\qquad\qquad\subfigure[Nonlocal stabilizer in Model-IVb\label{ns_ivb_1}]{\begin{tikzpicture}[scale=0.75,every node/.style={scale=0.75}]
\draw[help lines,step=1] (0,0) grid (3,10);
\draw[-latex] (0,0) -- (3,0);
\draw[-latex] (0,0) -- (0,10);
\draw (3,0)coordinate (A)node[below] {$i$};
\draw (0,10)coordinate (A)node[left] {$j$};
\draw[dashed] (0,0)rectangle(3,2.5);
\draw[dashed] (0,8.5)rectangle(3,10);
\filldraw[fill=blue,opacity=0.3] (0,0)rectangle(3,2.5);
    \draw[dashed,even odd rule] (0,2.5)--(0,8.5)--(0.5,8.5)--(0.5,7.5)--(1.5,7.5)--(1.5,2.5)--(0,2.5);
\filldraw[fill=teal,opacity=0.3,even odd rule] (0,2.5)--(0,8.5)--(0.5,8.5)--(0.5,7.5)--(1.5,7.5)--(1.5,2.5)--(0,2.5);
\filldraw[fill=red,opacity=0.3] (0,8.5)rectangle(3,10);
\node[greenop] (a) at (0,1)
{\color{white}$\bm{\mathsf{IX}}$};
\node[greenop] (a) at (0,8)
{\color{white}$\bm{\mathsf{IX}}$};
\node[blueop] (a) at (1,0)
{\color{white}$\bm{\mathsf{XX}}$};
\node[redop] (a) at (1,1)
{\color{white}$\bm{\mathsf{XI}}$};
\node[blueop] (a) at (1,2)
{\color{white}$\bm{\mathsf{XX}}$};
\node[redop] (a) at (1,3)
{\color{white}$\bm{\mathsf{XI}}$};
\node[redop] (a) at (1,4)
{\color{white}$\bm{\mathsf{XI}}$};
\node[redop] (a) at (1,5)
{\color{white}$\bm{\mathsf{XI}}$};
\node[redop] (a) at (1,6)
{\color{white}$\bm{\mathsf{XI}}$};
\node[greenop] (a) at (1,7)
{\color{white}$\bm{\mathsf{IX}}$};
\node[greenop] (a) at (1,9)
{\color{white}$\bm{\mathsf{IX}}$};
\node[greenop] (a) at (2,2)
{\color{white}$\bm{\mathsf{IX}}$};
\node[greenop] (a) at (2,9)
{\color{white}$\bm{\mathsf{IX}}$};
\end{tikzpicture}}
    \subfigure[Nonlocal stabilizer in Model-IVa\label{ns_iva_2}]{\begin{tikzpicture}[scale=0.75,every node/.style={scale=0.75}]
\draw[help lines,step=1] (-1,-1) grid (9,3);
\draw[-latex] (-1,-1) -- (9,-1);
\draw[-latex] (-1,-1) -- (-1,3);
\draw (9,-1)coordinate (A)node[below] {$i$};
\draw (-1,3)coordinate (A)node[left] {$j$};
\draw[dashed] (-1,-1)rectangle(0.5,3);
\draw[dashed] (0.5,1.5)rectangle(7.5,2.5);
\draw[dashed] (0.5,-0.5)rectangle(7.5,0.5);
\draw[dashed] (7.5,3)rectangle(9,-1);
\filldraw[fill=blue,opacity=0.3](-1,-1)rectangle(0.5,3);
\filldraw[fill=teal,opacity=0.3](0.5,1.5)rectangle(7.5,2.5);
\filldraw[fill=teal,opacity=0.3](0.5,-0.5)rectangle(7.5,0.5);
\filldraw[fill=red,opacity=0.3](7.5,3)rectangle(9,-1);

\node[greenop] (a) at (0,1)
{\color{white}$\bm{\mathsf{IX}}$};
\node[blueop] (a) at (1,0)
{\color{white}$\bm{\mathsf{XX}}$};
\node[greenop] (a) at (1,2)
{\color{white}$\bm{\mathsf{IX}}$};
\node[blueop] (a) at (3,0)
{\color{white}$\bm{\mathsf{XX}}$};
\node[greenop] (a) at (3,2)
{\color{white}$\bm{\mathsf{IX}}$};
\node[blueop] (a) at (5,0)
{\color{white}$\bm{\mathsf{XX}}$};
\node[greenop] (a) at (5,2)
{\color{white}$\bm{\mathsf{IX}}$};
\node[blueop] (a) at (7,0)
{\color{white}$\bm{\mathsf{XX}}$};
\node[greenop] (a) at (7,2)
{\color{white}$\bm{\mathsf{IX}}$};
\node[greenop] (a) at (8,1)
{\color{white}$\bm{\mathsf{IX}}$};
\end{tikzpicture}}
    \subfigure[Nonlocal stabilizer in Model-Vb\label{ns_vb_2}]{\begin{tikzpicture}[scale=0.75,every node/.style={scale=0.75}]
\draw[help lines,step=1] (0,-1) grid (9,2);
\draw[-latex] (0,-1) -- (9,-1);
\draw[-latex] (0,-1) -- (0,2);
\draw[dashed] (0,-1)rectangle(1.5,2);
\draw[dashed] (1.5,-1)--(1.5,0.5)--(2.5,0.5)--(2.5,1.5)--(3.5,1.5)--(3.5,0.5)--(4.5,0.5)--(4.5,1.5)--(5.5,1.5)--(5.5,0.5)--(6.5,0.5)--(6.5,1.5)--(7.5,1.5)--(7.5,-1)--(1.5,-1);
\draw[dashed] (7.5,-1)rectangle(9,2);
\filldraw[fill=blue,opacity=0.3] (0,-1)rectangle(1.5,2);
\filldraw[fill=teal,opacity=0.3] (1.5,-1)--(1.5,0.5)--(2.5,0.5)--(2.5,1.5)--(3.5,1.5)--(3.5,0.5)--(4.5,0.5)--(4.5,1.5)--(5.5,1.5)--(5.5,0.5)--(6.5,0.5)--(6.5,1.5)--(7.5,1.5)--(7.5,-1)--(1.5,-1);
\filldraw[fill=red,opacity=0.3] (7.5,-1)rectangle(9,2);
\draw (9,-1)coordinate (A)node[below] {$i$};
\draw (0,2)coordinate (A)node[left] {$j$};\node[greenop] (a) at (0,1)
{\color{white}$\bm{\mathsf{IX}}$};
\node[blueop] (a) at (1,0)
{\color{white}$\bm{\mathsf{XX}}$};
\node[greenop] (a) at (1,1)
{\color{white}$\bm{\mathsf{IX}}$};
\node[blueop] (a) at (3,0)
{\color{white}$\bm{\mathsf{XX}}$};
\node[greenop] (a) at (3,1)
{\color{white}$\bm{\mathsf{IX}}$};
\node[blueop] (a) at (5,0)
{\color{white}$\bm{\mathsf{XX}}$};
\node[greenop] (a) at (5,1)
{\color{white}$\bm{\mathsf{IX}}$};
\node[blueop] (a) at (7,0)
{\color{white}$\bm{\mathsf{XX}}$};
\node[greenop] (a) at (7,1)
{\color{white}$\bm{\mathsf{IX}}$};
\node[greenop] (a) at (8,1)
{\color{white}$\bm{\mathsf{IX}}$};
\end{tikzpicture}}
    \caption{Nonlocal stabilizers that can contribute to $S_\text{topo}$ in tripartition given in the figure. Red, green, and blue area is respectively the partition $A,B,C$. Such products of operators all have the same configuration with configuration of local operators in some certain MPSC. Nonlocal stabilizers in this figure can contribute to more exotic boundary geometry (dashed line surrounding green area).}
    \label{fig:STEE2}
\end{figure*}
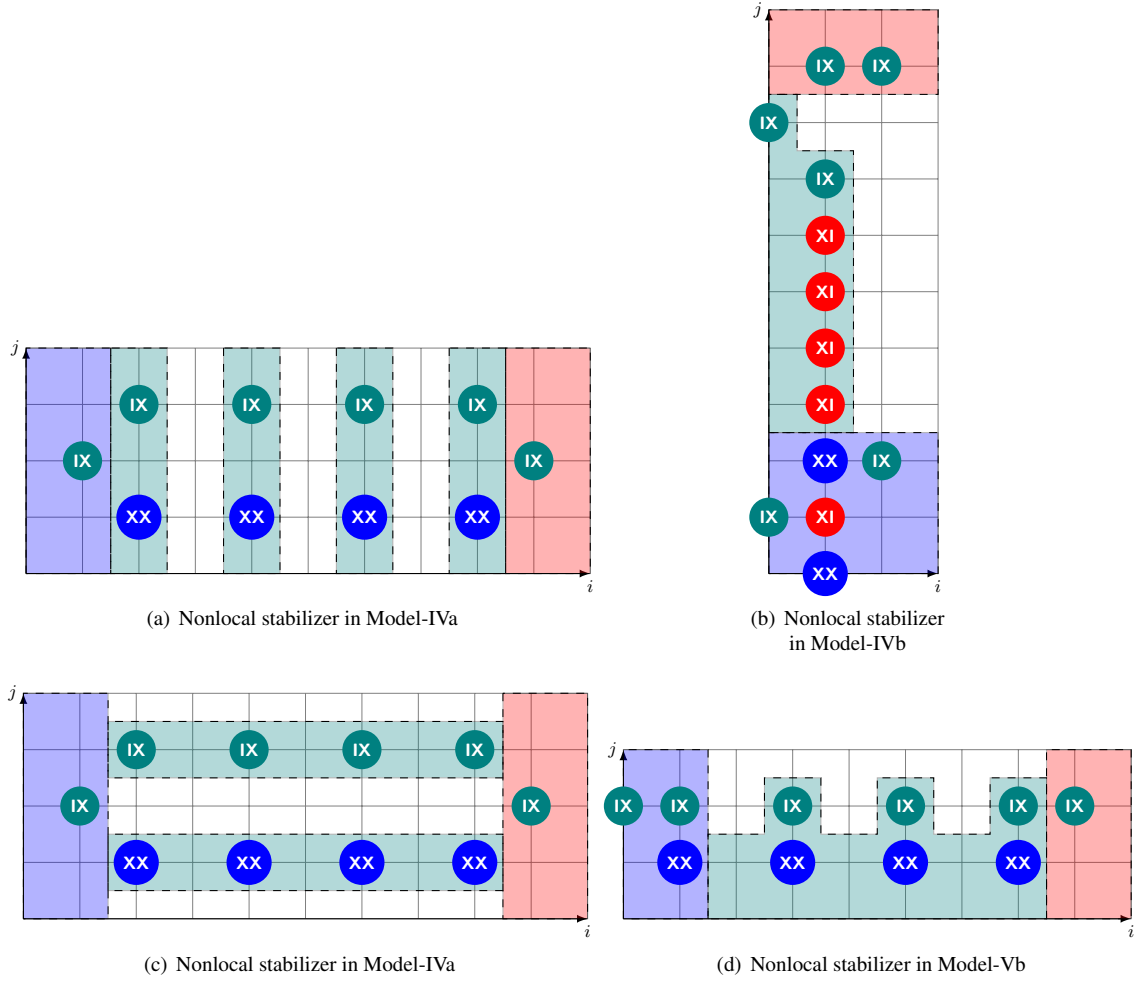

\begin{outline}[enumerate]
\1 \textbf{Nonlocal Stabilizers in Model-I}
Model-I (a I-MSPT model) possesses two types of subsystem symmetries, as mentioned in Fig.~\ref{mixCApic}. Each type of subsystem symmetry corresponds to a class of multi-point strange correlators, giving rise to a class of nonlocal stabilizers with the same geometry. An example of nonlocal stabilizers generated by the fractal-like symmetry (Fig.~\ref{p1}) is shown in Fig.~\ref{ns_I_1}, which gives a $\Delta S_{\text{topo}}=-1$ in the tripartite topological entropy. While the symmetry is fractal-like, the nonlocal stabilizer given by the symmetry has a string-like outlook, and can be prolonged in the $i$ direction by appropriately tuning the initial conditions that generates Fig.~\ref{ns_I_1}.

There also exists another type of nonlocal stabilizers with different directions, which is generated by the line-like symmetries (Fig.~\ref{p2}) in the model. 

The corresponding MPSC configuration $D_{L}(\mathbf{q},\mathbf{f};x,y)$ is shown in Table~\ref{tab:i}.

\1 \textbf{Nonlocal Stabilizers in Model-II}
Model-II (a II-MSPT model) possesses two types of subsystem symmetries, as mentioned in Fig.~\ref{mixCApic2}. Despite the lack of line-like symmetries, there are string-like nonlocal stabilizers that can contribute to $S_{\text{topo}}$ in this model. An example of nonlocal stabilizers generated by the fractal-like symmetry (Fig.~\ref{p7}) is shown in Fig.~\ref{ns_I_1}, which gives a $\Delta S_{\text{topo}}=-1$ in the tripartite topological entropy. The string-like stabilizer can be prolonged in the $i$ direction.

The corresponding MPSC configuration $D_{L}(\mathbf{q},\mathbf{f};x,y)$ is shown in Table~\ref{tab:i}.

\1 \textbf{Nonlocal Stabilizers in Model-III}
Model-III (a CSPT model) possesses chaotic-looking subsystem symmetries only, as mentioned in Fig.~\ref{otherpic}. So far we have not found any recognizable classes of nonlocal stabilizers that can contribute to the TEE due to the chaotic nature of the symmetry pattern.

\1 \textbf{Nonlocal Stabilizers in Model-IVa}
Model-IVa (an RSPT model) possesses regular subsystem symmetries, as mentioned in Fig.~\ref{periodic pic}. There are string-like nonlocal stabilizers that can run along a smooth boundary (Fig.~\ref{ns_iva_3}) and stabilizers that can fit into more peculiar boundary geometries (Fig.~\ref{ns_iva_1}, \ref{ns_iva_2}).

The corresponding MPSC configuration $D_{L}(\mathbf{q},\mathbf{f};x,y)$ is shown in Table~\ref{tab:i}.

\1 \textbf{Nonlocal Stabilizers in Model-IVb}
Model-IVb (an RSPT model) possesses regular subsystem symmetries, as mentioned in Fig.~\ref{pic_ivb}. There are string-like nonlocal stabilizers that can run along a smooth boundary (Fig.~\ref{ns_ivb_1}) and stabilizers that can fit into more peculiar boundary geometries (Fig.~\ref{ns_iva_1}, \ref{ns_iva_2}).

The corresponding MPSC configuration $D_{L}(\mathbf{q},\mathbf{f};x,y)$ is shown in Table~\ref{tab:i}.

\1 \textbf{Nonlocal Stabilizers in Model-Va}
Model-Va (Eq.~(\ref{sierrule}), an FSPT model) possesses fractal-like symmetries. Despite the lack of line-like symmetries, there are string-like nonlocal stabilizers that can contribute to $S_{\text{topo}}$ in this model. An example of nonlocal stabilizers generated by the fractal-like symmetry is shown in Fig.~\ref{ns_va_1}, which gives a $\Delta S_{\text{topo}}=-1$ in the tripartite topological entropy. The string-like stabilizer can be prolonged in the $i$ direction.

The corresponding MPSC configuration $D_{L}(\mathbf{q},\mathbf{f};x,y)$ is shown in Table~\ref{tab:i}.

\1 \textbf{Nonlocal Stabilizers in Model-Vb}
Model-Vb (an FSPT model) possesses fractal-like symmetries. Model-Vb is named Fibonacci FSPT in \cite{devakul_fractal_2019}, with the update rule
\begin{equation}\label{fiborule}
    f(x)=x^{-1}+1+x.
\end{equation}
Despite the lack of line-like symmetries\footnote{Generally, there will be row configurations in fractal symmetry patterns that look like a line. We are always referring to the whole time evolution when we are talking about the types of the symmetries. For example, for the fractal symmetry presented in Fig.~\ref{p1}, there are rows that look like a line in row 
 $1,3,7,15,...,2^k-1,...$, but these lines are actually slices of a fractal symmetry. In this paper, for consistency, we do not refer to these lines as ``line-like symmetries''.}, there are string-like nonlocal stabilizers that can contribute to $S_{\text{topo}}$ in this model. An example of nonlocal stabilizers generated by the fractal-like symmetry is shown in Fig.~\ref{ns_vb_1}, which gives a $\Delta S_{\text{topo}}=-1$ in the tripartite topological entropy. The string-like stabilizer can be prolonged in the $i$ direction. It is worth noticing that there are stabilizers in this model that can fit into a staggered boundary geometry, as shown in Fig.~\ref{ns_vb_2}.

\end{outline}

The corresponding MPSC configuration $D_{L}(\mathbf{q},\mathbf{f};x,y)$ is shown in Table~\ref{tab:i}.


\section{Summary and outlook}\label{summary}
In this work, we find exotic SPT phases protected by a variety of HOCA-generated symmetries. We identify HGSPT models with both fractal and line-like symmetries (e.g., Eq.~(\ref{MSPT1})), models with two distinct fractal symmetries (e.g., Eq.~(\ref{MSPT2})), and models with chaotic subsystem symmetries (e.g., Eq.~(\ref{MSPT3})). These models are derived from the HOCA rule, as explained in Section \ref{hgps}. We show that the framework of HOCA naturally encompasses these SPT phases protected by exotic symmetries and previously studied SSPT and FSPT phases, and we introduce labels that classify these SPT phases into different categories. To detect the nontrivial ground HGSPT order, we show the necessity of introducing multi-point strange correlators, which are a generalization of the strange correlator that involves more than two operators. The necessity is demonstrated by proving that all 2-point onsite strange correlators are trivial in the Sierpinski FSPT model, which is a fractal SPT model with a Sierpinski triangle symmetry (see Eq.~(\ref{FSPT})). This model can be recognized as an SPT phase generated by an order-1 CA, which can also be regarded as the simplest case of HOCA.  By examining the multi-point strange correlator of the given phase, we can determine the class of the phase. Also, we have found the relation between the multi-point strange correlator and the nonlocal stabilizers resulting in spurious topological entanglement entropy, revealing the connection between these two quantities showing long range behaviors in a short range entangleed state.

There are many interesting topics that remain unsolved. For example, while HOCA can be used to construct SPT models, the symmetries supported on HOCA-generated configurations can also be utilized to build other phases, including symmetry-breaking phases and symmetry-enriched topological (SET) orders, where the order-1 CA case is done in \cite{SET_CA}. Such future directions may require us to use more than one HOCA rule or use HOCA in higher dimensions, and HOCA patterns in these generalized conditions remain a future direction to explore, where the order-1 CA case is discussed in \cite{Devakul2021fractalizing}.  
 Moreover, the MPSC may be mapped to the membrane-like order parameters in the symmetry breaking models, of which some examples have previously shown in \cite{devakul_fractal_2019,zhou2021fractalquantumphasetransitions,doherty_identifying_2009}, showing the potential to probe new kinds of quantum criticality in symmetry breaking models. Another interesting topic is whether all types of subsystem symmetries can be generated by the HOCA framework, and how to develop a unified notation system to label miscellaneous subsystem symmetries. Finally, as chaotic patterns are realized as symmetries in CSPT ordered states, we may expect such quantum states to have diverse applications in computer science involving chaotic systems, such as for encryption \cite{HOCA-C}. Moreover, the multi-point strange correlators in HGSPT models can be studied by the Monte Carlo method, where the strange correlators of RSPT models have been studied in Ref.~\cite{zhou_detecting_2022}. Entanglements in HGSPT models are also intriguing. For example, whether there are spurious topological entanglement entropy \cite{spurious} in HGSPT models is an intriguing problem. Whether we can probe the HGSPT order via non-Hermitian perturbation \cite{nh1} and study the non-Hermitian entanglement in HGSPT order \cite{nh2,nh3} are also promising future directions. It will be interesting to probe the HGSPT in the realm of average symmetry-protected topological (ASPT) order \cite{ASPT1,ASPT2}, searching for peculiar behaviors of HGSPT models. Besides, using the relation between MPSC and STEE, we may probe a new way of detecting phases of matter via entanglements with geometric properties, paving the way for a general way to characterize exactly solvable models by analyzing their Hamiltonian terms without actually solving the model. Finally, because of the duality between the symmetry breaking  model and the HGSPT model, it will be interesting to map the MPSC back to the symmetry breaking models to detect new kinds of quantum criticality.

	\acknowledgements
This work was in part supported by National Natural Science Foundation of China (NSFC) Grant No. 12074438. The calculations reported were performed on resources provided by the Guangdong Provincial Key Laboratory of Magnetoelectric Physics and Devices, No. 2022B1212010008.


\begin{thebibliography}{177}%
\makeatletter
\providecommand \@ifxundefined [1]{%
 \@ifx{#1\undefined}
}%
\providecommand \@ifnum [1]{%
 \ifnum #1\expandafter \@firstoftwo
 \else \expandafter \@secondoftwo
 \fi
}%
\providecommand \@ifx [1]{%
 \ifx #1\expandafter \@firstoftwo
 \else \expandafter \@secondoftwo
 \fi
}%
\providecommand \natexlab [1]{#1}%
\providecommand \enquote  [1]{``#1''}%
\providecommand \bibnamefont  [1]{#1}%
\providecommand \bibfnamefont [1]{#1}%
\providecommand \citenamefont [1]{#1}%
\providecommand \href@noop [0]{\@secondoftwo}%
\providecommand \href [0]{\begingroup \@sanitize@url \@href}%
\providecommand \@href[1]{\@@startlink{#1}\@@href}%
\providecommand \@@href[1]{\endgroup#1\@@endlink}%
\providecommand \@sanitize@url [0]{\catcode `\\12\catcode `\$12\catcode `\&12\catcode `\#12\catcode `\^12\catcode `\_12\catcode `\%12\relax}%
\providecommand \@@startlink[1]{}%
\providecommand \@@endlink[0]{}%
\providecommand \url  [0]{\begingroup\@sanitize@url \@url }%
\providecommand \@url [1]{\endgroup\@href {#1}{\urlprefix }}%
\providecommand \urlprefix  [0]{URL }%
\providecommand \Eprint [0]{\href }%
\providecommand \doibase [0]{https://doi.org/}%
\providecommand \selectlanguage [0]{\@gobble}%
\providecommand \bibinfo  [0]{\@secondoftwo}%
\providecommand \bibfield  [0]{\@secondoftwo}%
\providecommand \translation [1]{[#1]}%
\providecommand \BibitemOpen [0]{}%
\providecommand \bibitemStop [0]{}%
\providecommand \bibitemNoStop [0]{.\EOS\space}%
\providecommand \EOS [0]{\spacefactor3000\relax}%
\providecommand \BibitemShut  [1]{\csname bibitem#1\endcsname}%
\let\auto@bib@innerbib\@empty
\bibitem [{\citenamefont {Wolfram}(1984{\natexlab{a}})}]{wolfram_cellular_1984}%
  \BibitemOpen
  \bibfield  {author} {\bibinfo {author} {\bibfnamefont {S.}~\bibnamefont {Wolfram}},\ }\bibfield  {title} {\bibinfo {title} {Cellular automata as models of complexity},\ }\href {https://doi.org/10.1038/311419a0} {\bibfield  {journal} {\bibinfo  {journal} {Nature}\ }\textbf {\bibinfo {volume} {311}},\ \bibinfo {pages} {419} (\bibinfo {year} {1984}{\natexlab{a}})}\BibitemShut {NoStop}%
\bibitem [{\citenamefont {Blecic}\ \emph {et~al.}(2013)\citenamefont {Blecic}, \citenamefont {Cecchini},\ and\ \citenamefont {Trunfio}}]{Blecic2013CellularAS}%
  \BibitemOpen
  \bibfield  {author} {\bibinfo {author} {\bibfnamefont {I.}~\bibnamefont {Blecic}}, \bibinfo {author} {\bibfnamefont {A.}~\bibnamefont {Cecchini}},\ and\ \bibinfo {author} {\bibfnamefont {G.~A.}\ \bibnamefont {Trunfio}},\ }\bibfield  {title} {\bibinfo {title} {Cellular automata simulation of urban dynamics through gpgpu},\ }\href {https://api.semanticscholar.org/CorpusID:6362393} {\bibfield  {journal} {\bibinfo  {journal} {The Journal of Supercomputing}\ }\textbf {\bibinfo {volume} {65}},\ \bibinfo {pages} {614} (\bibinfo {year} {2013})}\BibitemShut {NoStop}%
\bibitem [{\citenamefont {Guan}\ and\ \citenamefont {Clarke}(2010)}]{Guan2010AGP}%
  \BibitemOpen
  \bibfield  {author} {\bibinfo {author} {\bibfnamefont {Q.}~\bibnamefont {Guan}}\ and\ \bibinfo {author} {\bibfnamefont {K.~C.}\ \bibnamefont {Clarke}},\ }\bibfield  {title} {\bibinfo {title} {A general-purpose parallel raster processing programming library test application using a geographic cellular automata model},\ }\href {https://api.semanticscholar.org/CorpusID:17065636} {\bibfield  {journal} {\bibinfo  {journal} {International Journal of Geographical Information Science}\ }\textbf {\bibinfo {volume} {24}},\ \bibinfo {pages} {695 } (\bibinfo {year} {2010})}\BibitemShut {NoStop}%
\bibitem [{\citenamefont {Zhou}\ \emph {et~al.}(2022)\citenamefont {Zhou}, \citenamefont {Li}, \citenamefont {Yan}, \citenamefont {Ye},\ and\ \citenamefont {Meng}}]{zhou_detecting_2022}%
  \BibitemOpen
  \bibfield  {author} {\bibinfo {author} {\bibfnamefont {C.}~\bibnamefont {Zhou}}, \bibinfo {author} {\bibfnamefont {M.-Y.}\ \bibnamefont {Li}}, \bibinfo {author} {\bibfnamefont {Z.}~\bibnamefont {Yan}}, \bibinfo {author} {\bibfnamefont {P.}~\bibnamefont {Ye}},\ and\ \bibinfo {author} {\bibfnamefont {Z.~Y.}\ \bibnamefont {Meng}},\ }\bibfield  {title} {\bibinfo {title} {Detecting subsystem symmetry protected topological order through strange correlators},\ }\href {https://doi.org/10.1103/PhysRevB.106.214428} {\bibfield  {journal} {\bibinfo  {journal} {Phys. Rev. B}\ }\textbf {\bibinfo {volume} {106}},\ \bibinfo {pages} {214428} (\bibinfo {year} {2022})}\BibitemShut {NoStop}%
\bibitem [{\citenamefont {You}\ \emph {et~al.}(2018)\citenamefont {You}, \citenamefont {Devakul}, \citenamefont {Burnell},\ and\ \citenamefont {Sondhi}}]{you2018a}%
  \BibitemOpen
  \bibfield  {author} {\bibinfo {author} {\bibfnamefont {Y.}~\bibnamefont {You}}, \bibinfo {author} {\bibfnamefont {T.}~\bibnamefont {Devakul}}, \bibinfo {author} {\bibfnamefont {F.~J.}\ \bibnamefont {Burnell}},\ and\ \bibinfo {author} {\bibfnamefont {S.~L.}\ \bibnamefont {Sondhi}},\ }\bibfield  {title} {\bibinfo {title} {Subsystem symmetry protected topological order},\ }\href {https://link.aps.org/doi/10.1103/PhysRevB.98.035112} {\bibfield  {journal} {\bibinfo  {journal} {Phys. Rev. B}\ }\textbf {\bibinfo {volume} {98}},\ \bibinfo {pages} {035112} (\bibinfo {year} {2018})}\BibitemShut {NoStop}%
\bibitem [{\citenamefont {Devakul}\ \emph {et~al.}(2019)\citenamefont {Devakul}, \citenamefont {You}, \citenamefont {Burnell},\ and\ \citenamefont {Sondhi}}]{devakul_fractal_2019}%
  \BibitemOpen
  \bibfield  {author} {\bibinfo {author} {\bibfnamefont {T.}~\bibnamefont {Devakul}}, \bibinfo {author} {\bibfnamefont {Y.}~\bibnamefont {You}}, \bibinfo {author} {\bibfnamefont {F.~J.}\ \bibnamefont {Burnell}},\ and\ \bibinfo {author} {\bibfnamefont {S.~L.}\ \bibnamefont {Sondhi}},\ }\bibfield  {title} {\bibinfo {title} {{Fractal Symmetric Phases of Matter}},\ }\href {https://scipost.org/10.21468/SciPostPhys.6.1.007} {\bibfield  {journal} {\bibinfo  {journal} {SciPost Phys.}\ }\textbf {\bibinfo {volume} {6}},\ \bibinfo {pages} {007} (\bibinfo {year} {2019})}\BibitemShut {NoStop}%
\bibitem [{\citenamefont {Feldmeier}\ \emph {et~al.}(2020)\citenamefont {Feldmeier}, \citenamefont {Sala}, \citenamefont {De~Tomasi}, \citenamefont {Pollmann},\ and\ \citenamefont {Knap}}]{CA1}%
  \BibitemOpen
  \bibfield  {author} {\bibinfo {author} {\bibfnamefont {J.}~\bibnamefont {Feldmeier}}, \bibinfo {author} {\bibfnamefont {P.}~\bibnamefont {Sala}}, \bibinfo {author} {\bibfnamefont {G.}~\bibnamefont {De~Tomasi}}, \bibinfo {author} {\bibfnamefont {F.}~\bibnamefont {Pollmann}},\ and\ \bibinfo {author} {\bibfnamefont {M.}~\bibnamefont {Knap}},\ }\bibfield  {title} {\bibinfo {title} {Anomalous diffusion in dipole- and higher-moment-conserving systems},\ }\href {https://doi.org/10.1103/PhysRevLett.125.245303} {\bibfield  {journal} {\bibinfo  {journal} {Phys. Rev. Lett.}\ }\textbf {\bibinfo {volume} {125}},\ \bibinfo {pages} {245303} (\bibinfo {year} {2020})}\BibitemShut {NoStop}%
\bibitem [{\citenamefont {Neumann}(1966)}]{neumann1966a}%
  \BibitemOpen
  \bibfield  {author} {\bibinfo {author} {\bibfnamefont {J.}~\bibnamefont {Neumann}},\ }\href@noop {} {\emph {\bibinfo {title} {Theory of Self-Reproducing Automata}}}\ (\bibinfo  {publisher} {University of Illinois Press},\ \bibinfo {year} {1966})\BibitemShut {NoStop}%
\bibitem [{\citenamefont {Feynman}(1982)}]{feynman1982a}%
  \BibitemOpen
  \bibfield  {author} {\bibinfo {author} {\bibfnamefont {R.}~\bibnamefont {Feynman}},\ }\bibfield  {title} {\bibinfo {title} {Simulating physics with computers},\ }\href@noop {} {\bibfield  {journal} {\bibinfo  {journal} {International Journal of Theoretical Physics}\ }\textbf {\bibinfo {volume} {21}},\ \bibinfo {pages} {467–488} (\bibinfo {year} {1982})}\BibitemShut {NoStop}%
\bibitem [{\citenamefont {Feynman}(1986)}]{feynman1986a}%
  \BibitemOpen
  \bibfield  {author} {\bibinfo {author} {\bibfnamefont {R.}~\bibnamefont {Feynman}},\ }\bibfield  {title} {\bibinfo {title} {Quantum mechanical computers},\ }\href@noop {} {\bibfield  {journal} {\bibinfo  {journal} {Foundations of Physics (Historical Archive}\ }\textbf {\bibinfo {volume} {16}},\ \bibinfo {pages} {507–531} (\bibinfo {year} {1986})}\BibitemShut {NoStop}%
\bibitem [{\citenamefont {Arrighi}(2019)}]{arrighi_overview_2019}%
  \BibitemOpen
  \bibfield  {author} {\bibinfo {author} {\bibfnamefont {P.}~\bibnamefont {Arrighi}},\ }\href@noop {} {\bibinfo {title} {An overview of quantum cellular automata}} (\bibinfo {year} {2019}),\ \Eprint {https://arxiv.org/abs/1904.12956} {arXiv:1904.12956} \BibitemShut {NoStop}%
\bibitem [{\citenamefont {Arnault}\ \emph {et~al.}(2016)\citenamefont {Arnault}, \citenamefont {Di~Molfetta}, \citenamefont {Brachet},\ and\ \citenamefont {Debbasch}}]{arnault2016a}%
  \BibitemOpen
  \bibfield  {author} {\bibinfo {author} {\bibfnamefont {P.}~\bibnamefont {Arnault}}, \bibinfo {author} {\bibfnamefont {G.}~\bibnamefont {Di~Molfetta}}, \bibinfo {author} {\bibfnamefont {M.}~\bibnamefont {Brachet}},\ and\ \bibinfo {author} {\bibfnamefont {F.}~\bibnamefont {Debbasch}},\ }\bibfield  {title} {\bibinfo {title} {Quantum walks and non-abelian discrete gauge theory},\ }\href {https://link.aps.org/doi/10.1103/PhysRevA.94.012335} {\bibfield  {journal} {\bibinfo  {journal} {Phys. Rev. A}\ }\textbf {\bibinfo {volume} {94}},\ \bibinfo {pages} {012335} (\bibinfo {year} {2016})}\BibitemShut {NoStop}%
\bibitem [{\citenamefont {M\'arquez-Martin}\ \emph {et~al.}(2017)\citenamefont {M\'arquez-Martin}, \citenamefont {Di~Molfetta},\ and\ \citenamefont {P\'erez}}]{i2017a}%
  \BibitemOpen
  \bibfield  {author} {\bibinfo {author} {\bibfnamefont {I.}~\bibnamefont {M\'arquez-Martin}}, \bibinfo {author} {\bibfnamefont {G.}~\bibnamefont {Di~Molfetta}},\ and\ \bibinfo {author} {\bibfnamefont {A.}~\bibnamefont {P\'erez}},\ }\bibfield  {title} {\bibinfo {title} {Fermion confinement via quantum walks in (2+1)-dimensional and (3+1)-dimensional space-time},\ }\href {https://link.aps.org/doi/10.1103/PhysRevA.95.042112} {\bibfield  {journal} {\bibinfo  {journal} {Phys. Rev. A}\ }\textbf {\bibinfo {volume} {95}},\ \bibinfo {pages} {042112} (\bibinfo {year} {2017})}\BibitemShut {NoStop}%
\bibitem [{\citenamefont {Molfetta}\ and\ \citenamefont {Pérez}(2016)}]{giuseppe2016a}%
  \BibitemOpen
  \bibfield  {author} {\bibinfo {author} {\bibfnamefont {G.~D.}\ \bibnamefont {Molfetta}}\ and\ \bibinfo {author} {\bibfnamefont {A.}~\bibnamefont {Pérez}},\ }\bibfield  {title} {\bibinfo {title} {Quantum walks as simulators of neutrino oscillations in a vacuum and matter},\ }\href {https://dx.doi.org/10.1088/1367-2630/18/10/103038} {\bibfield  {journal} {\bibinfo  {journal} {New Journal of Physics}\ }\textbf {\bibinfo {volume} {18}},\ \bibinfo {pages} {103038} (\bibinfo {year} {2016})}\BibitemShut {NoStop}%
\bibitem [{\citenamefont {{Di Molfetta}}\ \emph {et~al.}(2014)\citenamefont {{Di Molfetta}}, \citenamefont {Brachet},\ and\ \citenamefont {Debbasch}}]{molfetta2014a}%
  \BibitemOpen
  \bibfield  {author} {\bibinfo {author} {\bibfnamefont {G.}~\bibnamefont {{Di Molfetta}}}, \bibinfo {author} {\bibfnamefont {M.}~\bibnamefont {Brachet}},\ and\ \bibinfo {author} {\bibfnamefont {F.}~\bibnamefont {Debbasch}},\ }\bibfield  {title} {\bibinfo {title} {Quantum walks in artificial electric and gravitational fields},\ }\href {https://www.sciencedirect.com/science/article/pii/S0378437113011059} {\bibfield  {journal} {\bibinfo  {journal} {Physica A: Statistical Mechanics and its Applications}\ }\textbf {\bibinfo {volume} {397}},\ \bibinfo {pages} {157} (\bibinfo {year} {2014})}\BibitemShut {NoStop}%
\bibitem [{\citenamefont {Stephen}\ \emph {et~al.}(2024)\citenamefont {Stephen}, \citenamefont {Dua}, \citenamefont {Lavasani},\ and\ \citenamefont {Nandkishore}}]{QCA1}%
  \BibitemOpen
  \bibfield  {author} {\bibinfo {author} {\bibfnamefont {D.~T.}\ \bibnamefont {Stephen}}, \bibinfo {author} {\bibfnamefont {A.}~\bibnamefont {Dua}}, \bibinfo {author} {\bibfnamefont {A.}~\bibnamefont {Lavasani}},\ and\ \bibinfo {author} {\bibfnamefont {R.}~\bibnamefont {Nandkishore}},\ }\bibfield  {title} {\bibinfo {title} {Nonlocal finite-depth circuits for constructing symmetry-protected topological states and quantum cellular automata},\ }\href {https://doi.org/10.1103/PRXQuantum.5.010304} {\bibfield  {journal} {\bibinfo  {journal} {PRX Quantum}\ }\textbf {\bibinfo {volume} {5}},\ \bibinfo {pages} {010304} (\bibinfo {year} {2024})}\BibitemShut {NoStop}%
\bibitem [{\citenamefont {Haldane}(1983)}]{haldane1983a}%
  \BibitemOpen
  \bibfield  {author} {\bibinfo {author} {\bibfnamefont {F.}~\bibnamefont {Haldane}},\ }\bibfield  {title} {\bibinfo {title} {Continuum dynamics of the 1-d heisenberg antiferromagnet: Identification with the o(3) nonlinear sigma model},\ }\href {https://doi.org/10.1016/0375-9601(83)90631-X.} {\bibfield  {journal} {\bibinfo  {journal} {Phys. Lett. A}\ }\textbf {\bibinfo {volume} {93}},\ \bibinfo {pages} {464} (\bibinfo {year} {1983})}\BibitemShut {NoStop}%
\bibitem [{\citenamefont {Affleck}\ \emph {et~al.}(1987)\citenamefont {Affleck}, \citenamefont {Kennedy}, \citenamefont {Lieb},\ and\ \citenamefont {Tasaki}}]{affleck1987a}%
  \BibitemOpen
  \bibfield  {author} {\bibinfo {author} {\bibfnamefont {I.}~\bibnamefont {Affleck}}, \bibinfo {author} {\bibfnamefont {T.}~\bibnamefont {Kennedy}}, \bibinfo {author} {\bibfnamefont {E.}~\bibnamefont {Lieb}},\ and\ \bibinfo {author} {\bibfnamefont {H.}~\bibnamefont {Tasaki}},\ }\bibfield  {title} {\bibinfo {title} {Rigorous results on valencebond ground states in antiferromagnets},\ }\href {https://doi.org/10.1103/PhysRevLett.59.799.} {\bibfield  {journal} {\bibinfo  {journal} {Phys. Rev. Lett}\ }\textbf {\bibinfo {volume} {59}},\ \bibinfo {pages} {799} (\bibinfo {year} {1987})}\BibitemShut {NoStop}%
\bibitem [{\citenamefont {Su}\ \emph {et~al.}(1979)\citenamefont {Su}, \citenamefont {Schrieffer},\ and\ \citenamefont {Heeger}}]{su1979a}%
  \BibitemOpen
  \bibfield  {author} {\bibinfo {author} {\bibfnamefont {W.}~\bibnamefont {Su}}, \bibinfo {author} {\bibfnamefont {J.}~\bibnamefont {Schrieffer}},\ and\ \bibinfo {author} {\bibfnamefont {A.}~\bibnamefont {Heeger}},\ }\bibfield  {title} {\bibinfo {title} {Solitons in polyacetylene},\ }\href {https://doi.org/10.1103/PhysRevLett.42.1698.} {\bibfield  {journal} {\bibinfo  {journal} {Phys. Rev. Lett}\ }\textbf {\bibinfo {volume} {42}},\ \bibinfo {pages} {1698} (\bibinfo {year} {1979})}\BibitemShut {NoStop}%
\bibitem [{\citenamefont {Schuch}\ \emph {et~al.}(2011)\citenamefont {Schuch}, \citenamefont {P\'erez-Garcia},\ and\ \citenamefont {Cirac}}]{schuch2011a}%
  \BibitemOpen
  \bibfield  {author} {\bibinfo {author} {\bibfnamefont {N.}~\bibnamefont {Schuch}}, \bibinfo {author} {\bibfnamefont {D.}~\bibnamefont {P\'erez-Garcia}},\ and\ \bibinfo {author} {\bibfnamefont {I.}~\bibnamefont {Cirac}},\ }\bibfield  {title} {\bibinfo {title} {Classifying quantum phases using matrix product states and projected entangled pair states},\ }\href {https://doi.org/10.1103/PhysRevB.84.165139.} {\bibfield  {journal} {\bibinfo  {journal} {Phys. Rev. B}\ }\textbf {\bibinfo {volume} {84}},\ \bibinfo {pages} {165139} (\bibinfo {year} {2011})}\BibitemShut {NoStop}%
\bibitem [{\citenamefont {Turner}\ \emph {et~al.}(2011)\citenamefont {Turner}, \citenamefont {Pollmann},\ and\ \citenamefont {Berg}}]{turner2011a}%
  \BibitemOpen
  \bibfield  {author} {\bibinfo {author} {\bibfnamefont {A.}~\bibnamefont {Turner}}, \bibinfo {author} {\bibfnamefont {F.}~\bibnamefont {Pollmann}},\ and\ \bibinfo {author} {\bibfnamefont {E.}~\bibnamefont {Berg}},\ }\bibfield  {title} {\bibinfo {title} {Topological phases of one-dimensional fermions: An entanglement point of view},\ }\href {https://doi.org/10.1103/PhysRevB.83.075102.} {\bibfield  {journal} {\bibinfo  {journal} {Phys. Rev. B}\ }\textbf {\bibinfo {volume} {83}},\ \bibinfo {pages} {075102} (\bibinfo {year} {2011})}\BibitemShut {NoStop}%
\bibitem [{\citenamefont {Fidkowski}\ and\ \citenamefont {Kitaev}(2011)}]{fidkowski2011a}%
  \BibitemOpen
  \bibfield  {author} {\bibinfo {author} {\bibfnamefont {L.}~\bibnamefont {Fidkowski}}\ and\ \bibinfo {author} {\bibfnamefont {A.}~\bibnamefont {Kitaev}},\ }\bibfield  {title} {\bibinfo {title} {Topological phases of fermions in one dimension},\ }\href {https://doi.org/10.1103/PhysRevB.83.075103} {\bibfield  {journal} {\bibinfo  {journal} {Phys. Rev. B}\ }\textbf {\bibinfo {volume} {83}},\ \bibinfo {pages} {075103} (\bibinfo {year} {2011})}\BibitemShut {NoStop}%
\bibitem [{\citenamefont {Pollmann}\ \emph {et~al.}(2012)\citenamefont {Pollmann}, \citenamefont {Berg}, \citenamefont {Turner},\ and\ \citenamefont {Oshikawa}}]{pollmann2012a}%
  \BibitemOpen
  \bibfield  {author} {\bibinfo {author} {\bibfnamefont {F.}~\bibnamefont {Pollmann}}, \bibinfo {author} {\bibfnamefont {E.}~\bibnamefont {Berg}}, \bibinfo {author} {\bibfnamefont {A.}~\bibnamefont {Turner}},\ and\ \bibinfo {author} {\bibfnamefont {M.}~\bibnamefont {Oshikawa}},\ }\bibfield  {title} {\bibinfo {title} {Symmetry protection of topological phases in one-dimensional quantum spin systems},\ }\href {https://doi.org/10.1103/PhysRevB.85.075125.} {\bibfield  {journal} {\bibinfo  {journal} {Phys. Rev. B}\ }\textbf {\bibinfo {volume} {85}},\ \bibinfo {pages} {075125} (\bibinfo {year} {2012})}\BibitemShut {NoStop}%
\bibitem [{\citenamefont {Chen}\ \emph {et~al.}(2011{\natexlab{a}})\citenamefont {Chen}, \citenamefont {Gu},\ and\ \citenamefont {Wen}}]{chen2011a}%
  \BibitemOpen
  \bibfield  {author} {\bibinfo {author} {\bibfnamefont {X.}~\bibnamefont {Chen}}, \bibinfo {author} {\bibfnamefont {Z.}~\bibnamefont {Gu}},\ and\ \bibinfo {author} {\bibfnamefont {X.}~\bibnamefont {Wen}},\ }\bibfield  {title} {\bibinfo {title} {Classification of gapped symmetric phases in one-dimensional spin systems},\ }\href {https://doi.org/10.1103/PhysRevB.83.035107.} {\bibfield  {journal} {\bibinfo  {journal} {Phys. Rev. B}\ }\textbf {\bibinfo {volume} {83}},\ \bibinfo {pages} {035107} (\bibinfo {year} {2011}{\natexlab{a}})}\BibitemShut {NoStop}%
\bibitem [{\citenamefont {Chen}\ \emph {et~al.}(2011{\natexlab{b}})\citenamefont {Chen}, \citenamefont {Liu},\ and\ \citenamefont {Wen}}]{chen2011b}%
  \BibitemOpen
  \bibfield  {author} {\bibinfo {author} {\bibfnamefont {X.}~\bibnamefont {Chen}}, \bibinfo {author} {\bibfnamefont {Z.}~\bibnamefont {Liu}},\ and\ \bibinfo {author} {\bibfnamefont {X.}~\bibnamefont {Wen}},\ }\bibfield  {title} {\bibinfo {title} {Two-dimensional symmetry-protected topological orders and their protected gapless edge excitations},\ }\href {https://doi.org/10.1103/PhysRevB.84.235141.} {\bibfield  {journal} {\bibinfo  {journal} {Phys. Rev. B}\ }\textbf {\bibinfo {volume} {84}},\ \bibinfo {pages} {235141} (\bibinfo {year} {2011}{\natexlab{b}})}\BibitemShut {NoStop}%
\bibitem [{\citenamefont {Kane}\ and\ \citenamefont {Mele}(2005)}]{kane2005a}%
  \BibitemOpen
  \bibfield  {author} {\bibinfo {author} {\bibfnamefont {C.}~\bibnamefont {Kane}}\ and\ \bibinfo {author} {\bibfnamefont {E.}~\bibnamefont {Mele}},\ }\bibfield  {title} {\bibinfo {title} {Quantum spin hall effect in graphene},\ }\href {https://doi.org/10.1103/PhysRevLett.95.226801.} {\bibfield  {journal} {\bibinfo  {journal} {Phys. Rev. Lett}\ }\textbf {\bibinfo {volume} {95}},\ \bibinfo {pages} {226801} (\bibinfo {year} {2005})}\BibitemShut {NoStop}%
\bibitem [{\citenamefont {Moore}\ and\ \citenamefont {Balents}(2007)}]{moore2007a}%
  \BibitemOpen
  \bibfield  {author} {\bibinfo {author} {\bibfnamefont {J.~E.}\ \bibnamefont {Moore}}\ and\ \bibinfo {author} {\bibfnamefont {L.}~\bibnamefont {Balents}},\ }\bibfield  {title} {\bibinfo {title} {Topological invariants of time-reversal-invariant band structures},\ }\href {https://doi.org/10.1103/PhysRevB.75.121306} {\bibfield  {journal} {\bibinfo  {journal} {Phys. Rev. B}\ }\textbf {\bibinfo {volume} {75}},\ \bibinfo {pages} {121306} (\bibinfo {year} {2007})}\BibitemShut {NoStop}%
\bibitem [{\citenamefont {Fu}\ \emph {et~al.}(2007)\citenamefont {Fu}, \citenamefont {Kane},\ and\ \citenamefont {Mele}}]{fu2007a}%
  \BibitemOpen
  \bibfield  {author} {\bibinfo {author} {\bibfnamefont {L.}~\bibnamefont {Fu}}, \bibinfo {author} {\bibfnamefont {C.~L.}\ \bibnamefont {Kane}},\ and\ \bibinfo {author} {\bibfnamefont {E.~J.}\ \bibnamefont {Mele}},\ }\bibfield  {title} {\bibinfo {title} {Topological insulators in three dimensions},\ }\href {https://doi.org/10.1103/PhysRevLett.98.106803} {\bibfield  {journal} {\bibinfo  {journal} {Phys. Rev. Lett.}\ }\textbf {\bibinfo {volume} {98}},\ \bibinfo {pages} {106803} (\bibinfo {year} {2007})}\BibitemShut {NoStop}%
\bibitem [{\citenamefont {Chen}\ \emph {et~al.}(2012)\citenamefont {Chen}, \citenamefont {Gu}, \citenamefont {Liu},\ and\ \citenamefont {Wen}}]{chen2012a}%
  \BibitemOpen
  \bibfield  {author} {\bibinfo {author} {\bibfnamefont {X.}~\bibnamefont {Chen}}, \bibinfo {author} {\bibfnamefont {Z.}~\bibnamefont {Gu}}, \bibinfo {author} {\bibfnamefont {Z.}~\bibnamefont {Liu}},\ and\ \bibinfo {author} {\bibfnamefont {X.}~\bibnamefont {Wen}},\ }\bibfield  {title} {\bibinfo {title} {Symmetry-protected topological orders in interacting bosonic systems},\ }\href {https://doi.org/10.1126/science.1227224.} {\bibfield  {journal} {\bibinfo  {journal} {Science}\ }\textbf {\bibinfo {volume} {338}},\ \bibinfo {pages} {1604} (\bibinfo {year} {2012})}\BibitemShut {NoStop}%
\bibitem [{\citenamefont {Levin}\ and\ \citenamefont {Gu}(2012{\natexlab{a}})}]{levin_braiding_2012}%
  \BibitemOpen
  \bibfield  {author} {\bibinfo {author} {\bibfnamefont {M.}~\bibnamefont {Levin}}\ and\ \bibinfo {author} {\bibfnamefont {Z.-C.}\ \bibnamefont {Gu}},\ }\bibfield  {title} {\bibinfo {title} {Braiding statistics approach to symmetry-protected topological phases},\ }\href {https://doi.org/10.1103/PhysRevB.86.115109} {\bibfield  {journal} {\bibinfo  {journal} {Physical Review B}\ }\textbf {\bibinfo {volume} {86}},\ \bibinfo {pages} {115109} (\bibinfo {year} {2012}{\natexlab{a}})}\BibitemShut {NoStop}%
\bibitem [{\citenamefont {Vishwanath}\ and\ \citenamefont {Senthil}(2013)}]{vishwanath2013a}%
  \BibitemOpen
  \bibfield  {author} {\bibinfo {author} {\bibfnamefont {A.}~\bibnamefont {Vishwanath}}\ and\ \bibinfo {author} {\bibfnamefont {T.}~\bibnamefont {Senthil}},\ }\bibfield  {title} {\bibinfo {title} {Physics of three-dimensional bosonic topological insulators: Surface-deconfined criticality and quantized magnetoelectric effect},\ }\href {https://doi.org/10.1103/PhysRevX.3.011016.} {\bibfield  {journal} {\bibinfo  {journal} {Phys. Rev}\ }\textbf {\bibinfo {volume} {X 3}},\ \bibinfo {pages} {011016} (\bibinfo {year} {2013})}\BibitemShut {NoStop}%
\bibitem [{\citenamefont {Yao}\ and\ \citenamefont {Ryu}(2013)}]{yao2013a}%
  \BibitemOpen
  \bibfield  {author} {\bibinfo {author} {\bibfnamefont {H.}~\bibnamefont {Yao}}\ and\ \bibinfo {author} {\bibfnamefont {S.}~\bibnamefont {Ryu}},\ }\bibfield  {title} {\bibinfo {title} {Interaction effect on topological classification of superconductors in two dimensions},\ }\href {https://doi.org/10.1103/PhysRevB.88.064507.} {\bibfield  {journal} {\bibinfo  {journal} {Phys. Rev. B}\ }\textbf {\bibinfo {volume} {88}},\ \bibinfo {pages} {064507} (\bibinfo {year} {2013})}\BibitemShut {NoStop}%
\bibitem [{\citenamefont {Gu}\ and\ \citenamefont {Wen}(2014)}]{gu2014a}%
  \BibitemOpen
  \bibfield  {author} {\bibinfo {author} {\bibfnamefont {Z.}~\bibnamefont {Gu}}\ and\ \bibinfo {author} {\bibfnamefont {X.}~\bibnamefont {Wen}},\ }\bibfield  {title} {\bibinfo {title} {Symmetry-protected topological orders for interacting fermions: Fermionic topological nonlinear $\sigma$ models and a special group supercohomology theory},\ }\href {https://doi.org/10.1103/PhysRevB.90.115141.} {\bibfield  {journal} {\bibinfo  {journal} {Phys. Rev. B}\ }\textbf {\bibinfo {volume} {90}},\ \bibinfo {pages} {115141} (\bibinfo {year} {2014})}\BibitemShut {NoStop}%
\bibitem [{\citenamefont {Qi}(2013)}]{qi2013a}%
  \BibitemOpen
  \bibfield  {author} {\bibinfo {author} {\bibfnamefont {X.}~\bibnamefont {Qi}},\ }\bibfield  {title} {\bibinfo {title} {A new class of (2 + 1)-dimensional topological superconductors with z8 topological classification},\ }\href {https://doi.org/10.1088/1367-2630/15/6/065002.} {\bibfield  {journal} {\bibinfo  {journal} {New J. Phys}\ }\textbf {\bibinfo {volume} {15}},\ \bibinfo {pages} {065002} (\bibinfo {year} {2013})}\BibitemShut {NoStop}%
\bibitem [{\citenamefont {Wang}\ and\ \citenamefont {Gu}(2018)}]{cheng2018a}%
  \BibitemOpen
  \bibfield  {author} {\bibinfo {author} {\bibfnamefont {Q.-R.}\ \bibnamefont {Wang}}\ and\ \bibinfo {author} {\bibfnamefont {Z.-C.}\ \bibnamefont {Gu}},\ }\bibfield  {title} {\bibinfo {title} {Towards a complete classification of symmetry-protected topological phases for interacting fermions in three dimensions and a general group supercohomology theory},\ }\href {https://doi.org/10.1103/PhysRevX.8.011055} {\bibfield  {journal} {\bibinfo  {journal} {Phys. Rev. X}\ }\textbf {\bibinfo {volume} {8}},\ \bibinfo {pages} {011055} (\bibinfo {year} {2018})}\BibitemShut {NoStop}%
\bibitem [{\citenamefont {Gaiotto}\ and\ \citenamefont {Kapustin}(2016)}]{gaiotto2016a}%
  \BibitemOpen
  \bibfield  {author} {\bibinfo {author} {\bibfnamefont {D.}~\bibnamefont {Gaiotto}}\ and\ \bibinfo {author} {\bibfnamefont {A.}~\bibnamefont {Kapustin}},\ }\bibfield  {title} {\bibinfo {title} {Spin tqfts and fermionic phases of matter},\ }\href {http://dx.doi.org/10.1142/S0217751X16450445} {\bibfield  {journal} {\bibinfo  {journal} {International Journal of Modern Physics A}\ }\textbf {\bibinfo {volume} {31}},\ \bibinfo {pages} {1645044} (\bibinfo {year} {2016})}\BibitemShut {NoStop}%
\bibitem [{\citenamefont {Thorngren}\ and\ \citenamefont {Else}(2018)}]{thorngren2018a}%
  \BibitemOpen
  \bibfield  {author} {\bibinfo {author} {\bibfnamefont {R.}~\bibnamefont {Thorngren}}\ and\ \bibinfo {author} {\bibfnamefont {D.}~\bibnamefont {Else}},\ }\bibfield  {title} {\bibinfo {title} {Gauging spatial symmetries and the classification of topological crystalline phases},\ }\href {https://doi.org/10.1103/PhysRevX.8.011040.} {\bibfield  {journal} {\bibinfo  {journal} {Phys. Rev}\ }\textbf {\bibinfo {volume} {X 8}},\ \bibinfo {pages} {011040} (\bibinfo {year} {2018})}\BibitemShut {NoStop}%
\bibitem [{\citenamefont {Else}\ and\ \citenamefont {Nayak}(2014)}]{else2014a}%
  \BibitemOpen
  \bibfield  {author} {\bibinfo {author} {\bibfnamefont {D.}~\bibnamefont {Else}}\ and\ \bibinfo {author} {\bibfnamefont {C.}~\bibnamefont {Nayak}},\ }\bibfield  {title} {\bibinfo {title} {Classifying symmetry-protected topological phases through the anomalous action of the symmetry on the edge},\ }\href {https://doi.org/10.1103/PhysRevB.90.235137.} {\bibfield  {journal} {\bibinfo  {journal} {Phys. Rev. B}\ }\textbf {\bibinfo {volume} {90}},\ \bibinfo {pages} {235137} (\bibinfo {year} {2014})}\BibitemShut {NoStop}%
\bibitem [{\citenamefont {Yoshida}(2015)}]{yoshida2016a}%
  \BibitemOpen
  \bibfield  {author} {\bibinfo {author} {\bibfnamefont {B.}~\bibnamefont {Yoshida}},\ }\bibfield  {title} {\bibinfo {title} {Topological phases with generalized global symmetries},\ }\href {https://api.semanticscholar.org/CorpusID:7702848} {\bibfield  {journal} {\bibinfo  {journal} {Physical Review B}\ }\textbf {\bibinfo {volume} {93}},\ \bibinfo {pages} {155131} (\bibinfo {year} {2015})}\BibitemShut {NoStop}%
\bibitem [{\citenamefont {Chen}\ \emph {et~al.}(2013)\citenamefont {Chen}, \citenamefont {Gu}, \citenamefont {Liu},\ and\ \citenamefont {Wen}}]{chen_symmetry_2013}%
  \BibitemOpen
  \bibfield  {author} {\bibinfo {author} {\bibfnamefont {X.}~\bibnamefont {Chen}}, \bibinfo {author} {\bibfnamefont {Z.-C.}\ \bibnamefont {Gu}}, \bibinfo {author} {\bibfnamefont {Z.-X.}\ \bibnamefont {Liu}},\ and\ \bibinfo {author} {\bibfnamefont {X.-G.}\ \bibnamefont {Wen}},\ }\bibfield  {title} {\bibinfo {title} {Symmetry protected topological orders and the group cohomology of their symmetry group},\ }\href {https://link.aps.org/doi/10.1103/PhysRevB.87.155114} {\bibfield  {journal} {\bibinfo  {journal} {Phys. Rev. B}\ }\textbf {\bibinfo {volume} {87}},\ \bibinfo {pages} {155114} (\bibinfo {year} {2013})}\BibitemShut {NoStop}%
\bibitem [{\citenamefont {Kapustin}\ \emph {et~al.}(2014)\citenamefont {Kapustin}, \citenamefont {Thorngren}, \citenamefont {Turzillo},\ and\ \citenamefont {Wang}}]{Kapustin2014FermionicSP}%
  \BibitemOpen
  \bibfield  {author} {\bibinfo {author} {\bibfnamefont {A.}~\bibnamefont {Kapustin}}, \bibinfo {author} {\bibfnamefont {R.}~\bibnamefont {Thorngren}}, \bibinfo {author} {\bibfnamefont {A.}~\bibnamefont {Turzillo}},\ and\ \bibinfo {author} {\bibfnamefont {Z.}~\bibnamefont {Wang}},\ }\bibfield  {title} {\bibinfo {title} {Fermionic symmetry protected topological phases and cobordisms},\ }\href {https://api.semanticscholar.org/CorpusID:42613274} {\bibfield  {journal} {\bibinfo  {journal} {Journal of High Energy Physics}\ }\textbf {\bibinfo {volume} {2015}},\ \bibinfo {pages} {1 } (\bibinfo {year} {2014})}\BibitemShut {NoStop}%
\bibitem [{\citenamefont {Kapustin}(2014)}]{Kapustin2014SymmetryPT}%
  \BibitemOpen
  \bibfield  {author} {\bibinfo {author} {\bibfnamefont {A.}~\bibnamefont {Kapustin}},\ }\href@noop {} {\bibinfo {title} {Symmetry protected topological phases, anomalies, and cobordisms: Beyond group cohomology}} (\bibinfo {year} {2014}),\ \Eprint {https://arxiv.org/abs/1403.1467} {arXiv:1403.1467} \BibitemShut {NoStop}%
\bibitem [{\citenamefont {Bi}\ \emph {et~al.}(2015)\citenamefont {Bi}, \citenamefont {Rasmussen}, \citenamefont {Slagle},\ and\ \citenamefont {Xu}}]{PhysRevB.91.134404}%
  \BibitemOpen
  \bibfield  {author} {\bibinfo {author} {\bibfnamefont {Z.}~\bibnamefont {Bi}}, \bibinfo {author} {\bibfnamefont {A.}~\bibnamefont {Rasmussen}}, \bibinfo {author} {\bibfnamefont {K.}~\bibnamefont {Slagle}},\ and\ \bibinfo {author} {\bibfnamefont {C.}~\bibnamefont {Xu}},\ }\bibfield  {title} {\bibinfo {title} {Classification and description of bosonic symmetry protected topological phases with semiclassical nonlinear sigma models},\ }\href {https://doi.org/10.1103/PhysRevB.91.134404} {\bibfield  {journal} {\bibinfo  {journal} {Phys. Rev. B}\ }\textbf {\bibinfo {volume} {91}},\ \bibinfo {pages} {134404} (\bibinfo {year} {2015})}\BibitemShut {NoStop}%
\bibitem [{\citenamefont {You}\ and\ \citenamefont {You}(2016)}]{PhysRevB.93.245135}%
  \BibitemOpen
  \bibfield  {author} {\bibinfo {author} {\bibfnamefont {Y.}~\bibnamefont {You}}\ and\ \bibinfo {author} {\bibfnamefont {Y.-Z.}\ \bibnamefont {You}},\ }\bibfield  {title} {\bibinfo {title} {Geometry defects in bosonic symmetry-protected topological phases},\ }\href {https://doi.org/10.1103/PhysRevB.93.245135} {\bibfield  {journal} {\bibinfo  {journal} {Phys. Rev. B}\ }\textbf {\bibinfo {volume} {93}},\ \bibinfo {pages} {245135} (\bibinfo {year} {2016})}\BibitemShut {NoStop}%
\bibitem [{\citenamefont {Lu}\ and\ \citenamefont {Vishwanath}(2012)}]{FT1}%
  \BibitemOpen
  \bibfield  {author} {\bibinfo {author} {\bibfnamefont {Y.-M.}\ \bibnamefont {Lu}}\ and\ \bibinfo {author} {\bibfnamefont {A.}~\bibnamefont {Vishwanath}},\ }\bibfield  {title} {\bibinfo {title} {Theory and classification of interacting integer topological phases in two dimensions: A chern-simons approach},\ }\href {https://doi.org/10.1103/PhysRevB.86.125119} {\bibfield  {journal} {\bibinfo  {journal} {Phys. Rev. B}\ }\textbf {\bibinfo {volume} {86}},\ \bibinfo {pages} {125119} (\bibinfo {year} {2012})}\BibitemShut {NoStop}%
\bibitem [{\citenamefont {Ye}\ and\ \citenamefont {Gu}(2015)}]{FT2}%
  \BibitemOpen
  \bibfield  {author} {\bibinfo {author} {\bibfnamefont {P.}~\bibnamefont {Ye}}\ and\ \bibinfo {author} {\bibfnamefont {Z.-C.}\ \bibnamefont {Gu}},\ }\bibfield  {title} {\bibinfo {title} {Vortex-line condensation in three dimensions: A physical mechanism for bosonic topological insulators},\ }\href {https://doi.org/10.1103/PhysRevX.5.021029} {\bibfield  {journal} {\bibinfo  {journal} {Phys. Rev. X}\ }\textbf {\bibinfo {volume} {5}},\ \bibinfo {pages} {021029} (\bibinfo {year} {2015})}\BibitemShut {NoStop}%
\bibitem [{\citenamefont {Gu}\ \emph {et~al.}(2016)\citenamefont {Gu}, \citenamefont {Wang},\ and\ \citenamefont {Wen}}]{FT3}%
  \BibitemOpen
  \bibfield  {author} {\bibinfo {author} {\bibfnamefont {Z.-C.}\ \bibnamefont {Gu}}, \bibinfo {author} {\bibfnamefont {J.~C.}\ \bibnamefont {Wang}},\ and\ \bibinfo {author} {\bibfnamefont {X.-G.}\ \bibnamefont {Wen}},\ }\bibfield  {title} {\bibinfo {title} {Multikink topological terms and charge-binding domain-wall condensation induced symmetry-protected topological states: Beyond chern-simons/bf field theories},\ }\href {https://doi.org/10.1103/PhysRevB.93.115136} {\bibfield  {journal} {\bibinfo  {journal} {Phys. Rev. B}\ }\textbf {\bibinfo {volume} {93}},\ \bibinfo {pages} {115136} (\bibinfo {year} {2016})}\BibitemShut {NoStop}%
\bibitem [{\citenamefont {Wang}\ \emph {et~al.}(2018)\citenamefont {Wang}, \citenamefont {Ohmori}, \citenamefont {Putrov}, \citenamefont {Zheng}, \citenamefont {Wan}, \citenamefont {Guo}, \citenamefont {Lin}, \citenamefont {Gao},\ and\ \citenamefont {Yau}}]{FT4}%
  \BibitemOpen
  \bibfield  {author} {\bibinfo {author} {\bibfnamefont {J.}~\bibnamefont {Wang}}, \bibinfo {author} {\bibfnamefont {K.}~\bibnamefont {Ohmori}}, \bibinfo {author} {\bibfnamefont {P.}~\bibnamefont {Putrov}}, \bibinfo {author} {\bibfnamefont {Y.}~\bibnamefont {Zheng}}, \bibinfo {author} {\bibfnamefont {Z.}~\bibnamefont {Wan}}, \bibinfo {author} {\bibfnamefont {M.}~\bibnamefont {Guo}}, \bibinfo {author} {\bibfnamefont {H.}~\bibnamefont {Lin}}, \bibinfo {author} {\bibfnamefont {P.}~\bibnamefont {Gao}},\ and\ \bibinfo {author} {\bibfnamefont {S.-T.}\ \bibnamefont {Yau}},\ }\bibfield  {title} {\bibinfo {title} {{Tunneling topological vacua via extended operators: (Spin-)TQFT spectra and boundary deconfinement in various dimensions}},\ }\href {https://doi.org/10.1093/ptep/pty051} {\bibfield  {journal} {\bibinfo  {journal} {Progress of Theoretical and Experimental Physics}\ }\textbf {\bibinfo {volume} {2018}},\ \bibinfo {pages} {053A01} (\bibinfo {year} {2018})}\BibitemShut {NoStop}%
\bibitem [{\citenamefont {Hsieh}\ \emph {et~al.}(2014)\citenamefont {Hsieh}, \citenamefont {Sule}, \citenamefont {Cho}, \citenamefont {Ryu},\ and\ \citenamefont {Leigh}}]{CFT1}%
  \BibitemOpen
  \bibfield  {author} {\bibinfo {author} {\bibfnamefont {C.-T.}\ \bibnamefont {Hsieh}}, \bibinfo {author} {\bibfnamefont {O.~M.}\ \bibnamefont {Sule}}, \bibinfo {author} {\bibfnamefont {G.~Y.}\ \bibnamefont {Cho}}, \bibinfo {author} {\bibfnamefont {S.}~\bibnamefont {Ryu}},\ and\ \bibinfo {author} {\bibfnamefont {R.~G.}\ \bibnamefont {Leigh}},\ }\bibfield  {title} {\bibinfo {title} {Symmetry-protected topological phases, generalized laughlin argument, and orientifolds},\ }\href {https://doi.org/10.1103/PhysRevB.90.165134} {\bibfield  {journal} {\bibinfo  {journal} {Phys. Rev. B}\ }\textbf {\bibinfo {volume} {90}},\ \bibinfo {pages} {165134} (\bibinfo {year} {2014})}\BibitemShut {NoStop}%
\bibitem [{\citenamefont {Hsieh}\ \emph {et~al.}(2016)\citenamefont {Hsieh}, \citenamefont {Cho},\ and\ \citenamefont {Ryu}}]{CFT2}%
  \BibitemOpen
  \bibfield  {author} {\bibinfo {author} {\bibfnamefont {C.-T.}\ \bibnamefont {Hsieh}}, \bibinfo {author} {\bibfnamefont {G.~Y.}\ \bibnamefont {Cho}},\ and\ \bibinfo {author} {\bibfnamefont {S.}~\bibnamefont {Ryu}},\ }\bibfield  {title} {\bibinfo {title} {Global anomalies on the surface of fermionic symmetry-protected topological phases in (3+1) dimensions},\ }\href {https://doi.org/10.1103/PhysRevB.93.075135} {\bibfield  {journal} {\bibinfo  {journal} {Phys. Rev. B}\ }\textbf {\bibinfo {volume} {93}},\ \bibinfo {pages} {075135} (\bibinfo {year} {2016})}\BibitemShut {NoStop}%
\bibitem [{\citenamefont {Han}\ \emph {et~al.}(2017)\citenamefont {Han}, \citenamefont {Tiwari}, \citenamefont {Hsieh},\ and\ \citenamefont {Ryu}}]{CFT3}%
  \BibitemOpen
  \bibfield  {author} {\bibinfo {author} {\bibfnamefont {B.}~\bibnamefont {Han}}, \bibinfo {author} {\bibfnamefont {A.}~\bibnamefont {Tiwari}}, \bibinfo {author} {\bibfnamefont {C.-T.}\ \bibnamefont {Hsieh}},\ and\ \bibinfo {author} {\bibfnamefont {S.}~\bibnamefont {Ryu}},\ }\bibfield  {title} {\bibinfo {title} {Boundary conformal field theory and symmetry-protected topological phases in $2+1$ dimensions},\ }\href {https://doi.org/10.1103/PhysRevB.96.125105} {\bibfield  {journal} {\bibinfo  {journal} {Phys. Rev. B}\ }\textbf {\bibinfo {volume} {96}},\ \bibinfo {pages} {125105} (\bibinfo {year} {2017})}\BibitemShut {NoStop}%
\bibitem [{\citenamefont {Chen}\ \emph {et~al.}(2014)\citenamefont {Chen}, \citenamefont {Lu},\ and\ \citenamefont {Vishwanath}}]{chen2014a}%
  \BibitemOpen
  \bibfield  {author} {\bibinfo {author} {\bibfnamefont {X.}~\bibnamefont {Chen}}, \bibinfo {author} {\bibfnamefont {Y.-M.}\ \bibnamefont {Lu}},\ and\ \bibinfo {author} {\bibfnamefont {A.}~\bibnamefont {Vishwanath}},\ }\bibfield  {title} {\bibinfo {title} {Symmetry-protected topological phases from decorated domain walls},\ }\href {https://doi.org/10.1038/ncomms4507.} {\bibfield  {journal} {\bibinfo  {journal} {Nat. Commun}\ }\textbf {\bibinfo {volume} {5}},\ \bibinfo {pages} {3507} (\bibinfo {year} {2014})}\BibitemShut {NoStop}%
\bibitem [{\citenamefont {Cheng}\ and\ \citenamefont {Gu}(2014)}]{tr1}%
  \BibitemOpen
  \bibfield  {author} {\bibinfo {author} {\bibfnamefont {M.}~\bibnamefont {Cheng}}\ and\ \bibinfo {author} {\bibfnamefont {Z.-C.}\ \bibnamefont {Gu}},\ }\bibfield  {title} {\bibinfo {title} {Topological response theory of abelian symmetry-protected topological phases in two dimensions},\ }\href {https://doi.org/10.1103/PhysRevLett.112.141602} {\bibfield  {journal} {\bibinfo  {journal} {Phys. Rev. Lett.}\ }\textbf {\bibinfo {volume} {112}},\ \bibinfo {pages} {141602} (\bibinfo {year} {2014})}\BibitemShut {NoStop}%
\bibitem [{\citenamefont {Hung}\ and\ \citenamefont {Wen}(2013)}]{tr2}%
  \BibitemOpen
  \bibfield  {author} {\bibinfo {author} {\bibfnamefont {L.-Y.}\ \bibnamefont {Hung}}\ and\ \bibinfo {author} {\bibfnamefont {X.-G.}\ \bibnamefont {Wen}},\ }\bibfield  {title} {\bibinfo {title} {Quantized topological terms in weak-coupling gauge theories with a global symmetry and their connection to symmetry-enriched topological phases},\ }\href {https://doi.org/10.1103/PhysRevB.87.165107} {\bibfield  {journal} {\bibinfo  {journal} {Phys. Rev. B}\ }\textbf {\bibinfo {volume} {87}},\ \bibinfo {pages} {165107} (\bibinfo {year} {2013})}\BibitemShut {NoStop}%
\bibitem [{\citenamefont {Wen}(2013)}]{tr3}%
  \BibitemOpen
  \bibfield  {author} {\bibinfo {author} {\bibfnamefont {X.-G.}\ \bibnamefont {Wen}},\ }\bibfield  {title} {\bibinfo {title} {Classifying gauge anomalies through symmetry-protected trivial orders and classifying gravitational anomalies through topological orders},\ }\href {https://doi.org/10.1103/PhysRevD.88.045013} {\bibfield  {journal} {\bibinfo  {journal} {Phys. Rev. D}\ }\textbf {\bibinfo {volume} {88}},\ \bibinfo {pages} {045013} (\bibinfo {year} {2013})}\BibitemShut {NoStop}%
\bibitem [{\citenamefont {Ye}\ and\ \citenamefont {Wang}(2013)}]{tr4}%
  \BibitemOpen
  \bibfield  {author} {\bibinfo {author} {\bibfnamefont {P.}~\bibnamefont {Ye}}\ and\ \bibinfo {author} {\bibfnamefont {J.}~\bibnamefont {Wang}},\ }\bibfield  {title} {\bibinfo {title} {Symmetry-protected topological phases with charge and spin symmetries: Response theory and dynamical gauge theory in two and three dimensions},\ }\href {https://doi.org/10.1103/PhysRevB.88.235109} {\bibfield  {journal} {\bibinfo  {journal} {Phys. Rev. B}\ }\textbf {\bibinfo {volume} {88}},\ \bibinfo {pages} {235109} (\bibinfo {year} {2013})}\BibitemShut {NoStop}%
\bibitem [{\citenamefont {Lapa}\ \emph {et~al.}(2017)\citenamefont {Lapa}, \citenamefont {Jian}, \citenamefont {Ye},\ and\ \citenamefont {Hughes}}]{tr5}%
  \BibitemOpen
  \bibfield  {author} {\bibinfo {author} {\bibfnamefont {M.~F.}\ \bibnamefont {Lapa}}, \bibinfo {author} {\bibfnamefont {C.-M.}\ \bibnamefont {Jian}}, \bibinfo {author} {\bibfnamefont {P.}~\bibnamefont {Ye}},\ and\ \bibinfo {author} {\bibfnamefont {T.~L.}\ \bibnamefont {Hughes}},\ }\bibfield  {title} {\bibinfo {title} {Topological electromagnetic responses of bosonic quantum hall, topological insulator, and chiral semimetal phases in all dimensions},\ }\href {https://doi.org/10.1103/PhysRevB.95.035149} {\bibfield  {journal} {\bibinfo  {journal} {Phys. Rev. B}\ }\textbf {\bibinfo {volume} {95}},\ \bibinfo {pages} {035149} (\bibinfo {year} {2017})}\BibitemShut {NoStop}%
\bibitem [{\citenamefont {Wang}\ \emph {et~al.}(2015{\natexlab{a}})\citenamefont {Wang}, \citenamefont {Gu},\ and\ \citenamefont {Wen}}]{tr6}%
  \BibitemOpen
  \bibfield  {author} {\bibinfo {author} {\bibfnamefont {J.~C.}\ \bibnamefont {Wang}}, \bibinfo {author} {\bibfnamefont {Z.-C.}\ \bibnamefont {Gu}},\ and\ \bibinfo {author} {\bibfnamefont {X.-G.}\ \bibnamefont {Wen}},\ }\bibfield  {title} {\bibinfo {title} {Field-theory representation of gauge-gravity symmetry-protected topological invariants, group cohomology, and beyond},\ }\href {https://doi.org/10.1103/PhysRevLett.114.031601} {\bibfield  {journal} {\bibinfo  {journal} {Phys. Rev. Lett.}\ }\textbf {\bibinfo {volume} {114}},\ \bibinfo {pages} {031601} (\bibinfo {year} {2015}{\natexlab{a}})}\BibitemShut {NoStop}%
\bibitem [{\citenamefont {Han}\ \emph {et~al.}(2019)\citenamefont {Han}, \citenamefont {Wang},\ and\ \citenamefont {Ye}}]{tr7}%
  \BibitemOpen
  \bibfield  {author} {\bibinfo {author} {\bibfnamefont {B.}~\bibnamefont {Han}}, \bibinfo {author} {\bibfnamefont {H.}~\bibnamefont {Wang}},\ and\ \bibinfo {author} {\bibfnamefont {P.}~\bibnamefont {Ye}},\ }\bibfield  {title} {\bibinfo {title} {Generalized wen-zee terms},\ }\href {https://doi.org/10.1103/PhysRevB.99.205120} {\bibfield  {journal} {\bibinfo  {journal} {Phys. Rev. B}\ }\textbf {\bibinfo {volume} {99}},\ \bibinfo {pages} {205120} (\bibinfo {year} {2019})}\BibitemShut {NoStop}%
\bibitem [{\citenamefont {Ye}\ and\ \citenamefont {Wen}(2013)}]{pc1}%
  \BibitemOpen
  \bibfield  {author} {\bibinfo {author} {\bibfnamefont {P.}~\bibnamefont {Ye}}\ and\ \bibinfo {author} {\bibfnamefont {X.-G.}\ \bibnamefont {Wen}},\ }\bibfield  {title} {\bibinfo {title} {Projective construction of two-dimensional symmetry-protected topological phases with u(1), so(3), or su(2) symmetries},\ }\href {https://doi.org/10.1103/PhysRevB.87.195128} {\bibfield  {journal} {\bibinfo  {journal} {Phys. Rev. B}\ }\textbf {\bibinfo {volume} {87}},\ \bibinfo {pages} {195128} (\bibinfo {year} {2013})}\BibitemShut {NoStop}%
\bibitem [{\citenamefont {Lu}\ and\ \citenamefont {Lee}(2014)}]{pc2}%
  \BibitemOpen
  \bibfield  {author} {\bibinfo {author} {\bibfnamefont {Y.-M.}\ \bibnamefont {Lu}}\ and\ \bibinfo {author} {\bibfnamefont {D.-H.}\ \bibnamefont {Lee}},\ }\bibfield  {title} {\bibinfo {title} {Spin quantum hall effects in featureless nonfractionalized spin-1 magnets},\ }\href {https://doi.org/10.1103/PhysRevB.89.184417} {\bibfield  {journal} {\bibinfo  {journal} {Phys. Rev. B}\ }\textbf {\bibinfo {volume} {89}},\ \bibinfo {pages} {184417} (\bibinfo {year} {2014})}\BibitemShut {NoStop}%
\bibitem [{\citenamefont {Liu}\ \emph {et~al.}(2014)\citenamefont {Liu}, \citenamefont {Mei}, \citenamefont {Ye},\ and\ \citenamefont {Wen}}]{pc3}%
  \BibitemOpen
  \bibfield  {author} {\bibinfo {author} {\bibfnamefont {Z.-X.}\ \bibnamefont {Liu}}, \bibinfo {author} {\bibfnamefont {J.-W.}\ \bibnamefont {Mei}}, \bibinfo {author} {\bibfnamefont {P.}~\bibnamefont {Ye}},\ and\ \bibinfo {author} {\bibfnamefont {X.-G.}\ \bibnamefont {Wen}},\ }\bibfield  {title} {\bibinfo {title} {$u(1)\ifmmode\times\else\texttimes\fi{}u(1)$ symmetry-protected topological order in gutzwiller wave functions},\ }\href {https://doi.org/10.1103/PhysRevB.90.235146} {\bibfield  {journal} {\bibinfo  {journal} {Phys. Rev. B}\ }\textbf {\bibinfo {volume} {90}},\ \bibinfo {pages} {235146} (\bibinfo {year} {2014})}\BibitemShut {NoStop}%
\bibitem [{\citenamefont {Ye}\ and\ \citenamefont {Wen}(2014)}]{pc4}%
  \BibitemOpen
  \bibfield  {author} {\bibinfo {author} {\bibfnamefont {P.}~\bibnamefont {Ye}}\ and\ \bibinfo {author} {\bibfnamefont {X.-G.}\ \bibnamefont {Wen}},\ }\bibfield  {title} {\bibinfo {title} {Constructing symmetric topological phases of bosons in three dimensions via fermionic projective construction and dyon condensation},\ }\href {https://doi.org/10.1103/PhysRevB.89.045127} {\bibfield  {journal} {\bibinfo  {journal} {Phys. Rev. B}\ }\textbf {\bibinfo {volume} {89}},\ \bibinfo {pages} {045127} (\bibinfo {year} {2014})}\BibitemShut {NoStop}%
\bibitem [{\citenamefont {Ye}\ \emph {et~al.}(2016)\citenamefont {Ye}, \citenamefont {Hughes}, \citenamefont {Maciejko},\ and\ \citenamefont {Fradkin}}]{pc5}%
  \BibitemOpen
  \bibfield  {author} {\bibinfo {author} {\bibfnamefont {P.}~\bibnamefont {Ye}}, \bibinfo {author} {\bibfnamefont {T.~L.}\ \bibnamefont {Hughes}}, \bibinfo {author} {\bibfnamefont {J.}~\bibnamefont {Maciejko}},\ and\ \bibinfo {author} {\bibfnamefont {E.}~\bibnamefont {Fradkin}},\ }\bibfield  {title} {\bibinfo {title} {Composite particle theory of three-dimensional gapped fermionic phases: Fractional topological insulators and charge-loop excitation symmetry},\ }\href {https://doi.org/10.1103/PhysRevB.94.115104} {\bibfield  {journal} {\bibinfo  {journal} {Phys. Rev. B}\ }\textbf {\bibinfo {volume} {94}},\ \bibinfo {pages} {115104} (\bibinfo {year} {2016})}\BibitemShut {NoStop}%
\bibitem [{\citenamefont {Wang}\ \emph {et~al.}(2015{\natexlab{b}})\citenamefont {Wang}, \citenamefont {Nahum},\ and\ \citenamefont {Senthil}}]{pc6}%
  \BibitemOpen
  \bibfield  {author} {\bibinfo {author} {\bibfnamefont {C.}~\bibnamefont {Wang}}, \bibinfo {author} {\bibfnamefont {A.}~\bibnamefont {Nahum}},\ and\ \bibinfo {author} {\bibfnamefont {T.}~\bibnamefont {Senthil}},\ }\bibfield  {title} {\bibinfo {title} {Topological paramagnetism in frustrated spin-1 mott insulators},\ }\href {https://doi.org/10.1103/PhysRevB.91.195131} {\bibfield  {journal} {\bibinfo  {journal} {Phys. Rev. B}\ }\textbf {\bibinfo {volume} {91}},\ \bibinfo {pages} {195131} (\bibinfo {year} {2015}{\natexlab{b}})}\BibitemShut {NoStop}%
\bibitem [{\citenamefont {Levin}\ and\ \citenamefont {Gu}(2012{\natexlab{b}})}]{bs1}%
  \BibitemOpen
  \bibfield  {author} {\bibinfo {author} {\bibfnamefont {M.}~\bibnamefont {Levin}}\ and\ \bibinfo {author} {\bibfnamefont {Z.-C.}\ \bibnamefont {Gu}},\ }\bibfield  {title} {\bibinfo {title} {Braiding statistics approach to symmetry-protected topological phases},\ }\href {https://doi.org/10.1103/PhysRevB.86.115109} {\bibfield  {journal} {\bibinfo  {journal} {Phys. Rev. B}\ }\textbf {\bibinfo {volume} {86}},\ \bibinfo {pages} {115109} (\bibinfo {year} {2012}{\natexlab{b}})}\BibitemShut {NoStop}%
\bibitem [{\citenamefont {Wang}\ and\ \citenamefont {Levin}(2014)}]{bs2}%
  \BibitemOpen
  \bibfield  {author} {\bibinfo {author} {\bibfnamefont {C.}~\bibnamefont {Wang}}\ and\ \bibinfo {author} {\bibfnamefont {M.}~\bibnamefont {Levin}},\ }\bibfield  {title} {\bibinfo {title} {Braiding statistics of loop excitations in three dimensions},\ }\href {https://doi.org/10.1103/PhysRevLett.113.080403} {\bibfield  {journal} {\bibinfo  {journal} {Phys. Rev. Lett.}\ }\textbf {\bibinfo {volume} {113}},\ \bibinfo {pages} {080403} (\bibinfo {year} {2014})}\BibitemShut {NoStop}%
\bibitem [{\citenamefont {Putrov}\ \emph {et~al.}(2017)\citenamefont {Putrov}, \citenamefont {Wang},\ and\ \citenamefont {Yau}}]{bs3}%
  \BibitemOpen
  \bibfield  {author} {\bibinfo {author} {\bibfnamefont {P.}~\bibnamefont {Putrov}}, \bibinfo {author} {\bibfnamefont {J.}~\bibnamefont {Wang}},\ and\ \bibinfo {author} {\bibfnamefont {S.-T.}\ \bibnamefont {Yau}},\ }\bibfield  {title} {\bibinfo {title} {Braiding statistics and link invariants of bosonic/fermionic topological quantum matter in 2+1 and 3+1 dimensions},\ }\href {https://doi.org/https://doi.org/10.1016/j.aop.2017.06.019} {\bibfield  {journal} {\bibinfo  {journal} {Annals of Physics}\ }\textbf {\bibinfo {volume} {384}},\ \bibinfo {pages} {254} (\bibinfo {year} {2017})}\BibitemShut {NoStop}%
\bibitem [{\citenamefont {Chan}\ \emph {et~al.}(2018)\citenamefont {Chan}, \citenamefont {Ye},\ and\ \citenamefont {Ryu}}]{bs4}%
  \BibitemOpen
  \bibfield  {author} {\bibinfo {author} {\bibfnamefont {A.~P.~O.}\ \bibnamefont {Chan}}, \bibinfo {author} {\bibfnamefont {P.}~\bibnamefont {Ye}},\ and\ \bibinfo {author} {\bibfnamefont {S.}~\bibnamefont {Ryu}},\ }\bibfield  {title} {\bibinfo {title} {Braiding with borromean rings in ($3+1$)-dimensional spacetime},\ }\href {https://doi.org/10.1103/PhysRevLett.121.061601} {\bibfield  {journal} {\bibinfo  {journal} {Phys. Rev. Lett.}\ }\textbf {\bibinfo {volume} {121}},\ \bibinfo {pages} {061601} (\bibinfo {year} {2018})}\BibitemShut {NoStop}%
\bibitem [{\citenamefont {Pretko}\ \emph {et~al.}(2020)\citenamefont {Pretko}, \citenamefont {Chen},\ and\ \citenamefont {You}}]{f1}%
  \BibitemOpen
  \bibfield  {author} {\bibinfo {author} {\bibfnamefont {M.}~\bibnamefont {Pretko}}, \bibinfo {author} {\bibfnamefont {X.}~\bibnamefont {Chen}},\ and\ \bibinfo {author} {\bibfnamefont {Y.}~\bibnamefont {You}},\ }\bibfield  {title} {\bibinfo {title} {Fracton phases of matter},\ }\href {http://dx.doi.org/10.1142/S0217751X20300033} {\bibfield  {journal} {\bibinfo  {journal} {International Journal of Modern Physics A}\ }\textbf {\bibinfo {volume} {35}},\ \bibinfo {pages} {2030003} (\bibinfo {year} {2020})}\BibitemShut {NoStop}%
\bibitem [{\citenamefont {Chamon}(2005)}]{f2}%
  \BibitemOpen
  \bibfield  {author} {\bibinfo {author} {\bibfnamefont {C.}~\bibnamefont {Chamon}},\ }\bibfield  {title} {\bibinfo {title} {Quantum glassiness in strongly correlated clean systems: An example of topological overprotection},\ }\href {https://doi.org/10.1103/PhysRevLett.94.040402} {\bibfield  {journal} {\bibinfo  {journal} {Phys. Rev. Lett.}\ }\textbf {\bibinfo {volume} {94}},\ \bibinfo {pages} {040402} (\bibinfo {year} {2005})}\BibitemShut {NoStop}%
\bibitem [{\citenamefont {Haah}(2011)}]{f3}%
  \BibitemOpen
  \bibfield  {author} {\bibinfo {author} {\bibfnamefont {J.}~\bibnamefont {Haah}},\ }\bibfield  {title} {\bibinfo {title} {Local stabilizer codes in three dimensions without string logical operators},\ }\href {https://doi.org/10.1103/PhysRevA.83.042330} {\bibfield  {journal} {\bibinfo  {journal} {Phys. Rev. A}\ }\textbf {\bibinfo {volume} {83}},\ \bibinfo {pages} {042330} (\bibinfo {year} {2011})}\BibitemShut {NoStop}%
\bibitem [{\citenamefont {Yoshida}(2013)}]{f4}%
  \BibitemOpen
  \bibfield  {author} {\bibinfo {author} {\bibfnamefont {B.}~\bibnamefont {Yoshida}},\ }\bibfield  {title} {\bibinfo {title} {Exotic topological order in fractal spin liquids},\ }\href {https://doi.org/10.1103/PhysRevB.88.125122} {\bibfield  {journal} {\bibinfo  {journal} {Phys. Rev. B}\ }\textbf {\bibinfo {volume} {88}},\ \bibinfo {pages} {125122} (\bibinfo {year} {2013})}\BibitemShut {NoStop}%
\bibitem [{\citenamefont {Vijay}\ \emph {et~al.}(2016)\citenamefont {Vijay}, \citenamefont {Haah},\ and\ \citenamefont {Fu}}]{f5}%
  \BibitemOpen
  \bibfield  {author} {\bibinfo {author} {\bibfnamefont {S.}~\bibnamefont {Vijay}}, \bibinfo {author} {\bibfnamefont {J.}~\bibnamefont {Haah}},\ and\ \bibinfo {author} {\bibfnamefont {L.}~\bibnamefont {Fu}},\ }\bibfield  {title} {\bibinfo {title} {Fracton topological order, generalized lattice gauge theory, and duality},\ }\href {https://doi.org/10.1103/PhysRevB.94.235157} {\bibfield  {journal} {\bibinfo  {journal} {Phys. Rev. B}\ }\textbf {\bibinfo {volume} {94}},\ \bibinfo {pages} {235157} (\bibinfo {year} {2016})}\BibitemShut {NoStop}%
\bibitem [{\citenamefont {Ma}\ \emph {et~al.}(2017)\citenamefont {Ma}, \citenamefont {Lake}, \citenamefont {Chen},\ and\ \citenamefont {Hermele}}]{f6}%
  \BibitemOpen
  \bibfield  {author} {\bibinfo {author} {\bibfnamefont {H.}~\bibnamefont {Ma}}, \bibinfo {author} {\bibfnamefont {E.}~\bibnamefont {Lake}}, \bibinfo {author} {\bibfnamefont {X.}~\bibnamefont {Chen}},\ and\ \bibinfo {author} {\bibfnamefont {M.}~\bibnamefont {Hermele}},\ }\bibfield  {title} {\bibinfo {title} {Fracton topological order via coupled layers},\ }\href {https://doi.org/10.1103/PhysRevB.95.245126} {\bibfield  {journal} {\bibinfo  {journal} {Phys. Rev. B}\ }\textbf {\bibinfo {volume} {95}},\ \bibinfo {pages} {245126} (\bibinfo {year} {2017})}\BibitemShut {NoStop}%
\bibitem [{\citenamefont {Vijay}(2017)}]{f7}%
  \BibitemOpen
  \bibfield  {author} {\bibinfo {author} {\bibfnamefont {S.}~\bibnamefont {Vijay}},\ }\href@noop {} {\bibinfo {title} {Isotropic layer construction and phase diagram for fracton topological phases}} (\bibinfo {year} {2017}),\ \Eprint {https://arxiv.org/abs/1701.00762} {arXiv:1701.00762} \BibitemShut {NoStop}%
\bibitem [{\citenamefont {Shirley}\ \emph {et~al.}(2018)\citenamefont {Shirley}, \citenamefont {Slagle}, \citenamefont {Wang},\ and\ \citenamefont {Chen}}]{f8}%
  \BibitemOpen
  \bibfield  {author} {\bibinfo {author} {\bibfnamefont {W.}~\bibnamefont {Shirley}}, \bibinfo {author} {\bibfnamefont {K.}~\bibnamefont {Slagle}}, \bibinfo {author} {\bibfnamefont {Z.}~\bibnamefont {Wang}},\ and\ \bibinfo {author} {\bibfnamefont {X.}~\bibnamefont {Chen}},\ }\bibfield  {title} {\bibinfo {title} {Fracton models on general three-dimensional manifolds},\ }\href {https://doi.org/10.1103/PhysRevX.8.031051} {\bibfield  {journal} {\bibinfo  {journal} {Phys. Rev. X}\ }\textbf {\bibinfo {volume} {8}},\ \bibinfo {pages} {031051} (\bibinfo {year} {2018})}\BibitemShut {NoStop}%
\bibitem [{\citenamefont {Ma}\ \emph {et~al.}(2022)\citenamefont {Ma}, \citenamefont {Shirley}, \citenamefont {Cheng}, \citenamefont {Levin}, \citenamefont {McGreevy},\ and\ \citenamefont {Chen}}]{f9}%
  \BibitemOpen
  \bibfield  {author} {\bibinfo {author} {\bibfnamefont {X.}~\bibnamefont {Ma}}, \bibinfo {author} {\bibfnamefont {W.}~\bibnamefont {Shirley}}, \bibinfo {author} {\bibfnamefont {M.}~\bibnamefont {Cheng}}, \bibinfo {author} {\bibfnamefont {M.}~\bibnamefont {Levin}}, \bibinfo {author} {\bibfnamefont {J.}~\bibnamefont {McGreevy}},\ and\ \bibinfo {author} {\bibfnamefont {X.}~\bibnamefont {Chen}},\ }\bibfield  {title} {\bibinfo {title} {Fractonic order in infinite-component chern-simons gauge theories},\ }\href {https://doi.org/10.1103/PhysRevB.105.195124} {\bibfield  {journal} {\bibinfo  {journal} {Phys. Rev. B}\ }\textbf {\bibinfo {volume} {105}},\ \bibinfo {pages} {195124} (\bibinfo {year} {2022})}\BibitemShut {NoStop}%
\bibitem [{\citenamefont {Aasen}\ \emph {et~al.}(2020)\citenamefont {Aasen}, \citenamefont {Bulmash}, \citenamefont {Prem}, \citenamefont {Slagle},\ and\ \citenamefont {Williamson}}]{f10}%
  \BibitemOpen
  \bibfield  {author} {\bibinfo {author} {\bibfnamefont {D.}~\bibnamefont {Aasen}}, \bibinfo {author} {\bibfnamefont {D.}~\bibnamefont {Bulmash}}, \bibinfo {author} {\bibfnamefont {A.}~\bibnamefont {Prem}}, \bibinfo {author} {\bibfnamefont {K.}~\bibnamefont {Slagle}},\ and\ \bibinfo {author} {\bibfnamefont {D.~J.}\ \bibnamefont {Williamson}},\ }\bibfield  {title} {\bibinfo {title} {Topological defect networks for fractons of all types},\ }\href {https://doi.org/10.1103/PhysRevResearch.2.043165} {\bibfield  {journal} {\bibinfo  {journal} {Phys. Rev. Res.}\ }\textbf {\bibinfo {volume} {2}},\ \bibinfo {pages} {043165} (\bibinfo {year} {2020})}\BibitemShut {NoStop}%
\bibitem [{\citenamefont {Slagle}(2021)}]{f11}%
  \BibitemOpen
  \bibfield  {author} {\bibinfo {author} {\bibfnamefont {K.}~\bibnamefont {Slagle}},\ }\bibfield  {title} {\bibinfo {title} {Foliated quantum field theory of fracton order},\ }\href {https://doi.org/10.1103/PhysRevLett.126.101603} {\bibfield  {journal} {\bibinfo  {journal} {Phys. Rev. Lett.}\ }\textbf {\bibinfo {volume} {126}},\ \bibinfo {pages} {101603} (\bibinfo {year} {2021})}\BibitemShut {NoStop}%
\bibitem [{\citenamefont {Li}\ and\ \citenamefont {Ye}(2020)}]{f12}%
  \BibitemOpen
  \bibfield  {author} {\bibinfo {author} {\bibfnamefont {M.-Y.}\ \bibnamefont {Li}}\ and\ \bibinfo {author} {\bibfnamefont {P.}~\bibnamefont {Ye}},\ }\bibfield  {title} {\bibinfo {title} {Fracton physics of spatially extended excitations},\ }\href {https://doi.org/10.1103/PhysRevB.101.245134} {\bibfield  {journal} {\bibinfo  {journal} {Phys. Rev. B}\ }\textbf {\bibinfo {volume} {101}},\ \bibinfo {pages} {245134} (\bibinfo {year} {2020})}\BibitemShut {NoStop}%
\bibitem [{\citenamefont {Li}\ and\ \citenamefont {Ye}(2021{\natexlab{a}})}]{f13}%
  \BibitemOpen
  \bibfield  {author} {\bibinfo {author} {\bibfnamefont {M.-Y.}\ \bibnamefont {Li}}\ and\ \bibinfo {author} {\bibfnamefont {P.}~\bibnamefont {Ye}},\ }\bibfield  {title} {\bibinfo {title} {Fracton physics of spatially extended excitations. ii. polynomial ground state degeneracy of exactly solvable models},\ }\href {https://doi.org/10.1103/PhysRevB.104.235127} {\bibfield  {journal} {\bibinfo  {journal} {Phys. Rev. B}\ }\textbf {\bibinfo {volume} {104}},\ \bibinfo {pages} {235127} (\bibinfo {year} {2021}{\natexlab{a}})}\BibitemShut {NoStop}%
\bibitem [{\citenamefont {Xu}\ and\ \citenamefont {Wu}(2008)}]{f14}%
  \BibitemOpen
  \bibfield  {author} {\bibinfo {author} {\bibfnamefont {C.}~\bibnamefont {Xu}}\ and\ \bibinfo {author} {\bibfnamefont {C.}~\bibnamefont {Wu}},\ }\bibfield  {title} {\bibinfo {title} {Resonating plaquette phases in su(4) heisenberg antiferromagnet},\ }\href {https://doi.org/10.1103/PhysRevB.77.134449} {\bibfield  {journal} {\bibinfo  {journal} {Phys. Rev. B}\ }\textbf {\bibinfo {volume} {77}},\ \bibinfo {pages} {134449} (\bibinfo {year} {2008})}\BibitemShut {NoStop}%
\bibitem [{\citenamefont {Pretko}(2017)}]{f15}%
  \BibitemOpen
  \bibfield  {author} {\bibinfo {author} {\bibfnamefont {M.}~\bibnamefont {Pretko}},\ }\bibfield  {title} {\bibinfo {title} {Subdimensional particle structure of higher rank $u(1)$ spin liquids},\ }\href {https://doi.org/10.1103/PhysRevB.95.115139} {\bibfield  {journal} {\bibinfo  {journal} {Phys. Rev. B}\ }\textbf {\bibinfo {volume} {95}},\ \bibinfo {pages} {115139} (\bibinfo {year} {2017})}\BibitemShut {NoStop}%
\bibitem [{\citenamefont {Ma}\ \emph {et~al.}(2018)\citenamefont {Ma}, \citenamefont {Hermele},\ and\ \citenamefont {Chen}}]{f16}%
  \BibitemOpen
  \bibfield  {author} {\bibinfo {author} {\bibfnamefont {H.}~\bibnamefont {Ma}}, \bibinfo {author} {\bibfnamefont {M.}~\bibnamefont {Hermele}},\ and\ \bibinfo {author} {\bibfnamefont {X.}~\bibnamefont {Chen}},\ }\bibfield  {title} {\bibinfo {title} {Fracton topological order from the higgs and partial-confinement mechanisms of rank-two gauge theory},\ }\href {https://doi.org/10.1103/PhysRevB.98.035111} {\bibfield  {journal} {\bibinfo  {journal} {Phys. Rev. B}\ }\textbf {\bibinfo {volume} {98}},\ \bibinfo {pages} {035111} (\bibinfo {year} {2018})}\BibitemShut {NoStop}%
\bibitem [{\citenamefont {Bulmash}\ and\ \citenamefont {Barkeshli}(2018)}]{f17}%
  \BibitemOpen
  \bibfield  {author} {\bibinfo {author} {\bibfnamefont {D.}~\bibnamefont {Bulmash}}\ and\ \bibinfo {author} {\bibfnamefont {M.}~\bibnamefont {Barkeshli}},\ }\bibfield  {title} {\bibinfo {title} {Higgs mechanism in higher-rank symmetric u(1) gauge theories},\ }\href {https://doi.org/10.1103/PhysRevB.97.235112} {\bibfield  {journal} {\bibinfo  {journal} {Phys. Rev. B}\ }\textbf {\bibinfo {volume} {97}},\ \bibinfo {pages} {235112} (\bibinfo {year} {2018})}\BibitemShut {NoStop}%
\bibitem [{\citenamefont {Seiberg}\ and\ \citenamefont {Shao}(2021)}]{f18}%
  \BibitemOpen
  \bibfield  {author} {\bibinfo {author} {\bibfnamefont {N.}~\bibnamefont {Seiberg}}\ and\ \bibinfo {author} {\bibfnamefont {S.-H.}\ \bibnamefont {Shao}},\ }\bibfield  {title} {\bibinfo {title} {{Exotic $\mathbb{Z}_N$ symmetries, duality, and fractons in 3+1-dimensional quantum field theory}},\ }\href {https://scipost.org/10.21468/SciPostPhys.10.1.003} {\bibfield  {journal} {\bibinfo  {journal} {SciPost Phys.}\ }\textbf {\bibinfo {volume} {10}},\ \bibinfo {pages} {003} (\bibinfo {year} {2021})}\BibitemShut {NoStop}%
\bibitem [{\citenamefont {Pretko}(2018{\natexlab{a}})}]{f19}%
  \BibitemOpen
  \bibfield  {author} {\bibinfo {author} {\bibfnamefont {M.}~\bibnamefont {Pretko}},\ }\bibfield  {title} {\bibinfo {title} {The fracton gauge principle},\ }\href {https://doi.org/10.1103/PhysRevB.98.115134} {\bibfield  {journal} {\bibinfo  {journal} {Phys. Rev. B}\ }\textbf {\bibinfo {volume} {98}},\ \bibinfo {pages} {115134} (\bibinfo {year} {2018}{\natexlab{a}})}\BibitemShut {NoStop}%
\bibitem [{\citenamefont {Gromov}(2019{\natexlab{a}})}]{f20}%
  \BibitemOpen
  \bibfield  {author} {\bibinfo {author} {\bibfnamefont {A.}~\bibnamefont {Gromov}},\ }\bibfield  {title} {\bibinfo {title} {Towards classification of fracton phases: The multipole algebra},\ }\href {https://doi.org/10.1103/PhysRevX.9.031035} {\bibfield  {journal} {\bibinfo  {journal} {Phys. Rev. X}\ }\textbf {\bibinfo {volume} {9}},\ \bibinfo {pages} {031035} (\bibinfo {year} {2019}{\natexlab{a}})}\BibitemShut {NoStop}%
\bibitem [{\citenamefont {Seiberg}(2020{\natexlab{a}})}]{f21}%
  \BibitemOpen
  \bibfield  {author} {\bibinfo {author} {\bibfnamefont {N.}~\bibnamefont {Seiberg}},\ }\bibfield  {title} {\bibinfo {title} {{Field theories with a vector global symmetry}},\ }\href {https://scipost.org/10.21468/SciPostPhys.8.4.050} {\bibfield  {journal} {\bibinfo  {journal} {SciPost Phys.}\ }\textbf {\bibinfo {volume} {8}},\ \bibinfo {pages} {050} (\bibinfo {year} {2020}{\natexlab{a}})}\BibitemShut {NoStop}%
\bibitem [{\citenamefont {Yuan}\ \emph {et~al.}(2020)\citenamefont {Yuan}, \citenamefont {Chen},\ and\ \citenamefont {Ye}}]{f22}%
  \BibitemOpen
  \bibfield  {author} {\bibinfo {author} {\bibfnamefont {J.-K.}\ \bibnamefont {Yuan}}, \bibinfo {author} {\bibfnamefont {S.~A.}\ \bibnamefont {Chen}},\ and\ \bibinfo {author} {\bibfnamefont {P.}~\bibnamefont {Ye}},\ }\bibfield  {title} {\bibinfo {title} {Fractonic superfluids},\ }\href {https://doi.org/10.1103/PhysRevResearch.2.023267} {\bibfield  {journal} {\bibinfo  {journal} {Phys. Rev. Res.}\ }\textbf {\bibinfo {volume} {2}},\ \bibinfo {pages} {023267} (\bibinfo {year} {2020})}\BibitemShut {NoStop}%
\bibitem [{\citenamefont {Chen}\ \emph {et~al.}(2021{\natexlab{a}})\citenamefont {Chen}, \citenamefont {Yuan},\ and\ \citenamefont {Ye}}]{f23}%
  \BibitemOpen
  \bibfield  {author} {\bibinfo {author} {\bibfnamefont {S.~A.}\ \bibnamefont {Chen}}, \bibinfo {author} {\bibfnamefont {J.-K.}\ \bibnamefont {Yuan}},\ and\ \bibinfo {author} {\bibfnamefont {P.}~\bibnamefont {Ye}},\ }\bibfield  {title} {\bibinfo {title} {Fractonic superfluids. ii. condensing subdimensional particles},\ }\href {https://doi.org/10.1103/PhysRevResearch.3.013226} {\bibfield  {journal} {\bibinfo  {journal} {Phys. Rev. Res.}\ }\textbf {\bibinfo {volume} {3}},\ \bibinfo {pages} {013226} (\bibinfo {year} {2021}{\natexlab{a}})}\BibitemShut {NoStop}%
\bibitem [{\citenamefont {Li}\ and\ \citenamefont {Ye}(2021{\natexlab{b}})}]{f24}%
  \BibitemOpen
  \bibfield  {author} {\bibinfo {author} {\bibfnamefont {H.}~\bibnamefont {Li}}\ and\ \bibinfo {author} {\bibfnamefont {P.}~\bibnamefont {Ye}},\ }\bibfield  {title} {\bibinfo {title} {Renormalization group analysis on emergence of higher rank symmetry and higher moment conservation},\ }\href {https://doi.org/10.1103/PhysRevResearch.3.043176} {\bibfield  {journal} {\bibinfo  {journal} {Phys. Rev. Res.}\ }\textbf {\bibinfo {volume} {3}},\ \bibinfo {pages} {043176} (\bibinfo {year} {2021}{\natexlab{b}})}\BibitemShut {NoStop}%
\bibitem [{\citenamefont {Yuan}\ \emph {et~al.}(2023)\citenamefont {Yuan}, \citenamefont {Chen},\ and\ \citenamefont {Ye}}]{f25}%
  \BibitemOpen
  \bibfield  {author} {\bibinfo {author} {\bibfnamefont {J.-K.}\ \bibnamefont {Yuan}}, \bibinfo {author} {\bibfnamefont {S.~A.}\ \bibnamefont {Chen}},\ and\ \bibinfo {author} {\bibfnamefont {P.}~\bibnamefont {Ye}},\ }\bibfield  {title} {\bibinfo {title} {Hierarchical proliferation of higher-rank symmetry defects in fractonic superfluids},\ }\href {https://doi.org/10.1103/PhysRevB.107.205134} {\bibfield  {journal} {\bibinfo  {journal} {Phys. Rev. B}\ }\textbf {\bibinfo {volume} {107}},\ \bibinfo {pages} {205134} (\bibinfo {year} {2023})}\BibitemShut {NoStop}%
\bibitem [{\citenamefont {Zhu}\ \emph {et~al.}(2023)\citenamefont {Zhu}, \citenamefont {Chen}, \citenamefont {Ye},\ and\ \citenamefont {Trebst}}]{f26}%
  \BibitemOpen
  \bibfield  {author} {\bibinfo {author} {\bibfnamefont {G.-Y.}\ \bibnamefont {Zhu}}, \bibinfo {author} {\bibfnamefont {J.-Y.}\ \bibnamefont {Chen}}, \bibinfo {author} {\bibfnamefont {P.}~\bibnamefont {Ye}},\ and\ \bibinfo {author} {\bibfnamefont {S.}~\bibnamefont {Trebst}},\ }\bibfield  {title} {\bibinfo {title} {Topological fracton quantum phase transitions by tuning exact tensor network states},\ }\href {https://doi.org/10.1103/PhysRevLett.130.216704} {\bibfield  {journal} {\bibinfo  {journal} {Phys. Rev. Lett.}\ }\textbf {\bibinfo {volume} {130}},\ \bibinfo {pages} {216704} (\bibinfo {year} {2023})}\BibitemShut {NoStop}%
\bibitem [{\citenamefont {Stahl}\ \emph {et~al.}(2022)\citenamefont {Stahl}, \citenamefont {Lake},\ and\ \citenamefont {Nandkishore}}]{f27}%
  \BibitemOpen
  \bibfield  {author} {\bibinfo {author} {\bibfnamefont {C.}~\bibnamefont {Stahl}}, \bibinfo {author} {\bibfnamefont {E.}~\bibnamefont {Lake}},\ and\ \bibinfo {author} {\bibfnamefont {R.}~\bibnamefont {Nandkishore}},\ }\bibfield  {title} {\bibinfo {title} {Spontaneous breaking of multipole symmetries},\ }\href {https://doi.org/10.1103/PhysRevB.105.155107} {\bibfield  {journal} {\bibinfo  {journal} {Phys. Rev. B}\ }\textbf {\bibinfo {volume} {105}},\ \bibinfo {pages} {155107} (\bibinfo {year} {2022})}\BibitemShut {NoStop}%
\bibitem [{\citenamefont {Argurio}\ \emph {et~al.}(2021)\citenamefont {Argurio}, \citenamefont {Hoyos}, \citenamefont {Musso},\ and\ \citenamefont {Naegels}}]{f28}%
  \BibitemOpen
  \bibfield  {author} {\bibinfo {author} {\bibfnamefont {R.}~\bibnamefont {Argurio}}, \bibinfo {author} {\bibfnamefont {C.}~\bibnamefont {Hoyos}}, \bibinfo {author} {\bibfnamefont {D.}~\bibnamefont {Musso}},\ and\ \bibinfo {author} {\bibfnamefont {D.}~\bibnamefont {Naegels}},\ }\bibfield  {title} {\bibinfo {title} {Fractons in effective field theories for spontaneously broken translations},\ }\href {https://doi.org/10.1103/PhysRevD.104.105001} {\bibfield  {journal} {\bibinfo  {journal} {Phys. Rev. D}\ }\textbf {\bibinfo {volume} {104}},\ \bibinfo {pages} {105001} (\bibinfo {year} {2021})}\BibitemShut {NoStop}%
\bibitem [{\citenamefont {Bidussi}\ \emph {et~al.}(2022)\citenamefont {Bidussi}, \citenamefont {Hartong}, \citenamefont {Have}, \citenamefont {Musaeus},\ and\ \citenamefont {Prohazka}}]{f29}%
  \BibitemOpen
  \bibfield  {author} {\bibinfo {author} {\bibfnamefont {L.}~\bibnamefont {Bidussi}}, \bibinfo {author} {\bibfnamefont {J.}~\bibnamefont {Hartong}}, \bibinfo {author} {\bibfnamefont {E.}~\bibnamefont {Have}}, \bibinfo {author} {\bibfnamefont {J.}~\bibnamefont {Musaeus}},\ and\ \bibinfo {author} {\bibfnamefont {S.}~\bibnamefont {Prohazka}},\ }\bibfield  {title} {\bibinfo {title} {{Fractons, dipole symmetries and curved spacetime}},\ }\href {https://scipost.org/10.21468/SciPostPhys.12.6.205} {\bibfield  {journal} {\bibinfo  {journal} {SciPost Phys.}\ }\textbf {\bibinfo {volume} {12}},\ \bibinfo {pages} {205} (\bibinfo {year} {2022})}\BibitemShut {NoStop}%
\bibitem [{\citenamefont {Jain}\ and\ \citenamefont {Jensen}(2022)}]{f30}%
  \BibitemOpen
  \bibfield  {author} {\bibinfo {author} {\bibfnamefont {A.}~\bibnamefont {Jain}}\ and\ \bibinfo {author} {\bibfnamefont {K.}~\bibnamefont {Jensen}},\ }\bibfield  {title} {\bibinfo {title} {{Fractons in curved space}},\ }\href {https://scipost.org/10.21468/SciPostPhys.12.4.142} {\bibfield  {journal} {\bibinfo  {journal} {SciPost Phys.}\ }\textbf {\bibinfo {volume} {12}},\ \bibinfo {pages} {142} (\bibinfo {year} {2022})}\BibitemShut {NoStop}%
\bibitem [{\citenamefont {Angus}\ \emph {et~al.}(2022)\citenamefont {Angus}, \citenamefont {Kim},\ and\ \citenamefont {Park}}]{f31}%
  \BibitemOpen
  \bibfield  {author} {\bibinfo {author} {\bibfnamefont {S.}~\bibnamefont {Angus}}, \bibinfo {author} {\bibfnamefont {M.}~\bibnamefont {Kim}},\ and\ \bibinfo {author} {\bibfnamefont {J.-H.}\ \bibnamefont {Park}},\ }\bibfield  {title} {\bibinfo {title} {Fractons, non-riemannian geometry, and double field theory},\ }\href {https://doi.org/10.1103/PhysRevResearch.4.033186} {\bibfield  {journal} {\bibinfo  {journal} {Phys. Rev. Res.}\ }\textbf {\bibinfo {volume} {4}},\ \bibinfo {pages} {033186} (\bibinfo {year} {2022})}\BibitemShut {NoStop}%
\bibitem [{\citenamefont {Grosvenor}\ \emph {et~al.}(2021)\citenamefont {Grosvenor}, \citenamefont {Hoyos}, \citenamefont {Peña-Benitez},\ and\ \citenamefont {Surówka}}]{f32}%
  \BibitemOpen
  \bibfield  {author} {\bibinfo {author} {\bibfnamefont {K.~T.}\ \bibnamefont {Grosvenor}}, \bibinfo {author} {\bibfnamefont {C.}~\bibnamefont {Hoyos}}, \bibinfo {author} {\bibfnamefont {F.}~\bibnamefont {Peña-Benitez}},\ and\ \bibinfo {author} {\bibfnamefont {P.}~\bibnamefont {Surówka}},\ }\href@noop {} {\bibinfo {title} {Space-dependent symmetries and fractons}} (\bibinfo {year} {2021}),\ \Eprint {https://arxiv.org/abs/2112.00531} {arXiv:2112.00531} \BibitemShut {NoStop}%
\bibitem [{\citenamefont {Banerjee}(2022)}]{f33}%
  \BibitemOpen
  \bibfield  {author} {\bibinfo {author} {\bibfnamefont {R.}~\bibnamefont {Banerjee}},\ }\href@noop {} {\bibinfo {title} {Noether type formulation for space dependent polynomial symmetries}} (\bibinfo {year} {2022}),\ \Eprint {https://arxiv.org/abs/2202.00326} {arXiv:2202.00326} \BibitemShut {NoStop}%
\bibitem [{\citenamefont {Pretko}(2018{\natexlab{b}})}]{f34}%
  \BibitemOpen
  \bibfield  {author} {\bibinfo {author} {\bibfnamefont {M.}~\bibnamefont {Pretko}},\ }\bibfield  {title} {\bibinfo {title} {The fracton gauge principle},\ }\href {https://doi.org/10.1103/PhysRevB.98.115134} {\bibfield  {journal} {\bibinfo  {journal} {Phys. Rev. B}\ }\textbf {\bibinfo {volume} {98}},\ \bibinfo {pages} {115134} (\bibinfo {year} {2018}{\natexlab{b}})}\BibitemShut {NoStop}%
\bibitem [{\citenamefont {Gromov}(2019{\natexlab{b}})}]{f35}%
  \BibitemOpen
  \bibfield  {author} {\bibinfo {author} {\bibfnamefont {A.}~\bibnamefont {Gromov}},\ }\bibfield  {title} {\bibinfo {title} {Towards classification of fracton phases: The multipole algebra},\ }\href {https://doi.org/10.1103/PhysRevX.9.031035} {\bibfield  {journal} {\bibinfo  {journal} {Phys. Rev. X}\ }\textbf {\bibinfo {volume} {9}},\ \bibinfo {pages} {031035} (\bibinfo {year} {2019}{\natexlab{b}})}\BibitemShut {NoStop}%
\bibitem [{\citenamefont {Seiberg}(2020{\natexlab{b}})}]{f36}%
  \BibitemOpen
  \bibfield  {author} {\bibinfo {author} {\bibfnamefont {N.}~\bibnamefont {Seiberg}},\ }\bibfield  {title} {\bibinfo {title} {{Field theories with a vector global symmetry}},\ }\href {https://scipost.org/10.21468/SciPostPhys.8.4.050} {\bibfield  {journal} {\bibinfo  {journal} {SciPost Phys.}\ }\textbf {\bibinfo {volume} {8}},\ \bibinfo {pages} {050} (\bibinfo {year} {2020}{\natexlab{b}})}\BibitemShut {NoStop}%
\bibitem [{\citenamefont {Gorantla}\ \emph {et~al.}(2022)\citenamefont {Gorantla}, \citenamefont {Lam}, \citenamefont {Seiberg},\ and\ \citenamefont {Shao}}]{f37}%
  \BibitemOpen
  \bibfield  {author} {\bibinfo {author} {\bibfnamefont {P.}~\bibnamefont {Gorantla}}, \bibinfo {author} {\bibfnamefont {H.~T.}\ \bibnamefont {Lam}}, \bibinfo {author} {\bibfnamefont {N.}~\bibnamefont {Seiberg}},\ and\ \bibinfo {author} {\bibfnamefont {S.-H.}\ \bibnamefont {Shao}},\ }\bibfield  {title} {\bibinfo {title} {Global dipole symmetry, compact lifshitz theory, tensor gauge theory, and fractons},\ }\href {https://doi.org/10.1103/PhysRevB.106.045112} {\bibfield  {journal} {\bibinfo  {journal} {Phys. Rev. B}\ }\textbf {\bibinfo {volume} {106}},\ \bibinfo {pages} {045112} (\bibinfo {year} {2022})}\BibitemShut {NoStop}%
\bibitem [{\citenamefont {You}\ \emph {et~al.}(2020)\citenamefont {You}, \citenamefont {Devakul}, \citenamefont {Sondhi},\ and\ \citenamefont {Burnell}}]{you2020a}%
  \BibitemOpen
  \bibfield  {author} {\bibinfo {author} {\bibfnamefont {Y.}~\bibnamefont {You}}, \bibinfo {author} {\bibfnamefont {T.}~\bibnamefont {Devakul}}, \bibinfo {author} {\bibfnamefont {S.~L.}\ \bibnamefont {Sondhi}},\ and\ \bibinfo {author} {\bibfnamefont {F.~J.}\ \bibnamefont {Burnell}},\ }\bibfield  {title} {\bibinfo {title} {Fractonic chern-simons and bf theories},\ }\href {https://doi.org/10.1103/PhysRevResearch.2.023249} {\bibfield  {journal} {\bibinfo  {journal} {Phys. Rev. Res.}\ }\textbf {\bibinfo {volume} {2}},\ \bibinfo {pages} {023249} (\bibinfo {year} {2020})}\BibitemShut {NoStop}%
\bibitem [{\citenamefont {Stephen}\ \emph {et~al.}(2019{\natexlab{a}})\citenamefont {Stephen}, \citenamefont {Nautrup}, \citenamefont {Bermejo-Vega}, \citenamefont {Eisert},\ and\ \citenamefont {Raussendorf}}]{stephen2019a}%
  \BibitemOpen
  \bibfield  {author} {\bibinfo {author} {\bibfnamefont {D.~T.}\ \bibnamefont {Stephen}}, \bibinfo {author} {\bibfnamefont {H.~P.}\ \bibnamefont {Nautrup}}, \bibinfo {author} {\bibfnamefont {J.}~\bibnamefont {Bermejo-Vega}}, \bibinfo {author} {\bibfnamefont {J.}~\bibnamefont {Eisert}},\ and\ \bibinfo {author} {\bibfnamefont {R.}~\bibnamefont {Raussendorf}},\ }\bibfield  {title} {\bibinfo {title} {Subsystem symmetries, quantum cellular automata, and computational phases of quantum matter},\ }\href {https://doi.org/10.22331/q-2019-05-20-142} {\bibfield  {journal} {\bibinfo  {journal} {{Quantum}}\ }\textbf {\bibinfo {volume} {3}},\ \bibinfo {pages} {142} (\bibinfo {year} {2019}{\natexlab{a}})}\BibitemShut {NoStop}%
\bibitem [{\citenamefont {Schmitz}\ \emph {et~al.}(2019)\citenamefont {Schmitz}, \citenamefont {Huang},\ and\ \citenamefont {Prem}}]{schmitz2019a}%
  \BibitemOpen
  \bibfield  {author} {\bibinfo {author} {\bibfnamefont {A.~T.}\ \bibnamefont {Schmitz}}, \bibinfo {author} {\bibfnamefont {S.-J.}\ \bibnamefont {Huang}},\ and\ \bibinfo {author} {\bibfnamefont {A.}~\bibnamefont {Prem}},\ }\bibfield  {title} {\bibinfo {title} {Entanglement spectra of stabilizer codes: A window into gapped quantum phases of matter},\ }\href {https://doi.org/10.1103/PhysRevB.99.205109} {\bibfield  {journal} {\bibinfo  {journal} {Phys. Rev. B}\ }\textbf {\bibinfo {volume} {99}},\ \bibinfo {pages} {205109} (\bibinfo {year} {2019})}\BibitemShut {NoStop}%
\bibitem [{\citenamefont {San~Miguel}\ \emph {et~al.}(2021)\citenamefont {San~Miguel}, \citenamefont {Dua},\ and\ \citenamefont {Williamson}}]{miguel2021a}%
  \BibitemOpen
  \bibfield  {author} {\bibinfo {author} {\bibfnamefont {J.~F.}\ \bibnamefont {San~Miguel}}, \bibinfo {author} {\bibfnamefont {A.}~\bibnamefont {Dua}},\ and\ \bibinfo {author} {\bibfnamefont {D.~J.}\ \bibnamefont {Williamson}},\ }\bibfield  {title} {\bibinfo {title} {Bifurcating subsystem symmetric entanglement renormalization in two dimensions},\ }\href {https://doi.org/10.1103/PhysRevB.103.035148} {\bibfield  {journal} {\bibinfo  {journal} {Phys. Rev. B}\ }\textbf {\bibinfo {volume} {103}},\ \bibinfo {pages} {035148} (\bibinfo {year} {2021})}\BibitemShut {NoStop}%
\bibitem [{\citenamefont {Burnell}\ \emph {et~al.}(2022)\citenamefont {Burnell}, \citenamefont {Devakul}, \citenamefont {Gorantla}, \citenamefont {Lam},\ and\ \citenamefont {Shao}}]{burnell2022a}%
  \BibitemOpen
  \bibfield  {author} {\bibinfo {author} {\bibfnamefont {F.~J.}\ \bibnamefont {Burnell}}, \bibinfo {author} {\bibfnamefont {T.}~\bibnamefont {Devakul}}, \bibinfo {author} {\bibfnamefont {P.}~\bibnamefont {Gorantla}}, \bibinfo {author} {\bibfnamefont {H.~T.}\ \bibnamefont {Lam}},\ and\ \bibinfo {author} {\bibfnamefont {S.-H.}\ \bibnamefont {Shao}},\ }\bibfield  {title} {\bibinfo {title} {Anomaly inflow for subsystem symmetries},\ }\href {https://doi.org/10.1103/PhysRevB.106.085113} {\bibfield  {journal} {\bibinfo  {journal} {Phys. Rev. B}\ }\textbf {\bibinfo {volume} {106}},\ \bibinfo {pages} {085113} (\bibinfo {year} {2022})}\BibitemShut {NoStop}%
\bibitem [{\citenamefont {Devakul}\ \emph {et~al.}(2018)\citenamefont {Devakul}, \citenamefont {Parameswaran},\ and\ \citenamefont {Sondhi}}]{devakul2018a}%
  \BibitemOpen
  \bibfield  {author} {\bibinfo {author} {\bibfnamefont {T.}~\bibnamefont {Devakul}}, \bibinfo {author} {\bibfnamefont {S.}~\bibnamefont {Parameswaran}},\ and\ \bibinfo {author} {\bibfnamefont {S.}~\bibnamefont {Sondhi}},\ }\bibfield  {title} {\bibinfo {title} {Correlation function diagnostics for typei fracton phases},\ }\href {https://doi.org/10.1103/PhysRevB.97.041110.} {\bibfield  {journal} {\bibinfo  {journal} {Phys. Rev. B}\ }\textbf {\bibinfo {volume} {97}},\ \bibinfo {pages} {041110} (\bibinfo {year} {2018})}\BibitemShut {NoStop}%
\bibitem [{\citenamefont {Devakul}\ and\ \citenamefont {Williamson}(2018)}]{FSPT1}%
  \BibitemOpen
  \bibfield  {author} {\bibinfo {author} {\bibfnamefont {T.}~\bibnamefont {Devakul}}\ and\ \bibinfo {author} {\bibfnamefont {D.~J.}\ \bibnamefont {Williamson}},\ }\bibfield  {title} {\bibinfo {title} {Universal quantum computation using fractal symmetry-protected cluster phases},\ }\href {https://doi.org/10.1103/PhysRevA.98.022332} {\bibfield  {journal} {\bibinfo  {journal} {Phys. Rev. A}\ }\textbf {\bibinfo {volume} {98}},\ \bibinfo {pages} {022332} (\bibinfo {year} {2018})}\BibitemShut {NoStop}%
\bibitem [{\citenamefont {Devakul}(2019)}]{FSPT2}%
  \BibitemOpen
  \bibfield  {author} {\bibinfo {author} {\bibfnamefont {T.}~\bibnamefont {Devakul}},\ }\bibfield  {title} {\bibinfo {title} {Classifying local fractal subsystem symmetry-protected topological phases},\ }\href {https://doi.org/10.1103/PhysRevB.99.235131} {\bibfield  {journal} {\bibinfo  {journal} {Phys. Rev. B}\ }\textbf {\bibinfo {volume} {99}},\ \bibinfo {pages} {235131} (\bibinfo {year} {2019})}\BibitemShut {NoStop}%
\bibitem [{\citenamefont {Zhou}\ and\ \citenamefont {Ye}(2023)}]{fr1}%
  \BibitemOpen
  \bibfield  {author} {\bibinfo {author} {\bibfnamefont {Y.}~\bibnamefont {Zhou}}\ and\ \bibinfo {author} {\bibfnamefont {P.}~\bibnamefont {Ye}},\ }\href@noop {} {\bibinfo {title} {Entanglement fractalization}} (\bibinfo {year} {2023}),\ \Eprint {https://arxiv.org/abs/2311.01199} {arXiv:2311.01199} \BibitemShut {NoStop}%
\bibitem [{\citenamefont {Gefen}\ \emph {et~al.}(1980)\citenamefont {Gefen}, \citenamefont {Mandelbrot},\ and\ \citenamefont {Aharony}}]{fr2}%
  \BibitemOpen
  \bibfield  {author} {\bibinfo {author} {\bibfnamefont {Y.}~\bibnamefont {Gefen}}, \bibinfo {author} {\bibfnamefont {B.~B.}\ \bibnamefont {Mandelbrot}},\ and\ \bibinfo {author} {\bibfnamefont {A.}~\bibnamefont {Aharony}},\ }\bibfield  {title} {\bibinfo {title} {Critical phenomena on fractal lattices},\ }\href {https://doi.org/10.1103/PhysRevLett.45.855} {\bibfield  {journal} {\bibinfo  {journal} {Phys. Rev. Lett.}\ }\textbf {\bibinfo {volume} {45}},\ \bibinfo {pages} {855} (\bibinfo {year} {1980})}\BibitemShut {NoStop}%
\bibitem [{\citenamefont {Kaufman}\ and\ \citenamefont {Griffiths}(1981)}]{fr3}%
  \BibitemOpen
  \bibfield  {author} {\bibinfo {author} {\bibfnamefont {M.}~\bibnamefont {Kaufman}}\ and\ \bibinfo {author} {\bibfnamefont {R.~B.}\ \bibnamefont {Griffiths}},\ }\bibfield  {title} {\bibinfo {title} {Exactly soluble ising models on hierarchical lattices},\ }\href {https://doi.org/10.1103/PhysRevB.24.496} {\bibfield  {journal} {\bibinfo  {journal} {Phys. Rev. B}\ }\textbf {\bibinfo {volume} {24}},\ \bibinfo {pages} {496} (\bibinfo {year} {1981})}\BibitemShut {NoStop}%
\bibitem [{\citenamefont {Gefen}\ \emph {et~al.}(1983)\citenamefont {Gefen}, \citenamefont {Meir}, \citenamefont {Mandelbrot},\ and\ \citenamefont {Aharony}}]{fr4}%
  \BibitemOpen
  \bibfield  {author} {\bibinfo {author} {\bibfnamefont {Y.}~\bibnamefont {Gefen}}, \bibinfo {author} {\bibfnamefont {Y.}~\bibnamefont {Meir}}, \bibinfo {author} {\bibfnamefont {B.~B.}\ \bibnamefont {Mandelbrot}},\ and\ \bibinfo {author} {\bibfnamefont {A.}~\bibnamefont {Aharony}},\ }\bibfield  {title} {\bibinfo {title} {Geometric implementation of hypercubic lattices with noninteger dimensionality by use of low lacunarity fractal lattices},\ }\href {https://doi.org/10.1103/PhysRevLett.50.145} {\bibfield  {journal} {\bibinfo  {journal} {Phys. Rev. Lett.}\ }\textbf {\bibinfo {volume} {50}},\ \bibinfo {pages} {145} (\bibinfo {year} {1983})}\BibitemShut {NoStop}%
\bibitem [{\citenamefont {Gefen}\ \emph {et~al.}(1984)\citenamefont {Gefen}, \citenamefont {Aharony},\ and\ \citenamefont {Mandelbrot}}]{fr5}%
  \BibitemOpen
  \bibfield  {author} {\bibinfo {author} {\bibfnamefont {Y.}~\bibnamefont {Gefen}}, \bibinfo {author} {\bibfnamefont {A.}~\bibnamefont {Aharony}},\ and\ \bibinfo {author} {\bibfnamefont {B.~B.}\ \bibnamefont {Mandelbrot}},\ }\bibfield  {title} {\bibinfo {title} {Phase transitions on fractals. iii. infinitely ramified lattices},\ }\href {https://doi.org/10.1088/0305-4470/17/6/024} {\bibfield  {journal} {\bibinfo  {journal} {Journal of Physics A: Mathematical and General}\ }\textbf {\bibinfo {volume} {17}},\ \bibinfo {pages} {1277} (\bibinfo {year} {1984})}\BibitemShut {NoStop}%
\bibitem [{\citenamefont {Tasaki}(1987)}]{fr6}%
  \BibitemOpen
  \bibfield  {author} {\bibinfo {author} {\bibfnamefont {H.}~\bibnamefont {Tasaki}},\ }\bibfield  {title} {\bibinfo {title} {Critical phenomena in fractal spin systems},\ }\href {https://doi.org/10.1088/0305-4470/20/13/050} {\bibfield  {journal} {\bibinfo  {journal} {Journal of Physics A: Mathematical and General}\ }\textbf {\bibinfo {volume} {20}},\ \bibinfo {pages} {4521} (\bibinfo {year} {1987})}\BibitemShut {NoStop}%
\bibitem [{\citenamefont {Koma}\ and\ \citenamefont {Tasaki}(1995)}]{fr7}%
  \BibitemOpen
  \bibfield  {author} {\bibinfo {author} {\bibfnamefont {T.}~\bibnamefont {Koma}}\ and\ \bibinfo {author} {\bibfnamefont {H.}~\bibnamefont {Tasaki}},\ }\bibfield  {title} {\bibinfo {title} {Classical $\mathit{XY}$ model in 1.99 dimensions},\ }\href {https://doi.org/10.1103/PhysRevLett.74.3916} {\bibfield  {journal} {\bibinfo  {journal} {Phys. Rev. Lett.}\ }\textbf {\bibinfo {volume} {74}},\ \bibinfo {pages} {3916} (\bibinfo {year} {1995})}\BibitemShut {NoStop}%
\bibitem [{\citenamefont {Yoshida}\ and\ \citenamefont {Kubica}(2014)}]{fr8}%
  \BibitemOpen
  \bibfield  {author} {\bibinfo {author} {\bibfnamefont {B.}~\bibnamefont {Yoshida}}\ and\ \bibinfo {author} {\bibfnamefont {A.}~\bibnamefont {Kubica}},\ }\href@noop {} {\bibinfo {title} {Quantum criticality from ising model on fractal lattices}} (\bibinfo {year} {2014}),\ \Eprint {https://arxiv.org/abs/1404.6311} {arXiv:1404.6311} \BibitemShut {NoStop}%
\bibitem [{\citenamefont {Liu}\ \emph {et~al.}(2021)\citenamefont {Liu}, \citenamefont {Zhou}, \citenamefont {Wang}, \citenamefont {Yin}, \citenamefont {Li}, \citenamefont {Huang}, \citenamefont {Guan}, \citenamefont {Li}, \citenamefont {Wang}, \citenamefont {Zheng}, \citenamefont {Liu}, \citenamefont {Han}, \citenamefont {Evans}, \citenamefont {Liu},\ and\ \citenamefont {Jia}}]{fr9}%
  \BibitemOpen
  \bibfield  {author} {\bibinfo {author} {\bibfnamefont {C.}~\bibnamefont {Liu}}, \bibinfo {author} {\bibfnamefont {Y.}~\bibnamefont {Zhou}}, \bibinfo {author} {\bibfnamefont {G.}~\bibnamefont {Wang}}, \bibinfo {author} {\bibfnamefont {Y.}~\bibnamefont {Yin}}, \bibinfo {author} {\bibfnamefont {C.}~\bibnamefont {Li}}, \bibinfo {author} {\bibfnamefont {H.}~\bibnamefont {Huang}}, \bibinfo {author} {\bibfnamefont {D.}~\bibnamefont {Guan}}, \bibinfo {author} {\bibfnamefont {Y.}~\bibnamefont {Li}}, \bibinfo {author} {\bibfnamefont {S.}~\bibnamefont {Wang}}, \bibinfo {author} {\bibfnamefont {H.}~\bibnamefont {Zheng}}, \bibinfo {author} {\bibfnamefont {C.}~\bibnamefont {Liu}}, \bibinfo {author} {\bibfnamefont {Y.}~\bibnamefont {Han}}, \bibinfo {author} {\bibfnamefont {J.~W.}\ \bibnamefont {Evans}}, \bibinfo {author} {\bibfnamefont {F.}~\bibnamefont {Liu}},\ and\ \bibinfo {author} {\bibfnamefont {J.}~\bibnamefont {Jia}},\ }\bibfield  {title} {\bibinfo {title} {Sierpi\ifmmode \acute{n}\else \'{n}\fi{}ski structure and
  electronic topology in bi thin films on insb(111)b surfaces},\ }\href {https://doi.org/10.1103/PhysRevLett.126.176102} {\bibfield  {journal} {\bibinfo  {journal} {Phys. Rev. Lett.}\ }\textbf {\bibinfo {volume} {126}},\ \bibinfo {pages} {176102} (\bibinfo {year} {2021})}\BibitemShut {NoStop}%
\bibitem [{\citenamefont {van Veen}\ \emph {et~al.}(2017)\citenamefont {van Veen}, \citenamefont {Tomadin}, \citenamefont {Polini}, \citenamefont {Katsnelson},\ and\ \citenamefont {Yuan}}]{fr10}%
  \BibitemOpen
  \bibfield  {author} {\bibinfo {author} {\bibfnamefont {E.}~\bibnamefont {van Veen}}, \bibinfo {author} {\bibfnamefont {A.}~\bibnamefont {Tomadin}}, \bibinfo {author} {\bibfnamefont {M.}~\bibnamefont {Polini}}, \bibinfo {author} {\bibfnamefont {M.~I.}\ \bibnamefont {Katsnelson}},\ and\ \bibinfo {author} {\bibfnamefont {S.}~\bibnamefont {Yuan}},\ }\bibfield  {title} {\bibinfo {title} {Optical conductivity of a quantum electron gas in a sierpinski carpet},\ }\href {https://doi.org/10.1103/PhysRevB.96.235438} {\bibfield  {journal} {\bibinfo  {journal} {Phys. Rev. B}\ }\textbf {\bibinfo {volume} {96}},\ \bibinfo {pages} {235438} (\bibinfo {year} {2017})}\BibitemShut {NoStop}%
\bibitem [{\citenamefont {Pai}\ and\ \citenamefont {Prem}(2019)}]{fr11}%
  \BibitemOpen
  \bibfield  {author} {\bibinfo {author} {\bibfnamefont {S.}~\bibnamefont {Pai}}\ and\ \bibinfo {author} {\bibfnamefont {A.}~\bibnamefont {Prem}},\ }\bibfield  {title} {\bibinfo {title} {Topological states on fractal lattices},\ }\href {https://doi.org/10.1103/PhysRevB.100.155135} {\bibfield  {journal} {\bibinfo  {journal} {Phys. Rev. B}\ }\textbf {\bibinfo {volume} {100}},\ \bibinfo {pages} {155135} (\bibinfo {year} {2019})}\BibitemShut {NoStop}%
\bibitem [{\citenamefont {Iliasov}\ \emph {et~al.}(2020)\citenamefont {Iliasov}, \citenamefont {Katsnelson},\ and\ \citenamefont {Yuan}}]{fr12}%
  \BibitemOpen
  \bibfield  {author} {\bibinfo {author} {\bibfnamefont {A.~A.}\ \bibnamefont {Iliasov}}, \bibinfo {author} {\bibfnamefont {M.~I.}\ \bibnamefont {Katsnelson}},\ and\ \bibinfo {author} {\bibfnamefont {S.}~\bibnamefont {Yuan}},\ }\bibfield  {title} {\bibinfo {title} {Hall conductivity of a sierpi\ifmmode \acute{n}\else \'{n}\fi{}ski carpet},\ }\href {https://doi.org/10.1103/PhysRevB.101.045413} {\bibfield  {journal} {\bibinfo  {journal} {Phys. Rev. B}\ }\textbf {\bibinfo {volume} {101}},\ \bibinfo {pages} {045413} (\bibinfo {year} {2020})}\BibitemShut {NoStop}%
\bibitem [{\citenamefont {Biesenthal}\ \emph {et~al.}(2022)\citenamefont {Biesenthal}, \citenamefont {Maczewsky}, \citenamefont {Yang}, \citenamefont {Kremer}, \citenamefont {Segev}, \citenamefont {Szameit},\ and\ \citenamefont {Heinrich}}]{fr13}%
  \BibitemOpen
  \bibfield  {author} {\bibinfo {author} {\bibfnamefont {T.}~\bibnamefont {Biesenthal}}, \bibinfo {author} {\bibfnamefont {L.~J.}\ \bibnamefont {Maczewsky}}, \bibinfo {author} {\bibfnamefont {Z.}~\bibnamefont {Yang}}, \bibinfo {author} {\bibfnamefont {M.}~\bibnamefont {Kremer}}, \bibinfo {author} {\bibfnamefont {M.}~\bibnamefont {Segev}}, \bibinfo {author} {\bibfnamefont {A.}~\bibnamefont {Szameit}},\ and\ \bibinfo {author} {\bibfnamefont {M.}~\bibnamefont {Heinrich}},\ }\bibfield  {title} {\bibinfo {title} {Fractal photonic topological insulators},\ }\href {https://www.science.org/doi/abs/10.1126/science.abm2842} {\bibfield  {journal} {\bibinfo  {journal} {Science}\ }\textbf {\bibinfo {volume} {376}},\ \bibinfo {pages} {1114} (\bibinfo {year} {2022})}\BibitemShut {NoStop}%
\bibitem [{\citenamefont {Yang}\ \emph {et~al.}(2022)\citenamefont {Yang}, \citenamefont {Zhou}, \citenamefont {Yao}, \citenamefont {Lv}, \citenamefont {Wang},\ and\ \citenamefont {Yuan}}]{fr14}%
  \BibitemOpen
  \bibfield  {author} {\bibinfo {author} {\bibfnamefont {X.}~\bibnamefont {Yang}}, \bibinfo {author} {\bibfnamefont {W.}~\bibnamefont {Zhou}}, \bibinfo {author} {\bibfnamefont {Q.}~\bibnamefont {Yao}}, \bibinfo {author} {\bibfnamefont {P.}~\bibnamefont {Lv}}, \bibinfo {author} {\bibfnamefont {Y.}~\bibnamefont {Wang}},\ and\ \bibinfo {author} {\bibfnamefont {S.}~\bibnamefont {Yuan}},\ }\bibfield  {title} {\bibinfo {title} {Electronic properties and quantum transport in functionalized graphene sierpinski-carpet fractals},\ }\href {https://doi.org/10.1103/PhysRevB.105.205433} {\bibfield  {journal} {\bibinfo  {journal} {Phys. Rev. B}\ }\textbf {\bibinfo {volume} {105}},\ \bibinfo {pages} {205433} (\bibinfo {year} {2022})}\BibitemShut {NoStop}%
\bibitem [{\citenamefont {Manna}\ \emph {et~al.}(2022)\citenamefont {Manna}, \citenamefont {Nandy},\ and\ \citenamefont {Roy}}]{fr15}%
  \BibitemOpen
  \bibfield  {author} {\bibinfo {author} {\bibfnamefont {S.}~\bibnamefont {Manna}}, \bibinfo {author} {\bibfnamefont {S.}~\bibnamefont {Nandy}},\ and\ \bibinfo {author} {\bibfnamefont {B.}~\bibnamefont {Roy}},\ }\bibfield  {title} {\bibinfo {title} {Higher-order topological phases on fractal lattices},\ }\href {http://dx.doi.org/10.1103/PhysRevB.105.L201301} {\bibfield  {journal} {\bibinfo  {journal} {Physical Review B}\ }\textbf {\bibinfo {volume} {105}} (\bibinfo {year} {2022})}\BibitemShut {NoStop}%
\bibitem [{\citenamefont {Li}\ \emph {et~al.}(2022)\citenamefont {Li}, \citenamefont {Mo}, \citenamefont {Jiang},\ and\ \citenamefont {Yang}}]{fr16}%
  \BibitemOpen
  \bibfield  {author} {\bibinfo {author} {\bibfnamefont {J.}~\bibnamefont {Li}}, \bibinfo {author} {\bibfnamefont {Q.}~\bibnamefont {Mo}}, \bibinfo {author} {\bibfnamefont {J.-H.}\ \bibnamefont {Jiang}},\ and\ \bibinfo {author} {\bibfnamefont {Z.}~\bibnamefont {Yang}},\ }\bibfield  {title} {\bibinfo {title} {Higher-order topological phase in an acoustic fractal lattice},\ }\href {https://doi.org/https://doi.org/10.1016/j.scib.2022.09.024} {\bibfield  {journal} {\bibinfo  {journal} {Science Bulletin}\ }\textbf {\bibinfo {volume} {67}},\ \bibinfo {pages} {2040} (\bibinfo {year} {2022})}\BibitemShut {NoStop}%
\bibitem [{\citenamefont {Zhu}\ \emph {et~al.}(2022)\citenamefont {Zhu}, \citenamefont {Jochym-O'Connor},\ and\ \citenamefont {Dua}}]{fr17}%
  \BibitemOpen
  \bibfield  {author} {\bibinfo {author} {\bibfnamefont {G.}~\bibnamefont {Zhu}}, \bibinfo {author} {\bibfnamefont {T.}~\bibnamefont {Jochym-O'Connor}},\ and\ \bibinfo {author} {\bibfnamefont {A.}~\bibnamefont {Dua}},\ }\bibfield  {title} {\bibinfo {title} {Topological order, quantum codes, and quantum computation on fractal geometries},\ }\href {https://doi.org/10.1103/PRXQuantum.3.030338} {\bibfield  {journal} {\bibinfo  {journal} {PRX Quantum}\ }\textbf {\bibinfo {volume} {3}},\ \bibinfo {pages} {030338} (\bibinfo {year} {2022})}\BibitemShut {NoStop}%
\bibitem [{\citenamefont {Toffoli}(1977)}]{toffoli1977a}%
  \BibitemOpen
  \bibfield  {author} {\bibinfo {author} {\bibfnamefont {T.}~\bibnamefont {Toffoli}},\ }\bibfield  {title} {\bibinfo {title} {Computation and construction universality},\ }\href@noop {} {\bibfield  {journal} {\bibinfo  {journal} {Journal of Computer and Systems Sciences}\ }\textbf {\bibinfo {volume} {15}},\ \bibinfo {pages} {213–231} (\bibinfo {year} {1977})}\BibitemShut {NoStop}%
\bibitem [{\citenamefont {Gu}\ and\ \citenamefont {Shuai}(2023)}]{gu2000a}%
  \BibitemOpen
  \bibfield  {author} {\bibinfo {author} {\bibfnamefont {J.}~\bibnamefont {Gu}}\ and\ \bibinfo {author} {\bibfnamefont {D.}~\bibnamefont {Shuai}},\ }\bibfield  {title} {\bibinfo {title} {The faster higher-order cellular automaton for hyper-parallel undistorted data compression},\ }\href {https://doi.org/10.1007/BF02948796} {\bibfield  {journal} {\bibinfo  {journal} {J. Comput. Sci. Technol.}\ }\textbf {\bibinfo {volume} {15}},\ \bibinfo {pages} {126–135} (\bibinfo {year} {2023})}\BibitemShut {NoStop}%
\bibitem [{\citenamefont {{Martin del Rey}}\ \emph {et~al.}(2005)\citenamefont {{Martin del Rey}}, \citenamefont {Mateus},\ and\ \citenamefont {S\'{a}nchez}}]{rey2005a}%
  \BibitemOpen
  \bibfield  {author} {\bibinfo {author} {\bibfnamefont {A.}~\bibnamefont {{Martin del Rey}}}, \bibinfo {author} {\bibfnamefont {J.~P.}\ \bibnamefont {Mateus}},\ and\ \bibinfo {author} {\bibfnamefont {G.~R.}\ \bibnamefont {S\'{a}nchez}},\ }\bibfield  {title} {\bibinfo {title} {A secret sharing scheme based on cellular automata},\ }\href {https://doi.org/https://doi.org/10.1016/j.amc.2005.01.026} {\bibfield  {journal} {\bibinfo  {journal} {Applied Mathematics and Computation}\ }\textbf {\bibinfo {volume} {170}},\ \bibinfo {pages} {1356} (\bibinfo {year} {2005})}\BibitemShut {NoStop}%
\bibitem [{\citenamefont {Bruyn}\ and\ \citenamefont {Bergh}(1991)}]{bruyn1991a}%
  \BibitemOpen
  \bibfield  {author} {\bibinfo {author} {\bibfnamefont {L.}~\bibnamefont {Bruyn}}\ and\ \bibinfo {author} {\bibfnamefont {M.}~\bibnamefont {Bergh}},\ }\bibfield  {title} {\bibinfo {title} {Algebraic properties of linear cellular automata},\ }\href@noop {} {\bibfield  {journal} {\bibinfo  {journal} {Linear algebra and its applications}\ }\textbf {\bibinfo {volume} {157}},\ \bibinfo {pages} {217–234} (\bibinfo {year} {1991})}\BibitemShut {NoStop}%
\bibitem [{\citenamefont {Chai}\ \emph {et~al.}(2005)\citenamefont {Chai}, \citenamefont {Cao},\ and\ \citenamefont {Zhou}}]{HOCA-C}%
  \BibitemOpen
  \bibfield  {author} {\bibinfo {author} {\bibfnamefont {Z.}~\bibnamefont {Chai}}, \bibinfo {author} {\bibfnamefont {Z.}~\bibnamefont {Cao}},\ and\ \bibinfo {author} {\bibfnamefont {Y.}~\bibnamefont {Zhou}},\ }\bibfield  {title} {\bibinfo {title} {Encryption based on reversible second-order cellular automata},\ }in\ \href@noop {} {\emph {\bibinfo {booktitle} {Parallel and Distributed Processing and Applications - ISPA 2005 Workshops}}},\ \bibinfo {editor} {edited by\ \bibinfo {editor} {\bibfnamefont {G.}~\bibnamefont {Chen}}, \bibinfo {editor} {\bibfnamefont {Y.}~\bibnamefont {Pan}}, \bibinfo {editor} {\bibfnamefont {M.}~\bibnamefont {Guo}},\ and\ \bibinfo {editor} {\bibfnamefont {J.}~\bibnamefont {Lu}}}\ (\bibinfo  {publisher} {Springer Berlin Heidelberg},\ \bibinfo {address} {Berlin, Heidelberg},\ \bibinfo {year} {2005})\ pp.\ \bibinfo {pages} {350--358}\BibitemShut {NoStop}%
\bibitem [{\citenamefont {You}\ \emph {et~al.}(2014)\citenamefont {You}, \citenamefont {Bi}, \citenamefont {Rasmussen}, \citenamefont {Slagle},\ and\ \citenamefont {Xu}}]{you_wave_2014}%
  \BibitemOpen
  \bibfield  {author} {\bibinfo {author} {\bibfnamefont {Y.~Z.}\ \bibnamefont {You}}, \bibinfo {author} {\bibfnamefont {Z.}~\bibnamefont {Bi}}, \bibinfo {author} {\bibfnamefont {A.}~\bibnamefont {Rasmussen}}, \bibinfo {author} {\bibfnamefont {K.}~\bibnamefont {Slagle}},\ and\ \bibinfo {author} {\bibfnamefont {C.}~\bibnamefont {Xu}},\ }\bibfield  {title} {\bibinfo {title} {Wave function and strange correlator of short-range entangled states},\ }\href {https://doi.org/10.1103/PhysRevLett.112.247202} {\bibfield  {journal} {\bibinfo  {journal} {Phys Rev Lett}\ }\textbf {\bibinfo {volume} {112}},\ \bibinfo {pages} {247202} (\bibinfo {year} {2014})}\BibitemShut {NoStop}%
\bibitem [{\citenamefont {Wu}\ \emph {et~al.}(2015)\citenamefont {Wu}, \citenamefont {He}, \citenamefont {You}, \citenamefont {Xu}, \citenamefont {Meng},\ and\ \citenamefont {Lu}}]{sc2}%
  \BibitemOpen
  \bibfield  {author} {\bibinfo {author} {\bibfnamefont {H.-Q.}\ \bibnamefont {Wu}}, \bibinfo {author} {\bibfnamefont {Y.-Y.}\ \bibnamefont {He}}, \bibinfo {author} {\bibfnamefont {Y.-Z.}\ \bibnamefont {You}}, \bibinfo {author} {\bibfnamefont {C.}~\bibnamefont {Xu}}, \bibinfo {author} {\bibfnamefont {Z.~Y.}\ \bibnamefont {Meng}},\ and\ \bibinfo {author} {\bibfnamefont {Z.-Y.}\ \bibnamefont {Lu}},\ }\bibfield  {title} {\bibinfo {title} {Quantum monte carlo study of strange correlator in interacting topological insulators},\ }\href {https://doi.org/10.1103/PhysRevB.92.165123} {\bibfield  {journal} {\bibinfo  {journal} {Phys. Rev. B}\ }\textbf {\bibinfo {volume} {92}},\ \bibinfo {pages} {165123} (\bibinfo {year} {2015})}\BibitemShut {NoStop}%
\bibitem [{\citenamefont {Wierschem}\ and\ \citenamefont {Sengupta}(2014)}]{sc3}%
  \BibitemOpen
  \bibfield  {author} {\bibinfo {author} {\bibfnamefont {K.}~\bibnamefont {Wierschem}}\ and\ \bibinfo {author} {\bibfnamefont {P.}~\bibnamefont {Sengupta}},\ }\bibfield  {title} {\bibinfo {title} {Strange correlations in spin-1 heisenberg antiferromagnets},\ }\href {https://doi.org/10.1103/PhysRevB.90.115157} {\bibfield  {journal} {\bibinfo  {journal} {Phys. Rev. B}\ }\textbf {\bibinfo {volume} {90}},\ \bibinfo {pages} {115157} (\bibinfo {year} {2014})}\BibitemShut {NoStop}%
\bibitem [{\citenamefont {Wierschem}\ and\ \citenamefont {Beach}(2016)}]{sc4}%
  \BibitemOpen
  \bibfield  {author} {\bibinfo {author} {\bibfnamefont {K.}~\bibnamefont {Wierschem}}\ and\ \bibinfo {author} {\bibfnamefont {K.~S.~D.}\ \bibnamefont {Beach}},\ }\bibfield  {title} {\bibinfo {title} {Detection of symmetry-protected topological order in aklt states by exact evaluation of the strange correlator},\ }\href {https://doi.org/10.1103/PhysRevB.93.245141} {\bibfield  {journal} {\bibinfo  {journal} {Phys. Rev. B}\ }\textbf {\bibinfo {volume} {93}},\ \bibinfo {pages} {245141} (\bibinfo {year} {2016})}\BibitemShut {NoStop}%
\bibitem [{\citenamefont {He}\ \emph {et~al.}(2016)\citenamefont {He}, \citenamefont {Wu}, \citenamefont {You}, \citenamefont {Xu}, \citenamefont {Meng},\ and\ \citenamefont {Lu}}]{sc5}%
  \BibitemOpen
  \bibfield  {author} {\bibinfo {author} {\bibfnamefont {Y.-Y.}\ \bibnamefont {He}}, \bibinfo {author} {\bibfnamefont {H.-Q.}\ \bibnamefont {Wu}}, \bibinfo {author} {\bibfnamefont {Y.-Z.}\ \bibnamefont {You}}, \bibinfo {author} {\bibfnamefont {C.}~\bibnamefont {Xu}}, \bibinfo {author} {\bibfnamefont {Z.~Y.}\ \bibnamefont {Meng}},\ and\ \bibinfo {author} {\bibfnamefont {Z.-Y.}\ \bibnamefont {Lu}},\ }\bibfield  {title} {\bibinfo {title} {Bona fide interaction-driven topological phase transition in correlated symmetry-protected topological states},\ }\href {https://doi.org/10.1103/PhysRevB.93.115150} {\bibfield  {journal} {\bibinfo  {journal} {Phys. Rev. B}\ }\textbf {\bibinfo {volume} {93}},\ \bibinfo {pages} {115150} (\bibinfo {year} {2016})}\BibitemShut {NoStop}%
\bibitem [{\citenamefont {Zhong}\ \emph {et~al.}(2016)\citenamefont {Zhong}, \citenamefont {Liu},\ and\ \citenamefont {Luo}}]{sc6}%
  \BibitemOpen
  \bibfield  {author} {\bibinfo {author} {\bibfnamefont {Y.}~\bibnamefont {Zhong}}, \bibinfo {author} {\bibfnamefont {Y.}~\bibnamefont {Liu}},\ and\ \bibinfo {author} {\bibfnamefont {H.-G.}\ \bibnamefont {Luo}},\ }\bibfield  {title} {\bibinfo {title} {Topological phase in 1d topological kondo insulator: Z2 topological insulator, haldane-like phase and kondo breakdown},\ }\href {https://api.semanticscholar.org/CorpusID:119451597} {\bibfield  {journal} {\bibinfo  {journal} {The European Physical Journal B}\ }\textbf {\bibinfo {volume} {90}} (\bibinfo {year} {2016})}\BibitemShut {NoStop}%
\bibitem [{\citenamefont {Niu}\ and\ \citenamefont {Qi}(2023)}]{sc7}%
  \BibitemOpen
  \bibfield  {author} {\bibinfo {author} {\bibfnamefont {Y.}~\bibnamefont {Niu}}\ and\ \bibinfo {author} {\bibfnamefont {Y.}~\bibnamefont {Qi}},\ }\href@noop {} {\bibinfo {title} {Strange correlator for 1d fermionic symmetry-protected topological phases}} (\bibinfo {year} {2023}),\ \Eprint {https://arxiv.org/abs/2312.01310} {arXiv:2312.01310} \BibitemShut {NoStop}%
\bibitem [{\citenamefont {Takayoshi}\ \emph {et~al.}(2016)\citenamefont {Takayoshi}, \citenamefont {Pujol},\ and\ \citenamefont {Tanaka}}]{c2}%
  \BibitemOpen
  \bibfield  {author} {\bibinfo {author} {\bibfnamefont {S.}~\bibnamefont {Takayoshi}}, \bibinfo {author} {\bibfnamefont {P.}~\bibnamefont {Pujol}},\ and\ \bibinfo {author} {\bibfnamefont {A.}~\bibnamefont {Tanaka}},\ }\bibfield  {title} {\bibinfo {title} {Field theory of symmetry-protected valence bond solid states in (2+1) dimensions},\ }\href {https://doi.org/10.1103/PhysRevB.94.235159} {\bibfield  {journal} {\bibinfo  {journal} {Phys. Rev. B}\ }\textbf {\bibinfo {volume} {94}},\ \bibinfo {pages} {235159} (\bibinfo {year} {2016})}\BibitemShut {NoStop}%
\bibitem [{\citenamefont {Vanhove}\ \emph {et~al.}(2018)\citenamefont {Vanhove}, \citenamefont {Bal}, \citenamefont {Williamson}, \citenamefont {Bultinck}, \citenamefont {Haegeman},\ and\ \citenamefont {Verstraete}}]{c5}%
  \BibitemOpen
  \bibfield  {author} {\bibinfo {author} {\bibfnamefont {R.}~\bibnamefont {Vanhove}}, \bibinfo {author} {\bibfnamefont {M.}~\bibnamefont {Bal}}, \bibinfo {author} {\bibfnamefont {D.~J.}\ \bibnamefont {Williamson}}, \bibinfo {author} {\bibfnamefont {N.}~\bibnamefont {Bultinck}}, \bibinfo {author} {\bibfnamefont {J.}~\bibnamefont {Haegeman}},\ and\ \bibinfo {author} {\bibfnamefont {F.}~\bibnamefont {Verstraete}},\ }\bibfield  {title} {\bibinfo {title} {Mapping topological to conformal field theories through strange correlators},\ }\href {https://doi.org/10.1103/PhysRevLett.121.177203} {\bibfield  {journal} {\bibinfo  {journal} {Phys. Rev. Lett.}\ }\textbf {\bibinfo {volume} {121}},\ \bibinfo {pages} {177203} (\bibinfo {year} {2018})}\BibitemShut {NoStop}%
\bibitem [{\citenamefont {Lootens}\ \emph {et~al.}(2020)\citenamefont {Lootens}, \citenamefont {Vanhove}, \citenamefont {Haegeman},\ and\ \citenamefont {Verstraete}}]{c7}%
  \BibitemOpen
  \bibfield  {author} {\bibinfo {author} {\bibfnamefont {L.}~\bibnamefont {Lootens}}, \bibinfo {author} {\bibfnamefont {R.}~\bibnamefont {Vanhove}}, \bibinfo {author} {\bibfnamefont {J.}~\bibnamefont {Haegeman}},\ and\ \bibinfo {author} {\bibfnamefont {F.}~\bibnamefont {Verstraete}},\ }\bibfield  {title} {\bibinfo {title} {Galois conjugated tensor fusion categories and nonunitary conformal field theory},\ }\href {https://doi.org/10.1103/PhysRevLett.124.120601} {\bibfield  {journal} {\bibinfo  {journal} {Phys. Rev. Lett.}\ }\textbf {\bibinfo {volume} {124}},\ \bibinfo {pages} {120601} (\bibinfo {year} {2020})}\BibitemShut {NoStop}%
\bibitem [{\citenamefont {McMahon}\ \emph {et~al.}(2018)\citenamefont {McMahon}, \citenamefont {Singh},\ and\ \citenamefont {Brennen}}]{c8}%
  \BibitemOpen
  \bibfield  {author} {\bibinfo {author} {\bibfnamefont {N.~A.}\ \bibnamefont {McMahon}}, \bibinfo {author} {\bibfnamefont {S.}~\bibnamefont {Singh}},\ and\ \bibinfo {author} {\bibfnamefont {G.~K.}\ \bibnamefont {Brennen}},\ }\bibfield  {title} {\bibinfo {title} {A holographic duality from lifted tensor networks},\ }\href {https://doi.org/10.1038/s41534-020-0255-7} {\bibfield  {journal} {\bibinfo  {journal} {npj Quantum Information}\ }\textbf {\bibinfo {volume} {6}},\ \bibinfo {pages} {36} (\bibinfo {year} {2018})}\BibitemShut {NoStop}%
\bibitem [{\citenamefont {Wu}\ \emph {et~al.}(2021)\citenamefont {Wu}, \citenamefont {Jian},\ and\ \citenamefont {Xu}}]{c10}%
  \BibitemOpen
  \bibfield  {author} {\bibinfo {author} {\bibfnamefont {X.-C.}\ \bibnamefont {Wu}}, \bibinfo {author} {\bibfnamefont {C.-M.}\ \bibnamefont {Jian}},\ and\ \bibinfo {author} {\bibfnamefont {C.}~\bibnamefont {Xu}},\ }\bibfield  {title} {\bibinfo {title} {{Universal features of higher-form symmetries at phase transitions}},\ }\href {https://scipost.org/10.21468/SciPostPhys.11.2.033} {\bibfield  {journal} {\bibinfo  {journal} {SciPost Phys.}\ }\textbf {\bibinfo {volume} {11}},\ \bibinfo {pages} {033} (\bibinfo {year} {2021})}\BibitemShut {NoStop}%
\bibitem [{\citenamefont {Vanhove}\ \emph {et~al.}(2022{\natexlab{a}})\citenamefont {Vanhove}, \citenamefont {Lootens}, \citenamefont {Van~Damme}, \citenamefont {Wolf}, \citenamefont {Osborne}, \citenamefont {Haegeman},\ and\ \citenamefont {Verstraete}}]{c13}%
  \BibitemOpen
  \bibfield  {author} {\bibinfo {author} {\bibfnamefont {R.}~\bibnamefont {Vanhove}}, \bibinfo {author} {\bibfnamefont {L.}~\bibnamefont {Lootens}}, \bibinfo {author} {\bibfnamefont {M.}~\bibnamefont {Van~Damme}}, \bibinfo {author} {\bibfnamefont {R.}~\bibnamefont {Wolf}}, \bibinfo {author} {\bibfnamefont {T.~J.}\ \bibnamefont {Osborne}}, \bibinfo {author} {\bibfnamefont {J.}~\bibnamefont {Haegeman}},\ and\ \bibinfo {author} {\bibfnamefont {F.}~\bibnamefont {Verstraete}},\ }\bibfield  {title} {\bibinfo {title} {Critical lattice model for a haagerup conformal field theory},\ }\href {https://doi.org/10.1103/PhysRevLett.128.231602} {\bibfield  {journal} {\bibinfo  {journal} {Phys. Rev. Lett.}\ }\textbf {\bibinfo {volume} {128}},\ \bibinfo {pages} {231602} (\bibinfo {year} {2022}{\natexlab{a}})}\BibitemShut {NoStop}%
\bibitem [{\citenamefont {Vanhove}\ \emph {et~al.}(2022{\natexlab{b}})\citenamefont {Vanhove}, \citenamefont {Lootens}, \citenamefont {Tu},\ and\ \citenamefont {Verstraete}}]{c14}%
  \BibitemOpen
  \bibfield  {author} {\bibinfo {author} {\bibfnamefont {R.}~\bibnamefont {Vanhove}}, \bibinfo {author} {\bibfnamefont {L.}~\bibnamefont {Lootens}}, \bibinfo {author} {\bibfnamefont {H.-H.}\ \bibnamefont {Tu}},\ and\ \bibinfo {author} {\bibfnamefont {F.}~\bibnamefont {Verstraete}},\ }\bibfield  {title} {\bibinfo {title} {Topological aspects of the critical three-state potts model},\ }\href {http://dx.doi.org/10.1088/1751-8121/ac68b1} {\bibfield  {journal} {\bibinfo  {journal} {Journal of Physics A: Mathematical and Theoretical}\ }\textbf {\bibinfo {volume} {55}},\ \bibinfo {pages} {235002} (\bibinfo {year} {2022}{\natexlab{b}})}\BibitemShut {NoStop}%
\bibitem [{\citenamefont {Lepori}\ \emph {et~al.}(2023)\citenamefont {Lepori}, \citenamefont {Burrello}, \citenamefont {Trombettoni},\ and\ \citenamefont {Paganelli}}]{c15}%
  \BibitemOpen
  \bibfield  {author} {\bibinfo {author} {\bibfnamefont {L.}~\bibnamefont {Lepori}}, \bibinfo {author} {\bibfnamefont {M.}~\bibnamefont {Burrello}}, \bibinfo {author} {\bibfnamefont {A.}~\bibnamefont {Trombettoni}},\ and\ \bibinfo {author} {\bibfnamefont {S.}~\bibnamefont {Paganelli}},\ }\bibfield  {title} {\bibinfo {title} {Strange correlators for topological quantum systems from bulk-boundary correspondence},\ }\href {http://dx.doi.org/10.1103/PhysRevB.108.035110} {\bibfield  {journal} {\bibinfo  {journal} {Physical Review B}\ }\textbf {\bibinfo {volume} {108}} (\bibinfo {year} {2023})}\BibitemShut {NoStop}%
\bibitem [{\citenamefont {Williamson}\ \emph {et~al.}(2019)\citenamefont {Williamson}, \citenamefont {Dua},\ and\ \citenamefont {Cheng}}]{spurious}%
  \BibitemOpen
  \bibfield  {author} {\bibinfo {author} {\bibfnamefont {D.~J.}\ \bibnamefont {Williamson}}, \bibinfo {author} {\bibfnamefont {A.}~\bibnamefont {Dua}},\ and\ \bibinfo {author} {\bibfnamefont {M.}~\bibnamefont {Cheng}},\ }\bibfield  {title} {\bibinfo {title} {Spurious topological entanglement entropy from subsystem symmetries},\ }\href {https://doi.org/10.1103/PhysRevLett.122.140506} {\bibfield  {journal} {\bibinfo  {journal} {Phys. Rev. Lett.}\ }\textbf {\bibinfo {volume} {122}},\ \bibinfo {pages} {140506} (\bibinfo {year} {2019})}\BibitemShut {NoStop}%
\bibitem [{\citenamefont {Levin}\ and\ \citenamefont {Wen}(2006)}]{LW}%
  \BibitemOpen
  \bibfield  {author} {\bibinfo {author} {\bibfnamefont {M.}~\bibnamefont {Levin}}\ and\ \bibinfo {author} {\bibfnamefont {X.-G.}\ \bibnamefont {Wen}},\ }\bibfield  {title} {\bibinfo {title} {Detecting topological order in a ground state wave function},\ }\href {https://doi.org/10.1103/PhysRevLett.96.110405} {\bibfield  {journal} {\bibinfo  {journal} {Phys. Rev. Lett.}\ }\textbf {\bibinfo {volume} {96}},\ \bibinfo {pages} {110405} (\bibinfo {year} {2006})}\BibitemShut {NoStop}%
\bibitem [{\citenamefont {Wolfram}(1984{\natexlab{b}})}]{wolfram1984d}%
  \BibitemOpen
  \bibfield  {author} {\bibinfo {author} {\bibfnamefont {S.}~\bibnamefont {Wolfram}},\ }\bibfield  {title} {\bibinfo {title} {Universality and complexity in cellular automata},\ }\href {https://doi.org/https://doi.org/10.1016/0167-2789(84)90245-8} {\bibfield  {journal} {\bibinfo  {journal} {Physica D: Nonlinear Phenomena}\ }\textbf {\bibinfo {volume} {10}},\ \bibinfo {pages} {1} (\bibinfo {year} {1984}{\natexlab{b}})}\BibitemShut {NoStop}%
\bibitem [{\citenamefont {Wolfram}(2002)}]{wolfram2002a}%
  \BibitemOpen
  \bibfield  {author} {\bibinfo {author} {\bibfnamefont {S.}~\bibnamefont {Wolfram}},\ }\href {https://doi.org/10.24097/wolfram.15062.data.} {\emph {\bibinfo {title} {A new kind of science}}}\ (\bibinfo  {publisher} {Wolfram Media Inc. Champaign},\ \bibinfo {year} {2002})\BibitemShut {NoStop}%
\bibitem [{\citenamefont {Dennunzio}\ \emph {et~al.}(2019)\citenamefont {Dennunzio}, \citenamefont {Formenti}, \citenamefont {Manzoni}, \citenamefont {Margara},\ and\ \citenamefont {Porreca}}]{dennunzio_dynamical_2019}%
  \BibitemOpen
  \bibfield  {author} {\bibinfo {author} {\bibfnamefont {A.}~\bibnamefont {Dennunzio}}, \bibinfo {author} {\bibfnamefont {E.}~\bibnamefont {Formenti}}, \bibinfo {author} {\bibfnamefont {L.}~\bibnamefont {Manzoni}}, \bibinfo {author} {\bibfnamefont {L.}~\bibnamefont {Margara}},\ and\ \bibinfo {author} {\bibfnamefont {A.~E.}\ \bibnamefont {Porreca}},\ }\bibfield  {title} {\bibinfo {title} {On the dynamical behaviour of linear higher-order cellular automata and its decidability},\ }\href {https://doi.org/https://doi.org/10.1016/j.ins.2019.02.023} {\bibfield  {journal} {\bibinfo  {journal} {Information Sciences}\ }\textbf {\bibinfo {volume} {486}},\ \bibinfo {pages} {73} (\bibinfo {year} {2019})}\BibitemShut {NoStop}%
\bibitem [{\citenamefont {Wikipedia}(2023)}]{noauthor_freshmans_2023}%
  \BibitemOpen
  \bibfield  {author} {\bibinfo {author} {\bibnamefont {Wikipedia}},\ }\href {https://en.wikipedia.org/w/index.php?title=Freshman%27s_dream&oldid=1174791494} {\bibinfo {title} {Freshman's dream}} (\bibinfo {year} {2023})\BibitemShut {NoStop}%
\bibitem [{\citenamefont {Zhou}\ \emph {et~al.}(2021)\citenamefont {Zhou}, \citenamefont {Zhang}, \citenamefont {Pollmann},\ and\ \citenamefont {You}}]{zhou2021fractalquantumphasetransitions}%
  \BibitemOpen
  \bibfield  {author} {\bibinfo {author} {\bibfnamefont {Z.}~\bibnamefont {Zhou}}, \bibinfo {author} {\bibfnamefont {X.-F.}\ \bibnamefont {Zhang}}, \bibinfo {author} {\bibfnamefont {F.}~\bibnamefont {Pollmann}},\ and\ \bibinfo {author} {\bibfnamefont {Y.}~\bibnamefont {You}},\ }\href {https://arxiv.org/abs/2105.05851} {\bibinfo {title} {Fractal quantum phase transitions: Critical phenomena beyond renormalization}} (\bibinfo {year} {2021}),\ \Eprint {https://arxiv.org/abs/2105.05851} {arXiv:2105.05851 [cond-mat.str-el]} \BibitemShut {NoStop}%
\bibitem [{\citenamefont {Doherty}\ and\ \citenamefont {Bartlett}(2009)}]{doherty_identifying_2009}%
  \BibitemOpen
  \bibfield  {author} {\bibinfo {author} {\bibfnamefont {A.~C.}\ \bibnamefont {Doherty}}\ and\ \bibinfo {author} {\bibfnamefont {S.~D.}\ \bibnamefont {Bartlett}},\ }\bibfield  {title} {\bibinfo {title} {Identifying {Phases} of {Quantum} {Many}-{Body} {Systems} {That} {Are} {Universal} for {Quantum} {Computation}},\ }\href {https://doi.org/10.1103/PhysRevLett.103.020506} {\bibfield  {journal} {\bibinfo  {journal} {Physical Review Letters}\ }\textbf {\bibinfo {volume} {103}},\ \bibinfo {pages} {020506} (\bibinfo {year} {2009})}\BibitemShut {NoStop}%
\bibitem [{\citenamefont {Zou}\ and\ \citenamefont {Haah}(2016)}]{STEEa}%
  \BibitemOpen
  \bibfield  {author} {\bibinfo {author} {\bibfnamefont {L.}~\bibnamefont {Zou}}\ and\ \bibinfo {author} {\bibfnamefont {J.}~\bibnamefont {Haah}},\ }\bibfield  {title} {\bibinfo {title} {Spurious long-range entanglement and replica correlation length},\ }\href {https://doi.org/10.1103/PhysRevB.94.075151} {\bibfield  {journal} {\bibinfo  {journal} {Phys. Rev. B}\ }\textbf {\bibinfo {volume} {94}},\ \bibinfo {pages} {075151} (\bibinfo {year} {2016})}\BibitemShut {NoStop}%
\bibitem [{\citenamefont {Kitaev}\ and\ \citenamefont {Preskill}(2006)}]{KP}%
  \BibitemOpen
  \bibfield  {author} {\bibinfo {author} {\bibfnamefont {A.}~\bibnamefont {Kitaev}}\ and\ \bibinfo {author} {\bibfnamefont {J.}~\bibnamefont {Preskill}},\ }\bibfield  {title} {\bibinfo {title} {Topological entanglement entropy},\ }\href {https://doi.org/10.1103/PhysRevLett.96.110404} {\bibfield  {journal} {\bibinfo  {journal} {Phys. Rev. Lett.}\ }\textbf {\bibinfo {volume} {96}},\ \bibinfo {pages} {110404} (\bibinfo {year} {2006})}\BibitemShut {NoStop}%
\bibitem [{\citenamefont {Kato}\ and\ \citenamefont {Brand\~ao}(2020)}]{STEEb}%
  \BibitemOpen
  \bibfield  {author} {\bibinfo {author} {\bibfnamefont {K.}~\bibnamefont {Kato}}\ and\ \bibinfo {author} {\bibfnamefont {F.~G. S.~L.}\ \bibnamefont {Brand\~ao}},\ }\bibfield  {title} {\bibinfo {title} {Toy model of boundary states with spurious topological entanglement entropy},\ }\href {https://doi.org/10.1103/PhysRevResearch.2.032005} {\bibfield  {journal} {\bibinfo  {journal} {Phys. Rev. Res.}\ }\textbf {\bibinfo {volume} {2}},\ \bibinfo {pages} {032005} (\bibinfo {year} {2020})}\BibitemShut {NoStop}%
\bibitem [{\citenamefont {Stephen}\ \emph {et~al.}(2019{\natexlab{b}})\citenamefont {Stephen}, \citenamefont {Dreyer}, \citenamefont {Iqbal},\ and\ \citenamefont {Schuch}}]{STEEc}%
  \BibitemOpen
  \bibfield  {author} {\bibinfo {author} {\bibfnamefont {D.~T.}\ \bibnamefont {Stephen}}, \bibinfo {author} {\bibfnamefont {H.}~\bibnamefont {Dreyer}}, \bibinfo {author} {\bibfnamefont {M.}~\bibnamefont {Iqbal}},\ and\ \bibinfo {author} {\bibfnamefont {N.}~\bibnamefont {Schuch}},\ }\bibfield  {title} {\bibinfo {title} {Detecting subsystem symmetry protected topological order via entanglement entropy},\ }\href {https://doi.org/10.1103/PhysRevB.100.115112} {\bibfield  {journal} {\bibinfo  {journal} {Phys. Rev. B}\ }\textbf {\bibinfo {volume} {100}},\ \bibinfo {pages} {115112} (\bibinfo {year} {2019}{\natexlab{b}})}\BibitemShut {NoStop}%
\bibitem [{\citenamefont {Kim}\ \emph {et~al.}(2023)\citenamefont {Kim}, \citenamefont {Levin}, \citenamefont {Lin}, \citenamefont {Ranard},\ and\ \citenamefont {Shi}}]{STEEd}%
  \BibitemOpen
  \bibfield  {author} {\bibinfo {author} {\bibfnamefont {I.~H.}\ \bibnamefont {Kim}}, \bibinfo {author} {\bibfnamefont {M.}~\bibnamefont {Levin}}, \bibinfo {author} {\bibfnamefont {T.-C.}\ \bibnamefont {Lin}}, \bibinfo {author} {\bibfnamefont {D.}~\bibnamefont {Ranard}},\ and\ \bibinfo {author} {\bibfnamefont {B.}~\bibnamefont {Shi}},\ }\bibfield  {title} {\bibinfo {title} {Universal lower bound on topological entanglement entropy},\ }\href {https://doi.org/10.1103/PhysRevLett.131.166601} {\bibfield  {journal} {\bibinfo  {journal} {Phys. Rev. Lett.}\ }\textbf {\bibinfo {volume} {131}},\ \bibinfo {pages} {166601} (\bibinfo {year} {2023})}\BibitemShut {NoStop}%
\bibitem [{\citenamefont {Stephen}\ \emph {et~al.}(2022)\citenamefont {Stephen}, \citenamefont {Dua}, \citenamefont {Garre-Rubio}, \citenamefont {Williamson},\ and\ \citenamefont {Hermele}}]{SET_CA}%
  \BibitemOpen
  \bibfield  {author} {\bibinfo {author} {\bibfnamefont {D.~T.}\ \bibnamefont {Stephen}}, \bibinfo {author} {\bibfnamefont {A.}~\bibnamefont {Dua}}, \bibinfo {author} {\bibfnamefont {J.}~\bibnamefont {Garre-Rubio}}, \bibinfo {author} {\bibfnamefont {D.~J.}\ \bibnamefont {Williamson}},\ and\ \bibinfo {author} {\bibfnamefont {M.}~\bibnamefont {Hermele}},\ }\bibfield  {title} {\bibinfo {title} {Fractionalization of subsystem symmetries in two dimensions},\ }\href {https://doi.org/10.1103/PhysRevB.106.085104} {\bibfield  {journal} {\bibinfo  {journal} {Phys. Rev. B}\ }\textbf {\bibinfo {volume} {106}},\ \bibinfo {pages} {085104} (\bibinfo {year} {2022})}\BibitemShut {NoStop}%
\bibitem [{\citenamefont {Devakul}\ and\ \citenamefont {Williamson}(2021)}]{Devakul2021fractalizing}%
  \BibitemOpen
  \bibfield  {author} {\bibinfo {author} {\bibfnamefont {T.}~\bibnamefont {Devakul}}\ and\ \bibinfo {author} {\bibfnamefont {D.~J.}\ \bibnamefont {Williamson}},\ }\bibfield  {title} {\bibinfo {title} {Fractalizing quantum codes},\ }\href {https://doi.org/10.22331/q-2021-04-22-438} {\bibfield  {journal} {\bibinfo  {journal} {Quantum}\ }\textbf {\bibinfo {volume} {5}},\ \bibinfo {pages} {438} (\bibinfo {year} {2021})}\BibitemShut {NoStop}%
\bibitem [{\citenamefont {Liang}\ \emph {et~al.}(2023)\citenamefont {Liang}, \citenamefont {Fang},\ and\ \citenamefont {Hu}}]{nh1}%
  \BibitemOpen
  \bibfield  {author} {\bibinfo {author} {\bibfnamefont {J.}~\bibnamefont {Liang}}, \bibinfo {author} {\bibfnamefont {C.}~\bibnamefont {Fang}},\ and\ \bibinfo {author} {\bibfnamefont {J.}~\bibnamefont {Hu}},\ }\href@noop {} {\bibinfo {title} {Probing topological phase transition with non-hermitian perturbations}} (\bibinfo {year} {2023}),\ \Eprint {https://arxiv.org/abs/2401.00530} {arXiv:2401.00530} \BibitemShut {NoStop}%
\bibitem [{\citenamefont {Chen}\ \emph {et~al.}(2021{\natexlab{b}})\citenamefont {Chen}, \citenamefont {Chen},\ and\ \citenamefont {Ye}}]{nh2}%
  \BibitemOpen
  \bibfield  {author} {\bibinfo {author} {\bibfnamefont {L.-M.}\ \bibnamefont {Chen}}, \bibinfo {author} {\bibfnamefont {S.~A.}\ \bibnamefont {Chen}},\ and\ \bibinfo {author} {\bibfnamefont {P.}~\bibnamefont {Ye}},\ }\bibfield  {title} {\bibinfo {title} {{Entanglement, non-hermiticity, and duality}},\ }\href {https://scipost.org/10.21468/SciPostPhys.11.1.003} {\bibfield  {journal} {\bibinfo  {journal} {SciPost Phys.}\ }\textbf {\bibinfo {volume} {11}},\ \bibinfo {pages} {003} (\bibinfo {year} {2021}{\natexlab{b}})}\BibitemShut {NoStop}%
\bibitem [{\citenamefont {Chen}\ \emph {et~al.}(2022)\citenamefont {Chen}, \citenamefont {Zhou}, \citenamefont {Chen},\ and\ \citenamefont {Ye}}]{nh3}%
  \BibitemOpen
  \bibfield  {author} {\bibinfo {author} {\bibfnamefont {L.-M.}\ \bibnamefont {Chen}}, \bibinfo {author} {\bibfnamefont {Y.}~\bibnamefont {Zhou}}, \bibinfo {author} {\bibfnamefont {S.~A.}\ \bibnamefont {Chen}},\ and\ \bibinfo {author} {\bibfnamefont {P.}~\bibnamefont {Ye}},\ }\bibfield  {title} {\bibinfo {title} {Quantum entanglement of non-hermitian quasicrystals},\ }\href {https://doi.org/10.1103/PhysRevB.105.L121115} {\bibfield  {journal} {\bibinfo  {journal} {Phys. Rev. B}\ }\textbf {\bibinfo {volume} {105}},\ \bibinfo {pages} {L121115} (\bibinfo {year} {2022})}\BibitemShut {NoStop}%
\bibitem [{\citenamefont {Ma}\ and\ \citenamefont {Wang}(2023)}]{ASPT1}%
  \BibitemOpen
  \bibfield  {author} {\bibinfo {author} {\bibfnamefont {R.}~\bibnamefont {Ma}}\ and\ \bibinfo {author} {\bibfnamefont {C.}~\bibnamefont {Wang}},\ }\bibfield  {title} {\bibinfo {title} {Average symmetry-protected topological phases},\ }\href {https://doi.org/10.1103/PhysRevX.13.031016} {\bibfield  {journal} {\bibinfo  {journal} {Phys. Rev. X}\ }\textbf {\bibinfo {volume} {13}},\ \bibinfo {pages} {031016} (\bibinfo {year} {2023})}\BibitemShut {NoStop}%
\bibitem [{\citenamefont {Ma}\ \emph {et~al.}(2023)\citenamefont {Ma}, \citenamefont {Zhang}, \citenamefont {Bi}, \citenamefont {Cheng},\ and\ \citenamefont {Wang}}]{ASPT2}%
  \BibitemOpen
  \bibfield  {author} {\bibinfo {author} {\bibfnamefont {R.}~\bibnamefont {Ma}}, \bibinfo {author} {\bibfnamefont {J.-H.}\ \bibnamefont {Zhang}}, \bibinfo {author} {\bibfnamefont {Z.}~\bibnamefont {Bi}}, \bibinfo {author} {\bibfnamefont {M.}~\bibnamefont {Cheng}},\ and\ \bibinfo {author} {\bibfnamefont {C.}~\bibnamefont {Wang}},\ }\href@noop {} {\bibinfo {title} {Topological phases with average symmetries: the decohered, the disordered, and the intrinsic}} (\bibinfo {year} {2023}),\ \Eprint {https://arxiv.org/abs/2305.16399} {arXiv:2305.16399} \BibitemShut {NoStop}%
\bibitem [{\citenamefont {Stephen}\ \emph {et~al.}(2019{\natexlab{c}})\citenamefont {Stephen}, \citenamefont {Nautrup}, \citenamefont {Bermejo-Vega}, \citenamefont {Eisert},\ and\ \citenamefont {Raussendorf}}]{CQCA1}%
  \BibitemOpen
  \bibfield  {author} {\bibinfo {author} {\bibfnamefont {D.~T.}\ \bibnamefont {Stephen}}, \bibinfo {author} {\bibfnamefont {H.~P.}\ \bibnamefont {Nautrup}}, \bibinfo {author} {\bibfnamefont {J.}~\bibnamefont {Bermejo-Vega}}, \bibinfo {author} {\bibfnamefont {J.}~\bibnamefont {Eisert}},\ and\ \bibinfo {author} {\bibfnamefont {R.}~\bibnamefont {Raussendorf}},\ }\bibfield  {title} {\bibinfo {title} {Subsystem symmetries, quantum cellular automata, and computational phases of quantum matter},\ }\href {https://doi.org/10.22331/q-2019-05-20-142} {\bibfield  {journal} {\bibinfo  {journal} {Quantum}\ }\textbf {\bibinfo {volume} {3}},\ \bibinfo {pages} {142} (\bibinfo {year} {2019}{\natexlab{c}})}\BibitemShut {NoStop}%
\bibitem [{\citenamefont {Daniel}\ \emph {et~al.}(2020)\citenamefont {Daniel}, \citenamefont {Alexander},\ and\ \citenamefont {Miyake}}]{CQCA2}%
  \BibitemOpen
  \bibfield  {author} {\bibinfo {author} {\bibfnamefont {A.~K.}\ \bibnamefont {Daniel}}, \bibinfo {author} {\bibfnamefont {R.~N.}\ \bibnamefont {Alexander}},\ and\ \bibinfo {author} {\bibfnamefont {A.}~\bibnamefont {Miyake}},\ }\bibfield  {title} {\bibinfo {title} {Computational universality of symmetry-protected topologically ordered cluster phases on 2d archimedean lattices},\ }\href {https://doi.org/10.22331/q-2020-02-10-228} {\bibfield  {journal} {\bibinfo  {journal} {Quantum}\ }\textbf {\bibinfo {volume} {4}},\ \bibinfo {pages} {228} (\bibinfo {year} {2020})}\BibitemShut {NoStop}%
\bibitem [{\citenamefont {Biswas}\ \emph {et~al.}(2022)\citenamefont {Biswas}, \citenamefont {Kwan},\ and\ \citenamefont {Parameswaran}}]{MCA}%
  \BibitemOpen
  \bibfield  {author} {\bibinfo {author} {\bibfnamefont {S.}~\bibnamefont {Biswas}}, \bibinfo {author} {\bibfnamefont {Y.~H.}\ \bibnamefont {Kwan}},\ and\ \bibinfo {author} {\bibfnamefont {S.~A.}\ \bibnamefont {Parameswaran}},\ }\bibfield  {title} {\bibinfo {title} {Beyond the freshman's dream: Classical fractal spin liquids from matrix cellular automata in three-dimensional lattice models},\ }\href {https://doi.org/10.1103/PhysRevB.105.224410} {\bibfield  {journal} {\bibinfo  {journal} {Phys. Rev. B}\ }\textbf {\bibinfo {volume} {105}},\ \bibinfo {pages} {224410} (\bibinfo {year} {2022})}\BibitemShut {NoStop}%
\bibitem [{\citenamefont {Devaney}(1989)}]{Devaney1989}%
  \BibitemOpen
  \bibfield  {author} {\bibinfo {author} {\bibfnamefont {R.~L.}\ \bibnamefont {Devaney}},\ }\href@noop {} {\emph {\bibinfo {title} {An Introduction to Chaotic Dynamical Systems}}}\ (\bibinfo  {publisher} {Addison-Wesley Advanced Book Program, Addison-Wesley},\ \bibinfo {year} {1989})\BibitemShut {NoStop}%
\bibitem [{\citenamefont {Dennunzio}\ \emph {et~al.}(2024)\citenamefont {Dennunzio}, \citenamefont {Formenti},\ and\ \citenamefont {Margara}}]{Dennunzio2024}%
  \BibitemOpen
  \bibfield  {author} {\bibinfo {author} {\bibfnamefont {A.}~\bibnamefont {Dennunzio}}, \bibinfo {author} {\bibfnamefont {E.}~\bibnamefont {Formenti}},\ and\ \bibinfo {author} {\bibfnamefont {L.}~\bibnamefont {Margara}},\ }\bibfield  {title} {\bibinfo {title} {An efficient algorithm deciding chaos for linear cellular automata over (z/mz)n with applications to data encryption},\ }\href {https://doi.org/https://doi.org/10.1016/j.ins.2023.119942} {\bibfield  {journal} {\bibinfo  {journal} {Information Sciences}\ }\textbf {\bibinfo {volume} {657}},\ \bibinfo {pages} {119942} (\bibinfo {year} {2024})}\BibitemShut {NoStop}%
\bibitem [{\citenamefont {Dennunzio}\ \emph {et~al.}(2020)\citenamefont {Dennunzio}, \citenamefont {Formenti}, \citenamefont {Grinberg},\ and\ \citenamefont {Margara}}]{Dennunzio2020}%
  \BibitemOpen
  \bibfield  {author} {\bibinfo {author} {\bibfnamefont {A.}~\bibnamefont {Dennunzio}}, \bibinfo {author} {\bibfnamefont {E.}~\bibnamefont {Formenti}}, \bibinfo {author} {\bibfnamefont {D.}~\bibnamefont {Grinberg}},\ and\ \bibinfo {author} {\bibfnamefont {L.}~\bibnamefont {Margara}},\ }\bibfield  {title} {\bibinfo {title} {Chaos and ergodicity are decidable for linear cellular automata over (z/mz)n},\ }\href {https://doi.org/https://doi.org/10.1016/j.ins.2020.05.123} {\bibfield  {journal} {\bibinfo  {journal} {Information Sciences}\ }\textbf {\bibinfo {volume} {539}},\ \bibinfo {pages} {136} (\bibinfo {year} {2020})}\BibitemShut {NoStop}%
\end{thebibliography}

%

	\appendix

 \section{Inability of order-1 CA in producing RSPT phases}\label{order1}
 While an HGSPT phase generated by order-1 CA is always protected by symmetries with exact self-similarity, the symmetry pattern is not always fractal. This claim is proved by the \textit{Freshman's Dream} theorem:
 \begin{equation}\label{freshman}
     (a_1)^p=a^p+b^p \mod p,~p\text{ is prime}.
 \end{equation}
 In this paper we focus on the CA rule with alphabet $\mathbb F_2$, meaning $p=2$. According to Eq.~(\ref{freshman}), for an update rule $f(x)=\sum_i {\lambda_i x^i}$, we have
 \begin{equation}
     f(x)^{2^k}=\sum_{i} \lambda_i x^{i2^k},~k\in  \mathbb N,
 \end{equation}
 which means
 \begin{equation}
     r_{2^k}(x)=r_0(x)f(x)^{2^k}=r_0(x)\left(\sum_{i} \lambda_i x^{i2^k}\right).
 \end{equation}
 , where $r_j(x)$ is defined in Eq.~(\ref{rx}). By picking the initial condition $r_0(x)=1$, we always get a series of self-similar rows:
 \begin{equation}
     r_{2^k}(x)=\sum_{i} \lambda_i x^{i2^k}.
 \end{equation}
 The self-similarity can be confirmed by observing that each term in $r_{2^{k_1}}(x)$ can be 1-to-1 mapped to $r_{2^{k_2}}(x)$ by a scaling transformation $\mathscr S$:
 \begin{equation}
     \mathscr S: x^{2^k_1}\to x^{2^k_2},
 \end{equation}
 proving the self-similarity of the CA pattern.
 
\begin{figure*}[htbp]
     \subfigure[$q=-1$]{\includegraphics[width=0.32\linewidth]{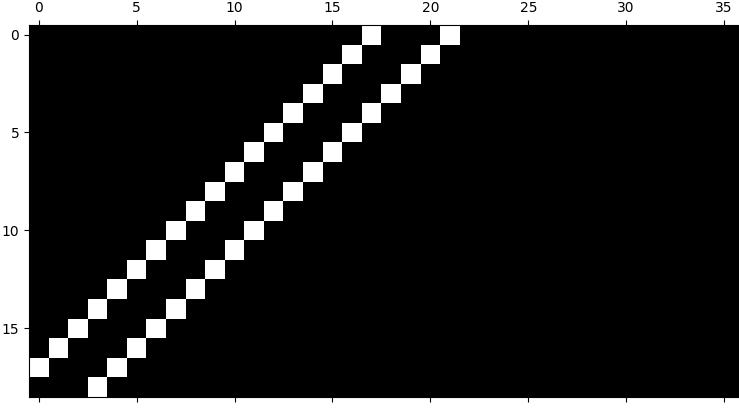}}
     \subfigure[$q=0$]{\includegraphics[width=0.32\linewidth]{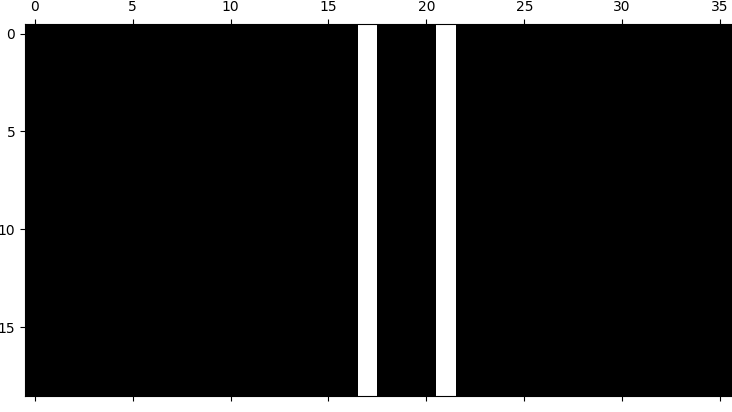}}
     \subfigure[$q=1$]{\includegraphics[width=0.32\linewidth]{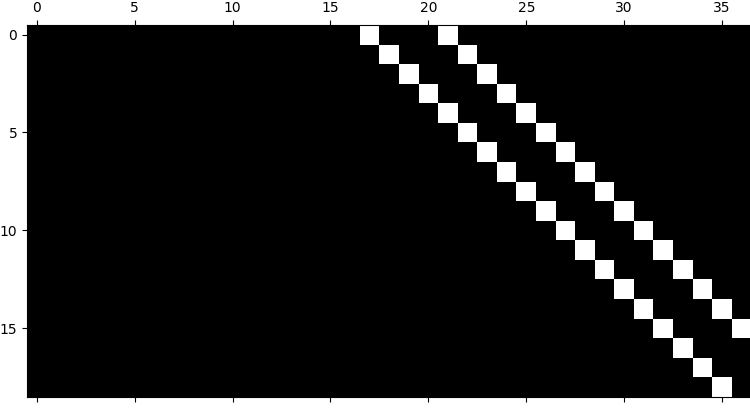}}
     \label{wr}
     \caption{2 different symmetry elements in 3 different weak RSPT models with $q=-1,0,1$ respectively. It can be observed that the symmetries in weak RSPT are always non-overlapping. White cubes are the spins that  the symmetry acts nontrivially on. The figure can be compared with Fig.~\ref{periodic pic}, which shows the symmetries of a genuine RSPT phase. Notice that 3 subfigures above are  symmetries from 3 different models while Fig.~\ref{periodic pic} shows 4 symmetries in the same model.}
 \end{figure*}

 However, self-similarity does not always mean fractal. If there is only one term in the update rule of the order-1 CA (i.e. the update rule is given by a monimial with respect to $x$), the resulting SPT will be protected by symmetries aligned along a line. However, these symmetries do not overlap with each other, and the whole HGSPT model can be decoupled into a set of $1D$ cluster models with $\mathbb Z_2\times \mathbb Z_2$ symmetry, which can be dubbed as weak RSPT\footnote{We refer to weak SSPT in Ref.~\cite{you2018a} as weak RSPT.}, being fundamentally different from a ``genuine'' RSPT mentioned in this paper, or strong SSPT mentioned in Ref.~\cite{you2018a}. In a genuine RSPT phase, each spin should be acted on by symmetries facing different directions. Under such context, we can safely claim that order-1 CA cannot generate RSPT phases. To demonstrate this in detail, we can write down the general form of update rule of weak RSPT:
 \begin{equation}
     f(x)=x^q,~q\in\mathbb Z.
 \end{equation}
 Symmetries in weak RSPT are always non-overlapping, and are all aligning along lines parallel to each other. We explicitly draw 3 cases where $q=-1,0,1$ in Fig.~\ref{wr}.

\section{Comparison of model-IVa and 2D cluster model}\label{diff}
Model-IVa (Eq.~(\ref{periodic_rule})) may seem equivalent to the 2D cluster model up to some shift of the sublattice at the first glance, we will show in the following text that they are actually two models with different symmetries.

\begin{figure*}[htb]
	\centering
	\subfigure[Original cluster model\label{clusterp1}]{\includegraphics[width=0.4\linewidth]{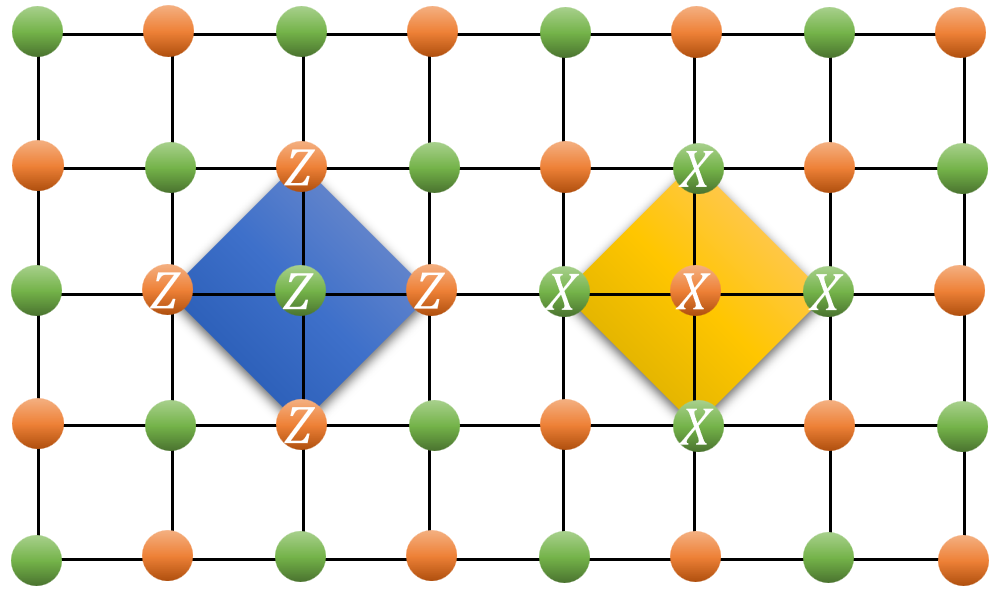}}
	\subfigure[2 sublattices\label{clusterp2}]{\includegraphics[width=0.4\linewidth]{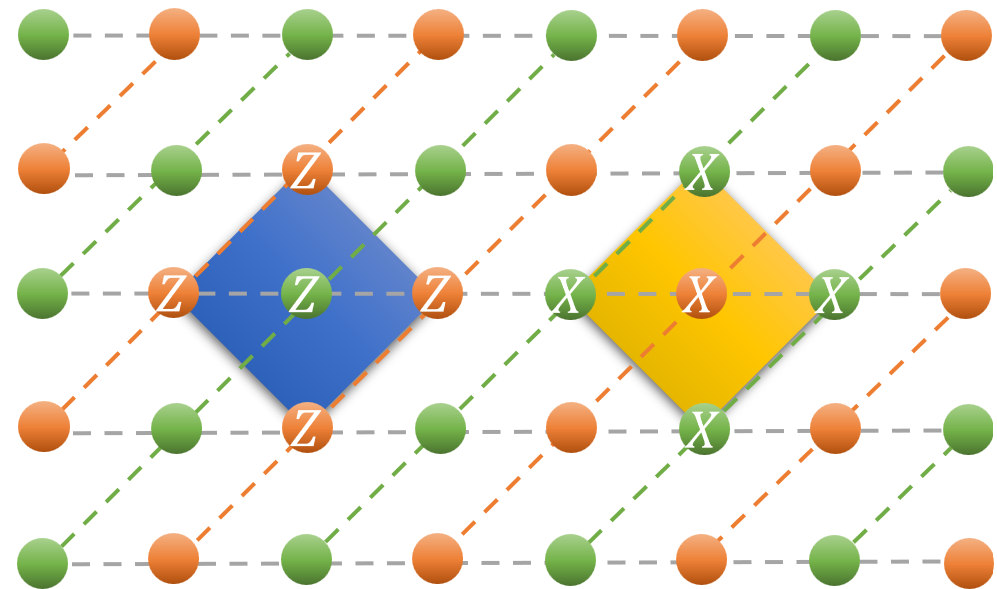}}
	\subfigure[Deformed sublattices\label{clusterp3}]{\includegraphics[width=0.4\linewidth]{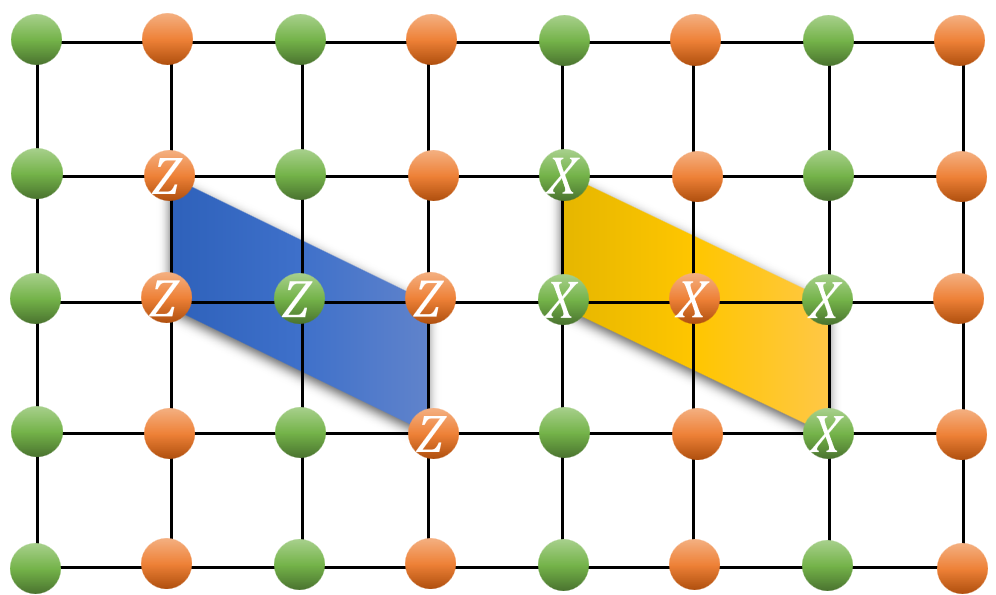}}
	\caption{Fig.~\ref{clusterp1} shows the Hamiltonian of the 2D cluster model up to some basis transformation. Two sublattices are shown in the Fig.~\ref{clusterp2}, where sublattice $(1)$ are drawn in orange and gray dashed lines and sublattice $(2)$ are drawn in green and gray dashed lines. There is 1 qubit living in each site. In Fig.~\ref{clusterp3} we deform 2 sublattices back to the form of a standard 2D square lattice, similar to the HGSPT model discussed in the main text.}
	\label{clusterH}
\end{figure*}

To better compare the difference between Model-IVa (Eq.~(\ref{periodic_rule})) and the 2D cluster model, we will start from the Hamiltonian of 2D cluster model, and make some basis transformation ($Z \leftrightarrow X$) to fit the general expression of HGSPT Hamiltonian:
\begin{equation}
	H_{\text{cluster}}=\sum_i A_i-\sum_j B_j,
\end{equation}
as shown in Fig.~\ref{clusterH}. Here, the $A_i$ term is products of 4 Pauli Z operators in orange sublattice and 1 Pauli Z operator in green sublattice, and $B_j$ term is products of 4 Pauli X operators in green sublattice and 1 Pauli X operator in orange sublattice.
The HGSPT model corresponds to the Hamiltonian shown in Fig.~\ref{clusterp3} are generated by HOCA rule (model-IVc)
\begin{equation}\label{clusterRule}
	\mathbf{f}(x)=\begin{pmatrix}
		x^{-1}+1\\
		x^{-1}
	\end{pmatrix}.
\end{equation} 

Now that the 2D cluster model is equivalent to the HGSPT model generated by rule Eq.~(\ref{clusterRule}), being different to the HOCA rule of Model-IVa (Eq.~(\ref{periodic_rule})), thus having different HOCA symmetries. Thus we conclude that 2D cluster model is not equivalent to Model-IVa (Eq.~(\ref{periodic_rule})).

\begin{figure*}[htbp]
		\centering
		\subfigure[$\mathbf{q}_1(x)$]{\includegraphics[width=0.45\linewidth]{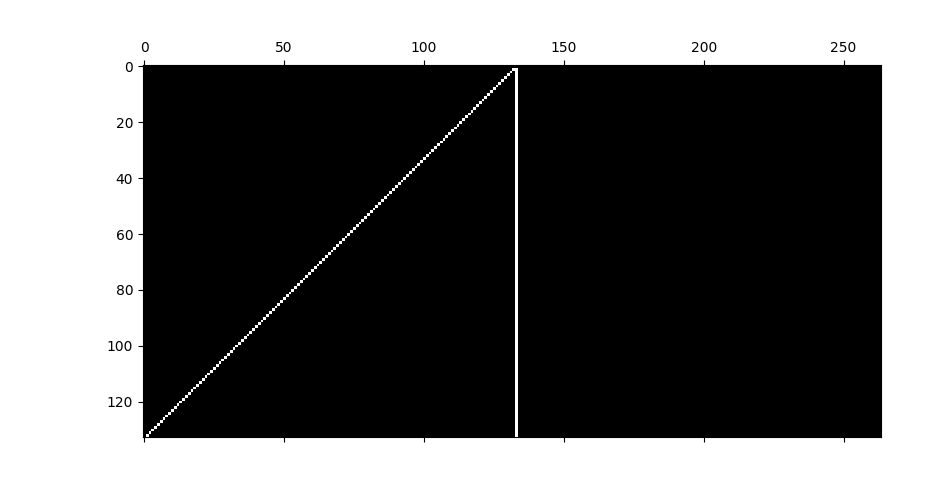}}
		\subfigure[$\mathbf{q}_2(x)$]{\includegraphics[width=0.45\linewidth]{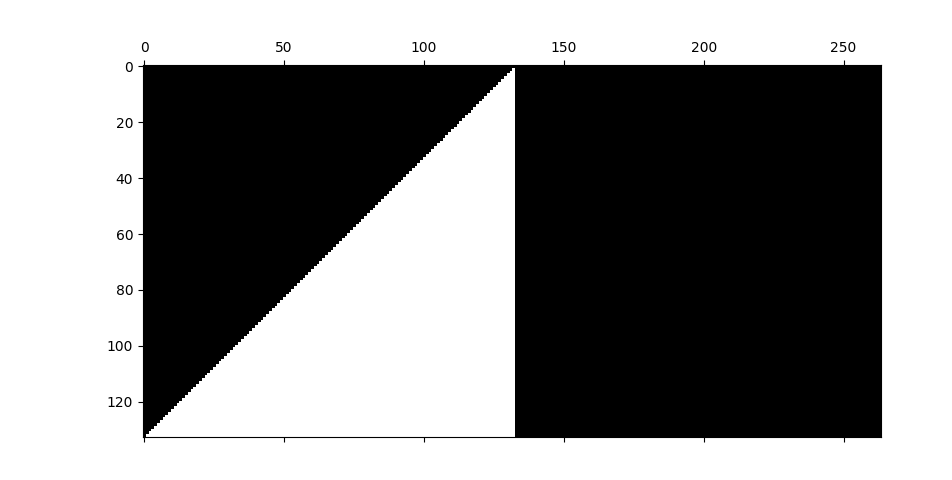}}\\
		\subfigure[$\mathbf{q}_3(x)$]{\includegraphics[width=0.45\linewidth]{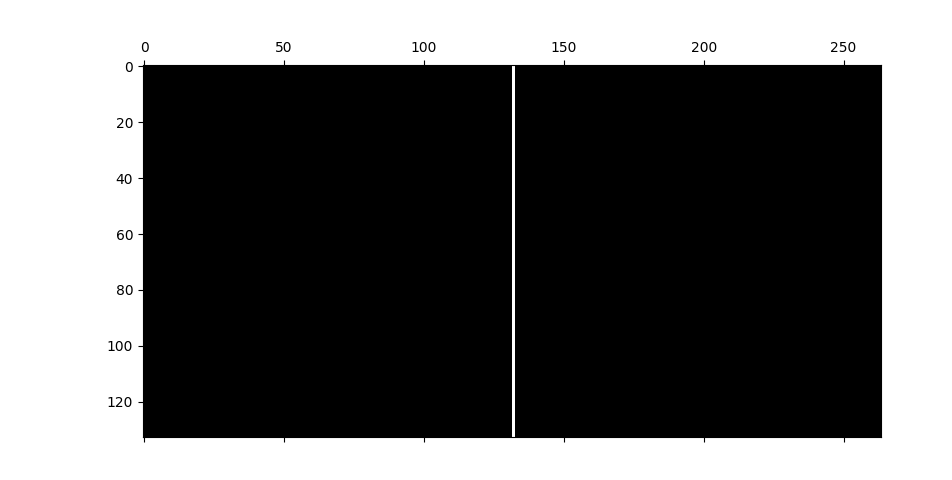}}
		\subfigure[$\mathbf{q}_4(x)$]{\includegraphics[width=0.45\linewidth]{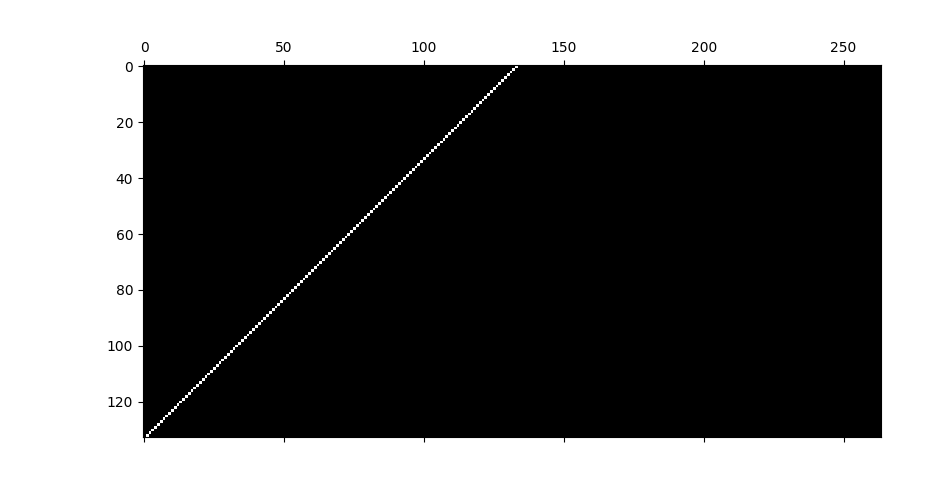}}
		\caption{4 patterns generated by order-2 CA (Eq.~(\ref{clusterRule})) in sublattice $(a)$. The initial condition are shown in Eq.~(\ref{ivc1}), Eq.~(\ref{ivc2}), Eq.~(\ref{ivc3}), Eq.~(\ref{ivc4}).  White pixels are spins that the Pauli-$X$ operator acts nontrivially on. The first 2 rows in each figure are determined by the initial condition, and the rest is determined by HOCA rule.}
		\label{fig:ivc}
	\end{figure*}

Given 4 initial conditions:
\begin{subequations}
		\begin{align}
			\mathbf{q}_1(x)&=\begin{pmatrix}
				0\\
				1+x
			\end{pmatrix},\label{ivc1}\\
			\mathbf{q}_2(x)&=\begin{pmatrix}
				0\\
				1
			\end{pmatrix},\label{ivc2}\\
			\mathbf{q}_3(x)&=\begin{pmatrix}
				1\\
				1
			\end{pmatrix},\label{ivc3}\\
			\mathbf{q}_4(x)&=\begin{pmatrix}
				x\\
				1
			\end{pmatrix}.\label{ivc4}
		\end{align}
	\end{subequations}
There are line-like and membrane-like symmetry elements present in the model, as shown in Fig.~\ref{fig:ivc}.

Another fundamental difference between two models is the topological transitivity of the HOCA rule, as shown in Table~\ref{chaostable}. Model-IVa have topological transitivity while model-IVc do not, indicating that two HOCA rules have different dynamical properties.

\section{2-point strange correlators in model-Va}
\label{calculation}
	
Now we want to diagnose whether the ground state of FSPT model (Eq.~(\ref{FSPT})) is trivial SRE state or not. Having known the degenerate edge modes of the model, it can be predicted that there have to be nontrivial results. First we introduce some terminologies:\\
	\begin{equation}
		\begin{aligned}
			\nabla_{ij}&:= \hat{Z}_{i j}^{(a)} \hat{Z}_{i, j-1}^{(a)} \hat{Z}_{i-1, j-1}^{(a)} \hat{Z}_{i j}^{(b)},\\
			\Delta_{ij}&:= \hat{X}_{i j}^{(b)} \hat{X}_{i, j+1}^{(b)} \hat{X}_{i+1, j+1}^{(b)} \hat{X}_{i j}^{(a)},
		\end{aligned}
	\end{equation}
	where the subscript  $ij$ denotes the coordinate of the operator, and superscript $(a/b)$ denotes the sublattice that the operator belongs to. The ground state to be diagnosed are taken as  \begin{equation}
		\ket{ \Psi}= \prod_{ij}\frac{1}{\sqrt 2}(1+\Delta_{ij})\ket{0\cdots 0} ,
	\end{equation}
	where $ | 0\cdots 0 \rangle $ denotes all qubits in the system are taken as $ +1 $ eigenstate of $ \sigma  _z $ operator. The trivial state should contain all symmetries in the ground state,  so the trivial state is taken as 
	\begin{equation}
		\ket{\Omega}=\ket{\hat X^{(a)}=\hat Z^{(b)}=1}.
	\end{equation}
	For FSPT states, the first problem is to specify the operator $ \phi $, we will try a set of different operators as candidates for $ \phi  $. 
	
	Overall results are shown in Table \ref{sierres}.
	\begin{table*}[ht]		\caption{Strange and normal correlator in the FSPT model (Eq.~(\ref{FSPT})) with correlation number 2}

		\centering
		\begin{tabular}{cccccccc}
			\hline
			
			\hline
			\rule[-1ex]{0pt}{2.5ex} Operator $ \phi(i,j)\phi  (i',j')  $  & $ \hat{X}_{ij}^{(a)}\xxa $ & $ \xa\xxb $ &$ \xb\xxb$ & $\za\zza$ &$ \za\zzb$&$\zb\zzb$&$\Y\YY$ \\
			\hline
			\rule[-1ex]{0pt}{2.5ex} Strange Correlator $ C(i,j;i',j') $ & 1 & 0 &0&0&0&1&0\\
			\hline
			\rule[-1ex]{0pt}{2.5ex} Normal Correlator $N(i,j;i',j')$ & 1 & 0 &0&0&0&1&0\\
			\hline
			
			\hline
		\end{tabular}
		\label{sierres}
	\end{table*}
	Candidates of local operators:
	
	\begin{outline}[enumerate]
		\1 $ \phi=\hat X^{(a/b)}_{ij} $;
		\1 $ \phi=\hat Z^{(a/b)}_{ij} $;
		\1 $ \phi  =\hat Y^{(a/b)}_{ij} $;
	\end{outline}
	
	Now we do a case-by-case calculation for all candidates, and notice that $i\ne j$ is always assumed.
	\begin{outline}[enumerate]
		\1 $ \phi=\hat X^{(a/b)}_{ij}  $: There are intuitively 3 cases for the choice above:
		\2 $\phi_{ij}=\hat{X}^{(a)}_{ij},\ \phi_{i'j'}=\hat{X}^{(a)}_{i'j'}$ : They are both symmetry elements of the model, resulting that \begin{equation}
			\bra{ \Omega}\phi_{ij}\phi_{i'j'}=\bra{ \Omega}.
		\end{equation}
		The strange correlator is \begin{equation}
			C(i,j;i',j')=1.
		\end{equation}
		\2 $\phi_{ij}=\hat{X}^{(a)}_{ij},\ \phi_{i'j'}=\hat{X}^{(b)}_{i'j'}$ : Only one spin in the $(b)$ sublattice is flipped. Such configuration cannot appear in the ground state, which makes \begin{equation}
			C(i,j;i',j')=0.
		\end{equation}
		\2 $\phi_{ij}=\hat{X}^{(b)}_{ij},\ \phi_{i'j'}=\hat{X}^{(b)}_{i'j'}$ :2 spins in the $(b)$ sublattice are flipped. Such configuration cannot appear in the ground state either, which makes \begin{equation}
			C(i,j;i',j')=0.
		\end{equation}

		\1 $ \phi  =\hat{Z}_{ij}^{(a/b)} $: There are similarly 3 cases here:
		\2 $\phi_{ij}=\hat{Z}^{(a)}_{ij},\ \phi_{i'j'}=\hat{Z}^{(a)}_{i'j'}$: 2 spins in the $(a)$ sublattice are flipped, which is not a configuration in the ground state.\begin{equation}
			C(i,j;i',j')=0.
		\end{equation}
		\2 $\phi_{ij}=\hat{Z}^{(a)}_{ij},\ \phi_{i'j'}=\hat{Z}^{(b)}_{i'j'}$: 1 spin in the $(a)$ sublattice is flipped, having zero overlap with the ground state. Therefore we have
		\begin{equation}
			C(i,j;i',j')=0.
		\end{equation}
		\2 $\phi_{ij}=\hat{Z}^{(b)}_{ij},\ \phi_{i'j'}=\hat{Z}^{(b)}_{i'j'}$: No changes are made to the trivial state $\ket{\Omega}$, so \begin{equation}
			C(i,j;i',j')=1.
		\end{equation}

		\1 $ \phi=\hat Y^{(a/b)}_{ij}  $: Using the data calculated above, we can easily obtain the result:
		\begin{equation}
			\begin{aligned}
				\ya \ket{\Omega}&=\frac{1}{2\i} \left[\za,\xa\right]\ket{\Omega}\\
				&=\frac{1}{2\i}\left(\za\xa-\xa\za\right)\ket{\Omega}\\
				&=\frac{1}{2\i}(\ket{+\cdots +-+\cdots+;0\cdots0}\\
				&\phantom{=}+\ket{+\cdots +-+\cdots+;0\cdots0})\\
				&=-\i\ket{+\cdots +-+\cdots+;0\cdots0},
			\end{aligned}
		\end{equation}
		and similarly:
		\begin{equation}
			\yb \ket{\Omega}=-\i \ket{+\cdots+;0\cdots010\cdots0}.
		\end{equation}
		
		There are overall 3 cases in the strange correlator:
		\2 $\phi_{ij}=\ya,\ \phi_{i'j'}=\ya$: \begin{equation}
			\begin{aligned}
				\bra{ \Omega}\phi  _{ij}\phi  _{i'j'}\ket{ \Psi}&=-\bra{+\cdots+-+\cdots+-+\cdots+;0\cdots 0}\ket{ \Psi}\\
				&=0.
			\end{aligned}
		\end{equation}
		There are 2 spins in the $(a)$ sublattice are flipped, and such configuration cannot be found in $\ket{ \Psi}$. So \begin{equation}
			C(i,j;i',j')=0.
		\end{equation}
		\2 $\phi_{ij}=\ya,\ \phi_{i'j'}=\yb$:\begin{equation}
			\begin{aligned}
				\bra{ \Omega}\phi  _{ij}\phi  _{i'j'}\ket{ \Psi}&=-\bra{+\cdots+-+\cdots+;0\cdots 010\cdots0}\ket{ \Psi}\\
				&=0.
			\end{aligned}
		\end{equation}
		There are 1 spin in the $(a)$ sublattice and 1 spin in the $(b)$ sublattice are flipped, and such configuration cannot be found in $\ket{ \Psi}$. So \begin{equation}
			C(i,j;i',j')=0.
		\end{equation}
		\2 $\phi_{ij}=\yb,\ \phi_{i'j'}=\yb$:\begin{equation}
			\begin{aligned}
				\bra{ \Omega}\phi  _{ij}\phi  _{i'j'}\ket{ \Psi}&=-\bra{+\cdots+;0\cdots 010\cdots010\cdots0}\ket{ \Psi}\\
				&=0.
			\end{aligned}
		\end{equation}
		There are 2 spins in the $(b)$ sublattice are flipped, and such configuration cannot be found in $\ket{ \Psi}$. So \begin{equation}
			C(i,j;i',j')=0.
		\end{equation}

	\end{outline}
	To determine whether the result of the strange correlator is trivial or not, we need to compare it with normal correlator. Now we examine the expectation value of operators in the paramagnetic phase $\ket{\Omega}=\ket{+\cdots+;0\cdots 0}$. In the calculation below $ \langle \Omega|\Omega \rangle=1$ and $i\neq j$ is always assumed.\\
	\begin{outline}[enumerate]
		
		\1 $ \phi=\hat X^{(a/b)}_{ij}  $: There are intuitively 3 cases for the choice above:
		\2 $\phi_{ij}=\hat{X}^{(a)}_{ij},\ \phi_{i'j'}=\hat{X}^{(a)}_{i'j'}$ : They are both symmetry elements of the model, resulting that \begin{equation}
			\bra{ \Omega}\phi_{ij}\phi_{i'j'}=\bra{ \Omega}.
		\end{equation}
		The normal correlator is \begin{equation}
			N(i,j;i',j')=1.
		\end{equation}
		\2 $\phi_{ij}=\hat{X}^{(a)}_{ij},\ \phi_{i'j'}=\hat{X}^{(b)}_{i'j'}$ : Only one spin in the $(b)$ sublattice is flipped.  \begin{equation}
			N(i,j;i',j')=0.
		\end{equation}
		\2 $\phi_{ij}=\hat{X}^{(b)}_{ij},\ \phi_{i'j'}=\hat{X}^{(b)}_{i'j'}$ : 2 spins in the $(b)$ sublattice are flipped. \begin{equation}
			N(i,j;i',j')=0.
		\end{equation}
		
		\1 $ \phi  =\hat{Z}_{ij}^{(a/b)} $: There are similarly 3 cases here:
		\2 $\phi_{ij}=\hat{Z}^{(a)}_{ij},\ \phi_{i'j'}=\hat{Z}^{(a)}_{i'j'}$: 2 spins in the $(a)$ sublattice are flipped.\begin{equation}
			N(i,j;i',j')=0.
		\end{equation}
		\2 $\phi_{ij}=\hat{Z}^{(a)}_{ij},\ \phi_{i'j'}=\hat{Z}^{(b)}_{i'j'}$: 1 spin in the $(a)$ sublattice is flipped. Therefore we have
		\begin{equation}
			N(i,j;i',j')=0.
		\end{equation}
		\2 $\phi_{ij}=\hat{Z}^{(b)}_{ij},\ \phi_{i'j'}=\hat{Z}^{(b)}_{i'j'}$: No changes are made to the trivial state $\ket{\Omega}$, so \begin{equation}
			N(i,j;i',j')=1.
		\end{equation}

		\1 $\phi  =\Y$: There will always spins flipped by the operator, by orthogonality we obtain
		\begin{equation}
			N(i,j;i',j')=0.
		\end{equation}
	\end{outline}

Here we have shown that all 2-point strange correlators is trivial in model-Va.
\section{Proof of minimal correlation number in model-Va}\label{proof}
	Now we want to prove that it will require at least 3 local Pauli operators to show nontrivial results in the model-Va (Eq.~(\ref{sierrule}))  , and 3 Pauli operators must locate at 3 corners of the fractal separately.
	
	\textbf{Theorem 1:}
	\begin{equation}\label{tm1}
		\begin{gathered}
			\text{Given a set of Hamiltonian terms } D=\{\Delta_{ij}\}, \\ \text{ the product of these Hamiltonian terms} \\ \text{act on $\ket\Omega$ equivalently as at least 3 onsite Pauli operators.}\\
   \text{The minimal case is}\\
			\left(\prod_{\Delta_{ij}\in D}\Delta_{ij}\right)\ket{\Omega}=\hat{X}^{(b)}_{i_0j_0}\hat{X}^{(b)}_{i_0,j_0+2^k}  \hat{X}^{(b)}_{i_0+2^k,j_0+2^k} \ket{\Omega},~k\in\mathbb N,\\
			\text{if and only if }  D \text{ form a Sierpinski fractal structure on the lattice. }\\
		\end{gathered}
	\end{equation}

	We will start with some notations. 
	\subsection{Notations}
\begin{figure}[htbp]
		\centering
		\begin{tikzpicture}[scale=0.9]
			\draw[help lines,step=1] (0,0) grid (3,3);
			\draw[-latex] (-1,0) -- (4,0);
			\draw[-latex] (0,-1) -- (0,4);
			\draw (4,0)coordinate (A)node[below] {$i$};
			\draw (0,4)coordinate (A)node[left] {$j$};
			\draw (0,0)coordinate (A)node[above right] {$s_{11}$};
			\draw (0,1)coordinate (B)node[above right] {$s_{21}$};
			\draw (0,2)coordinate (B)node[above right] {$s_{31}$};
			\draw (0,3)coordinate (B)node[above right] {$s_{41}$};
			\draw (1,0)coordinate (A)node[above right] {$s_{12}$};
			\draw (2,0)coordinate (A)node[above right] {$s_{13}$};
			\draw (3,0)coordinate (A)node[above right] {$s_{14}$};
			\draw (1,1)coordinate (B)node[above right] {$s_{22}$};
			\draw (1,2)coordinate (B)node[above right] {$s_{32}$};
			\draw (1,3)coordinate (B)node[above right] {$s_{42}$};
			\draw (2,1)coordinate (B)node[above right] {$s_{23}$};
			\draw (2,2)coordinate (B)node[above right] {$s_{33}$};
			\draw (2,3)coordinate (B)node[above right] {$s_{43}$};
			\draw (3,1)coordinate (B)node[above right] {$s_{24}$};
			\draw (3,2)coordinate (B)node[above right] {$s_{34}$};
			\draw (3,3)coordinate (B)node[above right] {$s_{44}$};
			\node (1) at(6.5,1.5) {$\to \mathsf{Spin}=\begin{bmatrix}
					s_{11} & s_{12} & s_{13} & s_{14} & \cdots\\
					s_{21} & s_{22} &  s_{23} & s_{24} &\cdots\\
					s_{31} & s_{32} &  s_{33} & s_{34} &\cdots\\
					s_{41} & s_{42} &  s_{43} & s_{44} &\cdots\\
					\vdots & \vdots &  \vdots & \vdots &\ddots\\
				\end{bmatrix}$};
		\end{tikzpicture}
		\caption{Illustration of the $\mathsf{Spin}$ matrix}
		\label{spin}
	\end{figure}
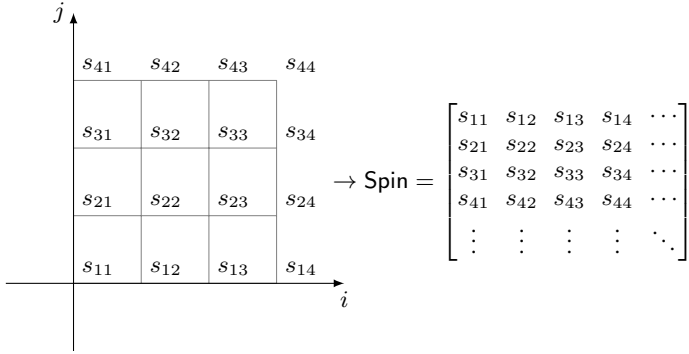

	In FSPT model defined above, $(a)$ and $(b)$ sublattice have different symmetry elements (made up of $\hat X$ or  $\hat Z$, separately), so the action of a Hamiltonian term $\nabla_{ij}$ or $\Delta_{ij}$ on trivial state $\ket{\Omega}$ will only flip spins in one sublattice. Therefore, it will be convenient to consider only one sublattice at a time, and here we will discuss $(b)$ sublattice, in which spins are all $\ket{0}$ in the trivial state $\ket{\Omega}$, and operators with nontrivial action on the sublattice is $\Delta_{ij}$. In the model, each sublattice is isomorphic to a square lattice. For simplicity, we denote the state of spin in $(b)$ sublattice ($\ket{0}$ or $\ket{1}$) and its corresponding lattice site $(i,j)$ by a matrix element $s_{i+1,j+1}=0 \text{ or } 1 $. All $s_{ij}$ form a matrix $\mathsf{Spin}$, shown in Fig.~\ref{spin}.
	
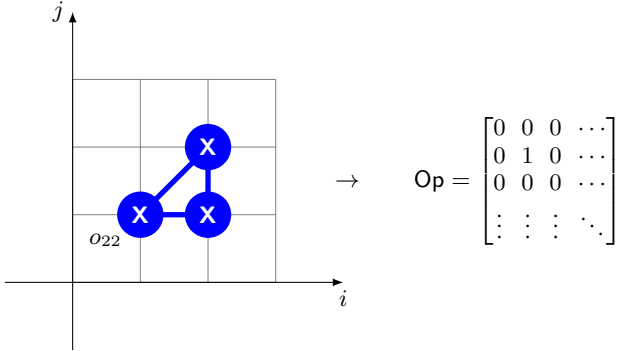
\begin{figure}[htbp]
		\centering
		\begin{tikzpicture}[scale=0.9]
			\draw[help lines,step=1] (0,0) grid (3,3);
			\draw[-latex] (-1,0) -- (4,0);
			\draw[-latex] (0,-1) -- (0,4);
			\draw (4,0)coordinate (A)node[below] {$i$};
			\draw (0,4)coordinate (A)node[left] {$j$};
			\draw[bold] (1,1)--(2,1)--(2,2)--cycle;
			\node[blueop] (a) at (1,1){\color{white}$\bm{\mathsf{X}}$};
			\node[blueop] (a) at (2,1){\color{white}$\bm{\mathsf{X}}$};
			\node[blueop] (a) at (2,2){\color{white}$\bm{\mathsf{X}}$};
			\draw (0.85,0.85)coordinate (A)node[below left] {$o_{22}$};
			\node (1) at(6,1.5) {$\to\qquad \mathsf{Op}=\begin{bmatrix}
					0 & 0 & 0 &\cdots\\
					0 & 1 & 0 &\cdots\\
					0&0&0&\cdots\\
					\vdots & \vdots &\vdots & \ddots
				\end{bmatrix}$};
		\end{tikzpicture}
		\caption{Illustration of the $\mathsf{Op}$ matrix}
		\label{operator}
	\end{figure}

	Now we represent whether a site is acted by Hamiltonian term $\Delta_{ij}$ or not by a matrix $\mathsf{Op}=\{o_{ij}\}$. Note that we only consider the action on $(b)$ sublattice only. The position of a Hamiltonian term is marked by the location of its local Pauli operator with the least $i,j$ value in  $(b)$ sublattice. For example, if $\Delta_{ij}=\hat{X}_{i j}^{(b)} \hat{X}_{i, j+1}^{(b)} \hat{X}_{i+1, j+1}^{(b)} \hat{X}_{i j}^{(a)}$ acts on our state, then we note this by $o_{ij}=1$. An example is shown in Fig.~\ref{operator}.
	
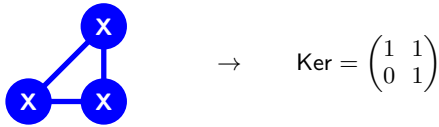
\begin{figure}[htbp]
		\centering
		\begin{tikzpicture}
			\draw[bold] (1,1)--(2,1)--(2,2)--cycle;
			\node[blueop] (a) at (1,1){\color{white}$\bm{\mathsf{X}}$};
			\node[blueop] (a) at (2,1){\color{white}$\bm{\mathsf{X}}$};
			\node[blueop] (a) at (2,2){\color{white}$\bm{\mathsf{X}}$};
			\node (1) at(5,1.5) {$\to\qquad \mathsf{Ker}=\begin{pmatrix}
					1&1\\
					0&1
				\end{pmatrix}$};
		\end{tikzpicture}
		\caption{Illustration of the $\mathsf{Ker}$ matrix}
		\label{ker}
	\end{figure}

	We express the shape of $\Delta_{ij}$ by a matrix $\mathsf{Ker}=\{k_{ij}\}=\begin{pmatrix}
	1&1\\
	0&1
\end{pmatrix}$, shown in Fig.~\ref{ker}.\par
	We can simply generate $\mathsf{Spin}$ by $\mathsf{Op}$ and $\mathsf{Ker}$ :
	\begin{equation}
		\begin{gathered}
			\mathsf{Spin}_{(a+1)\times (b+1)}\equiv\mathsf{Op}_{a\times b}\diamond \mathsf{Ker}_{2\times 2}, \text{ defined as } \\
			s_{ij}=\sum_{m}\sum_n o_{i-m+1,j-n+1}k_{mn}\mod 2,
			\\ \text{if } i-m+1 \text{ or } j-n+1 \le 0,\\ \text{or }  i-m+1>a \text{ or } j-n+1>b,\\ 
			\text{ then } o_{i-m+1,j-n+1}=0.
		\end{gathered}
	\end{equation}
	Our claim above is straightforward, which is basically a translation to the matrix language. \par
	Finally, we denote all elements in the $i$-th row of matrix $\mathbf{A}=\{a_{ij}\}$ by $a_{i*}$, and similarly $j$-th column by $a_{*j}$.
	
	\subsection{Proof of Theorem 1}
	\textbf{Lemma 1:} 
	\begin{equation}\label{lm1}
		\begin{gathered}
			\text{If } \mathsf{Op}\text{ is not null,}\\ \text{denote the number of nonzero values in matrix } \textbf{A} \text{ by } N(\textbf{A}),\\ \text{ then  }N(\mathsf{Spin})\ge 3,\text{ where }\mathsf{Spin}=\mathsf{Op}\diamond \mathsf{Ker}.
		\end{gathered}
	\end{equation}
	
	\textbf{Proof:}
	\begin{outline}[enumerate]
		\1 First we truncate $\mathsf{Op}$ by deleting all null columns and rows at the edge. We suppose that $\mathsf{Op}$ is a $a\times b$ matrix after the truncation. 
		\1 \textbf{Lemma 1.1} 
		\2 \textbf{Claim:} Changing the value of an element $o \in o_{a*},\ o\ne o_{ab}$ from 0 to 1 will increase $N(\mathsf{Spin})$ by $1$. 
		\2 \textbf{Proof:} Through direct calculation, $o_{ai}\ (1\le i<b)= s_{a+1,i+1}$. The claim above is obvious.
		\1 \textbf{Lemma 1.2}\label{1.2}
		\2 \textbf{Claim:} If $o_{*b}$ is not null, then $N(s_{*,b+1})\ge 2$.
		\2 \textbf{Proof:} We examine $o_{1b},o_{2b},\cdots, o_{ab} $ in order. Through direct calculation, each nonzero value $o_{ib}$ in the sequence will either increase $N(s_{*,b+1})$ by 2 (if $o_{i-1,b}=0$) or 0 (if $o_{i-1,b}\ne 0$). Since there are at least 1 nonzero elements, so it follows that $N(s_{*,b+1})\ge 2$. 
		
		\1 It follows that $o_{a*}$ and $o_{* b}$ should contain at least one nonzero value (otherwise it would have been truncated). Note that $s_{a+1,*}$ is completed determined by $o_{a* }$, and $s_{* ,b+1}$ by $o_{* b}$. There are 2 cases here:
		\2 $o_{ab}\ne 0$: This already meets the condition above. From \textbf{Lemma 1.2} we know that $N(\mathsf{Spin})\ge N(s_{*,b+1})\ge 2$.
		\2 $o_{ab}=0$: It means that there are at least 1 nonzero value in $\{o| o\in o_{a*},\ o\ne o_{ab}\}$ and $\{o| o\in o_{*b},\ o\ne o_{ab}\}$. From \textbf{Lemma 1.1} and \textbf{Lemma 1.2} we know that $N(\mathsf{Spin})\ge 3$. In this case, \textbf{Lemma 1} is proven.\\
		\1 \textbf{Lemma 1.3} 
		\2 \textbf{Claim:} $N(s_{1*}\cup s_{*1}-s_{a+1,1}-s_{1,b+1})\ge 1$.
		\2 \textbf{Proof:} $o_{1*}$ and $o_{*1}$ must be nontrivial because of the truncation. Notice that $s_{i1}=o_{i1}$, so $N(s_{*1})\ge1$.\\
		$\phantom{=}$
		\1 If case 4(a) is true, because of \textbf{Lemma 1.3}, we have $N(\mathsf{Spin})\ge2+1=3$. So we completed our proof for \textbf{Lemma 1}.
	\end{outline}
	\textbf{Lemma 2:}
	\begin{equation}
		\begin{gathered}
			\text{If $N(\mathsf{Spin})=3$,}\\
			\text{then $\forall o\in o_{1*}\cup o_{*b},\ o=1.$}  
		\end{gathered}
	\end{equation}
	\textbf{Proof:}
	\begin{outline}[enumerate]
		\1 We assume that $o_{ab}=0$, \textbf{Lemma 1.3} and the discussion 4(b) in the proof of \textbf{Lemma 1} we know that $N(\mathsf{Spin})\ge4$, so our assumption is false and $o_{ab}=1$.
		\1 \textbf{Lemma 2.1}
		\2 \textbf{Claim:} We call a set of nonzero matrix elements with one common index and one consecutive index a \textbf{string}, and adding any elements to this set should makes it no longer satisfy the definition of a string. For example, $\{a_{11}=1,a_{12}=1,a_{13}=1\}$ form a string. There are 2 strings in the set $\{a_{11}=1,a_{12}=1,a_{13}=0,a_{14}=1\}$, they are separately $\{a_{11},a_{12}\}$ and $\{a_{14}\}$. We denote the maximum number of strings that can be possibly defined in a set of matrix elements $A$ by $S(A)$. We claim that if $N(\mathsf{Spin})=3$, then $S(o_{*b})=1$.
		\2 \textbf{Proof:} Through direct calculation, 1 string correspond to 2 endpoints, flipping 2 spins. So we observe that
		\begin{equation}
			N(s_{*,b+1})=2S(o_{*b}).
		\end{equation}
		To ensure that $N(\mathsf{Spin})=3$, we require that $N(s_{*,b+1})=2$. So it follows that $S(o_{*b})=1$. 
		\1 \textbf{Lemma 2.2}
		\2 \textbf{Claim:} \begin{equation}
			o_{1b}=1.
		\end{equation}
		\2 \textbf{Proof:} We assume that $o_{1b}=0$. From \textbf{Lemma 2.1} we know that if $N(\mathsf{Spin})=3$, then $N(s_{*,b+1})= 2$. Since $o_{1*}$ is nontrivial and $o_{1b}=0$, we find that $S(o_{1*}-o_{1b})\ge1$ and $N(s_{1*}-s_{1,b+1})\ge 2$. In total we have $N(\mathsf{Spin})\ge4$, which is contradictory to our assumption. It follows that $o_{1b}=1$. 
		\1 From \textbf{Lemma 2.1} and \textbf{Lemma 2.2} we immediately see that\begin{equation}\label{res1}
			\forall o\in o_{*b}, \ o=1.
		\end{equation}
		\1 \textbf{Lemma 2.3}
		\2 \textbf{Claim:} If $N(\mathsf{Spin})=3,$ then $ \forall o\in o_{1*},\ o=1.$
		\2 \textbf{Proof:} Similarly we can prove $o_{11}=1$ by considering $S(o_{*1})$ and $S(o_{1*})$. If there are any null elements in $o_{1*}$, then it follows that $S(o_{1*})\ge 2$ and $N(s_{1*}-s_{1,b+1})\ge 3$, contradicting our assumption $N(\mathsf{Spin})=3$. Therefore, \begin{equation}
			\forall o\in o_{1*},\ o=1.
		\end{equation}
		\1 Combining Eq.~(\ref{res1}) and \textbf{Lemma 2.3}, we finished our proof for \textbf{Lemma 2}. 
 
	\end{outline}
	Now we move on to prove \textbf{Theorem 1}.\\
	\textbf{Proof:}
	\begin{outline}[enumerate]
		\1 According to \textbf{Lemma 2}, we already have $N(s_{1*}\cup s_{*,b+1})=3$, which means that there are already 3 spins flipped by the Hamiltonian terms. If we want to meet the condition of \textbf{Theorem 1}, the rest of the $\mathsf{Spin}$ matrix should not occur any nonzero elements. So the rest part of $\mathsf{Op}$ should be selected to ensure that there are no other spins flipped by the Hamiltonian term $\Delta_{ij}$.
		\1 To satisfy the condition above, we observe that condition
		\begin{equation}\label{res2}
			o_{ij}=o_{i-1,j}+o_{i,j+1}\mod 2
		\end{equation}
		should be satisfied by all $o_{ij},\ 1<i\le a,\ 1<j< b$. The proof is straightforward, Eq.~(\ref{res2}) is equivalent to $s_{i,j+1}=0$. We also notice that this is \textbf{intrinsically equivalent to the update rule of Sierpinski fractal} (Eq.~(\ref{sierrule})).
		\1 We also observe that \begin{equation}\label{cons1}
			o_{21}=o_{31}=\cdots=o_{a1}=o_{a2}=\cdots=o_{a,b-1}=0,
		\end{equation}
		otherwise extra flipped spin will be generated, and there are no way to cancel these extra spins.
		\1 We already know \begin{equation}
			o_{11}=o_{12}=\cdots=o_{1b}=o_{2b}=\cdots=o_{ab}=1
		\end{equation}
		from \textbf{Lemma 2}, so we can generate the rest part of $\mathsf{Op}$ by Eq.~(\ref{res2}). It follows that all $o_{ij},\ 1<i\le a,\ 1<j< b$ is uniquely determined. So the problem here is to select $a$ and $b$ to meet the constraint of Eq.~(\ref{cons1}).
		\1 Iterating Eq.~(\ref{res2}), we find a formula for a general term for $o_{ij},\ i>1,\ 1\le j<b$:
		\begin{equation}\label{res3}
			\begin{aligned}
				o_{ij}&=o_{i-1,j}+o_{i,j+1}\mod 2\\
				&=\sum_{l_1=j}^bo_{i-1,l_1}\mod 2\\
				&=\sum_{l_1=j}^b \sum_{l_2=l_1}^b o_{i-2,l_2}\mod 2\\
				&\phantom{=}\vdots\\
				&=\sum_{l_1=j}^b \sum_{l_2=l_1}^b\cdots \sum_{l_{i-1}=l_{i-2}}^b o_{1,l_{i-1}}\mod 2\\
				&=\sum_{l_1=j}^b \sum_{l_2=l_1}^b\cdots \sum_{l_{i-1}=l_{i-2}}^b 1\mod 2\\
				&=\frac{1}{(i-1)!}\prod_{k=1}^{i-1}(b-j+k)\mod 2\\
				&=\frac{(b-j+i-1)!}{(i-1)!(b-j)!}\mod2\\
				&=C_{b-j+i-1}^{i-1}\mod 2.
			\end{aligned}
		\end{equation}
		We can see that $o_{ij}$ is symmetric along the counter-diagonal by examining the variable substitution
		\begin{equation}
			\begin{gathered}
				(i,j)\to(i',j')\\
				i\to b+1-j'\\
				j\to b+1-i'
			\end{gathered}
		\end{equation}
		and equation Eq.~(\ref{res3}) is invariant under the substitution:
		\begin{equation}\label{res4}
			\begin{aligned}
				C^{i-1}_{b-j+i-1}&=C^{b+1-j'-1}_{b-b-1+i'+b+1-j'-1}\\
				&=C^{b-j'}_{b-j'+i'-1}\\
				&=C^{i'-1}_{b-j'+i'-1}.
			\end{aligned}
		\end{equation}
		\1 Now we can say \begin{equation}\label{cons2}
			a=b
		\end{equation}
		since
		\begin{equation}
			o_{21}=o_{31}=\cdots=o_{b1}=0 \iff o_{b2}=\cdots=o_{b,b-1}=0,
		\end{equation}
		according to Eq.~(\ref{res4}).
		\1 \textbf{Lemma 3}
		\2 \textbf{Claim:} Constraint Eq.~(\ref{cons1}) is satisfied if and only if\begin{equation}
			a=b=2^n,\ n\in \mathbb{N^*}.
		\end{equation}
		\2 \textbf{Proof:} 
		\3 We can consider $o_{21},\cdots,o_{a1}$ only because of the symmetry.
		\3 Using Eq.~(\ref{res3}) and Eq.~(\ref{cons2}), we know that
		\begin{equation}\label{res5}
			\begin{aligned}
				o_{i1}=C^{i-1}_{a+i-2}=C^{a-1}_{a+i-2}.
			\end{aligned}
		\end{equation}
		\3 Noticing that if $o_{2j}=o_{3j}=\cdots =o_{nj}=0$, using Eq.~(\ref{res2}) we know that
		\begin{equation}
			o_{1,j-1}=o_{2,j-1}=o_{3,j-1}=\cdots =o_{n,j-1}=1.
		\end{equation}
		So if $a=a_0$ satisfies
		\begin{equation}\label{cons3}
			o_{11}=o_{21}=\cdots =o_{a1}=1,
		\end{equation}
		then $a=a_0-1$ satisfies the Eq.~(\ref{cons1}).
			
			\3 \textbf{Theorem 2 (Lucas's Theorem)}: For non-negative integers $m$ and $n$ and a prime $p$, the following congruence relation holds:
			\begin{equation}
				\binom{m}{n}=\prod_{i=0}^k\binom{m_i}{n_i} \mod p,
			\end{equation}
			where
			\begin{equation}
				m=\sum_{i=0}^{k}m_i p^i,\quad n=\sum_{i=0}^{k}n_i p^i.
			\end{equation}
			\3 \textbf{Corollary 2.1}:
			\begin{equation}
				\begin{gathered}
					\text{For } m=\sum_{i=0}^{k}m_i 2^i,\ n=\sum_{i=0}^{k}n_i 2^i,\\
					\binom{m}{n} \text{ is odd } \iff (m_i,n_i)\ne (0,1),\ \forall i\in \{0,1,2,\ldots,k\} .
				\end{gathered}
			\end{equation}
			\3 To meet constraint Eq.~(\ref{cons3}), from \textbf{Corollary 2.1} we know
			\begin{equation}
				\begin{gathered}
					m=a_0+i-2,\ n=a_0-1,\\
					\text{For } m=\sum_{i=0}^{k}m_i 2^i,\ n=\sum_{i=0}^{k}n_i 2^i,\\
					(m_i,n_i)\ne (0,1).
				\end{gathered}
			\end{equation} 
			So we need $n_0=n_1=\cdots n_{k-1}=0$, therefore $n=a_0-1=2^k,\ a_0=2^k+1,\ k\in \mathbb{N}^*$.
			\3 Therefore, to satisfy Eq.~(\ref{cons1}), we have
			\begin{equation}
				a=b=a_0-1=2^k, \ k\in \mathbb{N}^*,
			\end{equation}
			which is what we wanted.  
			\1 Combining \textbf{Lemma 3} and Eq.~(\ref{res2}), $\mathsf{Op}$ indeed form a complete Sierpinski fractal, and there are 3 spins flipped by the configuration of Hamiltonian terms, lying at 3 corners of the fractal, which are separately $s_{11},s_{1,a+1},s_{a+1,a+1}$, finishing our proof for \textbf{Theorem 1}.
		\end{outline}

 \section{Review of various cellular automata in constructing subsystem symmetries}\label{ca_review}
In the realm of constructing subsystem symmetries via cellular automata (CA), there have been several approaches that has been discovered. This includes quantum cellular automata (QCA) \cite{CQCA1,CQCA2} and matrix cellular automata (MCA) \cite{MCA}. While all of these methods utilized CA to generate subsystem symmetries, there are distinctions among the symmetries generated by different CAs. In the sections following, we will give a brief review of different CA approaches and compare these methods in constructing subsystem symmetries. In short, they differ in the type and symmetry of the spatial configuration of the subsystem, and in the dimension and the geometry of the background lattice. The overall comparison of these CA approaches is given in Table~\ref{tab:ca_compare} based on the models constructed by these CA approaches in the literature. 

A \textbf{quantum cellular automata (QCA) }defines a map between locally supported operators, enlarging the sizes of their supports by an amount independent of the sizes of the original supports. When we restrict our sight onto qubit systems, it is natural to consider Clifford quantum cellular automata (CQCA), which is a QCA inducing an automorphism of $N$-qubit Pauli group. By means of \emph{projected entangled pair states} (PEPS), one can construct SPT phases protected by certain types of subsystem symmetries, including line-like and fractal symmetries. There are some differences between HOCA and CQCA in terms of constructing phases with subsystem symmetries:

\begin{enumerate}
	\item Types of subsystem symmetries: CQCA managed to generate SPT phases protected by line-like and fractal symmetries, which are all named as $L$-cycle symmetries. In our notation, the CQCA method can produce FSPT and RSPT phases. So far, CSPT phases and MSPT phases has not been constructed from CQCA method as far as our best knowledge.
	\item Symmetries of the spatial configuration of subsystem: As for the SPT orders protected by fractal symmetries in \cite{CQCA1}, there are SPT orders generated by HOCA protected by locally identical symmetry. However, the symmetry generated by HOCA can be considered as a part of the CQCA generated fractal symmetry. This is because of the unidirectional nature of the HOCA evolution, resulting in the lack of spatial symmetry of the HOCA generated symmetry. The CQCA discussed in \cite{CQCA1} are all symmetric, hence CQCA generated subsystems have spatial symmetries different  form HOCA generated subsystems.
\end{enumerate}

The \textbf{matrix cellular automata (MCA)} proposed in \cite{MCA} is actually an LCA over $\mathbb{Z}^2_2$ in the terminologies proposed in the realm of cellular automata. The authors in \cite{MCA} used MCA to make essential use of the crystallographic structure, i.e. two 3D lattices of corner sharing triangles: trillium and hyperhyperkagome (HHK). 


\begin{table*}
    \centering
    \begin{tabular}{c|ccc}
    \hline
    \hline
        CA Method & Lattice Dimension & Lattice Geometry & Discovered Subsystem Symmetries \\
        \hline
         CQCA& 2D \cite{CQCA1,CQCA2} & \makecell{Square \cite{CQCA1},\\ 11 Archimedean lattices \cite{CQCA2}} & \makecell{Fractal \cite{CQCA1,CQCA2}, Line-like \cite{CQCA1},\\Ribbon \cite{CQCA2}, Cone \cite{CQCA2}} \\
         \hline
         MCA& 3D\cite{MCA} & Trillium, HHK\cite{MCA} & Fractal \\
         \hline
         HOCA& 2D & Square & Fractal, Line-like, Chaotic, Mixed \\
         \hline
         \hline
    \end{tabular}
    \caption{Comparison of several CA approaches in constructing subsystem symmetries. }
    \label{tab:ca_compare}
\end{table*}

\section{Mathematical properties of HOCA}\label{HOCA_prop}

In this appendix, we give a brief review of some mathematical properties of HOCA, which are mainly obtained by considering a duality between linear HOCA and linear cellular automata (LCA) proposed in Ref.\cite{dennunzio_dynamical_2019}. We will show there is an algorithm that can decide whether a HOCA rule corresponding to a concrete model in the main body satisfy a certain criteria of chaos or not, and a subset of the HOCA rules does satisfy the criteria. Unless otherwise specified, we only consider one-dimensional CA in this appendix.

\subsection{Duality between linear HOCA and LCA over $\mathbb{Z}^n_2$}
\label{dualityCA}

At first, we review the duality between LCA and linear HOCA. Because the duality allows us to express all liear HOCA rules used in this paper as LCA over $\mathbb{Z}^n_m$, where $n$ is the order of the original HOCA, we mainly consider such LCA in this appendix to utilize relevant mathematical results.
\\ 

\noindent \textbf{Definition of linear HOCA (LHOCA)}:
A linear HOCA can be formally summarized as a structure $\mathcal H= \langle n,S,r,\mathbf{f}\rangle$, where $n \geq 1$ is the order (also refered as memory size), $S$ is the alphabet, $r$ is the radius\footnote{Note that here we use $r$ instead of $M$ as in the main body to represent the radius to avoid confusion with $\mathbf{M}$ representing matrices.}, $\mathbf{f}$ is the \emph{local rule} defined in Eq.~(\ref{eq:update_rule}). As discussed in the main body, we mainly focus on the case $S = \mathbb Z_{2}$. For latter convenience, we represent $\mathbf{f}$ by a series of coefficients $a_{i}^j \in \mathbb Z_{2}$, where $a_{i}^j$ is the coefficient of $x^i$ term in $f_{n+1-j}(x)$, $j=1,2,\cdots,n$ and $i=-r,-r+1,\cdots,r-1,r$.

\noindent  \textbf{Definition of linear cellular automata (LCA)}:
An LCA over alphabet $\mathbb{Z}^n_{m}$ of order-$1$ is a CA $\mathcal L=\langle \mathbb Z^n_{m},r,f\rangle$ where the alphabet $S$ has been taken as $\mathbb{Z}^n_{m}$, the local rule $f$ is defined by $2r+1$ matrices $\mathbf{M}_{-r},\dots,\mathbf{M}_{0},\dots,\mathbf{M}_{r}\in\text{Mat}(n,\mathbb Z_{m})$, such that the time evolution can be expressed as follows:

\begin{align}
\label{eq:def_linearf}
f\left(\mathbf{x}_{-r},\ldots,\mathbf{x}_0,\ldots,\mathbf{x}_r\right)=\left[\sum_{i=-r}^r\mathbf{M}_i\cdot \mathbf{x}_i\right]_m
\end{align}
for any $(\mathbf{x}_{-r},\ldots,\mathbf{x}_0,\ldots,\mathbf{x}_r)\in(\mathbb{Z}_m^n)^{2r+1}$. Here a $n$-row vector $\mathbf{x}_{i}$ denotes the state of site $i$, and $\text{Mat}(n,\mathbb Z_m)$ denotes the set of $n \times n$ matrices with coefficients in $\mathbb Z_{m}$. Note that we use $f$ \textit{without} any subscripts to denote the local rule of LCA rather than a component of $\mathbf{f}$.
\\

\noindent \textbf{Definition of Frobenius LCA}: An LCA is said to be a Frobenius LCA if the matrix associated to it $\mathbf{M}(x):=\sum_{i=-r}^{r}\mathbf{M}_i x^{-i}$ ($x$ is a formal variable) is in Frobenius normal form, that is to say, it has the following form:
\begin{equation}
    \mathbf{M}(x)=\begin{bmatrix}
        0&1&0&\cdots& 0&0\\
        0&0&1&\cdots& 0&0\\
        0&0&0&\cdots &0&0\\
        \vdots&\vdots&\vdots&\cdots&\vdots&\vdots\\
        0&0&0&\cdots&0&1\\
        m_0(x)&m_1(x)&m_2(x)&\cdots &m_{n-2}(x)&m_{n-1}(x)
    \end{bmatrix},
\end{equation}
where each $m_i(x)$ is a polynomial about the formal variable $x$.

\noindent  \textbf{Topological conjugacy of LHOCA and LCA}:
The LHOCA defined by $\mathcal H= \langle n,\mathbb Z_{2},r,\mathbf{f}\rangle$, where $\mathbf{f}$ is specified by $a^j_{i}$ , $j\in [1,n],i\in [-r,r]$, can be simulated by an LCA $\mathcal L=\langle \mathbb Z^n_{m},r,f\rangle$, with $f$ totally determined by $\mathbf{f}$:

\begin{align}
\label{eq:lca_m0}
\mathbf{M}_0=\begin{bmatrix}0&1&0&\ldots&0&0\\0&0&1&\ldots&0&0\\0&0&0&\ldots&0&0\\\vdots&\vdots&\vdots&\ddots&\ldots&\vdots\\0&0&0&\ldots&0&1\\a_0^1&a_0^2&a_0^3&\ldots&a_0^{n-1}&a_0^n\end{bmatrix}\:,
\end{align}

and for $i \in [-r,r], i \neq 0$,

\begin{align}
\label{eq:lca_mi}
\mathbf{M}_i=\begin{bmatrix}0&0&0&\ldots&0&0\\0&0&0&\ldots&0&0\\0&0&0&\ldots&0&0\\\vdots&\vdots&\vdots&\ddots&\vdots&\vdots\\0&0&0&\ldots&0&0\\a_i^1&a_i^2&a_i^3&\ldots&a_i^{n-1}&a_i^n\end{bmatrix}\mathrm{~.}
\end{align}

In Ref.\cite{dennunzio_dynamical_2019}, it has been shown the above correspondence is a topological conjugacy that preserves dynamical properties, thus we can investigate the dynamical properties of LHOCA by considering the corresponding LCA. 

\subsection{D-chaos of LCA}\label{D-chaos}

In this appendix, we consider the well-accepted notion of chaos of discrete time dynamical system (DTDS) proposed by Devaney (often abbreviated as D-chaos). In general, the criteria of D-chaos is composed of three components: \textit{topological transitivity},
\textit{sensitivity to initial conditions} and \textit{denseness of periodic orbits}\cite{Devaney1989}. In this subsection, we briefly review the definition of these properties for DTDS, and introduce an algorithm that can decide whether an 1D LCA is D-chaotic or not proposed in Ref.\cite{Dennunzio2024}.
\\

\noindent \textbf{Definition of discrete time dynamical system (DTDS)}: A discrete dynamic system is a pair $(\mathcal{X},\mathcal{F})$ where $\mathcal{X}$ is a space equipped with a metric $d$, and $\mathcal{F}$ is a transformation on $\mathcal{X}$ which is continuous with respect to that metric. The dynamical evolution of a DTDS is described by an initial state $x^{(0)} \in \mathcal{X}$ evolving as $x^{(t)}= \mathcal{F}^t(x^{(0)})$, $\forall t \in \mathbb{N}$.

In the realm of 1D CA, the space $\mathcal{X}$ is taken as $S^{\mathbb{Z}}$, where $S$ is the alphabet of the CA. Therefore the space $\mathcal{X}$ represents the space of configurations at a specific time step. $\mathcal{F}$ is naturally the global rule of the CA. Here we take the metric as the standard Cantor distance $$d(c,c')=
\begin{cases}
	\frac{1}{2^n},& c\neq c',\\
	0,& c=c',
\end{cases}
$$
where $$ n=\min \{i\geq 0: c_{i} \neq c_{i}' \text{ or } c_{-i}\neq c_{-i}'\}.$$ Here $c_{i}$ denotes the state of configuration $c$ at site $i$, which is an element of the alphabet $S$.
\\

\noindent \textbf{Definition of topological transitivity of DTDS\label{topo_trans}}: 
A DTDS $(\mathcal{X},\mathcal{F})$ is said to have \emph{topological transitivity} if for an arbitrary pair of open nonempty subsets of $\mathcal{X}$, $\exists n\in \mathbb{N}$, such that $\mathcal{F}^n(U) \cap V\neq \emptyset$. 
\\

\noindent \textbf{Definition of sensitivity to the initial conditions of DTDS}: A DTDS $(\mathcal{X},\mathcal{F})$ is said to be \emph{sensitive to the initial conditions} if there exists $\epsilon >0$ such that $\forall x \in \mathcal{ X},~ \delta>0$, then there exists $y \in \mathcal{X},~n\in \mathbb{N}$, such that $d(x,y)< \delta$ and $d(\mathcal{F}^n(x),\mathcal{F}^n(y))> \epsilon$.
\\

\noindent \textbf{Definition of denseness of periodic orbits of DTDS}:
An element $x \in \mathcal{X}$ is said to be a \textit{periodic point} if there exists a natural number $n>0$ such that $\mathcal{F}^n(x) = x$. The denseness of periodic orbits is the denseness of the set composed of all such periodic points.

By definition, as a DTDS, an 1D LCA simultaneously satisfying the above three properties is chaotic according to Devaney's notion\cite{Devaney1989}.

\subsection{Deciding chaos of linear HOCA}


In Ref.\cite{Dennunzio2024}, the authors proposed an efficient method to decide whether a 1D LCA is D-chaotic or not. Obviously, with the duality between linear HOCA and LCA, we can also use this method to decide the chaos of linear HOCA. 

Firstly, for a 1D LCA $\mathcal L=\langle \mathbb Z^n_{m},r,f\rangle$, where the local rule $f$ is specified by $2r+1$ matrices $\mathbf{M}_{-r},\dots,\mathbf{M}_{0},\dots,\mathbf{M}_{r}\in\text{Mat}(n,\mathbb Z_{m})$ according to Eq.~(\ref{eq:def_linearf}), we can write down the Laurent polynomial associated with $\mathcal L$:

\begin{equation}\label{mx_def}
    \mathbf{M}(X)=\sum^r_{i=-r} \mathbf{M}_i X^{-i},
\end{equation}
 
where $X$ is merely a formal variable.

And then we can define the characteristic polynomial of matrix $M(X)$:
$$\chi_{M(X)} (t)= \text{det} (tI_n - M(X)),$$
where $t$ is another formal variable, $I_n$ is the $n\times n$ identity matrix.

In Ref.\cite{Dennunzio2024,Dennunzio2020}, the authors proved that a 1D LCA $\mathcal{L}$ is chaotic if and only if its characteristic polynomial $\chi_{M(X)} (t)$ satisfy the following condition:
when we recognize $\chi_{M(X)} (t)$ as a polynomial $\xi(X,X^{-1})$ of $X,X^{-1}$, the greatest common divisor of the coefficients of all terms in $\xi(X,X^{-1})$ denoted as $\gamma(t)$ has a degree smaller than $1$. Note that for alphabet $S=\mathbb{Z}^n_2$, all coefficients can only be $0$ or $1$ after taking $\text{mod}\ 2$.

Finally, we obtain a general procedure to decide whether a linear HOCA $\mathcal H= \langle n,\mathbb{Z}_2,r,\mathbf{f}\rangle$ corresponding to a model studied in this paper is chaotic (according to Devaney's notion) or not as follows:

\begin{enumerate}
	\item Represent $\mathbf{f}$ by a series of coefficients $a_{i}^j \in \mathbb Z_{2}$, where $a_{i}^j$ is the coefficient of $x^i$ term in $f_{n+1-j}(x)$;
	\item Recompose all $a_{i}^j$ coefficients to $2r+1$ matrices $\mathbf{M}_{-r},\dots,\mathbf{M}_{0},\dots,\mathbf{M}_{r}$ according to Eq.~(\ref{eq:lca_m0}) and
	\linebreak
	Eq.~(\ref{eq:lca_mi}), which specifies a 1D LCA $\mathcal{L}$;
	\item Write down the Laurent polynomial associated with $\mathcal{L}$ and its characteristic polynomial $\chi_{M(X)} (t)= \text{det} (tI_n - M(X))$, then recognize $\chi_{M(X)} (t)$ as a polynomial $\xi(X,X^{-1})$ of $X,X^{-1}$;
	\item Compute $\gamma(t)$, the greatest common divisor of the coefficients of all terms in $\xi(X,X^{-1})$, the original HOCA $\mathcal{H}$ is chaotic when $\text{deg}(\gamma(t))<1$, otherwise it is not.
\end{enumerate}

With this procedure, we can determine the chaotic property of all linear HOCA corresponding to models studied in this paper, and the results are summarized in Table~\ref{chaostable}. Compared to our criterion of deciding a chaotic HOCA rule, the algorithm proposed here is looser since many rules whose patterns do not seem chaotic visually are decided to be chaotic by the procedure here.

\begin{table}
    \centering
    \begin{tabular}{c|cccccccc}
    \hline
    \hline
        Model No. & I & II & III & IVa & IVb & IVc & Va & Vb \\
        \hline
        $\gamma(t)$ & 1 & 1 & 1 & 1 & 1 & $t+1$ & 1 & 1\\
        \hline
        $\deg \gamma(t)$ &0&0&0&0&0&1&0&0\\
        \hline
        Topological Transitivity& Yes & Yes & Yes & Yes & Yes & No & Yes & Yes\\
    \hline
    \hline
    \end{tabular}
    \caption{Deciding whether there are topological transitivity (or equivalently, D-chaos) in the HGSPT models mentioned in the paper using characteristic polynomials.}
    \label{chaostable}
\end{table}

\subsection{Deciding the sensitivity to initial conditions of linear HOCA}\label{sensitivity}
Deciding the sensitivity to initial conditions of HOCA can be done by the following procedure \cite{dennunzio_dynamical_2019}:
\begin{outline}[enumerate]
    \1 Find the Frobenius LCA that is topologically conjugate to the linear HOCA;
    \1 Decide the sensitivity of Frobenius LCA.
\end{outline}

First, we will start with some notations.

\textbf{Definition ($\deg+$ and $\deg^-$, sensitivity of polynomial)}: Given any Laurent polynomial $P(X)$ with coefficients in $\mathbb{Z}_{p^k}$ where $p$ is a prime and $k \in \mathbb{N}_+$ (containing both positive and negative degree of $X$), $\deg^+ [P(X)]$ (resp. $\deg^- [P(X)]$) is the maximum degree among those of the monomials having both positive (resp. negative) degree and coefficient which is not multiple of $p$. If there is no monomial satisfying the required conditions, then $\deg^+[P(X)]=0$ (resp. $\deg^-[P(X)]=0$). If either  $\deg^+[P(X)]>0$ or  $\deg^-[P(X)]<0$, 

\textbf{Example:} Consider a polynomial $P(X)$ over $\mathbb{Z}_4=\mathbb{Z}_{2^2}$:
\begin{equation}
    P(X)=2X^{-3}+X^{-1}+1+X+3X^2+2X^3,
\end{equation}
then $\deg^+[P(X)]=2$ and $\deg^{-}[P(X)]=-1$.

\textbf{Theorem \cite{dennunzio_dynamical_2019}}: An LCA over $\mathbb{Z}_{p^k}^n$ is \textit{sensitive to initial conditions} iff there exists some $i \in \{0,1,...,n-1\}$, such that $m_i(X)$ is sensitive, where $m_i(X)$ is the polynomial at the $n$-th row, $i$-th column of the matrix $\mathbf{M}(X)$. $\mathbf{M}(X)$ is defined in Eq.~(\ref{mx_def}).

Using the theorem above and the topological conjugacy between LCA over $\mathbb{Z}_{p^k}^n$ and the HOCA over $\mathbb{Z}_{p^k}$, we can decide whether a given HOCA rule is sensitive to the initial condition. All HOCA rule mentioned in Table~\ref{table1} is sensitive to the initial condition. We want to further point out that HOCA rules that can create a self-similar fractal pattern is sensitive to the initial condition, argued as following:
\begin{outline}[enumerate]
    \1 A necessary condition of a self-similar pattern in 2D square lattice is the \textit{scaling behavior} of a single row: There exists an infinite sequence $\{t_i\}$, such that $r_{t_{i+1}}(x)=r_{t_{i}}(x^p), ~p \in \mathbb{N}+,~p>1$. Here $r(x)$ is defined in Eq.~(\ref{rx}).
    \1 Suppose that the HOCA rule is not sensitive to the initial condition, i.e. all $m_i(x)$ satisfy $\deg^\pm[m_i(x)]=0$, then we conclude that the \textit{radius} $R$ of the HOCA rule is $0$ immediately.
    \1 The \textit{radius} $R$ of a HOCA rule determines how fast the change of a site can propagate throughout the system. $R=0$ means that the time evolution of each site is governed by itself only, and cannot be affected by the sites nearby. 
    \1 Using the additivity of the HOCA rule we immediately obtain that if a site is at state $0$ at time $t$, then it will remain at state $0$ governed by a radius-0 HOCA rule, which contradicts our initial assumption that the rule can produce a self-similar fractal pattern.
    \1 We conclude that all HOCA rules that can generate a fractal pattern are sensitive to the initial condition.
\end{outline}

\section{Mathematical discussion on the validity of two criteria in section~\ref{detection}}\label{sc_proof}
 Now we give an brief explanation on two criteria in Sec.~\ref{detection}. For a ``regular" $M=0$ pattern, we can observe that $\mathfrak{n}_{\text{inf} }=\mathfrak{n}_{\text{sup} }$ regardless of the dimension of the pattern (for a regular pattern, $\mathfrak n$ grows linearly with $L$ or remains a constant, which can be recognized as a fundamental feature of regular patterns), explaining our criterion 1. We observe that for HOCA patterns with $M=1$ (i.e. fractal patterns) there are exact self-similarity in the pattern, making $\mathfrak{n}_{\text{inf} }=\text{Const.}$. While at the same time,  $\mathfrak{n}_{\text{sup} }=\infty$ are always true for a fractal pattern, explaining our criterion 2. The initial condition in the criteria above can be always fully enumerated if the model is defined on an open slab. The detailed proof are shown below.

As defined in Section~\ref{notation}, we label the class of a HGSPT model by two values (Eq.~(\ref{eq:class_label})):
\begin{equation}
	\begin{aligned}
		X_r&=1-\lceil \min\{M\}\rceil,\\
		X_f&=\lfloor\max\{M\}\rfloor,
	\end{aligned}
\end{equation}
where $\{M\}$ represents the set of all possible $M$ generated by the given HOCA rule, which at the same time determines all possible symmetry patterns for an HGSPT phase. As explained in Section~\ref{notation}, if $X_r=1$, then there are regular symmetry patterns (defined as HOCA configuration with $M=0$) in the model, and $X_r=0$ indicates otherwise. Similarly, if $X_f=1$, then there are fractal symmetry patterns (defined as HOCA configuration with $M=1$) in the model, and $X_f=0$ indicates otherwise. Now we want to show that:
\begin{outline}[enumerate]
    \1 For any HOCA configuration $\mathscr F(x,y)$ (Eq.~(\ref{eq:HOCA_config})) with $M=0$ generated by initial condition $\mathbf{q}$ and HOCA rule $\mathbf{f}$, we have $\mathscr{N}(\mathbf{q},\mathbf{f})=1$.
    \1 For any HOCA configuration $\mathscr F(x,y)$ (Eq.~(\ref{eq:HOCA_config})) with $M=1$ generated by initial condition $\mathbf{q}$ and HOCA rule $\mathbf{f}$, we have $\mathscr{N}(\mathbf{q},\mathbf{f})=0$.
\end{outline}
\textbf{Proof:} First we recall the definition of $M$ given  in Eq.~(\ref{M1}, \ref{M2}, \ref{M3}):
\begin{equation}
	M=\frac{S_u-S_d}{S_u},
\end{equation}
where
\begin{equation}
	S_u(n)=\limsup_{k\to\infty} \frac{\sum_{i=k}^{k+n}A(i)}{k}
\end{equation}
and
\begin{equation}
	S_d(n)=\liminf_{k\to\infty}\frac{\sum_{i=k}^{k+n}A(i)}{k}.
\end{equation}

The key point in this proof is to see the correspondence between $\mathfrak{n}[D_{L}(\mathbf{q},\mathbf{f})]$ and $A(t)$. $\mathfrak{n}[D_{L}(\mathbf{q},\mathbf{f})]$ stands for the number of local Pauli operators in a given MPSC, and $A(t)$ is the number of sites with state $1$ in the $t$-th row of a HOCA pattern. First, recall that an MPSC that can detect the nontrivial SRE ground state of HGSPT model acts as identity on the ground state, which can further be decomposed into products on Hamiltonian terms. If we put a Hamiltonian term (e.g. the one made up of Pauli $X$ operators) on each site of with state $1$ in a truncated HOCA pattern, then in the ``bulk'' (opposite to boundary, to be explained below) of the pattern all Pauli operators will cancel out by definition of the HOCA rule. On the boundary (i.e. the first and last $n$ rows of the pattern) there will be generally Pauli operators that have not been cancelled out. If we are to obtain a product of operators that acts effectively on the ground state as products of Hamiltonian terms from a truncated HOCA pattern by putting an Hamiltonian term on each site with state 1, then there will be some terms on the boundary of the pattern (Eq.~(\ref{eq:qx}, \ref{eq:mx})) that do not cancel out. The number of such terms we must add is exactly the number of operators of the MPSC that we want to construct. If we truncate the HOCA pattern at row $j$, then such terms will appear at row $j+1,j+2,...,j+n$. Because of the locality of HOCA rule (memory size $n$, radius $R$ limits the area of impact of flipping one single site), the number of such terms ($\mathfrak{n}$) scales linearly with $A(t)$, and can be mutually substituted while the time step grows to infinity. Therefore, $S_u (\text{resp. }S_d)=\text{Const.}$ is equivalent to $\mathfrak{n}_{\sup}(\text{resp. }\mathfrak{n}_{inf})=\infty$, and 
 $S_d=S_u$ will indicate $\mathfrak{n}_{\sup}=\mathfrak{n}_{\inf}$, and $S_d=0$ will be equivalent to $\mathfrak{n}_{\inf}=o(L)$. Using this connection, we see that the definition of $X_r$ and $X_f$ is actually equivalent to the two criteria raised in Section~\ref{detection}:
\begin{outline}[enumerate]
    \1 $M=1$ indicates that $$S_u-S_d=S_u,$$ which suggests $S_u=\text{Const.}$ and $S_d=0$. $S_d=0$ indicates that there are an infinite sequence $\{t_i\}$ such that $A(t_i)$ is a constant. Since the locality of the HOCA rule, if we plug the corresponding initial condition into $\mathfrak{n}_{\inf}[D_L(\mathbf{q},\mathbf{f})]$, we will obtain $\mathfrak{n}_{\inf}=\text{Const.}$ The claim above can be verified by choosing the sequence $\{L_i:L_i=t_i-n\}$, then by the locality of HOCA rule we have that $\{\mathfrak{n}[D_{L_i}(\mathbf{q},\mathbf{f})]\}$ will remain a constant. So we conclude that $\mathfrak{n}_{\inf}=\text{Const.}$. On the other hand, $S_u=\text{Const.}$ indicates that $\mathfrak{n}_{\sup}=\infty$. So we have
    $$\mathscr{N}=\frac{0}{\infty}=0.$$
    \1 $M=0$ indicates that $$S_u=S_d,$$ which suggests $S_u=S_d=\text{Const.}$. This indicates that if we plug the corresponding initial condition into $\mathfrak{n}_{\inf}[D_L(\mathbf{q},\mathbf{f})]$ and $\mathfrak{n}_{\sup}[D_L(\mathbf{q},\mathbf{f})]$, we will obtain $\mathfrak{n}_{\inf}=\mathfrak{n}_{\sup}=\text{Const.}$ So we have
    $$\mathscr{N}=1.$$
\end{outline}

\section{Mathematical discussion on 
 the universality of HGSPT phases}\label{universality}
As HOCA managed to produce a large variety of symmetry patterns, one may wonder if any kind of subsystem symmetry can be generated by HOCA approach mentioned in the main text. In this section we will show the ``completeness'' of HGSPT model.

\textbf{Proposition:} Given a pattern $S(x,y)$ defined on an open slab with size $L_x\times L_y$, there will be at least one HOCA rule $\mathbf{f}$ and initial condition $\mathbf{q}$, such that the HOCA configuration $\mathscr{F}(x,y)$ is identical to $S(x,y)$ in the open slab. We say any finite patterns can always be \textit{locally simulated} by an HOCA rule.

\textbf{Proof:} The proof of this proposition involves the topological transitivity of the HOCA rule (Appendix~\ref{topo_trans}):

A DTDS $(\mathcal{X},\mathcal{F})$ is said to have \emph{topological transitivity} if for an arbitrary pair of open nonempty subsets $U,V$ in $\mathcal{X}$, then there exists a positive natural number $n$, such that $\mathcal{F}^n(U) \cap V\neq \emptyset$. 

We start from a desired HOCA pattern $S(x,y)$, and maps it to a single-row Frobenius LCA configuration $\alpha$. The validity of this process is guaranteed by the topological conjugacy between an order-$n$ HOCA over $\mathbb{Z}_2$ and an LCA over $\mathbb{Z}^n_2$, which can be considered as a single row of $ \mathbb{Z}_2$ vectors with $n$ components. So this is a one-to-one map, without losing or adding any information. Then, we select an open set of configurations $V$ that contains $\alpha$ locally, i.e. each configuration $v\in V$ shares the same configuration with $\alpha$ within the domain of $\alpha$, and can be arbitrarily chosen outside of the domain of $\alpha$. Then, because of the topological transitivity of the LCA rule, if we pick an LCA rule with topological transitivity, then for any configuration subset $U$, there exists $N\in \mathbb N^*$ such that $\mathcal{F}^N(U)\cap V\ne\emptyset$, where $\mathcal F$ is the global rule of the LCA. Without lost of generality, we suppose that $\mathcal{F}^N(u)\in V,~u\in U$. Then we can map LCA configuration $u$ back to HOCA configuration $u_0$. By selecting $u_0$ as the initial condition of HOCA rule, we obtain the desired symmetry pattern $S(x,y)$ in an open slab after $N$ steps of evolution governed by HOCA. This finishes our proof of the proposition above.
\hfill{\textbf{Q.E.D.}}

\begin{figure*}[htb]
    \centering
    \includegraphics[width=0.8\linewidth]{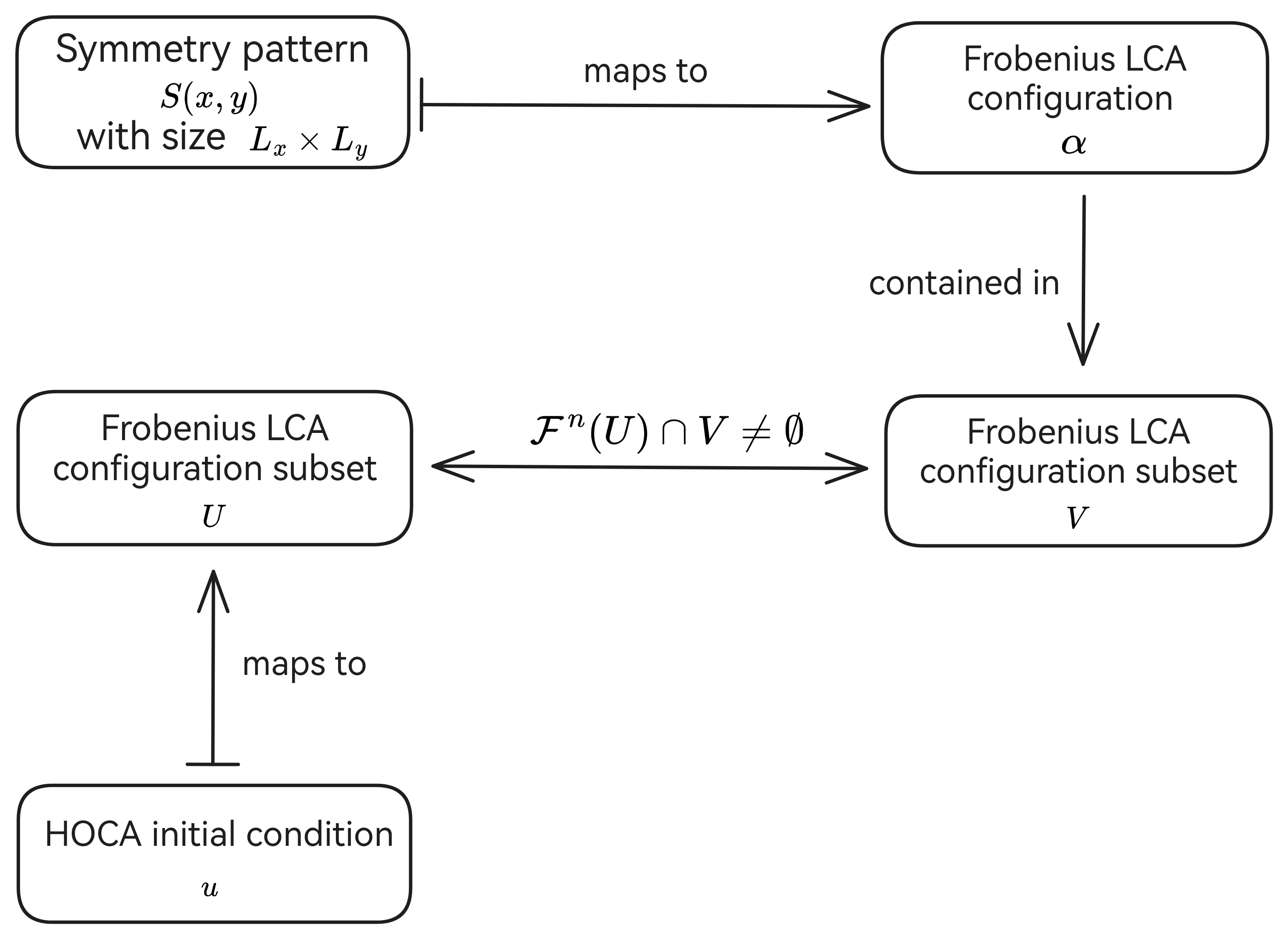}
    \caption{A pictorial illustration of main idea in proof in Appendix~\ref{universality}.}
    \label{fig:uni_proof}
\end{figure*}

The proof indicates that any order-$n$ HOCA rule with topological transitivity have the ability to simulate symmetry patterns within an $L_x\times L_y$ slab with $L_y\le n$. A pictorial illustration of the proof above is shown in Fig.~\ref{fig:uni_proof}.

\end{document}